\documentclass[rmp,
aps,
twocolumn
]{revtex4-1}
\usepackage{amsmath}
\usepackage{epsfig}
\usepackage{amssymb}
\usepackage{color}
\usepackage{graphicx}%

\newtheorem{theorem}{Theorem}

\newcommand{\ri}{\mathrm{i}}
\newcommand{\sech}{\mathrm{sech}}
\newcommand{\bq}{{\bf q}}
\newcommand{\bp}{{\bf p}}

\newcommand{\bw}{{\bf w}}

\newcommand{\tbw}{\tilde{\bf w}}

\newcommand{\tb}{\tilde{b}}

\newcommand{\tmu}{\tilde{\mu}}

\newcommand{\vep}{\varepsilon }

\newcommand{\br}{ {\bf r}}

\newcommand{\tpsi}{\tilde{\psi}}

\newcommand{\PT}{{\cal PT}}
\newcommand{\CPT}{{\cal CPT}}

\newcommand{\T}{{\cal T}}
\newcommand{\C}{{\cal C}}
\newcommand{\p}{{\cal P}}

\newcommand{\cH}{{\cal H}}
\newcommand{\IM}{\textrm{Im$\,$}}
\newcommand{\RE}{\textrm{Re$\,$}}
\newcommand{\Ai}{\textrm{Ai$\,$}}
\newcommand{\Bi}{\textrm{Bi$\,$}}
\newcommand{\diag}{\mbox{diag}}

\newcommand{\lan}{\langle}
\newcommand{\ran}{\rangle}
\newcommand{\tE}{\tilde{E}}
\newcommand{\tchi}{\tilde{\chi}}
\newcommand{\ua}{{\uparrow}}
\newcommand{\da}{{\downarrow}}
\newcommand{\bPsi}{{\bf \Psi }}
\newcommand{\bpsi}{ {\bf \psi} }

\begin{document}

\title{Nonlinear waves in $\PT$-symmetric systems}

\author{Vladimir V. Konotop}
\affiliation{Centro de F\'{\i}sica Te\'orica e Computacional, and Departamento de F\'{\i}sica, Faculdade de Ci\^encias,
 Universidade de Lisboa,
Campo Grande, Ed. C8, Lisboa  1749-016, Portugal}
\email{vvkonotop@fc.ul.pt}

\author{Jianke Yang}
\affiliation{Department of Mathematics and Statistics, University of
Vermont, Burlington, VT 05401, USA}
\email{jyang@math.uvm.edu}

\author{Dmitry A. Zezyulin}
\affiliation{Centro de F\'{\i}sica Te\'orica e Computacional, and Departamento de F\'isica, Faculdade de Ci\^encias,
 Universidade de Lisboa,
Campo Grande, Ed. C8, Lisboa  1749-016, Portugal}
\email{dzezyulin@fc.ul.pt}


\begin{abstract}
Recent progress on nonlinear properties of parity-time ($\PT$-)
symmetric systems is comprehensively reviewed in this article. $\PT$
symmetry started out in non-Hermitian quantum mechanics, where
complex potentials obeying $\PT$ symmetry could exhibit all-real
spectra. This concept later spread out to optics, Bose-Einstein
condensates, electronic circuits, and many other physical fields,
where a judicious balancing of gain and loss constitutes a
$\PT$-symmetric system. The natural inclusion of nonlinearity into
these $\PT$ systems then gave rise to a wide array of new phenomena
which have no counterparts in traditional dissipative systems.
Examples include the existence of continuous families of nonlinear
modes and integrals of motion, stabilization of nonlinear modes
above $\PT$-symmetry phase transition, symmetry breaking of
nonlinear modes, distinctive soliton dynamics, and many others. In
this article, nonlinear $\PT$-symmetric systems arising from various
physical disciplines are presented; nonlinear properties of these
systems are thoroughly elucidated;
and relevant experimental results are described. In addition,
emerging applications of $\PT$ symmetry are pointed out.
\end{abstract}

\pacs{11.30.Er, 03.65.Ge, 05.45.Yv, 42.65.Tg, 42.65.Wi, 42.25.Bs, 42.81.Dp, 02.30.Ik}

\maketitle

\tableofcontents

\section{Introduction}
\label{sec:intro}

Symmetries are the most fundamental properties of nature, which are
responsible for many physical phenomena we observe. Not long ago it
was suggested by \textcite{BendBoet} that parity ($\p$) and time
($\T$) symmetries can be responsible for purely real spectra of
non-Hermitian operators. While examples of such operators were known
for a long time, the discovery of \textcite{BendBoet} had profound
significance, because \textcolor{black}{it suggested a possibility
of $\PT$-symmetric modification of the conventional quantum
mechanics which considers observables as Hermitian operators in the
Hilbert space $L^2$}. This idea was further developed
by~\textcite{Most_a,Most_b} who introduced and explored a general
class of pseudo-Hermitian operators with special symmetries and
purely real spectra. These works have since stimulated intensive
research on $\PT$-symmetric operators. Developments on this front
are nicely covered in a series of reviews
\cite{Bender05,Bender07,Markis1,Most_QM} and special issues in
\textcite{JPA06,JPA08,JPA12}.

The concept of $\PT$ symmetry has gone far beyond quantum mechanics
and has spread to many branches of physics. \textcolor{black}{\textcite{Ruschhaupt}
noticed that if a medium where a light pulse
propagates has an even refractive index profile and odd gain-loss
landscape, then one can construct an optical analog of $\PT$-symmetric
quantum mechanics. The real explosion of the $\PT$-symmetric optics and photonics started after the works by \textcite{ElGan07,Musslimani_1,Markis} who suggested and elaborated  paraxial $\PT$-symmetric optics.  Moreover,~\textcite{ElGan07} established the concept of $\PT$-symmetric waveguide optics, by showing that discrete optics provides a simple but nontrivial
framework for the study of $\PT$-symmetric systems.}  The $\PT$-symmetric optical theories were soon confirmed in a series of experiments~\cite{Guo,Ruter,Regens,Feng11}. Extension of $\PT$
symmetry to other branches of physics then quickly followed.

These developments suggested further extension of the theory to
include {\em nonlinearity}, which is inherent in many fields of physics
and is responsible for a wide variety of new phenomena.
This study was initiated in nonlinear optics with linear
$\PT$-symmetric potentials by \textcite{Musslimani_1}.
Later on optical systems with nonlinear $\PT$-symmetric potentials
were also explored~\cite{AKKZ,Miroshnichenko}.
\textcolor{black}{Presently the first experimental studies of
    nonlinear $\PT$-symmetric physics are already
    available~\cite{Peng2014,Wimmer2015}. From a practical point of
    view, important applications of $\PT$ symmetry, such as single-mode
    $\PT$ lasers~\cite{Feng2014,Hodaei2014} and unidirectional
    reflectionless $\PT$-symmetric metamaterial at optical
    frequencies~\cite{Feng13} have also emerged.}

\textcolor{black}{Why are $\PT$ systems interesting for physics beyond
    quantum mechanics? There are a number of reasons. One reason is that
    $\PT$ systems, being dissipative in nature, exhibit many properties
    of conservative systems, such as all-real linear spectra and
    existence of nonlinear steady states with continuous ranges of
    energy values. Such hybrid properties make $\PT$ systems physically
    very novel. Another reason is that $\PT$ systems offer some exciting
    applications, such as those mentioned in the previous paragraph. A
    third reason is that, loss was always considered to be a detrimental
    physical effect in the past. $\PT$ symmetry makes loss useful, which
    is physically very enlightening. Finally, gain and loss can be
    varied in time, opening new possibilities for flexible control and
    steering of physical processes. }

In this review we describe recent developments on nonlinear
$\PT$-symmetric systems. Even though reviews on linear $\PT$
theories and non-Hermitian quantum mechanics have been written (see
\textcite{Bender07,Most_QM} for instance) a comprehensive review on
nonlinear $\PT$-symmetric systems is still lacking. More
importantly, the field of nonlinear $\PT$ systems has been
developing very rapidly, and a large body of knowledge has been
obtained just in the past few years. Thus it is timely to write a
review on this subject.\footnote{After submission of this review we became aware of the work of \textcite{SSHDLK16} which addresses $\PT$ symmetry in optical applications.}

\section{Non-hermitian operators with real spectra}
\label{sec:1}

In this section, we overview the main concepts in the theory of
$\PT$-symmetric (and, more generally, non-Hermitian) linear
operators.  We do not intend to cover all available results of this
extremely vast field, but  rather to systematize the material
relevant for description of nonlinear systems presented in the
subsequent sections. For comprehensive reviews on non-Hermitian
operators in physics and mathematics, in addition to the works
listed in the Introduction
we also mention the reviews by
\textcite{Cannata07,Muga04,Rotter2009,Garcia14, Daley14}, as well as
the monograph of~\textcite{Moisey11}.

\subsection{Definition and basic properties}
\label{subsubsec:PT}

Let $\psi({\br}, t)$ be a complex-valued wavefunction of a quantum
particle. Evolution of $\psi({\br}, t)$ in space $\bf r$ and time
$t$ is governed by the Schr\"odinger equation
\begin{equation}
\label{eq:Schrod_main}
i \frac{\partial \psi}{\partial t} = H \psi(\br, t),
\end{equation}
where the linear operator $H$ acts in a Hilbert space
\textcolor{black}{$L^2(\mathbb{R}^D)$} endowed with an inner product $\langle \psi,
\phi \rangle=\int_{\mathbb{R}^D}\psi^*(\br,t)\phi(\br,t)d\br$, where $D$ is
the space dimension, the asterisk stands for complex conjugation,
and (unless stated otherwise) we consider the units where
$\hbar=m=1$ with $m$ being the mass of the particle.

For a given linear operator $H$, the Hermitian conjugation $H^\dag$
is defined by the relation $\lan H^\dag \psi,\phi\ran=\lan  \psi,H
\phi\ran$ for any two functions $\psi$ and $\phi$ in
$\cH(\mathbb{R}^D)$. An operator $H$ is said to be Hermitian (or
self-adjoint) if $H^\dag=H$, i.e., $\lan H\psi, \phi\ran=\lan
\psi,H\phi\ran$ \textcolor{black}{(a mathematically rigorous
definition of the Hermiticity \cite{RS80} additionally requires the
operator $H$ to be densely defined, i.e., the domain of $H$ must be
a dense subset of $L^2(\mathbb{R}^D)$; for the sake of simplicity,
we assume that this requirement holds for any operator  we
consider).}

The spectrum of any Hermitian operator is purely
real, while the converse  is not true, i.e., Hermiticity is
sufficient for reality of the spectrum but not necessary.

The two fundamental discrete symmetries in physics are given by the parity operator, $\p$, defined as
\begin{eqnarray}
\label{p_reflect}
\p \psi(\br, t)=\psi(-\br,t),
\end{eqnarray}
and by the time reversal operator, $\T$, defined in Wigner's sense as \cite{Wigner}
\begin{eqnarray}
\label{t_Wigner}
\T \psi(\br, t)=\psi^*(\br, -t). 
\end{eqnarray}
The operator $\T$ is antilinear, i.e., \textcolor{black}{
$\T(\lambda \psi +\phi) = \lambda^* \T\psi + \T\phi$, for any
vectors $\psi$, $\phi$ and a complex number  $\lambda$}.
Additionally,
\begin{equation}
\label{PT_prop_1}
\p^2=\T^2=I, \quad [\p,\T]=0,
\end{equation}
where $I$ is the identity operator.

An operator $H$ is said to be \textit{$\PT$ symmetric} if
\begin{equation}
\label{PT_defin}
[\PT,H]=0.
\end{equation}
Using  (\ref{PT_prop_1}),  definition (\ref{PT_defin}) can be rewritten as $H=\PT H\PT$.


Rapidly growing interest in $\PT$-symmetric operators was triggered
by the seminal work of \textcite{BendBoet} where a connection
between $\PT$ symmetry and reality of the spectrum was pointed out.
To emphasize this connection, \textcite{BendBoet} introduced the
notion of \textit{unbroken $\PT$ symmetry}. $\PT$ symmetry of a
$\PT$-symmetric operator $H$ is said to be unbroken if any
eigenfunction of $H$ is at the same time an eigenfunction of the
$\PT$ operator. In this case, the relation $H \psi = E \psi $
implies the existence of $\lambda$ such that  $\PT \psi = \lambda
\psi$. From (\ref{PT_prop_1}) it follows that  there exists a real
constant $\varphi$ such that $\lambda = e^{i\varphi}$, i.e., any
eigenvalue of the $\PT$ operator is a pure
phase~\cite{BendBoetMeis}.

Unlike Hermiticity, $\PT$ symmetry is not sufficient for the
spectrum to be purely real. However, it becomes sufficient when
combined with the requirement for the $\PT$ symmetry to be unbroken.
Indeed, let $E$ be an eigenvalue of $H$ with the eigenfunction
$\psi$, i.e., $H\psi = E \psi$. Applying the $\PT$ operator to both
sides of this equation and utilizing (\ref{PT_prop_1}), one obtains
$H (\PT \psi) = E^* (\PT \psi)$. Then, if the $\PT$ symmetry of $H$
is unbroken, $H\psi = E^*\psi$, and hence the eigenvalue $E$ is
real. Since this procedure is applied to every eigenvalue of $H$ we
conclude that the spectrum of $H$ is purely real.

If the unbroken $\PT$ symmetry does not hold, then the $\PT$
symmetry is said to be {\em broken}.
The broken $\PT$ symmetry is typically associated with the presence of complex eigenvalues in the spectrum of $H$.

\textcolor{black}{Unlike Hermiticity, $\PT$ symmetry does
    not ensure the completeness of eigenvectors of the operator. Even if
    the spectrum of a $\PT$-symmetric operator $H$ is entirely real, the
    set of eigenfunctions of $H$ may not constitute a complete basis.
    The typical scenarios when
    the eigenvectors lose their completeness correspond to the presence of an \emph{exceptional point}   
    (see \textcite{Kato,MoFr} and Sec.~\ref{sec:EP}) or a \emph{spectral
        singularity} (see Sec.~\ref{se:lasing}). These
    features are not exclusive to $\PT$-symmetric operators and can be
    encountered for more general non-Hermitian operators as well. }

The described connection between the $\PT$ symmetry and reality of
the spectrum does not involve the definition (\ref{p_reflect}) of
the parity operator, but rather relies on properties
(\ref{PT_prop_1}) and the fact that $\T$ is antilinear. Therefore,
one can also consider the \emph{generalized} parity operator $\p$
\textcolor{black}{\cite{BenBerMan02, Most2003b, Most2008}}, with $\p$ being an arbitrary unitary linear
operator: $\p^\dag \p =  \p \p^\dag = I$.  Then properties
(\ref{PT_prop_1}) also imply that $\p$ is self-adjoint, i.e.,
$\p^\dag = \p$.

Now we consider a few examples.

\paragraph{$\PT$-symmetric parabolic potentials.}

A  Schr\"odinger operator with a complex potential $U(x)$,
\begin{equation}
\label{Schrod}
H = -\frac{d^2}{dx^2} +U(x), \quad U(x) = V(x)+i W(x),
\end{equation}
is $\PT$ symmetric if $U^*(x)=U(-x)$, i.e., its real and imaginary
parts are even and odd, respectively:
\begin{equation}
\label{VevenWodd}
V(x) = V(-x), \quad W(x) = -W(-x).
\end{equation}
The simplest example of such a potential is  the complex parabolic potential~\cite{Kato,Znozil1999,BenderJones08}
\begin{eqnarray}
\label{parabolic}
U(x)= (x-i\alpha)^2, \quad \alpha \in \mathbb{R}.
\end{eqnarray}
Its  eigenvalues and eigenfunctions are
\begin{equation}
\label{spect:parab}
E_n=2n+1, \quad \psi_n(x)=H_n(x-i\alpha)e^{-(x-i\alpha)^2/2},
\end{equation}
where $n=0,1,\ldots$, and $H_n(x)$ is the $n$th Hermite polynomial.
Thus the  $\PT$ symmetry of the parabolic potential  is unbroken for
any $\alpha$ (that is $E_n$ are real for all $\alpha$).

\paragraph{Bender--Boettcher potential.}

Generalizing a conjecture of Bessis and Zinn-Justin,
\textcite{BendBoet} investigated the spectrum of the potential
\begin{eqnarray}
\label{BB_potential}
U(x) = -(ix)^{N}.
\end{eqnarray}
For $1< N < 4$, the eigenvalue problem is posed on the real axis and the potential acquires the form  $U(x) = -|x|^N \exp\{i\,\textrm{sign}\,  (x)\, \pi N/2\}$. If   $0<N\leq 1$ or $N \geq 4$, then the problem  must be posed on a contour lying in the complex plane.

\begin{figure}
    \includegraphics[width=0.5\columnwidth]{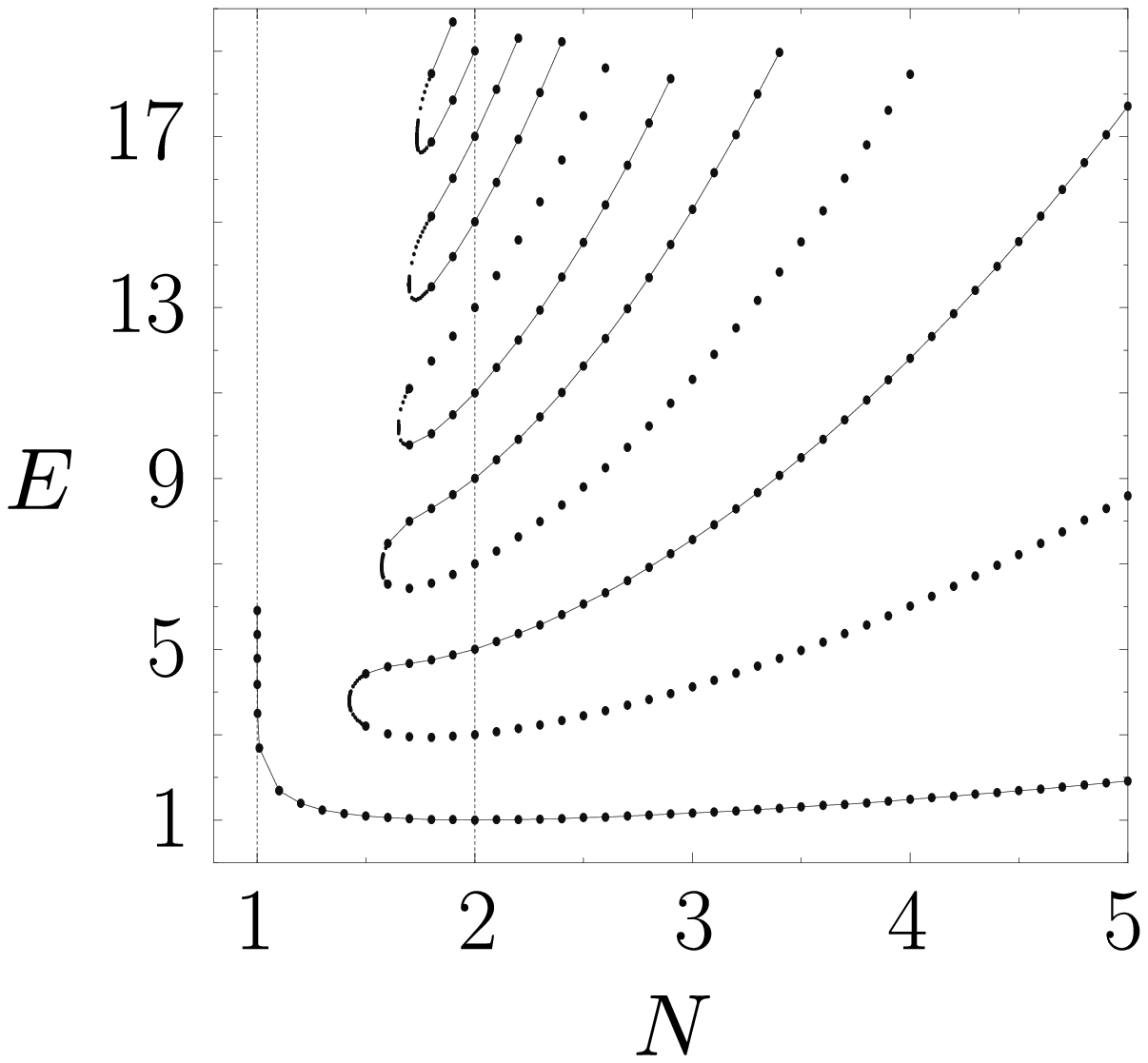}
    \caption{Real eigenvalues  of  potential (\ref{BB_potential}) for different $N$.   Adapted from \textcite{BendBoet}.
    }
    \label{fig:spectrum_Bender}
\end{figure}

Numerical results of~\textcite{BendBoet} on the spectrum of this
potential [Fig.~\ref{fig:spectrum_Bender}] show that for $N\geq 2$
the spectrum is real and positive [a rigorous proof of this fact
belongs to \textcite{DoreyProof}]. At the lower boundary of this
region, $N=2$, this potential becomes a real parabolic potential.
When $1<N<2$ one observes a finite number of real positive
eigenvalues and an infinite number of complex conjugate pairs of
eigenvalues. As $N$ approaches 1 from above, the lowest real eigenvalue
approaches infinity, and for $N<1$ there are no real eigenvalues.
Thus $\PT$ symmetry is unbroken for $N\geq  2$, but becomes
\textit{spontaneously broken} as parameter $N$ crosses the $\PT$-symmetry breaking threshold $N_{cr}=2$.

\paragraph{Two-level   $\PT$-symmetric   system.}

Consider now a Hamiltonian defined by a  $2\times 2$ matrix \cite{BenBerMan02}
\begin{equation}
\label{H2times2}
{H} = \left( \begin{array}{cc} i\gamma  & \kappa\\ \kappa  & -i\gamma \end{array} \right)=\kappa\sigma_1+i\gamma\sigma_3,
\end{equation}
where $\gamma\geq 0$ and $\kappa\geq 0$ are real parameters and hereafter we use the conventional notations for the Pauli matrices:
\begin{equation}
\label{Pauli}
\sigma_1=\left(\!\begin{array}{cc}
0 &1 \\ 1&0
\end{array}
\!\right),
\,\, \sigma_2=\left(\!\begin{array}{cc}
0 &-i \\ i&0
\end{array}\!\right),
\,\,\sigma_3=\left(\!\begin{array}{cc}
1 &0 \\ 0&-1
\end{array}\!\right).
\end{equation}

Hamiltonian (\ref{H2times2}) acts in a Hilbert space which consists of two-component column vectors $\psi=(\psi_1,\psi_2)^T$ (hereafter the superscript $T$ stands for the matrix transpose), with complex entries $\psi_{1,2}$, and the inner product is defined as  $\langle \psi, \phi \rangle = \psi_1^*\phi_1 + \psi_2^* \phi_2$. Hamiltonian   (\ref{H2times2}) is $\PT$ symmetric with $\p=\sigma_1$ and $\T$ being the complex conjugation.
The eigenvalues and eigenvectors  of $H$ are given by 
\begin{equation}
\label{eq:H_matrix_E}
E_{1,2}=\pm \sqrt{\kappa^2 - \gamma^2},\
\psi^{(1,2)}=
\left(\!\!
\begin{array}{c}
i{\gamma}/{\kappa}\pm \sqrt{1- {\gamma^2}/{\kappa^2}} \\ 1
\end{array}
\!\right).
\end{equation}
Thus $\PT$ symmetry is unbroken (the spectrum is all-real) if $\gamma  \leq \kappa$ and is broken (both eigenvalues are imaginary) if $\gamma > \kappa$. At $\gamma=\kappa$, $\PT$-symmetry breaking occurs. At this point,
the two eigenvalues collide, and the eigenvectors  become linearly
dependent. Thus $\PT$-symmetry breaking occurs at the point where
the Hamiltonian is a non-diagonal Jordan block. The respective
algebraic multiplicity of the eigenvalue is two, larger than its
geometric multiplicity of one. Such points in the parameter space
$(\gamma, \kappa)$ are called {\em exceptional} points \cite{Kato}
or branch points~\cite{MoFr}.

\subsection{Exceptional points}
\label{sec:EP}

Transition through an exceptional point is the most typical scenario
of $\PT$ symmetry breaking, which arises also in a more general
context of non-Hermitian physics \cite{MoFr,Heiss,Rotter2009}. Now
we take a closer look at what happens at an exceptional point by
considering a (not necessarily $\PT$-symmetric) Hamiltonian
\cite{Heiss}
\begin{eqnarray}
\label{eq:matrix}
H(\epsilon)=\left(
\begin{array}{cc}
E_{1} &  0
\\
0 & E_{2}
\end{array}
\right)+i\epsilon  \left(
\begin{array}{cc}
h_{11} & h_{12}
\\
h_{21} & h_{22}
\end{array}
\right),
\end{eqnarray}
where $E_1$ and $E_2$ are real. Eigenvalues of $H(\epsilon)$ are
\begin{eqnarray}
\label{eq:excep_spectr}
E_{1,2}(\epsilon)=\frac 12  \left[ E_1+E_2+i\epsilon(h_{11}-h_{22})\right]    \hspace{1.4cm}
\nonumber\\
\pm \frac 12 \sqrt{[(h_{11}-h_{22})^2+4h_{12}h_{21}](\epsilon-\epsilon_1)(\epsilon_2-\epsilon)},
\end{eqnarray}
where
\begin{eqnarray}
\label{eq:except}
\epsilon_{1,2}= [{E_1-E_2}]/[{2 \sqrt{h_{12}h_{21}} \mp i(h_{11}-h_{22}) }].
\end{eqnarray}
For arbitrary parameters the spectrum of $H(\epsilon)$ contains
\emph{two} distinct eigenvalues. However, at $\epsilon=\epsilon_1$
[or $\epsilon=\epsilon_2$] the two eigenvalues \emph{coalesce},
i.e., $E_1(\epsilon_1)=E_2(\epsilon_1)$ [or
$E_1(\epsilon_2)=E_2(\epsilon_2)$]. At these points
$H(\epsilon_{1,2})$ has  only \emph{one} linearly independent eigenvector, which means
that $\epsilon_1$ and $\epsilon_2$ are exceptional points.

These exceptional points may be complex numbers and transition
through an exceptional point requires variation of a complex
parameter $\epsilon$, i.e., is controlled by two real parameters. We
simplify the consideration by imposing the conditions
$h_{11}=h^*_{22}$, and $h_{12}h_{21}>0$ (each of $h_{12}$ and
$h_{21}$ may be complex). Then $\epsilon_{1,2}$ are real and we
consider real $\epsilon$. For the sake of definiteness we also set
$\epsilon_1<\epsilon_2$ and consider
$(h_{11}-h_{22})^2+4h_{12}h_{21}<0$. Then upon increase of
$\epsilon$ from zero, eigenvalues $E_{1,2}(\epsilon)$ move toward
each other along the real axis and collide at $\epsilon=\epsilon_1$.
At this instant phase transition occurs. After collision they move
to the complex plane, then become real again at
$\epsilon=\epsilon_2$ where they collide on the real axis a second
time. As $\epsilon$ approaches an exceptional point functions
$E_{1,2}(\epsilon)$ display a typical square root behavior,
$\sim\sqrt{\epsilon-\epsilon_{1,2}}$. The described restoration of
unbroken $\PT$ symmetry is referred to as {\em reentrant}
$\PT$ symmetry. It is noted that reentrant $\PT$ symmetry resembles
``bubbles of instability" in equilibria of Hamiltonian systems
\cite{MacKay1987}.

Exceptional points are inherently different from the
\emph{degeneracy} of eigenvalues, which corresponds to the situation
where two eigenvalues coalesce but their eigenvectors remain
linearly independent (i.e., the eigenvalue has a diagonal Jordan
block). In our case, the simplest example of degeneracy occurs when
$E_1=E_2$ and $\epsilon=0$.

\subsection{$\PT$ symmetry and pseudo-Hermiticity}
\label{subsec:pseudoHermit}

Although $\PT$ symmetry itself is not sufficient to guarantee the
reality of the spectrum of a Hamiltonian $H$, it ensures that
complex eigenvalues (if any) always exist in complex-conjugate
pairs. Indeed, if $E$ is a complex eigenvalue (with nonzero
imaginary part) and $\psi$ is the corresponding eigenvector, then
$E^*$ is also an eigenvalue with eigenvector $\PT\psi$. This, in
particular, implies that in the finite dimensional case $\PT$
symmetry of a linear operator results in reality of all coefficients
of the characteristic  equation of the Hamiltonian.
\textcite{BenderManheim}   proved that the converse is also correct:
if all the coefficients of the characteristic polynomial are real,
then the corresponding Hamiltonian is $\PT$ symmetric.

A necessary and sufficient condition for the spectrum of a
non-Hermitian Hamiltonian to be purely real can be formulated in
terms of a more general property called \textit{pseudo-Hermiticity}
{\cite{Most_a,{LW69}}}.  A Hamiltonian $H$ is said to be
$\eta$-pseudo-Hermitian if there exists a Hermitian invertible
linear operator $\eta$ such that
\begin{eqnarray}
\label{pH_defin}
H^\dagger=\eta H \eta^{-1}.
\end{eqnarray}
It is clear that if $\eta$ is the identity operator, then definition
(\ref{pH_defin}) is equivalent to Hermiticity, i.e.,
pseudo-Hermiticity is a generalization of Hermiticity. In many
cases, pseudo-Hermiticity can also be considered as a generalization
of $\PT$ symmetry. For example, if $H$ is a symmetric matrix
Hamiltonian, then $\PT$ symmetry implies $H\p-\p H^*=0$, and hence
$H^\dag = H^* = \p H \p$,
i.e., the pseudo-Hermiticity of $H$.
As another example,  the Schr\"odinger operator (\ref{Schrod}) with complex potential (\ref{VevenWodd}) is $\p$-pseudo-Hermitian.

An  immediate corollary of the pseudo-Hermiticity is that the
quantity $Q=\lan \eta \psi, \psi\ran$ is invariant under the time
evolution (\ref{eq:Schrod_main}) generated by the Hamiltonian $H$,
i.e., $d Q/d t \equiv 0$ \cite{Most_a}. In the case of the
Schr\"odinger operator (\ref{Schrod})--(\ref{VevenWodd}) this leads
to a conserved quantity~\cite{BQZ01}
\begin{equation}
\label{eq:int_Q}
Q= \int_{-\infty}^\infty \psi(x,t)\psi^*(-x, t)\ dx.
\end{equation}

\textcolor{black}{\textcite{Solombrino02} introduced a possibly more general concept of weak pseudo-Hermiticity which does not require the operator $\eta$ in (\ref{pH_defin}) to be Hermitian. Whenever one considers only  diagonalizable operators with discrete spectrum, the class of all pseudo-Hermitian operators coincides with the class of all weakly pseudo-Hermitian operators. Moreover, in this case (weak) pseudo-Hermititicy is equivalent to the presence of an antilinear symmetry, such as  $\PT$ symmetry:  a diagonalizable operator $H$ with discrete spectrum is (weakly) pseudo-Hermitian if and only if there exists an invertible antilinear operator $\Omega$ such that $\Omega^2=I$ and $[H, \Omega]=0$  \cite{Solombrino02, Most_c}.}

Notion  of the pseudo-Hermiticity allows one to formulate necessary and
sufficient conditions for a a Hamiltonian to possess a purely real
spectrum. Let us consider the case of the discrete spectrum, and let
a Hamiltonian have a complete set of biorthonormal eigenvectors
$\{ |\psi_n\rangle,|\phi_n\rangle\}$ defined by \cite{FM81}
\begin{eqnarray*}
    H|\psi_n\rangle=E_n|\psi_n\rangle, &\quad H^\dag|\phi_n\rangle=E_n^*|\phi_n\rangle,\\
    \langle\phi_n|\psi_n\rangle=\delta_{n,m}, &\quad \sum_n |\psi_n\rangle\langle\phi_n|=I.
\end{eqnarray*}
Then the following theorem holds.
\begin{theorem}[\textcite{Most_b}]
    \label{theo:most}
    Let $H$ be a Hamiltonian that  acts in a Hilbert space, has a discrete spectrum, and admits a complete set of biorthonormal eigenvectors $\{ |\psi_n\rangle,|\phi_n\rangle\}$. Then the spectrum of $H$ is real if and only if there is an invertible linear operator ${\cal O}$ such that $H$ is ${\cal O}{\cal O}^\dagger$-pseudo-Hermitian: $H=({\cal O}{\cal O}^\dag)H^\dag({\cal O}{\cal O}^\dag)^{-1}$.
\end{theorem}

To illustrate the application of Theorem~\ref{theo:most}, consider
the $\PT$-symmetric operator (\ref{H2times2}). It possesses a
complete set of biorthonormal eigenvectors unless  $\epsilon=\gamma
/\kappa = 1$.  Since at $\epsilon<1$ the spectrum of $H$ is real,
Theorem~\ref{theo:most} guarantees that there exists the operator
${\cal O}$ such that $H$ is  $\eta$-pseudo-Hermitian with
$\eta={\cal O}{\cal O}^\dagger$. Notice that although $H$ is
$\p$-pseudo-Hermitian, this cannot be used in
Theorem~\ref{theo:most}, because the parity operator $\p=\sigma_1$
does not admit the representation $\p = {\cal O}{\cal O}^\dagger$
(this can be verified straightforwardly). Therefore there must exist
{another} operator $\eta\not=\p$ when the spectrum of $H$ is purely
real. By straightforward algebra one finds that
\begin{equation*}
\eta  = \frac{1}{\epsilon^2}\left(\! \begin{array}{cc}
1 & i\epsilon \\ -i \epsilon & 1
\end{array} \!\right), \quad
\,\, {\cal O}  = \frac{1}{\epsilon}\left(\! \begin{array}{cc}
0 &i\\ \sqrt{1-\epsilon^2}  & \epsilon
\end{array} \!\right).
\end{equation*}
Theorem~\ref{theo:most} also indicates that no such operators exist
in the broken $\PT$-symmetry case of $\epsilon>1$.

It is noted that $\PT$ symmetry is not necessary for a non-Hermitian operator to have a real spectrum. \textcite{NY16a}  showed that if
an operator $H$ satisfies a weaker symmetry relation $H^\dagger \eta
=\eta H$ for some operator $\eta$ (not necessarily invertible), then
under some mild conditions on the kernel of $\eta$, complex
eigenvalues of $H$ (if any) always come in conjugate pairs, and a
real spectrum is often possible. Imposing this 
symmetry relation on the Schr\"odinger operator (\ref{Schrod}) for
differential operators $\eta$, wide classes of non-$\PT$-symmetric
complex potentials with  all-real spectra were constructed \cite{NY16a}. For an arbitrary real function $w(x)$, one such class of potentials is $U(x)=-w^2(x)-iw'(x)$ (see also (\ref{VW}) in Sec.
\ref{subsec:IST}), and another class is
\begin{equation}
\label{eq:And}
U(x) = -\left(\frac{1}{4} w^2+ \frac{w'^2-2 w''w +c}{4w^2}\right)-iw',
\end{equation}
where $c$ is a free
real parameter. The latter class of potentials generalizes the earlier result of  \textcite{Andrianov99} who discovered potentials (\ref{eq:And})  with negative $c$  using  the supersymmetry technique (addressed in Sec.~\ref{subsec:SUSY}).

\subsection{Real spectrum and effect of perturbations}
\label{subsec:perturb}

Since $\PT$ symmetry ensures that complex eigenvalues appear as
complex-conjugate pairs, one can expect that if $\PT$ symmetry is
unbroken and the real eigenvalues are ``well-separated'' from each
other, then the reality of  the spectrum is  ``robust'' against
sufficiently small perturbations.
While this intuitive expectation is not always correct, in many
situations it is indeed true.
In particular, this happens if a perturbed $\PT$-symmetric operator
is ``close'' to a self-adjoint operator~\cite{CaGraS,CaCaGra}. Let
us consider a Hermitian operator $H_0$ perturbed as
$H(\epsilon)=H_0+\epsilon H_1$, where $\epsilon$ is a  real
parameter. We also require operators $H_0$ and  $H_1$ to be
pseudo-Hermitian with the same operator $\eta$,
\begin{equation}
\label{eq:Calicetti}
H_{0}^\dag = \eta H_{0}\eta^{-1},\quad H_{1}^\dag =  \eta H_{1}\eta^{-1},
\end{equation}
where $\eta^2=I$ (for $\PT$-symmetric operators, $\eta$ is a parity
operator $\p$). Then according to the following theorem, the
spectrum of $H(\epsilon)$ is real provided that $\epsilon$ is small
enough.
\begin{theorem}[\textcite{CaGraS}]
    \label{theor:GaGraS} Let $H_0$ be a self-adjoint positive operator in a
    Hilbert space. Let $H_0$  have  only  discrete spectrum
    $\{0\leq  \lambda_0 < \lambda_1<\cdots< \lambda_n< \cdots\}$, where each
    eigenvalue $\lambda_j$ is simple, and $\delta=\inf_{j\geq
        0}[\lambda_{j+1}-\lambda_j]/2>0$. Let also $H_0$ and $H_1$ satisfy
    (\ref{eq:Calicetti}), and $H_1$ be continuous.
    Then the spectrum of $H(\epsilon)$ is real if
    $\epsilon\in\mathbb{R}$ and $|\epsilon|<\delta/\|H_1\|$.
\end{theorem}
Here the operator norm is defined in the usual way:
$\|H_1\|=\sup\{\|H_1f\|; \|f\|=1\}$, \textcolor{black}{and the operator $H_0$ is said to be
    positive if $\langle H_0\psi, \psi\rangle\geq 0$ for any $\psi$ from the Hilbert space \cite{RS80}.}

Theorem~\ref{theor:GaGraS} guarantees the existence of a large class
of pseudo-Hermitian operators with real spectra constructed as
perturbations of a given Hermitian operator, provided the spectrum
of the unperturbed operator is bounded below and its eigenvalues are
``well separated''. As a simple  example, consider the Schr\"odinger
operator with a harmonic potential $H_0=-{d^2}/{dx^2}+x^{2}$ and
$\PT$-symmetric perturbation $H_1=iW(x)$, with $W(x)=-W(-x)$ and
$W(x)\in L^\infty(\mathbb{R})$ (recall that the $L^\infty$-norm is
defined as
$\|W\|_{L^\infty}=\sup_{x\in \mathbb{R}}|W(x)|$).
Then  $\delta=1$ and the spectrum of $H_0+ \epsilon  H_1$ is real at
least for $|\epsilon|<1/\|W\|_{L^\infty}$. A similar result is
obtained for the power-law potentials $V(x)=x^{2m}$ with polynomial
perturbations { $iW(x)$},
provided that the odd degree $m^\prime$ of polynomials $W(x)$ is less than $m-1$ \cite{CaGra05}.

\subsection{Supersymmetry and real spectra}
\label{subsec:SUSY}

The concept of supersymmetry (SUSY) was first introduced in quantum
field theories and high-energy physics (see \textcite{Cooper95} and
the references therein). Subsequently, SUSY was utilized in quantum
mechanics to construct analytically solvable potentials. This
construction is based on the factorization of the Schr\"odinger
operator into the product of two first-order
operators~\cite{Infeld51}. Switching the order of these two
first-order operators gives another Schr\"odinger operator with a
new potential (called the {\em partner potential}) which shares the
same spectrum as the original potential (except possibly a single
discrete eigenvalue). {\color{black} Extending the idea of SUSY,
    parametric families of complex potentials with all-real spectra can
    be
    constructed~\cite{Khare89,Miri2013,Cannata98,Andrianov99,Bagchi01}.}

{\color{black} Let us employ the idea of SUSY to construct complex
    potentials with all-real spectra, following
    \textcite{Khare89,Miri2013,Yang14b}.} To this end, we consider the
Schr\"odinger operator (\ref{Schrod}) and assume that $U(x)$ has
purely real spectrum. Let $E_1$ and $\psi_1$ be an eigenvalue and
its eigenfunction of $H$, i.e., $(H-E_1)\psi_1=0$. We first
factorize the linear operator in this equation as
\begin{equation}
\label{e:factorV1}
H-E_1=A_-A_+, \quad
A_\pm=\pm \frac{d}{dx}+Y(x),
\end{equation}
where function $Y(x)$  is obtained from the requirement
$A_+\psi_1=0$, which yields  $Y(x) = -\psi_{1,x}/\psi_1$.

Now we switch operators $A_+$ and $A_-$ on the right side of
(\ref{e:factorV1}). This
leads to a new Schr\"odinger operator defined by
$ H_p-E_1=A_+A_-$,
i.e.,
\begin{equation}  \label{f:V2}
H_p=-\frac{d^2}{dx^2}+U_p(x)\quad \mbox{with $U_p=U+2Y_x$}.
\end{equation}
$U_p$ is the partner potential of
$U$ and has the same spectrum as $U$ (with the only possible
exception of $E_1$), since operators $A_+A_-$ and $A_-A_+$ share the
same spectrum.

The partner potential $U_p$ is   real or $\PT$ symmetric if
$U$ is so. In order to obtain a \textit{ non-$\PT$-symmetric} potential with all-real spectrum, we
build a new factorization for the partner potential:
$
H_p-E_1 = \widetilde{A}_+\widetilde{A}_-$,
with
$
\displaystyle{\widetilde{A}_\pm=\pm \frac{d}{dx}+\widetilde{Y}(x)}.
$
Equating both factorizations for $H_p-E_1$, we obtain the relation
$\widetilde{Y}_x+\widetilde{Y}^2=Y_x+Y^2$ which is  a Riccati
equation for $\widetilde{Y}$. Decomposition  $\widetilde{Y}=Y+1/f$
leads to a linear equation $f_x-2Yf=1$ which can be  readily solved.
This   yields
\begin{equation}
\label{f:Wtilde}
\widetilde{Y}(x)=-\frac{d}{dx} \ln
(\widetilde{\psi}_1),
\,\,
\widetilde{\psi}_1(x) = \frac{\psi_1(x)}{c+\int_0^x [\psi_1(\xi)]^2 d\xi},
\end{equation}
where $c$ is an arbitrary complex constant.
For this new $U_p$ factorization, its partner
potential is defined through $\widetilde{H}-E_1=A_-A_+$ and is given by
$
\widetilde{U}=U_p-2\widetilde{Y}_x.
$
Utilizing the $U_p$ and $\widetilde{Y}$ in formulas (\ref{f:V2}) and
(\ref{f:Wtilde}), this $\widetilde{U}$ potential is found to
be
\begin{equation}  \label{f:V1tilde}
\widetilde{U}(x)=U(x)-2\frac{d^2}{dx^2}\ln\left[c+\int_0^x
[\psi_1(\xi)]^2 d\xi\right].
\end{equation}
For generic values of the complex constant $c$,
$\widetilde{U}$  is complex and not $\PT$ symmetric. In
addition, its spectrum is identical to that of $U$. Hence if $U$
has an all-real spectrum, so does $\widetilde{U}$. The
potential $\widetilde{U}$ is referred to as the {\em superpotential};
it represents a family of potentials parametrized by $c$.

\begin{figure}
    \includegraphics[width=\columnwidth]{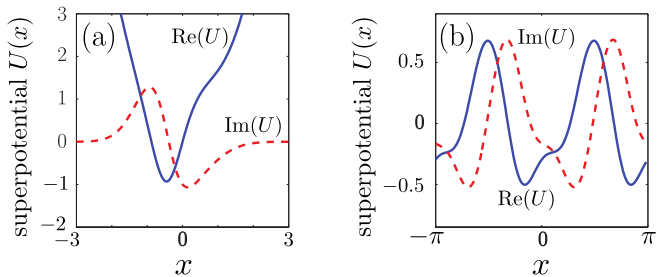}
    \caption{(a) Superpotential (\ref{f:Vexample1}) with $c=1+i$;
        (b) Periodic superpotential (\ref{f:Vexample3}) with $c=0.5-2i$
        and $V_0=1$.}  \label{fig_SUSY}
\end{figure}

Now we give two explicit examples of non-$\PT$-symmetric
superpotentials (\ref{f:V1tilde}) with all-real spectra. The first
one   is constructed from the parabolic potential $U(x)=x^2$ and its
first eigenmode of $E_1=1$ with $\psi_1=e^{-x^2/2}$. Then the
superpotential (\ref{f:V1tilde}) reads
\begin{equation}  \label{f:Vexample1}
U(x)=x^2-2\frac{d^2}{dx^2}\ln\left[c+\int_0^x e^{-\xi^2}d\xi\right],
\end{equation}
see Fig.~\ref{fig_SUSY}(a).
The spectrum of this superpotential
(for  any  $c$) is $\{1, 3, 5, \dots\}$, i.e.,  is all-real.

In the second example, the superpotential (\ref{f:V1tilde}) is built
from the $\PT$-symmetric periodic potential $U(x)=-V_0^2 e^{2ix}$
and its Bloch mode $\psi^{(1)}=I_1(V_0e^{ix})$ with eigenvalue
$E_1=1$. Here $V_0$ is a real constant, and $I_n$ is the
modified Bessel function. The resulting periodic superpotential
(\ref{f:V1tilde}) reads
\begin{equation}  \label{f:Vexample3}
U(x)=-V_0^2 e^{2ix}-2\frac{d^2}{dx^2}\ln\left[c+\int_0^x  I_1^2(V_0
e^{i\xi}) d\xi\right],
\end{equation}
see Fig.~\ref{fig_SUSY}(b).
The diffraction (dispersion) relation of this
superpotential (for all $c$ values) is the same as that of the
original potential $U(x)=-V_0^2 e^{2ix}$, i.e., $E=-(k+2m)^2$,
where $k$ is in the first Brillouin zone, $k\in [-1, 1]$, and $m$ is any
non-negative integer.

If $U(x)$ is  a localized real potential, then SUSY   allows to construct localized
complex superpotentials (\ref{f:V1tilde}) with all-real spectra  \cite{Miri2013,Yang14b}.

\subsection{Soliton theory and $\PT$-symmetric potentials}
\label{subsec:IST}

Nonlinear integrable equations solvable by the inverse scattering
transform technique play a special role in physics and mathematics.
Some of those equations, like the Korteweg-de Vries (KdV) equation
\begin{eqnarray}
u_t-6uu_x+u_{xxx}=0,
\end{eqnarray}
and the nonlinear Schr\"odinger (NLS) equation
\begin{eqnarray}
\label{eq:NLS}
i\psi_t+\psi_{xx}+g|\psi|^2\psi=0
\end{eqnarray}
with $g$ being a real constant, constitute fundamental models
describing a large variety of physical phenomena
\cite{Lamb,AS81,Dodd,NMPZ,FaTa87}. The starting point of the inverse
scattering transform is the representation of a nonlinear equation
as a compatibility condition for two linear equations (the so-called
Lax pair). \textcite{Wadati} noticed that the Lax representation
offers a way to construct a wide class of $\PT$-symmetric (as well
as complex asymmetric) potentials with purely real spectra.
Indeed,
let us consider the modified Korteweg-de Vries (mKdV) equation
\begin{eqnarray}
\label{mKdV}
w_t+6w^2w_{x}+w_{xxx}=0
\end{eqnarray}
for the real function $w(x,t)$, where $x\in \mathbb{R}$ is the
spatial coordinate, and $t>0$ is time. We will consider decaying functions: $\lim_{|x|\to\infty}w(x,t)=0$. Equation (\ref{mKdV}) is the
compatibility condition for the Zakharov-Shabat (ZS) spectral
problem  \cite{ZS71}
\begin{equation}
\label{ZS}
\phi_{1x}+i\zeta \phi_1=w(x,t)\phi_2, \,\,\, \phi_{2x}-i\zeta \phi_2=-w(x,t)\phi_1,
\end{equation}
where $\zeta$ is the spectral parameter, and the linear system
\begin{eqnarray*}
    \phi_{1t}=2i\zeta(w^2-2\zeta^2)\phi_1+(2i\zeta w_x-2w^3-w_{xx}+4\zeta^2 w)\phi_2,
    \\
    \phi_{2t}= (2i\zeta w_x+2w^3+w_{xx}-4\zeta^2 w)\phi_1-2i\zeta(w^2-2\zeta^2)\phi_2.
\end{eqnarray*}
Then the new function $\phi=\phi_2-i\phi_1$ solves the linear
Schr\"odinger equation  $H\phi=E\phi$,
where $H$ is given by (\ref{Schrod}) with potential \cite[Sec.~2.12]{Lamb}
\begin{eqnarray}
\label{VW}
U(x,t) = -w^2(x,t)-iw_x(x,t),
\end{eqnarray}
and  $E=-\zeta^2$. Here time $t$ plays the role of a parameter. If $w(x,t)$ is an even function, the potential $U(x,t)$ is $\PT$ symmetric; for general real $w(x,t)$, this potential is complex and asymmetric.

Discrete eigenvalues of the ZS problem (\ref{ZS}) are either purely imaginary or situated symmetrically with respect to the imaginary axis.
Its continuous spectrum is the real axis. Thus from {\em any} solution $w(x,t)$ of the mKdV equation (\ref{mKdV}) that
possesses purely imaginary discrete eigenvalues of (\ref{ZS}), one can obtain a complex potential $U(x,t)$ defined by (\ref{VW}), with purely real spectrum. Further, we notice that $w(x,t)$ depends on the parameter $t$,  
while the spectrum of the ZS problem does not depend on $t$. This means that $t$ can be considered as a ``deformation'' parameter, and $w(x,t)$  generates a family of deformable potentials $U(x,t)$ with real spectra. As an example,  we present a $\PT$-symmetric potential obtained from the two-soliton solution of the mKdV equation, with $\zeta_{1,2}=i\eta_{1,2}/2$, where $0<\eta_{1}<\eta_2$. It is generated by the function (here $t=0$, and not indicated)~\cite{WadatiOhkuma,Wadati}
$
w(x)=2\epsilon \Delta {g(x)}/{f(x)}$,
where $\Delta={(\eta_1+\eta_2)}/{(\eta_1-\eta_2)}$, $g=\eta_1\cosh(\eta_2 x)+\eta_2\cosh(\eta_1 x)$,
\begin{eqnarray*}
    f= \cosh[(\eta_2 +\eta_1)x]+ \frac{4\epsilon^2\eta_1\eta_2 }{(\eta_1-\eta_2)^2}
    +\Delta^2\cosh[(\eta_2 -\eta_1)x],
\end{eqnarray*}
and has the form
\begin{eqnarray}
\label{eq:two_sol_V}
U(x)=-4\Delta^2\left({g}/{f}\right)^2
-2i\epsilon\Delta \left({g}/{f}  \right)_x.
\end{eqnarray}
By changing $\eta_{1,2}$ or $\epsilon$ one can modify the potential shape without violating the reality of the spectrum.

In addition we  notice that   the   potentials of the form (\ref{VW}) were also discussed in the earlier literature in the context of supersymmetry \cite{Unanyan92, BSF91,Andrianov99}  and in application  to   
neutrino physics
\cite{Balantekin88, Notzold87}.

\section{$\PT$ symmetry in nonlinear physics}
\label{sec:2}

Rapidly growing interest in $\PT$ symmetry and particularly in its
interplay with other physical phenomena, such as periodicity,
discreteness, or nonlinearity, is stimulated by  possibilities of
extending the paradigm far beyond its quantum mechanical
applications. In this section, we review suggestions on
implementation of $\PT$ symmetry in physical systems of different
natures.

\subsection{Optics}
\label{subsec:optics}

\paragraph*{Paraxial optics {\it vs.} quantum mechanics.} The Schr\"{o}dinger equation (\ref{eq:Schrod_main}) with the Hamiltonian (\ref{Schrod})
\begin{equation}
\label{opt:parabolic}
i\Psi_t+ \Psi_{xx}  {-} U(x)\Psi=0
\end{equation}
has direct mathematical analogy with the theory of optical wave
propagation under the paraxial approximation~\textcolor{black}{\cite{ElGan07,Musslimani_1,Markis}}. To introduce
this analogy, we consider propagation of a linear monochromatic {TE}
wave $Ee^{i\omega_0 t}$ with frequency $\omega_0 $, in a
waveguide confined to the $(x,z)$ plane, i.e.,  bounded in the
domain $-\ell\leq y\leq \ell$, where $2\ell$ is the waveguide width,
by parallel claddings (say, by Bragg mirrors).
The field diffraction in such a waveguide is described by the Helmholtz equation
$\nabla^2E+k_0^2n^2(x)E=0$, where $k_0=\omega_0/c$.
Let 
the refractive index of the medium be weakly modulated along the $x$-direction, i.e.,
$n(x)=n_0+n_1(x)$, where $n_0$ is the constant component, and
$|n_1(x)|\ll n_0$ describes the modulation.
In the paraxial approximation (i.e., under small diffraction angles) the field can
be represented as $E=\psi(\xi,\zeta)\phi(y) e^{i\beta z}$, where
$\beta$ is the propagation constant, $\phi(y)$ describes the
transverse distribution of the field and solves the equation
$d^2\phi/dy^2 +k_0^2n_0^2\phi=\beta^2\phi$ subject to the continuity
boundary conditions at $y=\pm\ell$ (determined by the cladding), and $\psi(\xi,\zeta)$ solves Eq.
(\ref{opt:parabolic}) where $U(x)=n_0n_1(x)/k_0^2$ and the independent variables were re-named
as $\zeta=k_0 z\to t$ and $\xi=k_0 x\to x/\sqrt{2}$.

Thus variation of the dielectric permittivity 
 $-\varepsilon(x)=-n^2(x)$, in optical applications plays the role of a potential in
the Schr\"{o}dinger equation, and non-Hermitian quantum mechanics
can be emulated by optical media with the refractive index or permittivity
[\emph{cf.} (\ref{VevenWodd})]
\begin{eqnarray}
\label{nPT}
n(x)=n^*(-x), \quad\mbox{or}\quad  \varepsilon(x)=\varepsilon^*(-x).
\end{eqnarray}

\textcolor{black}{
\paragraph*{Modeling pulse propagation through a $\PT$-layer.}
A simple model for refractive index (\ref{nPT}) is provided by two
cells with a gas of two-level atoms described by the Lorentz model. It was used by~\textcite{Ruschhaupt} for
description of pulse propagation through a $\PT$-symmetric layer inserted in a waveguide confined by two metallic plates parallel to the $(x,y)$-plane and having distance $2a$ between them}. If the
atomic population in the left cell ($-\ell < x <0$) is inverted, and
in the right cell ($0<x<\ell$) the atoms are in the ground state,
then the dielectric permittivity for a monochromatic beam with
central frequency $\omega$ reads~\cite{Chiao}
{\color{black}
$\varepsilon=1-f(x)\omega_p^2/(\omega^2-\omega_0^2+2i\gamma\omega)$},
where $f(x)= -1$ for $x\in(-\ell,0)$ (cell with gain), $f(x)=1$ for
$x\in(0,\ell)$ (cell with loss) and zero for $|x|>\ell$, $\omega_p$
and $\omega_0$ are the plasma and resonance frequencies,
respectively, and
{\color{black} $\gamma$ is the damping constant.}
Assuming that the pulse frequency is close to the resonant one, i.e.
$\omega-\omega_0 \ll \gamma$,
and that the plasma frequency is small enough,
$\omega_p\ll\gamma$, the system can be shown to obey
Eq.~(\ref{opt:parabolic}) with the complex potential $U(x)\approx
-i\omega_p^2f(x)/4\gamma$ (where distance is measured in units of $a/\pi$).
Accounting also for the real part of the dielectric permittivity
such step-like potential can be re-written as
\begin{eqnarray}
\label{pot:step_like}
U=\left\{
\begin{array}{lll}
V_0+i\gamma & x\in (-\ell,0), & \mbox{(gain) }
\\
V_0-i\gamma & x\in (0,\ell), & \mbox{(absorption)}
\\
0 & |x|> \ell, & \mbox{(vacuum)}
\end{array}
\right.
\end{eqnarray}
with $V_0$ and $\gamma$ being real positive constants.

\paragraph*{Discrete optics.}
Interplay between gain and loss in wave scattering attracted
additional attention due to suggestions on constructing
nonreciprocal optical devices~\cite{Poladian}.
In~\textcite{GreOr,Kulishov1,Kulishov2} this problem was considered
for a guiding medium  with a periodic modulation of the real and
imaginary parts of the refractive index, $U=\delta_c \cos (2\beta_0
x)+i \delta_s \sin (2\beta_0 x )$, where $\delta_c$ and $\delta_s$
are the depths of modulaions, and $\beta_0=2\pi/\Lambda$ is defined
by the lattice period $\Lambda$. \textcite{LREKCC} considered this
phenomenon, termed as unidirectional invisibility, for the
particular $\PT$-symmetric configuration. In the presence of weak
constant absorption $\tilde{\gamma}$, i.e., when $U(x)\to
U(x)+i\tilde{\gamma}$, one can employ the two-mode approximation
$\psi(x,t)\approx [q_1(x)e^{i\beta x}+q_2(x)e^{-i\beta
    x}]e^{-i\beta^2 t}$, where  $q_1(x)$ and $q_2(x)$ are slowly varying
amplitudes of the forward and backward propagating waves. Using this
ansatz in the paraxial equation (\ref{opt:parabolic}), in the
leading order one obtains [\emph{cf.}~(\ref{H2times2})]
\begin{eqnarray}
\label{opt:coupled_ode}
i\frac{dq}{dx} =Hq, \quad
q=\left(
\begin{array}{c}
q_1   \\ q_2
\end{array}
\right),
\quad
H=\left(
\begin{array}{cc}
-i\gamma & \kappa_{12}
\\
\kappa_{21} & i\gamma
\end{array}
\right),
\end{eqnarray}
where $\gamma=\tilde{\gamma}/(2\beta)$,
$\kappa_{12}=(\delta_s-\delta_c)e^{2ib x}/4\beta$,
$\kappa_{21}=(\delta_s+\delta_c)e^{-2ib x}/4\beta$, and $b=\beta_0-\beta$ is
the phase mismatch.

System (\ref{opt:coupled_ode}) is also known as the simplest model
for stationary propagation of light in an optical coupler with gain
and loss~\cite{CSP}. Indeed, if one considers the real part of the
dielectric permittivity (\ref{nPT}) to have two localized and well
separated maxima, then $U(x)$ has the form of a double-well
potential in quantum mechanics. The field in (\ref{opt:parabolic})
can be searched in the form~\cite{LandauQM}
$\Psi=[q_1(z)E_1(x)+q_2(z)E_2(x)]e^{i\beta z}$, where $E_{1,2}(x)$
are the field distributions localized in the vicinity of the
potential minima. Then by straightforward algebra one can show that
$q_{1,2}$ solve the system (\ref{opt:coupled_ode}) with properly
defined matrix elements.

\begin{figure}
    \includegraphics[width=\columnwidth]{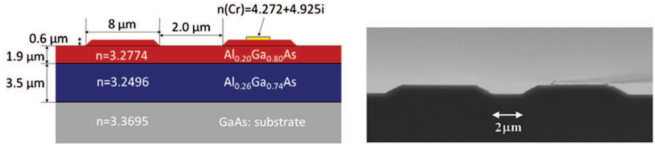}%
    \caption{(Color online) Left panel: details of two-waveguide layered structure which includes the Cr stripe and features complex refractive index;
        right panel: scanning electron microscopy picture of the finalized device. Adapted from \textcite{Guo}.}
    \label{fig:guo}
\end{figure}

The link between (\ref{opt:coupled_ode}) and (\ref{opt:parabolic}),
yet in a different setting where each of the two arms of a coupler
has balanced gain and loss, was established by~\textcite{ElGan07},
who recognized the relevance of the model for $\PT$-symmetric
optics. If $\PT$ symmetry is unbroken, a medium with balanced gain
and loss allows for stationary propagation of light, while the light
undergoes attenuation or amplification if the $\PT$ symmetry is
broken. Moreover, since the existence of an exceptional point, and
hence the transition between different propagation regimes, do not
require the exact balance between gain and loss [Sec.~\ref{sec:EP}],
one can introduce the concept of a {\em
    passive} $\PT$-symmetric coupler~\cite{Guo}.  Indeed, consider an imbalanced generalization of Eqs.~(\ref{opt:coupled_ode}),
\begin{eqnarray}
\label{opt:coupled_ode_gen}
i\dot{q}_1 =-i\gamma_1 q_1 +\kappa q_2 \quad
i\dot{q}_2 =i\gamma_2 q_2 +\kappa q_1,
\end{eqnarray}
where $\gamma_1 \ne \gamma_2$. Hereafter an overdot stands for the
derivative with respect to an evolution variable, which can be
either the propagation distance or time, depending on the context.
By substitution
$q_{1,2}(z)=\tilde{q}_{1,2}(z)\exp[(\gamma_2-\gamma_1) z/2]$ one
verifies that $\tilde{q}_{1,2}$ solves (\ref{opt:coupled_ode}) with
$\gamma=(\gamma_1+\gamma_2)/2$ and $\kappa_{12}=\kappa_{21}=\kappa$,
thus reducing the dissipative system (\ref{opt:coupled_ode_gen}) to
a $\PT$-symmetric one. Propagation of a 1.55$\mu$m--wavelength beam
in a passive coupler fabricated on a multilayer Al$_x$Ga$_{1-x}$As
hetero-structure with one nonlossy waveguide and another waveguide
with controlled absorption was used in the experiment
of~\textcite{Guo} (see Fig.~\ref{fig:guo}), where transition between
broken and unbroken $\PT$ symmetries was observed for the first
time. Later, $\PT$-symmetry breaking was observed experimentally in
different physical settings, such as the microwave billiard
(implemented in a microcavity)~\cite{Bittner2012} and
{\color{black} polarization}
of the electromagnetic radiation interacting with a
metasurface~\cite{Lawrence20014}.

\begin{figure}
    \includegraphics[width=\columnwidth]{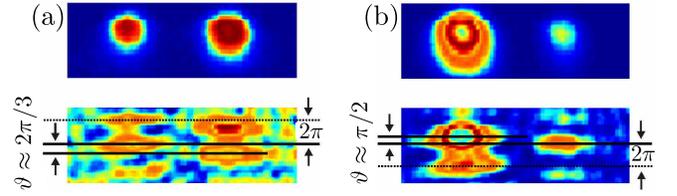}%
    \caption{(Color online)  Measured density distribution  (upper panels) and relative phase difference between the two components (lower panels) of the $\PT$-symmetric coupler in unbroken (a)  and broken (b) $\PT$ symmetries.  Below the $\PT$-symmetry breaking threshold  the phase difference lies in the
        interval $[0,\pi]$, depending on the magnitude of gain, while above the threshold   this value is fixed at $\pi/2$. 
        Reprinted by permission from Macmillan Publishers Ltd: [Nature Physics] (R\"uter, C. E., K. G. Makris, R. El-Ganainy,   D. N.  Christodoulides, M. Segev, and D.
    Kip, 2010, Observation of parity-time symmetry in optics.
    Nat. Phys. \textbf{6}, 192--195), copyright 2010.}
    \label{fig:PT_breaking}
\end{figure}

$\PT$-symmetry phase transition in a coupler with gain and loss was
observed by~\textcite{Ruter}. In the experimental setup two
waveguides were created in a photorefractive Fe-dopped Lithium
niobate substrate. The loss was determined by excitations of
electrons from Fe$^{2+}$-centers, while gain was created by the pump
light through the two-wave mixing determined by the concentration of
Fe$^{3+}$centers. The model was described by
Eqs.~(\ref{opt:coupled_ode_gen}). Having the loss coefficient
$\gamma_1$ fixed at the level $\gamma_1=3.3\,$cm$^{-1}$ and
increasing the gain coefficient $\gamma_2$, \textcite{Ruter} observed
spontaneous $\PT$-symmetry breaking as shown in
Fig.~\ref{fig:PT_breaking}.

System (\ref{opt:coupled_ode})  also reveals other important effects
observable in media with balanced gain and loss.
\textcite{Kulishov1} found that a finite medium with periodically modulated complex refractive index (or a coupler) of length $L$
possesses distinct transmission and reflection properties depending
on whether the light is applied at $z=0$ or at $z=L$. This unidirectional propagation, which was also studied for the $\PT$-symmetric configuration \cite{LREKCC},  is described by the entries $M_{ij}$ of the transfer matrix $M(L)$ defined through the relation $q(L)=M(L)q(0)$. In
particular, defining the left (right) transmission and reflection
coefficients $t_{L(R)}$ and $r_{L(R)}$ by the conditions $q_2(L)=0$
($q_1(0)=0$), one obtains
\begin{eqnarray}
\label{opt:tr}
t_L=t_R=\frac{1}{M_{22}},\quad r_L=-\frac{M_{21}}{M_{22}}, \quad r_R=\frac{M_{12}}{M_{22}}.
\end{eqnarray}
The difference in light propagation from the left and right can be
described by the contrast ratio
$C=(|r_L|^2-|r_R|^2)/(|r_L|^2+|r_R|^2)$~\cite{Feng13}. For the
system (\ref{opt:coupled_ode}) with $\beta_0=\beta$ one obtains
$C=2\delta/(1+\delta^2)$, where $\delta=\delta_s/\delta_c$. $C$
achieves unity at $\delta=1$. This phenomenon was observed in the
experiments of \textcite{Feng11,Feng13} at wavelength of 1.55 $\mu$m
using Si waveguides with periodic dissipation implemented by the
embedded Ge/Cr structures.
%
%
In order not to violate the Lorentz reciprocal theorem (see e.g.
\textcite{Haus}),  the prediction on differences in the left and
right propagations of a given mode is explained by excitation of the
orthogonal modes, when the input channel is changed to the opposite
one~\cite{FAN_comm}.

\paragraph*{Nonlinearity.} As soon as optical applications are considered, accounting for
nonlinearity becomes a natural step. Considering light propagation
in a Kerr-type medium, where the refractive index is a function of
the field intensity, $n(x, |\psi|^2)=n(x)+n_2|\psi|^2$,
Eq.~(\ref{opt:parabolic}) is generalized to the NLS equation with a
potential $U(x)$, i.e.,
\begin{eqnarray}
\label{opt:NLS}
i\Psi_t+ \Psi_{xx} {-}U(x)\Psi+g|\Psi|^2\Psi=0,
\end{eqnarray}
where $g=n_0n_2/k_0^2$ is the nonlinear coefficient which can be
either positive (focusing medium) or negative (defocusing medium).
Equation (\ref{opt:NLS}) with focusing nonlinearity and a $\PT$-symmetric
periodic potential was introduced by~\textcite{Musslimani_1}. The
nonlinear generalization of the coupler model
(\ref{opt:coupled_ode}), i.e., a nonlinear $\PT$-symmetric dimer
\begin{eqnarray}
\label{opt:coupled_ode_nl}
\begin{array}{l}
i\dot{q}_0 =-i\gamma q_0 +\kappa q_1+\chi|q_0|^2q_0,
\\
i\dot{q}_1 =\phantom{+}i\gamma q_1 +\kappa q_0+\chi|q_1|^2q_1,
\end{array}
\end{eqnarray}
was introduced by~\textcite{Ramezani,SXK10}.

\paragraph*{\textcolor{black}{Synthetic photonic lattices.}}

{An idea of experimental implementation of a fully discrete
(discrete ``time-space'') $\PT$-symmetric lattice was proposed
by~\textcite{Miri2012,Regens}. Such a synthetic lattice is created in the time-domain.
It is produced by two fiber loops having slightly different lengths and coupled by a $50$\% coupler
illustrated in Fig.~\ref{fig:synthetic1}. If the gain and loss are held constant in each loop,
then the lattice is locally $\PT$ symmetric, i.e., symmetric with respect to the $\p$ inversion ($n\to-n$)
combined with the complex conjugation 
for a fixed value of $m$. If gain and loss alternate on every other round trip the synthetic lattice obeys global
$\PT$ symmetry, with the above $\p$ operator and $\T$-operator
inverting $m\to-m$ with simultaneous complex conjugation.

\begin{figure}
        \includegraphics[width=\columnwidth]{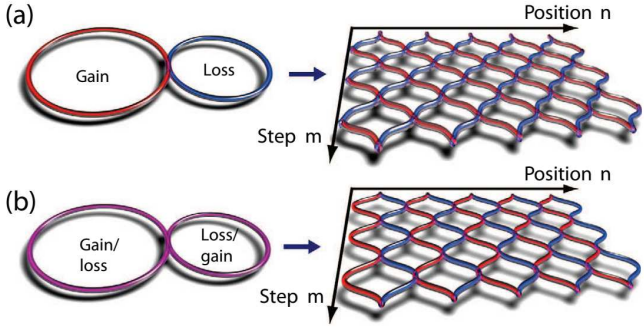}%
        \caption{(Color online) Mapping of the pulse propagation in loops consisting of dispersion compensating fibers into a 2D lattice corresponding to (a) locally $\PT$-symmetric and (b) globally $\PT$-symmetric settings. Red and blue segments correspond to pulses traveling during the paths with gain and with losses. Alternating gain and losses are  shown by purple rings in (b). From~\textcite{Wimmer2015}.}
        \label{fig:synthetic1}
\end{figure}
}

{ Pulses, whose fields are denoted by $u_n^m$ and $v_n^m$, travel in
    shorter and longer loops, respectively~\cite{Regens2011}.
    Here the upper index $m$ stands for the time interval as measured in round trips. The subindex $n$ denotes the position of a single pulse during one cycle. 
      The nonlinearity
    of the fibers  leads to phase accumulation proportional to the pulse
    intensities. The system evolution is modeled by the nonlinear
    map~\cite{Wimmer2015}
    \begin{equation}
    \label{eq:synthetic1}
    \begin{array}{l}
    u_n^{m+1}={\sqrt{G_u/2}}\left(u_{n+1}^{m}+iv_{n+1}^m\right)e^{\frac{i\Gamma}{2} \left|u_{n+1}^{m}+iv_{n+1}^m\right|^2}e^{i\phi_n}, \quad
    \\
    v_n^{m+1}={\sqrt{G_v/2}}\left(v_{n-1}^{m}+iu_{n-1}^m\right)e^{\frac{i\Gamma}{2} \left|v_{n-1}^{m}+iu_{n-1}^m\right|^2},\quad
    \end{array}
    \end{equation}
    where gain and loss factors $G_{u,v}$ characterize the pulse
    amplitudes at the coupler output.  The phase function $\phi_n$,
    controlled by the phase modulator in the experiment, is the imposed
    phase shift governing the $\PT$-symmetric potential (in addition to
    the phase shift $\pi/2$ generated by the coupler, which is also
    accounted for). The case $G_{u,v}=1$ and $\phi_n=0$ corresponds to
    the conservative case. The described synthetic network
    allowed~\textcite{Wimmer2015} to report the first observation of
    $\PT$-symmetric lattice solitons (see
    Sec.~\ref{sec:synthetic_nonlin} below). }

\paragraph*{$\PT$ symmetry introduced by time management.}

The idea of inducing $\PT$ symmetry by time management can be developed further by inclusion of non-autonomous gain-and-loss coefficients $\theta(t)$ in the classical linear oscillator~\cite{TL14}:
\begin{equation}
\label{eq:discr:0D}
\ddot{q} + 2\theta(t)\dot{q} + \omega_0^2 q = 0.
\end{equation}
If $\theta(t)$ is  periodic,   $\theta(t+T) = \theta(t)$, and acquires positive and negative  values, e.g., $\theta=\gamma$ for $0\leq t< T/2$ and    $\theta=-\gamma$ for $T/2\leq t<  T$, then the system 
can feature bounded or unbounded dynamics, depending on the choice
of parameters  $\omega_0$  and $\gamma$, which resembles
$\PT$-symmetric behavior.

\subsection{\textcolor{black}{$\PT$ lasers}}
\label{subsec:PTlaser}

One of the most important applications of $\PT$ symmetry is in the
design of new single-mode lasers. Laser cavities typically support a
large number of closely spaced modes, which is undesirable since it
leads to mode competition, random fluctuations, worse
monochromaticity and worse laser quality. Recently it was
demonstrated experimentally that utilizing the concepts of $\PT$
symmetry and $\PT$-symmetry breaking, new laser devices with
enhanced single-mode operations and greater tunability can be
realized \cite{Feng2014,Hodaei2014}. The basic idea is that, by
strategically designing gain and loss to obey $\PT$ symmetry, almost
all of the modes in the laser cavity can be neutralized, except for
a single lasing mode which amplifies. Hence single-mode operation is
achieved.

\begin{figure}
    \includegraphics[width=\columnwidth]{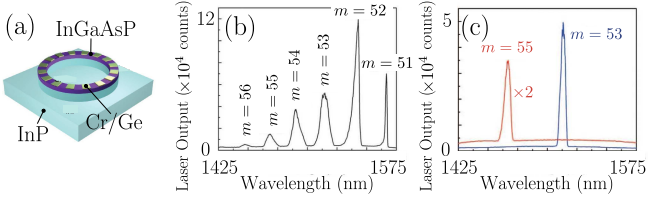}%
    \caption{(Color online) (a) Schematic of the $\PT$ microring laser. The diameter and width of the microring resonator are 8.9 $\mu$m and
        900 nm, respectively.
        (b) Multimode lasing spectrum observed from the typical microring
        WGM laser, showing a series of lasing modes corresponding to
        different azimuthal orders. (c) Single-mode lasing spectra of the
        $\PT$ microring lasers operating at the $m = 53$ and $m = 55$ azimuthal
        orders.  Adapted from Feng, L., Z. J. Wong, R. Ma, Y. Wang, and X.
    Zhang, 2014, Single-mode laser by parity-time symmetry breaking.
    Science \textbf{346}, 972--975.  Reprinted with permission from AAAS.
    }  \label{fig_PTlaser1}
\end{figure}

In the experiment by \textcite{Feng2014}, the $\PT$-symmetric
microring resonator was designed with 500-nm-thick InGaAsP multiple
quantum wells (MQWs) on an InP substrate [Fig. \ref{fig_PTlaser1}(a)].
InGaAsP MQWs have a high material gain coefficient around 1500~nm.
The gain/loss modulation, satisfying an exact $\PT$ symmetry
operation, was periodically introduced using additional Cr-Ge
structures on top of the InGaAsP MQW along the azimuthal direction
($\varphi$):
\begin{equation}  \label{e:Deltan}
\Delta n=\left\{\begin{array}{l} n_{gain}=-in''\left[\frac{l\pi}{m}<\varphi< \frac{(l+\frac{1}{2})\pi}{m}\right], \\
n_{loss}=in''\left[ \frac{(l+\frac{1}{2})\pi}{m} <\varphi< \frac{(l+1)\pi}{m}\right], \end{array}\right.
\end{equation}
where $n''$ denotes the index modulation in only the imaginary part;
$m$ is the azimuthal order of the desired whispering-gallery mode (WGM) in the microring; and
$l = 0, 1, 2, \dots, 2m-1$ divides the microring into $2m$ periods.
Due to the rotational symmetry of the microring, $\PT$-symmetry
breaking in this resonator is threshhold-less, i.e., it
occurs even if the strength of gain/loss modulation is
infinitesimal.

In the absence of the Cr/Ge gain-loss modulation (\ref{e:Deltan}), a
typical multimode lasing spectrum with different WGM azimuthal
orders was observed [Fig.~\ref{fig_PTlaser1}(b)]. But under this
$\PT$-symmetric index modulation, a single lasing mode was obtained
[Fig.~\ref{fig_PTlaser1}(c)]. The location of this single mode and its
power efficiency are close to those without the gain-loss modulation
[Fig.~\ref{fig_PTlaser1}(b)].

\begin{figure}
    \includegraphics[width=\columnwidth]{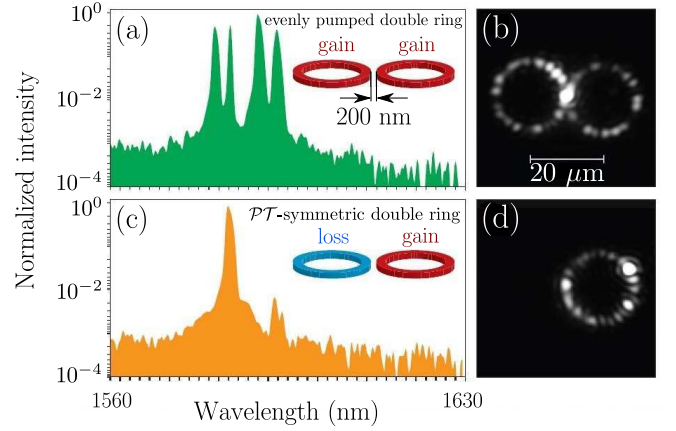}%
    \caption{(Color online) (a) Spectrum obtained from an evenly pumped pair of microrings. (b) The intensity
        pattern shows that both resonators equally contribute. (c) Single-moded spectrum under $\PT$-symmetric
        conditions. (d) Lasing exclusively occurs in the active resonator.  Adapted from Hodaei, H.,  M.-A. Miri, M. Heinrich, D. N. Christodoulides, and M.
    Khajavikhan, 2014,
    Parity-time-symmetric microring lasers.
    Science {\bf 346}, 975--978.  Reprinted with permission from AAAS.
        }  \label{fig_PTlaser2}
\end{figure}

In a different experiment by \textcite{Hodaei2014}, a single-mode
$\PT$ laser was demonstrated by utilizing two adjacent microrings,
one with gain and the other with loss (the active ring was based on
InGaAsP quantum wells), see Fig.~\ref{fig_PTlaser2}. In this case,
due to linear coupling between the two microrings, $\PT$-symmetry
breaking has a gain/loss threshold, which is equal to the coupling
constant between the rings. Thus when the gain-loss contrast is
increased beyond this coupling constant, $\PT$-symmetry breaking
occurs, and an amplifying lasing mode appears.

The experimental results are summarized in Fig.~\ref{fig_PTlaser2}.
When there is only one active ring, or both rings are active, a
familiar multimode lasing spectrum was observed [Fig.~\ref{fig_PTlaser2}(a)]. But when the two rings are placed in $\PT$
configuration, a single dominant spectral peak appears, resulting in
single-mode operation [Fig.~\ref{fig_PTlaser2}(c)].

In both experiments, the laser design was based on a linear model by
assuming a steady lasing state with a certain gain coefficient. But
it should be recognized that lasing itself is an intrinsically nonlinear
process. Nonlinear modeling of these $\PT$-laser devices in the broken phase is an
important open question. Below the $\PT$-symmetry breaking threshold such ring structures support stable nonlinear vortex modes~\cite{KarKonTor}.

\subsection{Atomic gasses}
\label{subsec:atomic}

Atomic media
are intrinsically dissipative. However, it was suggested
by~\textcite{Scully} and ~\textcite{FCSSUZ} and shown experimentally
by~\textcite{Zibrov} that by using destructive interference in the
imaginary part of the dielectric susceptibility it is possible to
obtain sufficiently large real refractive indices at small
absorption. This is achievable in a gas of multilevel atoms subject
to two far-off-resonant control fields~\cite{Yavuz,PUGY} or in a
mixture of isotopes of two $\Lambda$-atoms~\cite{BARK,Simmmons}.

The respective atomic schemes use two Raman resonances, one of which results in gain
and another leads to absorption. The imaginary part of probe-field
susceptibility becomes a non-monotonic function of the frequency
with positive (gain) and negative (absorbing) domains. Moreover, real and imaginary parts of the
susceptibility can be designed respectively as even and odd functions of  probe-field
frequency~\cite{Yavuz,PUGY,BARK,Simmmons}.
For a monochromatic beam the change $\omega\to -\omega$ can be
viewed as the time inversion which, when accompanied by the spatial
symmetry, can lead to the $\PT$-symmetric refractive
index.

\begin{figure}
    \includegraphics[width=\columnwidth]{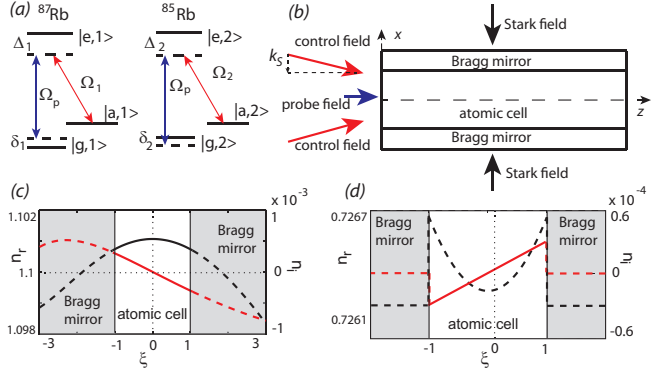}%
    \caption{(Color online) (a) Isotopes of $\Lambda$-atoms and Raman
        transitions.
        (b) A geometry for the atomic cell and fields applied. (c)  Locally parabolic~\cite{HZKH} and (d) two-channel~\cite{HZHKM} spatial distributions of the real (solid line) and imaginary (dashed line) parts of the refractive index.}
    \label{fig:lambda_atom}
\end{figure}

\textcite{HHK} considered a scheme shown in Fig.~\ref{fig:lambda_atom}(a,b). A ground ($|g,s\rangle$), lower ($|a,s\rangle$), and excited  ($|e,s\rangle$)  atomic states of two isotopes ($s=1,2$) with densities $N_{1,2}$ are coupled by two strong control fields and by a probe field with the half Rabi frequencies $\Omega_{1,2}$ and $\Omega_p$, respectively. 
All fields are far-off resonance, i.e., $\Delta_{s}\gg\Omega_{s}$,
where  $\Delta_{s}=\omega_{e,s}-\omega_{a,s}-\omega_c$ is the
one-photon  detuning, $\hbar \omega_{l,s}$ ($l=g,a,e$) is the energy
of the state $|l,s\rangle$ and $\omega_{p}$ ($\omega_{c}$) is the
center frequency of the probe (control) field. The first scheme
($s=1$, $\delta_1>0$) exhibits two-photon absorption for the probe
field, while the second one ($s=2$, $\delta_2<0$) provides
two-photon gain.

The mixture of isotopes is loaded in an atomic cell with Bragg
cladding [Fig.~\ref{fig:lambda_atom}(b)]. Spatial modulation of the
susceptibility is  achieved by a continuous-wave laser field (Stark
field) $E_{\rm S}(x)\cos(\omega_{\rm S} t)$ with the amplitude  $E_{\rm S}$
and frequency $\omega_{\rm S}$. Such field
originates $x$-dependent shifts of the one-photon detunings
$\Delta_{s}(x)=\Delta_{s}- (\alpha_{e,s}-\alpha_{g,s})E_{\rm
    S}^2(x)/(4\hbar)$.

The susceptibility for the probe field is
computed from the density-matrix formalism~\cite{BARK}:
\begin{eqnarray}
\label{chi}
\frac{\chi_p(x)}{\chi_0}=
\frac{\delta_1-i\gamma_{ag}}{(\delta_1+\Delta_1
    -i\gamma_{eg})(\delta_1-i\gamma_{ag})-|\Omega_1|^2}
\nonumber\\
-\eta \frac{|\Omega_2|^2
    (\Delta_2+i\gamma_{ag})^{-1}}{(\delta_2+\Delta_2
    -i\gamma_{eg})(\delta_2-i\gamma_{ag})-|\Omega_2|^2}
. \end{eqnarray}
Here $\chi_0=N_1d_{eg,1}^2/(\varepsilon_0\hbar)$, and $\eta=
N_2d_{eg,2}^2/N_1d_{eg,1}^2$ characterizes the ratio between the
densities, $\varepsilon_0$ is the vacuum
permittivity, $d_{eg,s}$ stands for the dipole moment of the
transition between the ground and excited states of the $s$th
system, and $\gamma_{ij}$ are dephasing rates at transitions
$i\leftrightarrow j$.

A refractive index satisfying $\PT$ symmetry conditions  (\ref{nPT}) can   be obtained numerically using an optimization procedure \cite{HHK}. As an example, implementation of this algorithm in a gas of rubidium isotopes yields $\PT$-symmetric permittivity  $\chi_{p}\approx 10^{-3} (7.5\cos\xi+i  0.394\sin\xi)$,
where $\xi= 2\pi x/\lambda_S$ and $\lambda_S$ is the Stark field wavelength. 

The cell confines atoms in space and can be used to cut undesirable
deviations from the $\PT$ symmetry. This allows for construction of
refractive indexes of different shapes, such as parabolic~\cite{HHK}
and double-hump~\cite{HZHKM}, see Fig.~\ref{fig:lambda_atom}(c,d).
The described ideas were further generalized through the use of more sophisticated atomic schemes, like four-level atoms~\cite{LDH13}, as well as involving nonlinear effects~\cite{HZKH,HZHKM}.
Effect of nonlinearity on the $\PT$-symmetry phase transition in finite-size systems with various profiles of the complex refractive index were studied by~\textcite{WaMaLi}.

\subsection{Plasmonic waveguides}
\label{sec:plasmonic}

\begin{figure}
    \includegraphics[width=\columnwidth]{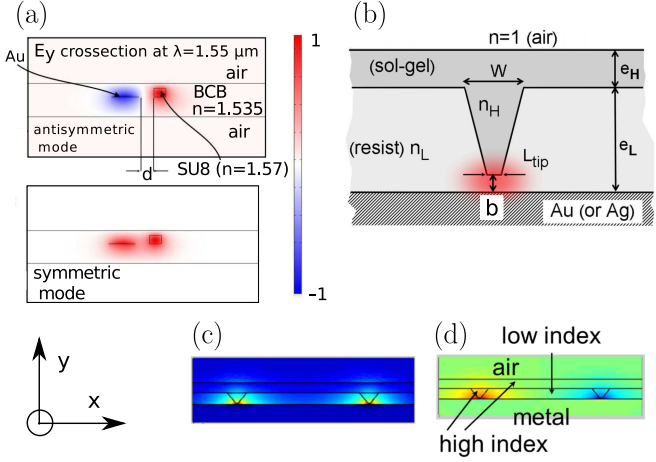}
    \caption{(Color online) (a) Antisymmetric and symmetric modes in a long-range plasmonic waveguide.
        The color map shows $y$-electric
        field components of the modes.  (b) Cross-section of a single hybrid waveguide.
        (c) and (d) Even and odd modes in coupled waveguides, respectively.
        Adapted from~\textcite{Benisty}.
    }
    \label{fig:plasmonic_waveguide}
\end{figure}

The model (\ref{opt:coupled_ode_gen}) appears to be suitable for description of plasmonic
waveguides. Such systems possess intrinsic Joule's loss due to
metallic components. On the other hand, they can be combined with
gain mechanisms achieved by plasmon amplification using stimulated
emission of radiation~\cite{BerSto03} or with dielectric active
media. These ideas were proposed by~\textcite{Benisty,LBD13}, who
suggested to  use   long range surface plasmon polariton waveguides
based on metallic layers.  Strongly confined guidance can be
achieved using schemes of hybrid dielectric-plasmonic waveguides
illustrated in Fig.~\ref{fig:plasmonic_waveguide}.

The first device [Fig.~\ref{fig:plasmonic_waveguide}(a)] is a long-range plasmonic waveguide working at the $1.55$ $\mu$m wavelength. The prototypical unit \cite{Degiron13}
has an Au lossy stripe $36\,$nm $\times$ $4.6\,\mu$m cross-section and SU8 stripe with gain of the   $1.5\,\mu$m$\times\,\,2\,\mu$m cross-section embedded in a transparent layer of benzocyclobutene-based polymer (BCB). The Au stripe and SU8 waveguide are separated by the distance $d=2.5\,{\mu}$m. The antisymmetric and symmetric modes, which are almost TM-polarized, are 
shown in the upper and lower panels, respectively.

The second hybrid model [Fig.~\ref{fig:plasmonic_waveguide}(b,c,d)]
consists of two waveguides, each representing a high-index ($n_H$)
sol-gel inverse rib optical waveguide linked  to a gold (or silver)
metallic plate through the low-refractive-index ($n_L$)
filling~\cite{BenBes}. The field (at $\lambda=633$~nm) in the
waveguide is concentrated at the tip end (the domain marked by red
color and the transverse width $b$). The figure shows calculations
performed  for the lossy structure. Gain can be introduced by pumping   the
high-index material in a limited region of the sole inverse rib  and is
expected to be on the order of 500~cm$^{-1}$. Alternately, the gain
can be provided by adding organic elements to the
structure~\cite{Benisty} such as optically pumped polymer with
dye~\cite{Noginov08}.

\textcolor{black}{ Multilayered plasmonic waveguides can be made of
    identical parallel metallic plates separated by dielectrics. To
    ensure $\PT$ symmetry, the dielectric layers should have alternating
    gain and loss (from the two sides of the metallic layer). Such a
    structure was theoretically studied by \textcite{AlDi2014a}, where
    the calculations were performed for Ag, as a lossless metal, whose
    dielectric permittivity is given by $\epsilon_{{\rm
            Ag}}=1-(\omega_p/\omega)^2$, and TiO$_2$ layers as dielectric slabs
    with $n=3.2\pm ik$, where $k$ is tunable gain or loss. A stack of
    five layers of length $150\,$nm and width 30$\,$nm was considered,
    operating at subwavelength frequencies of (transverse-magnetic) TM
    polarized plasmons. In the absence of gain and loss, the
    metamaterial exhibits negative index response resulting in negative
    diffraction. When $\PT$ symmetry is imposed, the authors numerically
    obtained several effects including double negative refraction,
    unidirectional invisibility, and reflection and transmission
    coefficients whose moduli simultaneously exceed unity. A detailed
    study of spectral characteristics of this stack was performed by
    \textcite{AlDi2014b}.}  A general analysis of  $\PT$ symmetry in subwavelength guiding
optical systems,
based on the full system of Maxwell's equations, rather than the paraxial
approximation, was given by \textcite{HYKMC14}. In particular, it was
found that,
on the subwavelength scale, the broken $\PT$ symmetry may be restored, while
the paraxial approximation misses this possibility.

 \subsection{Metamaterials and \textcolor{black}{Transformation Optics}}
 \label{sec:metadimer}

\begin{figure}
    \includegraphics[width=\columnwidth]{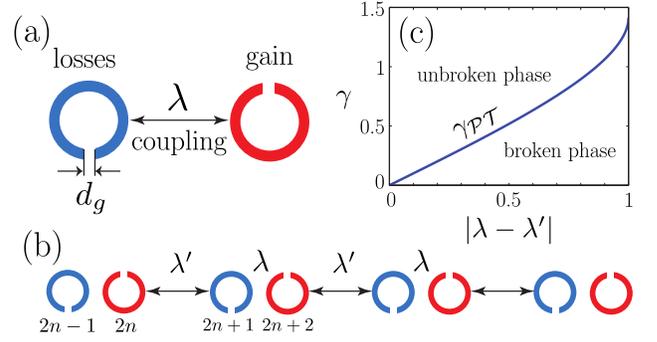}
    \caption{(Color online) (a) A $\PT$-symmetric dimer of SSRs;  (b) schematic illustration of a 1D $\PT$-metamaterial;
        (c) Domains of broken and unbroken $\PT$ symmetry  for the array shown in (b);  the phase transition line $\gamma_{\PT}$ is computed in Sec.~\ref{sec:1D_metamaterial}. Based on \textcite{LazTsir13}.}
    \label{fig:metamat}
 \end{figure}

 \paragraph*{One-dimensional $\PT$-symmetric metamaterial.} Now we turn to an idea  of \textcite{LazTsir13} on implementation of
 $\PT$ symmetry in metamaterials~\cite{Pendry2004}.
 The simplest building block for
 such systems, meta-atoms, is  a planar highly conductive split-ring
 resonator (SRR)~\cite{SarShal04}. A  typical SRR is characterized by
 losses. To compensate these losses, one can consider coupling of a
 lossy SRR with one incorporating gain elements
 [Fig.~\ref{fig:metamat}(a)].
 Assembling such SRR dimers in an array, one can build a $\PT$-symmetric metamaterial [Fig.~\ref{fig:metamat}(b)].

 Each SRR can be regarded as an RLC circuit with self-inductance $L$,
 Ohmic resistance $R$, and capacitance $C$. The nonlinearity can be
 introduced by a Kerr dielectric \cite{ZSK03} with $\varepsilon(|{\bf
    E}|^2)=\varepsilon_0\left(\varepsilon_{\ell}+\alpha |{\bf
    E}|^2/E_c^2\right)$, where $\varepsilon_0$ is the permittivity of
 the vacuum, $\varepsilon_{\ell}$ is the linear dielectric
 permittivity, $E_c$ is a characteristic electric field, and
 $\alpha=\pm 1$ is the sign of nonlinearity. The charge $Q_n$ of the
 $n$th SRR and capacitance $C=\varepsilon(|{\bf E}_g|^2) A/d_g$ [where
 $A$ is the area of SRR wire cross section and $d_g$ the size of the
 gap, see Fig.~\ref{fig:metamat}(a)] are related as
 $C(U_n)=dQ_n/dU_n$, where $U_n=d_gE_{gn}$ is the voltage across the
 SRR's gap~\cite{LET06}. In the weakly nonlinear limit, $U_n$ can be
 expressed by ${U_n}/{U_c}=q_n -\beta q_n^3+{\cal O}(q_n^5)$, where
 $q_n=Q_n/(C_{\ell} U_c)$, $C_{\ell}= \varepsilon_0\varepsilon_{\ell}
 A/d_g$ is the linear capacitance, $U_c=d_gE_c$, and
 $\beta={\alpha}/{(3\varepsilon_{\ell})}$ .

 The SRRs interact with each other through the near field due to magnetic dipole-dipole interactions
 [electric coupling in the geometry shown in Fig.~\ref{fig:metamat} can be neglected~\cite{HTZR}].  Then
 dynamical equations for the charges $Q_n$ and currents $I_n$  read  ($n=1,2$)
 \begin{equation*}
 \frac{dQ_n}{dt}=I_n,
 \quad
 L\frac{dI_n}{dt}-(-1)^{n}RI_n +U_n=M\frac{dI_{3-n}}{dt}+{\cal E}_n,
 \end{equation*}
 where $M$ is the mutual inductance of the  SRRs determining the strength of the coupling, ${\cal E}_n$ is the electromotive force induced in each SRR by the applied field, and $U_n$ is expressed through $q_n$ as indicated above.  
 It is also taken into account that the first SRR ($n=1$) has losses described by the resistance $R$, while the second SRR ($n=2$) has gain which has the same strength as losses, i.e., $-R$.

 Using the dimensionless variable $\tau=t/\sqrt{LC_{\ell}}$ and
 defining $\gamma=R\sqrt{C_{\ell}/L}>0$ and $\epsilon_n={\cal
    E}_n/U_c$, one arrives at the system
 \begin{eqnarray}
 \label{eq:meta_dimer}
 \begin{array}{l}
 \ddot{q}_{1}+\lambda\ddot{q}_{2}+ q_1+\gamma \dot{q}_1=\beta q_1^3 +\epsilon_1(\tau),
 \\
 \lambda\ddot{q}_{1}+\ddot{q}_{2}+ q_2-\gamma \dot{q}_2=\beta q_2^3 +\epsilon_2(\tau),
 \end{array}
 \end{eqnarray}
 where $\dot{q}_n=dq_n/d\tau$.

 In the linear limit ($\alpha=0$) and in the absence of the external
 driving [$\epsilon_n(\tau)\equiv 0$], eigenfrequencies $\Omega$
 ($q_{1,2}\propto e^{i\Omega \tau}$) of dimer (\ref{eq:meta_dimer})
 read
 \begin{equation}
 \label{eq:meta_eigen_dimer}
 \Omega_\pm^2=\frac {2-\gamma^2\pm \sqrt{4\lambda^2-4\gamma^2+ \gamma^4}}{2(1-\lambda^2)}.
 \end{equation}
 Considering  $\lambda^2<1$, which  corresponds to a typical physical
 setting and ensures stable dynamics in the conservative case (i.e.,
 at $\gamma=0$), one finds that unbroken $\PT$ symmetry (with real
 $\Omega_\pm$) corresponds to $0\leq\gamma\leq\gamma_\PT
 =\sqrt{2-2\sqrt{1-\lambda^2}}$, where $\gamma_{\PT}$ is the point of
 phase transition where the real eigen-frequencies $\Omega_+$ and
 $\Omega_-$ coalesce. At $\gamma>2$ the
 dimer is unstable for arbitrary coupling $\lambda$.

 \paragraph*{\textcolor{black}{Transformation Optics.}}
 Versatility of design of metamaterials allowed for development of a new area of the Transformation Optics. The idea \cite{Leonhardt06, Pendry2006}  consists  in designing a refractive index in a way to guide geometrical rays at will, in particular, avoiding a chosen domain thus making it invisible. The required refractive index can be constructed with help of an appropriate coordinate transformation $\br'={\bf F}(\br)$ to Cartesian coordinates $\br'$ where electric and magnetic fields $\{{\bf E}, {\bf H}\}$ are emitted by the given sources $\{{\bf J}, {\bf M}\}$.  Such coordinate transformation results in transformations of the electric and magnetic fields $\{{\bf E}, {\bf H}\}$, source current and magnetization $\{{\bf J}, {\bf M}\}$, and permittivity and permeability tensors  $\{\hat{\varepsilon},\hat{\mu}\}$, which can be obtained from the Maxwell equations \cite{Castaldi20013}:
 \begin{subequations}
    \begin{eqnarray}
    &&\{{\bf E}, {\bf H}\}(\br)=\Lambda^T(\br)\cdot \{{\bf E}', {\bf H}'\}(\br'),
    \\
    &&\{{\bf J}, {\bf M}\}(\br)=\det[\Lambda(\br)]\Lambda^{-1}(\br)\cdot \{{\bf E}, {\bf H}\}(\br'),
    \\
    &&\hat{\varepsilon}(\br)=\hat{\mu}(\br)=\det[\Lambda(\br)]\Lambda^{-1}(\br)\cdot[\Lambda^{-1}(\br)]^T,
    \end{eqnarray}
 \end{subequations}
 where $\Lambda=\partial(x',y',z')/\partial(x,y,z)$ is the Jacobian of the transformation. In practice the so designed device performs prescribed optical transformation of the rays in a flat real space $\br$ with inhomogeneous perimittivity and permeability to the homogeneous space $\br'$.

 \textcite{Castaldi20013} extended these ideas for designing $\PT$-symmetric metamaterials. In the vectorial problem  the requirement for $\PT$ symmetry can be reduced to  $\hat{\varepsilon}(\br)=\hat{\varepsilon}^*(-\br)$ [or $\hat{\mu}(\br)=\hat{\mu}^*(-\br)$], which can be fulfilled if the chosen transformation ensures $\Lambda(\br)=\Lambda^*(-\br)$. Thus the coordinate transformation must be complex. It turns out, however, that in the described procedure a continuous transformation $\br'={\bf F}(\br)$ leads to $\PT$-symmetric potentials having no spontaneous $\PT$-symmetry breaking. Potentials with exceptional points can be designed using suitable discontinuous transformations.

 As a simple but important example consider the transformation $x'=x$, $y'=y$ and $z'=ib(1\mp z/d)$ where Re$z\gtrless 0$, Im$z=0^+$, and $|z|\leq d$ \cite{Castaldi20013}. Then for the TM polarization the relevant nonzero components of the tensors $\hat\varepsilon$ and $\hat{\mu}$ are given by: $\varepsilon_{xx}=\mu_{xx}=\mp ib/d$, and $\varepsilon_{zz}=\pm id/b$. This transformation is particularly interesting for radiation emitted by a line-source $M_y^\prime=\delta(x')\delta(z'-i b)$ in Cartesian coordinates mapping it in the real-space source $M_y=\delta(x)\delta(z)$ in a medium with complex permittivity. The real axis $z'=0$ is transformed into the slab boundaries $z=\mp d$. The respective metamaterial slabs can be fabricated by periodic stacking of subwavelength layers of material constituents with opposite-signed permittivities and permeabilities. 

\begin{figure}
    \includegraphics[width=\columnwidth]{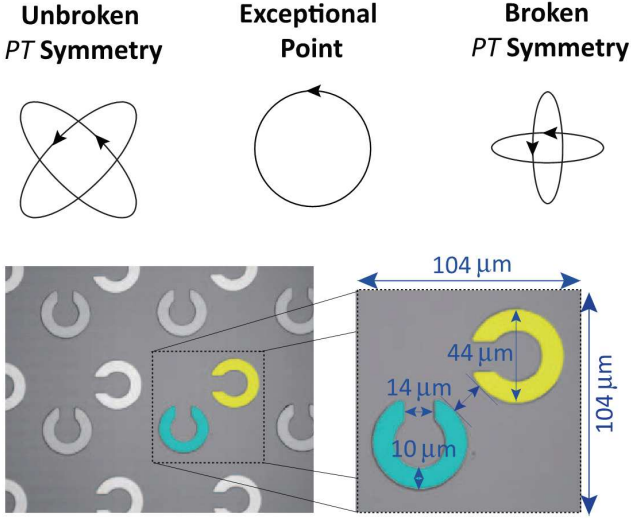}
    \caption{(Color online) Upper panels: Polarizations corresponding to unbroken and broken $\PT$-symmetric phases and to the exceptional point.  Lower panel: Photograph of $\PT$ symmetric metasurface on silicon substrate  composed of 300 nm thick silver (yellow in the color version or light gray in the grayscale version) and lead (turquoise in the color version or dark gray in the grayscale version) SRRs.  From \textcite{Lawrence20014}.}
    \label{fig:metasurf}
 \end{figure}

 \paragraph*{\textcolor{black}{$\PT$-symmetry breaking in polarization space.}}
 Turning now to 2D  metamaterials, we describe the direct observation of $\PT$-symmetry breaking by \textcite{Lawrence20014} using THz time domain spectroscopy of the metasurfaces as the one shown in Fig.~\ref{fig:metasurf}. The authors explored the dependence of the transmitted field polarization ${\bf E}=(E_x, E_y)^Te^{i\omega t}$ on the metasurface properties defined by the Lorentzian dipoles ${\bf p}=(p_x,p_y)^T$, $p_{x,y}\propto e^{i\omega t}$, oriented along perpendicular directions (corresponding to the geometry of the meta-molecules shown in Fig.~\ref{fig:metasurf}), resonating at the same frequency $\omega_0$, and characterized by the decay rates $\gamma_y<\gamma_x\ll \omega_0$. The link between the field and the dipoles is given by the polarizability matrix written as
 \begin{eqnarray}
 \label{eq:metasurface}
 S{\bf p}+G_{xy}\sigma_1{\bf p}-i\Gamma \sigma_3{\bf p}=g{\bf E}.
 \end{eqnarray}
Here $S=\delta+G_{xx}+i(\gamma_x+\gamma_y)/2$, $\Gamma=(\gamma_x-\gamma_y)/2$, the real coupling $G_{xy}$ is the summation of retarded fields from all $x$-oriented antennas acting on an $y$-oriented antenna, $G_{xx}$ is the summation of retarded coupling from all antennas oriented along the same direction,   $\delta=\omega-\omega_0$ is small detuning from the resonance ($\delta \ll 1$), and $g$ characterizes polarizability.

 The eigenstates of the polarizability matrix in the l.h.s. of (\ref{eq:metasurface}) are determined by the second matrix, which in its turn has a typical structure of the $\PT$-symmetric dimer (the first term scales out the net dissipation; the considered system is passive). In the unbroken ($2G_{xy}>\Gamma$) and broken ($2G_{xy}<\Gamma$) $\PT$-symmetric phases the field is elliptically polarized but has different orientation of the axes as illustrated in Fig.~\ref{fig:metasurf}, while at the exceptional point ($2G_{xy}=\Gamma$) the polarization is circular. These predictions were confirmed experimentally by~\textcite{Lawrence20014} on a number of metasurfaces, fabricated using photolithography, with the separation between the silver and lead SRRs in each unit cell varying from 2 to 20 $\mu$m.

\subsection{\textcolor{black}{Exciton-polariton condensates}}

Unlike atomic condensates existing at ultra-low temperatures, where gain and losses are usually avoided, and introducing $\PT$ symmetry requires special efforts (see Sec.~\ref{sec:BEC}, \ref{subsec:linear-CPT-BEC}), condensates of quasiparticles are obtained in the exited states (at relatively high temperatures) and must be supported by the pump since they are usually subject to appreciable losses. Thus, on the one hand, the balance between gain and losses is fundamental for supporting these condensates, and, on the other hand, they are intrinsically nonlinear systems due to interactions among quasiparticles. This readily suggests that condensates of quasiparticles are natural candidates for experimental implementation and exploration of  the nonlinear $\PT$-symmetric systems.

\begin{figure}
    \includegraphics[width=\columnwidth]{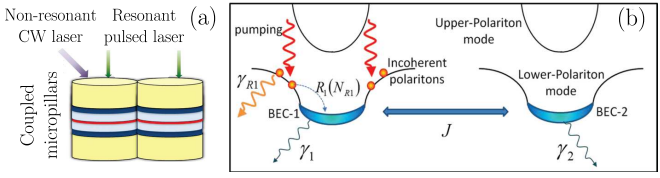}
    \caption{(Color online) (a) Semiconductor pillars pumped on the one side form a polariton Josephson junction.  (b) The model for the junction: a continuous--wave laser excites high--energy excitons pumping to the reservoir. A resonant laser is used to create the required initial conditions. Adapted from~\textcite{Lien2014}.
    }
    \label{fig:plasmonic}
\end{figure}

\textcite{Lien2014} suggested a setting for implementation of  $\PT$ symmetry with an exciton-polariton condensate.  The system consists of coupled micropillars as shown in Fig.~\ref{fig:plasmonic}(a,b). The junction is described by the   Gross-Pitaevskii equation for the vectorial wavefunction ${\boldsymbol \Psi}=\left(\Psi_1,\Psi_2\right)^T=\left(\sqrt{N_{1}}e^{i\varphi_1},\sqrt{N_2}e^{i\varphi_2}\right)^T$: $id{\boldsymbol \Psi}/dt=H{\boldsymbol \Psi}$, where
\begin{eqnarray}
\label{eq:polarit_2}
H\equiv\left(\begin{array}{cc}
E_1 & -J
\\
-J & E_2
\end{array}\right),\quad E_j=\epsilon_j+V_j+ U_j|\Psi_j|^2,
\end{eqnarray}
$j=1,2$, $\epsilon_j$ are the single particle ground states, $J$ characterizes tunneling between the two sides, $U_j$ is the strength of the nonlinear interactions of quasi-particles, and the local dispersion is ignored for the wave-functions of the ground state. The effective potentials in a simplified form are given by  $V_1=gN_{R}/A+{\cal G}P+\frac i2\left(RN_{R}-\gamma_1\right)$ and $V_2= -\frac i2\gamma_2$, where
$R$ is a constant, $\gamma_{1,2}$ are the decay rates of the condensates, ${\cal G}$ corresponds to the interaction of the condensate with high-energy excitons, the $g$ describes interaction between the condensate and reservoir polaritons, $P$ is the pump of the first reservoir, and $N_{R}$ is the population of the first reservoir whose dynamics is determined from the equation
\begin{eqnarray}
\label{eq:polarit_N}
\dot{N}_{R}=P-\gamma_{R1}N_{R}-RN_{R}|\psi_1|^2
\end{eqnarray}
with  $\gamma_{R}$ being the reservoir decay; the population of the second reservoir, which is not pumped, is neglected.

The model
(\ref{eq:polarit_2})-(\ref{eq:polarit_N}) admits stationary
solutions (i.e. $\boldsymbol{\Psi}\propto e^{-2i\Omega t}$ and
$N_{R}$ constant) with real frequencies
\begin{eqnarray}
\Omega_\pm= (E_1+E_2)\pm  \sqrt{(E_1+E_2)^2+4J^2},
\end{eqnarray}
if a nonzero dc Josephson current
$2J\sqrt{N_{1}N_{2}}\sin\left(\Delta\varphi\right)$, where
$\Delta\varphi=\varphi_2-\varphi_1$ is the relative phase, balances
pump from the one side and loss from the other side. If the pump
also compensates the total loss, $RN_{R}=\gamma_1+\gamma_2$, two
analytical solutions can be derived. They exist subject to the
condition $J^2\geq\gamma_2^2/4$, which determines the unbroken
$\PT$-symmetric phase, while $J^2=\gamma_2^2/4$ corresponds to the
exceptional point of the system.

\textcite{Chestnov2015}  showed that $\PT$ symmetry of the
coupled exciton-photon system, which can be implemented in a
specific regime of pumping the exciton state and depletion of the
reservoir,  enables permanent Rabi oscillations of the condensate in
the external magnetic field. For a particular case of the acoustic
phonon assisted pumping $RN_\alpha$, where $R$ is the pumping rate,
$\alpha=\pm$  corresponds to spin projection parallel (antiparallel)
to the external magnetic field, and $N$ is the reservoir population,
the system of Boltzmann kinetic equation governing the condensate,
which is supposed to be at zero momentum state, reads
\begin{subequations}
    \label{eq:condensate_2}
    \begin{eqnarray}
    &&2\dot{\chi} _\alpha= (RN_\alpha-\gamma_X)\chi_\alpha +2i\delta_\alpha\chi_\alpha-2i\Omega\phi_\alpha
    \\
    &&2\dot{\phi} _\alpha=-\gamma_P\phi_\alpha-2i\Omega\chi_\alpha
    \end{eqnarray}
\end{subequations}
where $\phi_\alpha$ and $\chi_\alpha$ are the photonic and excitonic components of the condensate wavefunction, $\gamma_X$ and $\gamma_P$,  are the decay rates of excitons and polaritons, $2\Omega$ is the Rabi frequency, $\delta_\pm= \omega_P-\omega_X-\Delta_Z$ is the effective detuning  of the cavity mode and exciton frequency in the presence of the Zeeman splitting $\hbar \Delta_Z$. The evolution of the reservoir populations is described by (\ref{eq:polarit_N}) with $N_R$ replaced by $N_\alpha$.

The system (\ref{eq:condensate_2}) reveals $\PT$ symmetry if the
photonic gain compensates total loss: $p_X=\gamma_X+\gamma_P$, and
the Zeeman splitting results in the zero effective detuning, i.e.
$\delta_\alpha=0$. Then, neglecting variation of the pump
$p_XN_\alpha$ on the time scale of the Rabi oscillations, for pump
below some threshold value $P<P_{th}$, one obtains two
eigenfrequencies of the condensate: $\pm \omega_0=\pm
\sqrt{4\Omega^2-\gamma_P^2/4}$ for the steady state solutions
$\chi,\phi\propto e^{\pm i\omega_0 t}$. On the other hand, at large
$P>P_{th}$ and provided the above  conditions for the $\PT$ symmetry
hold, one can obtain permanently oscillating regimes, where
$\chi(t)=\chi_1e^{i\omega_0 t}+\chi_2e^{-i\omega_0 t}$ and
$\phi(t)=\phi_1e^{i\omega_0 t}+\phi_2e^{-i\omega_0 t}$.

Finally, we mention that the non-Hermitian nature of exciton-polariton
condensates was explored experimentally by \textcite{Ostrovskaya}. A
chaotic exciton-polariton billiard was created, which exhibits
multiple exceptional points, crossing and anti-crossing of energy
levels, mode switching, and topological Berry phase, subject to
proper changes of system parameters.

\subsection{Bose-Einstein condensates}
\label{sec:BEC}

Bose-Einstein condensates (BECs) represent  another promising area
for theoretical and experimental study of interplay between
nonlinearity and   phenomena originated by $\PT$ symmetry.
{\textcite{KGM08}} suggested to consider a BEC in a double-well
potential with atoms injected into one well and removed from another
well. Removal of the atoms can be achieved in different ways: by
applying laser radiation or an electronic beam to ionize
atoms~\cite{GeWuReLaOt,BaLott}, using inelastic interactions of
atoms with the trap potential~\cite{Muga04,Cannata07}, or by
stimulating transitions to higher levels with subsequent removal of
the excited atoms from the trap. Loading of atoms ensuring exact
compensation of the losses  can be implemented  by an atomic
laser~\cite{laser} or by some more sophisticated technique such as
combination of tilted potential wells~\cite{Kreibich}. Nonlinearity
in BECs stems from two-body interactions and in the mean-field
approximation  results in the cubic term in the Gross-Pitaevskii
equation (GPE)~\cite{PitStrin03}. To describe the removal or loading
of atoms within the framework of the mean-field model, one can start
with the Master Equation in the Lindblad
form~\cite{SKon10,BaLott,WiTriHe11} which shows excellent agreement
with the available experimental data~\cite{BaLott}. In 3D, GPE with
loading and removal of atoms reads
\begin{eqnarray}
\label{BEC_3D}
i\Psi_t=-\Delta\Psi+[V(\br)+iW(\br)]\Psi-g|\Psi|^2\Psi,
\end{eqnarray}
where $g\sim -a_s$  characterizes the strength of two-body
interactions ($g>0$  and $g<0$ correspond to negative and positive scattering
length $a_s$), $V(\br)$ and $W(\br)$ are the real and
imaginary parts of the external potential, and the dimensionless
units ($\hbar=2m=1$) are used. Model (\ref{BEC_3D}) with a
$\PT$-symmetric double-well trap
\begin{equation}
\label{eq:double_well_3D}
V(\br)=\omega_x^2x^2+y^2+z^2+v_0e^{-\sigma x^2}\!, \,\,  W(\br)=\gamma xe^{-\rho x^2}
\end{equation}
with $\sigma=2\rho\ln(v_0\sigma/\omega_x^2)$ (all parameters are positive) was investigated
by~\textcite{Dast,Dast_b}. \textcite{cartar} considered a 1D model
with a double well potential composed of two Dirac $\delta$-functions.

An alternative description of $\PT$-symmetric BEC models can be
developed from a non-Hermitian Bose-Hubbard model, where the
gain-loss coefficient $\gamma$ is introduced
explicitly~\cite{GraKoNied2008,Graefe,GraefeLiverani}:
\begin{equation}
H=i\gamma(a_2^\dag a_2-a_1^\dag a_1) +v(a_1^\dag a_2+a_2^\dag a_1)
+c(a_1^\dag a_1-a_2^\dag a_2)^2.
\end{equation}
Here $a_j$ and $a_j^\dag$ are the bosonic annihilation and creation
operators for the $j$-th mode, $v$ is the tunneling rate,
and $c$ is the strength of two-body interactions.

\subsection{Spin-orbit coupled Bose-Einstein condensates}
\label{subsec:linear-CPT-BEC}

Now we consider a BEC of two states, $|\ua\rangle$ and $|\da\rangle$, belonging to
the ground manifold and coupled with an excited state by laser
beams, like in the $\Lambda$-schemes shown in
Fig.~\ref{fig:lambda_atom}(a). These can be hyperfine states $|F=1,
m_F=0\rangle$ and $|F=1, m_F=+1\rangle$   of $^{87}$Rb
atoms~\cite{LJ-GS11} or degenerate dark states~\cite{DGJO11}. Such a
system can emulate the phenomenon of spin-orbit (SO) coupling in
condensed-matter physics~\cite{SAG08,GaSpi13} and gives rise to a
SO-coupled BEC, which was produced experimentally
by~\textcite{LJ-GS11}.

Let atoms in the $|\da\rangle$ ( $|\ua\rangle$) state are removed from (loaded into) the system with the rate $\gamma>0$. In absence of two-body interactions the system Hamiltonian reads~\cite{KaKoZe14}
\begin{eqnarray}
\label{eq:SO_BEC_Hamilt}
H=-\frac{1}{2}\frac{\partial^2}{\partial x^2}  +\omega\sigma_1+
i\kappa\sigma_1\frac{\partial}{\partial x} +i\gamma\sigma_3
+V(x).
\end{eqnarray}
Here $\kappa$ is the strength of SO-coupling, $V(x)$ is the external
trap potential, $\omega$ is the strength of linear coupling from the
Zeeman field, and we have added the injection and removal of atoms
to the standard model~\cite{LJ-GS11}

According to definitions (\ref{p_reflect}) and (\ref{t_Wigner}), the
Hamiltomian (\ref{eq:SO_BEC_Hamilt}) acting in the Hilbert space of
the  vectors $\bPsi=(\Psi_{\ua},\Psi_{\da})^T$ is not $\PT$
symmetric. However, if we define a charge (or pseudo-spin) operator
$\C$:  $\C\bPsi(x,t)=\sigma_1\bPsi(x,t)$, then $[\CPT, H]=0$, i.e.,
$H$ is $\CPT$ symmetric [compare with (\ref{PT_defin})]. Interpreting the states
$|\ua\rangle$ and $|\da\rangle$ as having negative  and positive
energies  with respect to the average chemical potential $\mu$, one
finds that $\C$ indeed obeys properties of the charge operator: it
exchanges the states with positive and negative energies, $\C^2=1$,
$[\C,\PT]=0$; it has eigenvalues $\pm 1$ and changes the direction
of currents [see Fig.~\ref{fig:SO_PT_nonlin} and discussion in
Sec.~\ref{sec:CPT_BEC}].

In Sec.~\ref{subsec:optics}, $\sigma_1$ was interpreted as the
parity operator since it was obtained from $\p$ defined by
(\ref{p_reflect}). This reflects ambiguity in definitions of
symmetry operators. In particular, one can consider $H$ as
$\PT$-symmetric in a more general sense, where  $\tilde{\p}=\C\p$ is
a new parity operator. We also emphasize that the $\C$ operator used
here should not be confused with the $\C$ operator introduced
by~\textcite{BBJ2002} as an observable representing the signature of
the $\PT$-norm which allows the definition of an inner product
having positive-definite signatures. Recently $\C$-operator was discussed also in optical context~\cite{DaBaMa15}.

In the model (\ref{eq:SO_BEC_Hamilt}), $\PT$ symmetry (or $\CPT$ symmetry) is
determined by two parameters: $\omega/\gamma$ and
$\kappa/(\gamma\ell)$, where $\ell$ is the characteristic scale of
the wavefunction: $|\partial \Psi_{\ua,\da}/\partial
x|\sim|\Psi_{\ua,\da}|/\ell$. Thus a sufficiently broad mode
($\ell\gg\kappa/\gamma$) has effectively SO-decoupled components and
cannot balance gain and loss.  Since, however, $V(x)$ limits the size of the SO-BEC (i.e., makes $\ell$ bounded), the unbroken symmetry may exist even when the homogeneous condensate is unstable, i.e. an external potential can
control $\PT$-symmetry breaking.
\begin{figure}
    \includegraphics[width=\columnwidth]{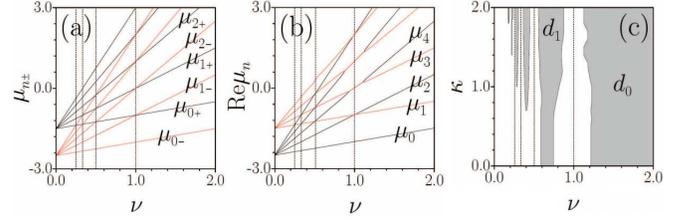}%
    \caption{ (Color online)
        (a) Spectrum (\ref{eq:CPT:gamma=0}) for $\omega=0.5$ and $\kappa=2$.
        Crossing eigenvalues occur at $\nu_m$, $m=1,2,\ldots$. (b) Real parts of eigenvalues
        for $\gamma=0.2$, $\omega=0.5$ and $\kappa=2$, which merge pairwise near $\nu_m$, and
        the corresponding imaginary parts become nonzero at the same time.
        (c) Domains of unbroken (shadowed regions $d_{0}$, $d_{1}$, ...) and broken (white regions) $\CPT$ symmetries
        shown in the plane $(\kappa,\nu)$ at $\gamma=0.2$ and $\omega=0.5$. Dashed lines correspond to $\nu=\nu_m$. From \textcite{KaKoZe14}.
    }
    \label{fig:SO_PT_lin}
\end{figure}

Properties of  Hamiltonian (\ref{eq:SO_BEC_Hamilt}) with a parabolic trap $V(x)=\nu^2 x^2/2$ are illustrated in Fig.~\ref{fig:SO_PT_lin}. At $\gamma=0$ the model is Hermitian and its the spectrum is real: 
\begin{eqnarray}
\label{eq:CPT:gamma=0} {\mu}_{n, \pm}= \nu(n+1/2) \pm \omega - \kappa^2/2, \quad n=0,1, \ldots
\end{eqnarray}
At $ \nu=\nu_m$, where $ \nu_m={2 {\omega}}/{m} $ with $
m=1,2,\ldots $, the spectrum  contains an {\em infinite} number of
double eigenvalues: $\mu_{n-m,+}=\mu_{n,-}$, $n=m, m+1,
\ldots$ [Fig.~\ref{fig:SO_PT_lin}(a)]. When $\gamma$ is nonzero,
these double eigenvalues become complex leading to broken $\PT$
symmetry [Fig.~\ref{fig:SO_PT_lin}(b)]. For   fixed $\kappa$
$\PT$ symmetry is always broken in a sufficiently broad trap (small
$\nu$)  [Fig.~\ref{fig:SO_PT_lin}(c)]. As $\nu$ grows, the domains
of unbroken symmetry (shadowed domains $d_0$, $d_1$, \ldots) appear
and alternate with the domains of broken symmetry. In a sufficiently
narrow trap (large $\nu$) the $\CPT$ symmetry is unbroken for all
$\kappa$ (domain $d_0$). This stripe-like structure  is related to
the presence of an infinite number of double eigenvalues in the
spectrum (\ref{eq:CPT:gamma=0}). In Fig.~\ref{fig:SO_PT_lin},
values of $\nu$ that correspond to the double eigenvalues are shown
by dashed vertical lines, which always belong to the region of broken $\CPT$ symmetry 
and  ``separate'' domains $d_{m-1}$ and $d_m$ of unbroken symmetry.

 \subsection{Superconductivity}
 \label{subsec:superconduct}

 The Bender--Boettcher potential (\ref{BB_potential}) with $N=1$
 turns out to be relevant for the pattern formation of phase slip
 centers in superconducting wires~\cite{RuSteMa} and in the theory of
 fluctuation superconductivity~\cite{CGBV12}.

 Considering a superconducting wire in the interval $[-L,L]$ along
 the $x$-axis, one can start with the time-dependent complex
 Ginzburg-Landau equation~\cite{GL50} for the order parameter
 {$\Psi$}. In the dimensionless form, the model reads
 \begin{equation}
 \label{eq:CGL}
 \Psi_t + H\Psi+|\Psi|^2\Psi=0,\quad H=-{\partial^2}/{\partial x^2}+ i\varphi-\Gamma.
 \end{equation}
 Here $\varphi$ is the electric potential,
 $1/\Gamma \propto (1-T/T_c)^{-1}$ is the characteristic relaxation
 time of the order parameter, $T$ is the temperature, and $T_c$ is
 the transition temperature. The order parameter is subject to the
 zero boundary conditions $\Psi(\pm L)=0$ (this choice is not
 crucial). Potential $\varphi$ induces constant current $I$, i.e.,
 $\varphi= -I x$ (the Ohmic resistivity is set to one).

 The normal state corresponds to $\Psi=0$, and thus the phase
 transition can be characterized by solutions of the linear version
 of Eq.~(\ref{eq:CGL}). Setting $L=1$ and
 $\Psi(x,t)=u(x)e^{(\Gamma-\lambda)t}$ one obtains the eigenvalue
 problem~\cite{RuSteMa}:
 \begin{equation}
 \label{eq:superc_eig}
 u_{xx}+ixIu=-\lambda u, \quad u(\pm 1)=0,
 \end{equation}
 which involves the $\PT$-symmetric potential
 (\ref{BB_potential}) with $N=1$.
 The normal state is unstable if  $\Gamma < \textrm{Re\ }[\lambda(I)]$.
 At $I=0$, the spectrum  is real and discrete: $\lambda_n=\pi^2n^2/4$, $n=1,2,\ldots$. The spectrum remains real for sufficiently small $I$,
 {which  is guaranteed by Theorem~\ref{theor:GaGraS}}.
 Increase of $I$ eventually results in collision of eigenvalues (with the first collision occurring at $I_{cr}\approx 12.31$),
 followed by emergence of complex eigenvalues. 

 The complex spectrum at $I>I_{cr}$ implies breaking of the Cooper
 pairs. On the other hand, it leads to the energy difference in the
 two lowest states resulting in Josephson oscillations between them
 and consequently in the symmetry breaking of the time averaged order
 parameter. These phenomena were described and experimentally
 validated by~\textcite{CGBV12}.

\subsection{Magnetics}
\label{se:magnet}

\textcite{Gaididei} suggested a way to implement a $\PT$-symmetric
configuration in a double-wire magnetic structure in which a
spin-polarized current $j$ propagates along the $z$-axis in positive and negative directions
in the first and second wires, respectively.
The spin-transfer torque efficiency function has the
form
$\varepsilon_{n,\alpha}={\eta\Lambda^2}/{\left[(\Lambda^2+1)-(-1)^\alpha
    (1-\Lambda^2)S_{n,\alpha}^{z}\right]}$, where $S_{n,\alpha}^{\nu}$
with $\nu=x,y,z$ is the $\nu$th component of the spin vector
$\mathbf{S}_{n,\alpha}$ of the $n$th site at the $\alpha$th wire,
$\eta$ is the degree of spin polarization,  $\alpha=1,2$, and the parameter
$\Lambda\geqslant1$ describes the mismatch between spacer and
ferromagnet resistance \cite{slon,sluk}.

The magnetic energy of the system amounts to
$
{\cal E}={\cal E}_1+{\cal E}_2+{\cal E}_{12},
$
where
\begin{eqnarray}
\label{energy}
{\cal E}_{\alpha}
=-J_1\,\sum_{n}\mathbf{S}_{n,\alpha}\,\mathbf{S}_{n+1,\alpha}-
\frac{1}{2}\,A\,\sum_{n}\Big(S^z_{n,\alpha}\Big)^2
\end{eqnarray}
is the magnetic energy of the $\alpha$th wire   with the last term being the easy axis anisotropy (characterized by the constant $A$),
$
{\cal E}_{12}=-J_2\,\sum_{n}\mathbf{S}_{n,1}\,\mathbf{S}_{n,2}
$
represents an interaction between the wires, $S_{n, \alpha}^{(z)}$
is the $z$-component of the spin $\mathbf{S}_{n,\alpha}$ in the
$\alpha$-th wire, and $J_{1,2}$ are the respective
exchange energies. The dynamics of the system is described by the
Landau-Lifshitz equation augmented with spin-torque terms:
\begin{equation}
\label{eq:LL}
\frac{d\mathbf{S}_{n,\alpha}}{dt}= \mathbf{S}_{n,\alpha}\times\frac{\delta \cal{E}}{\delta \mathbf{S}_{n,\alpha}}
+
(-1)^{\alpha}
j\varepsilon_{n,\alpha}\mathbf{S}_{n,\alpha}
\times \Big(\mathbf{S}_{n,\alpha}\times \hat{\mathbf{z}}\Big),
\end{equation}
where $\hat{\mathbf{z}}$ is a unitary vector along the
$z$-direction. The last term in Eq.~\eqref{eq:LL} represents
spin-torques which are due to interaction with a spin-polarized
current $j$.

Let us consider weak deviations of spins from the ferromagnetic stationary state $\mathbf{S}_{n,\alpha}=(0,0, 1)$. To this end we introduce complex amplitudes $\psi_{n,\alpha}$ defined by
\begin{eqnarray*}
    \label{HP}S^{x}_{n,\alpha }&=& \Big(\psi_{n,\alpha}+ \psi^*_{n,\alpha}\Big)\,\sqrt{1- \mid\psi_{n,\alpha}\mid^2}\;,\nonumber\\
    S^y_{n,\alpha}&=& -i\,\Big(\psi_{n,\alpha}-\psi^*_{n,\alpha}\Big)\,\sqrt{1- \mid\psi_{n,\alpha}\mid ^2}\;,\nonumber\\
    S^z_{n,\alpha}&=&1-2\mid\psi_{n,\alpha}\mid^2,
\end{eqnarray*}
and consider the small amplitude (linear) limit
$|\psi_{n,\alpha}|\ll 1$. Using the Fourier transform
$\psi_{n,j}=N^{-\frac{1}{2}}\sum\limits_k\,e^{i k n+i\omega_k
    t}\,\tpsi_{k,j}$, where $N$ is the number of spins in the chain and
$\omega_k=(A+J_2)+4\,J_1\,\sin^2(k/2)$ is the dispersion relation of
linear spin waves, one obtains a coupled $\PT$-symmetric system
(\ref{opt:coupled_ode}) with $q=(\tpsi_{k,1},\tpsi_{k,2})^T$,
$\gamma=j\eta/2$, and $\kappa_{12}=\kappa_{21}=J_2$.

The model can be generalized to the anti-ferromagnetic
case $\mathbf{S}_{n,\alpha}=(0,0, (-1)^n)$, as well as to the
continuum limit.

Another model of two coupled ferromagnetic films, one with gain and
the other with loss placed in an external magnetic field, was
introduced by \textcite{LeKoSha14}.

\subsection{Electronic circuits}
\label{subsec:electric}

\begin{figure}
    \includegraphics[width=\columnwidth]{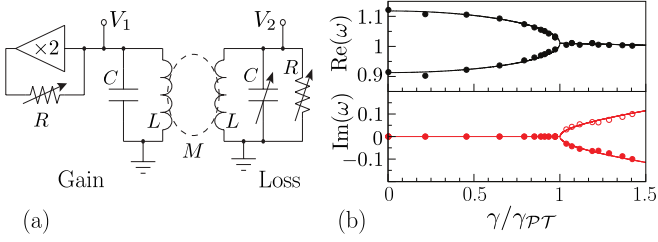}%
    \caption{(Color online)  (a) Electronic $\PT$-symmetric dimer. The left circuit contains the gain element due to feedback from a voltage-doubling buffer.
        (b) Theoretical (solid line) and experimental (circles) values of the eigenfrequency $\omega$ [open circles are reflections of the data with respect to the Im$(\omega) = 0$
        axis]. Here $\omega_0=2\times 10^5$ s$^{-1}$ and $\mu=0.2$. Adapted from~\textcite{Schindler}.  }
    \label{fig:RLC_cirquit}
\end{figure}

The concept of a passive $\PT$-symmetric system (see Sec.~\ref{subsec:optics}) relies on the existence of
an exceptional point which separates qualitatively different dynamical regimes. 
This phenomenon is well known for simplest mechanical systems.
Indeed, for a damped oscillator  $\ddot{x} +2\gamma \dot{x}+x=0$,
the exceptional point occurs at $\gamma=1$, and it separates
underdamped ($\gamma<1$) and overdamped ($\gamma>1$)
oscillations.
A $\PT$-symmetric
generalization of such a system corresponds to two coupled
oscillators with damping and gain:
\begin{eqnarray}
\label{mech:two_oscil}
\ddot{x} +2\gamma \dot{x}+x=-2\kappa y,
\quad
\ddot{y} -2\gamma \dot{y}+y=-2\kappa x.
\end{eqnarray}
This idea was suggested and experimentally implemented using the RLC
circuits by \textcite{Schindler} [Fig.~\ref{fig:RLC_cirquit}]. A
mechanical realization of a   dimer of oscillators was reported
by~\textcite{BBPS}.

The scheme in Fig.~\ref{fig:RLC_cirquit}(a)
obeys Kirchhoff's laws
\begin{eqnarray}
\label{eq:Kirchhoff}
\begin{array}{l}
I_n^{C}+I_n^{R}+I_n^{L}=0, \quad I_n^R=(-1)^n\Gamma\omega_0 Q_n^C,
\\[1mm]
\omega_0^2Q_1^C=\dot{I}_1^L + \mu   \dot{I}_2^L,    \quad \omega_0^2Q_2^C=\dot{I}_2^L + \mu   \dot{I}_1^L,
\end{array}
\end{eqnarray}
where $I$ are $Q$ are respectively  currents and charges in the
amplified ($n=1$) and lossy ($n=2$) circuits with capacitor
(``$C$''), resistor (``$R$'') and inductor (``$I$'');
$\omega_0=1/\sqrt{LC}$ is the natural frequency of each isolated
coil,
$\mu=M/L$ characterizes inductive coupling of the coils, and $\Gamma=\sqrt{L/C}/R$ is the effective gain-loss parameter. An overdot in (\ref{eq:Kirchhoff}) stands for the derivative with respect to time $t$. 

Model (\ref{mech:two_oscil}) follows directly from
(\ref{eq:Kirchhoff}) with $x=Q_2^C$, $y=Q_1^C$,
$2\gamma=\Gamma\sqrt{1-{\kappa^2}}$, $\kappa=-\mu$, and the overdot for derivative
with respect to $\tau=\omega_0 t/\sqrt{1-{\kappa^2}}$.
Eigenfrequencies $\omega$ of Eq. (\ref{mech:two_oscil}), with
$x,\,y\propto e^{i\omega t}$, are given by
\begin{eqnarray}
\label{eq:PT_phase_lin_oscil}
\omega_{\pm}^2=1-2\gamma^2\pm2\sqrt{\kappa^2-\gamma^2+\gamma^4}.
\end{eqnarray}
Thus $\PT$ symmetry is unbroken
if $0\leq \gamma\leq \gamma_\PT$, where
$\gamma_\PT=\sqrt{{1}/{2}-\sqrt{{1}/{4}-\kappa^2}} $
is the phase transition point.
Unlike the case of a Schr\"odinger-type $\PT$-symmetric dimer,
where increase in coupling favors unbroken symmetry [see
(\ref{eq:H_matrix_E})], here increase of $|\kappa|$ eventually breaks the
$\PT$ symmetry. Figure~\ref{fig:RLC_cirquit}(b) presents the
comparison of theoretical and experimental results of
\textcite{Schindler}. For further theoretical studies of the model see~\textcite{Bender13b,NBRMCK13}.

System (\ref{mech:two_oscil}) admits a Hamiltonian formulation with
the Hamiltonian~\cite{Bender13b}
\begin{equation}
\label{Hamilt_mecahnical}
H=pq+\gamma(yq-xp)+(1-\gamma^2)xy+\kappa(x^2+y^2).
\end{equation}
Time reversal and parity operators can be defined as~\cite{Bender13b}
\begin{eqnarray*}
    \label{PT_oscillators}
    \begin{array}{l}
        \T:\quad x\to x,\,\,y\to y,\,\,p \to-p,\,\,q\to-q,
        \\
        \p:\quad x\to-y,\,\,y\to -x,\,\,p\to-q,\,\,q\to-p.
    \end{array}
\end{eqnarray*}
Connection between the velocities and momenta is given by
\begin{equation}
\label{velocity_momenta}
\dot{x}={\partial H}/{\partial p}=q-\gamma x,\quad \dot{y}={\partial H}/{\partial q}=p+\gamma y,
\end{equation}
while the second pair of Hamilton equations, $\dot p=-\partial H/\partial x$ and $\dot q=-\partial H/\partial y$, yields Eqs.~(\ref{mech:two_oscil}).

Nonlinearity can be included in coupled circuits by taking into
account the internal currents in the electric-circuit dimer. The
resulting system is modeled by coupled Van der Pol oscillators and
features asymmetric transport observed by \textcite{Kottos2013}.

\subsection{Micro-cavities}
\label{subsec:mechan}

\begin{figure}
    \includegraphics[width=\columnwidth]{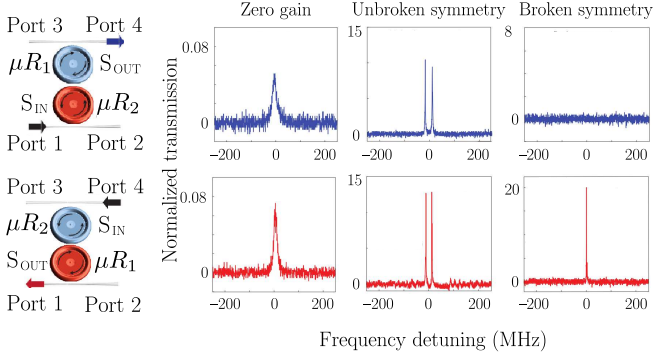}%
    \caption{(Color online) Left column: the system composed of two whispering-gallery-mode resonators
        coupled with each other and with two fiber-taper waveguides.
        The resonators are silica toroids of approximately $30\,\mu$m radius.
        One microcavity ($\mu R_1$) is active due to  E$^{3+}$ dopants while the other ($\mu R_2$) is passive.
        The ions in the active microcavity are pumped by the laser in 1.417 $\mu$m band providing gain in 1.55 $\mu$m band.
        The second, third and fourth columns show different operating regimes of the coupler.
        The upper and lower lines correspond to left and right incidence of light.
        The fourth column illustrates non-reciprocity of light propagation. 
        Reprinted by permission from Macmillan Publishers Ltd: [Nature Physics] (Peng, B., \c{S}. K. \"Ozdemir, F. Lei, F. Monifi, M. Gianfreda, G. L. Long, S. Fan, F. Nori, C. M. Bender, and L. Yang, 2014,   Parity-time-symmetric whispering-gallery microcavities.
   Nat. Phys. {\bf 10}, 394--398), copyright 2014.
        }
    \label{fig:coupled-microcavities}
\end{figure}

Coupled $\PT$-symmetric oscillators (\ref{PT_oscillators}) is a
fairly general model. In particular, it is possible to design an
optical scheme assuring asymmetric transport in analogy with the
coupled RLC-circuits considered above~\cite{Kottos2013}.  {Such a
    scheme extends the idea of all-silicon passive optical diode with
    two passive micro-cavities connected by an optical
    waveguide~\cite{FaWaV12}, to a configuration with mutually balanced
    active and passive cavities.} Experimental implementations of
$\PT$-symmetric micro-cavities were reported by
\textcite{Peng2014} and \textcite{CJHYWJLWX}. The prototypical
experimental setup is illustrated in
Fig.~\ref{fig:coupled-microcavities}.

\textcite{NBRMCK13} suggested that
the observed non-reciprocity may reside in nonlinear Fano resonances which can be  captured by a model of a linear conservative array interacting with two nonlinear sites with gain (``$g$'', placed at $n=0$)
and loss (``$l$'', placed at $n=N$):
\begin{equation*}
\begin{array}{lcr}
i\dot{\phi}_n&=&-C(\phi_{n-1}+\phi_{n+1})-V_g\phi_g\delta_{n,0}-V_l\phi_l\delta_{n,N},
\\
i\dot{\phi}_{g/l}&=&-(E\mp i\gamma) \phi_{g/l}-\chi|\phi_{g/l}|^2\phi_{g/l}-V_{g/l}\phi_{0/N}.
\end{array}
\end{equation*}
Here $\phi$ are field amplitudes, $C$ is the coupling constant between the neighboring sites in the linear chain,
$E\mp i\gamma$ models eigenmodes of the two micro-resonators, and $V_{g,l}$ are the coupling coefficients between the chain and the impurities.
The nonlinear system interacts with the linear one at sites $\phi_0$ and $\phi_N$.
The left incidence of a monochromatic wave, with dispersion relation $\omega(q)=2C\cos(q)$ ($\phi_{n,g,l}=A_{n,g,l} e^{i\omega t}$),
is characterized by the reflection and transmission coefficients 
\begin{equation*}
r_L=i\frac{V_gA_g+V_lA_le^{iqN}}{2C\sin q}, \,\, t_L= I +i\frac{V_gA_g+V_lA_le^{-iqN}}{2C\sin q},
\end{equation*}
where $I$ is the input amplitude, and amplitudes $A_{g,l}$ are found
from the system
\begin{eqnarray}
\label{eq:amp_Fano}
\begin{array}{l}
(E-\omega-i\gamma)A_g+\chi|A_g|^2A_g+V_g(I+r_L)=0,
\\
(E-\omega+i\gamma)A_l+\chi|A_l|^2A_l+V_le^{iqN}t_L=0.
\end{array}
\end{eqnarray}
For the right incident wave, $r_R$ and $t_R$ are obtained from the above  expressions 
by the substitution $\gamma\to-\gamma$ and $V_g\leftrightarrow V_l$.

The  coupled active and passive micro-cavities can also be studied
in optomechanics, where they have promising applications such as
phonon lasers~\cite{Nori14}. Micro-cavities driven by blue- and
red-detuned laser fields with mechanically connected movable mirrors
were studied in~\textcite{Xu14}.

 \subsection{Acoustics}
 \label{subsec:acoustics}

 \textcite{ZRSZZ14} extended the idea of $\PT$ symmetry to propagation of sound waves. A linear acoustic wave characterized by the pressure $P(z,t)$
 in a bulk medium with $z$-dependent density $\rho(z)$  and bulk modulus $K(z)$ is governed by the wave equation
 $ KP_{zz}-\rho P_{tt}=0. $ In a general situation, dissipation of
 sound waves can be described by a complex bulk modulus with negative
 imaginary part. In the meantime, by including piezoelectric elements
 connected with electric circuits~\cite{PopaCum14}, it is possible to
 implement gain elements described by the positive imaginary part of
 $K$. Thus, by combining the elements with gain and loss to ensure
 properties $K(z)=K^*(-z)$ and $\rho(z)=\rho^*(-z)$, for a
 monochromatic wave $P\propto e^{i\omega t}$ one obtains the linear
 Helmholtz equation
 \begin{equation}
 \label{eq:acoust_Helmholtz}
 \frac{d^2P}{dz^2}+\omega^2U(z)P=0,\quad  U(z)=\frac{\rho(z)}{K(z)}=U^*(-z).
 \end{equation}
 For a particular waveguide configuration of three gain and three
 loss sections separated by five passive   blocks,  such a medium can
 become unidirectionally transparent for some frequencies
 \cite{ZRSZZ14}.

 Nonlinear phenomena in the sound wave propagation can be accounted for by considering the complete strain tensor, which remains an open problem.

\section{$\PT$-symmetric discrete lattices}
\label{sec:3}

In this section, we present detailed analysis of $\PT$-symmetric
lattices. Motivated by physical applications in Sec.~\ref{sec:2},
our main attention will be on discrete nonlinear Schr\"odinger
(dNLS) -- type equations, although other nonlinear lattice models
are also touched upon.

\subsection{Formalism for discrete nonlinear systems}

We focus on the most studied nonlinear networks where a number of waveguides is either
even or infinite. For some details on arrays with an odd number of
waveguides see~\textcite{LKFRK,LiKev,LKMG}.

Mathematical description of the network is based on a  system
of ordinary differential equations
\begin{eqnarray}
\label{eq:discr:dyn_main}
i\dot \bq  =  H \bq + F(\bq)\bq,
\end{eqnarray}
where $\bq=\bq(z)=(q_{-N+1},...,q_N)^T$ is a column-vector of $2N$
elements, $H$ is a $2N\times 2N$ symmetric matrix accounting for the
linear coupling between sites, and $F(\bq)$ is a $2N\times 2N$
matrix whose elements depend on the field $\bq$.
In this section, we consider cubic nonlinearity where entries
of matrix $F(\bq)$ are given by
\begin{equation}
\label{eq:discr:F}
[F(\bq)]_{pj}  =
\sum_{l,m=1-N}^N
{f}_{pj}^{lm}q_l^*  q_m,  \quad p,j = -N+1, \ldots, N,
\end{equation}
and $f_{pj}^{lm}$ are constant coefficients.

$\PT$ symmetry of $H$ implies the following property: if $\bq(z)$ is
a solution of the underlying linear system  (i.e., the system with
$F(\bq)\equiv 0$), then $\PT \bq(z) = \p\bq^*(-z)$ is also a
solution. This is not true for the nonlinear case in general. If
however the nonlinearity obeys the constraint
\begin{equation}
\label{eq:discr:PTreqnonlin}
\p  F^*(\bq)  = F(\p \bq^*) \p
\mbox{\quad for all \ } \bq,
\end{equation}
then the mentioned property holds also in the nonlinear case
\cite{KevrSIAM,LKMG}, and we say that the nonlinear system
(\ref{eq:discr:dyn_main}) is $\PT$ symmetric.

\subsection{Discrete   configurations and their linear properties}
\label{sec:discrete}

\subsubsection{Arrays with nearest-neighbor linear interactions}
\label{subsec:open_closed}

\begin{figure}
    \includegraphics[width=1\columnwidth]{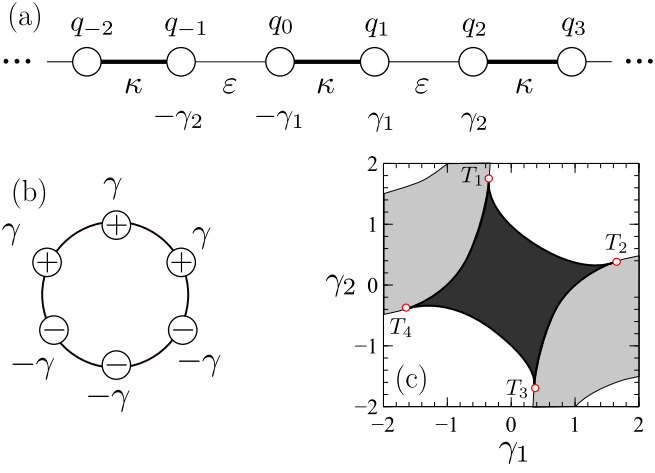}%
    \caption{(a) Infinite $\PT$-symmetric chain with alternating coupling defined by $\kappa$, $\epsilon$ and gain-loss coefficients $\gamma_1$, $\gamma_2$. (b) A clustered closed $\PT$-symmetric chain of six sites.
        (c)  $\PT$-symmetry breaking ``phase diagram'' of the open $\PT$-symmetric quadrimer \cite{ZK_12_PRL} which corresponds to the network shown in (a) truncated  at $q_{-1}$ and $q_2$, with $\kappa=\epsilon = 1$.
    }
    \label{fig-PD}
\end{figure}

Let us start with an infinite array in which a waveguide $n$  is
linked only with its two neighbors, waveguides $n-1$ and $n+1$. The simplest $\PT$-symmetric configuration that allows for the balance between gain and loss can be assembled using the following rule:
if any site of this network (say, $q_n$) has gain (loss) characterized
by $\gamma_n$, then the site $q_{-n+1}$ situated symmetrically with
respect to the ``center'' of the chain has loss (gain) characterized
by $-\gamma_n$, see  Fig.~\ref{fig-PD}(a).
Generally speaking, linear coupling between adjacent sites can also depend on the site number. In Fig.~\ref{fig-PD} we showcase a particular case where the network has \emph{alternating} coupling which is equal to $\kappa$ between $2n$-th and $(2n+1)$-th waveguides and equal to $\epsilon$ between $(2n-1)$-th and $2n$-th waveguides.
If $\kappa=\epsilon$, then the coupling becomes \emph{homogeneous}.

Linear properties of the described array are determined by a bi-infinite tridiagonal matrix $H$ with entries \cite{PZK}
\begin{equation}
H_{n,m} = c_n \delta_{n+1,m} +c_{n+1} \delta_{n-1, m} - (c_n + c_{n+1}  - i \gamma_n) \delta_{n,n},
\end{equation}
where $n$ and $m$ run through all integers from $-\infty$ to $\infty$,  $c_n=\kappa$ for even $n$ and $c_n=\epsilon$ for odd $n$, and  the gain and loss coefficients $\gamma_n$ obey the property $\gamma_n = -\gamma_{-n+1}$.
This lattice is $\PT$ symmetric with the parity operator defined by
$[\p \bq (t)]_n = q_{1-n}(t)$.

Let us set the nonlinearity $F(\bq)$ as
\begin{equation}
\label{F}
\left[F(\bq)\right]_{n,m}= \left[(1-\chi_n)|q_{n}|^2+\chi_n|q_{1-n}|^2\right]\delta_{n,m},
\end{equation}
where the real coefficients $\chi_n$ obey the relation
$\chi_n=\chi_{1-n}$.
The case with $\chi_n=0$ for all $n$ corresponds to on-site Kerr
nonlinearity, while the limit $\chi_n=1$ for all $n$ means nonlinear
coupling between the sites $n$ and $1-n$. The above nonlinearity
satisfies (\ref{eq:discr:PTreqnonlin}), i.e., the corresponding
nonlinear system is $\PT$ symmetric.

Problem  (\ref{eq:discr:dyn_main})   can also be rewritten in terms of
$q_n(t)$ as a generalized dNLS equation with gain and dissipation \cite{KPZ, Dmitriev}
\begin{eqnarray}
\label{eq:discr:dnls-general}
i \dot{q}_n = c_n(q_{n+1} - q_{n}) + c_{n+1}(q_{n-1}-q_n) + i \gamma_n q_n
\nonumber \\ +\left[(1-\chi_n)|q_{n}|^2+\chi_n|q_{1-n}|^2\right] q_n.
\end{eqnarray}

Imposing additional boundary conditions, one can transform the infinite system (\ref{eq:discr:dnls-general}) into a finite array.  Assuming that $n$ in (\ref{eq:discr:dimer_general}) runs from $-N+1$ to $N$ for a given $N$ and imposing zero boundary conditions, i.e., $q_{-N}(t)\equiv q_{N+1}(t)\equiv 0$, we define an \emph{open} chain of $2N$ sites.  Alternatively, one can impose periodic boundary conditions, i.e., $q_{-N}=q_N$ and $q_{-N+1}=q_{N+1}$, thus defining a  \emph{closed} chain (``necklace'') of $2N$ sites. An example of the latter configuration is shown in Fig.~\ref{fig-PD}(b).

The simplest case  $N=1$ corresponds to a nonlinear $\PT$-symmetric dimer
\begin{eqnarray}
\begin{array}{l}
i \dot{q}_0 = q_1 - i\gamma q_0  +  \left[(1-\chi) |q_0|^2 + \chi |q_1|^2\right] q_0, \\[1mm]
i \dot{q}_1 = q_0 + i \gamma q_1 + \left[\chi |q_{{0}}|^2 + (1 - \chi) |q_{ {1}}|^2\right] q_1,
\end{array}
\label{eq:discr:dimer-chi}
\end{eqnarray}
generalizing the model (\ref{opt:coupled_ode_nl}). In Eqs.~(\ref{eq:discr:dimer-chi}), the linear coupling is set to be equal to one,  and subscripts $1$ for $\gamma$ and $\chi$ are omitted. 

Setting $N=2$, we obtain an array of four waveguides -- a
$\PT$-symmetric quadrimer. Its properties are much richer than the
simplest dimer model (\ref{eq:discr:dimer-chi}). Indeed, quadrimers
allow one to distinguish between open and closed geometries, to
study effects of inhomogeneous couplings and gain-loss coefficients,
and to observe different types of broken $\PT$ symmetry (or
\emph{degrees} of $\PT$ symmetry breaking in the terminology
of~\textcite{Joglekar10,Scott11}), such as ``partially'' broken
symmetry (two real eigenvalues and two complex eigenvalues) and
``fully'' broken $\PT$ symmetry (all four eigenvalues are complex). These features
of $\PT$-symmetric quadrimers  were systematically addressed by
\textcite{Bendix10,ZK_12_PRL,LKMG,LiKev}.
As a particular example, we consider an open  quadrimer with homogeneous linear coupling $\kappa=\epsilon=1$ and arbitrary gain and losses coefficients $\gamma_{1,2}$ \textcite{ZK_12_PRL}, i.e.,
\begin{equation}
H =  \left(\begin {array}{cccc}
-i\gamma_2-2 & 1 & 0 & 0\\%
1& -i\gamma_1-2 & 1 & 0\\%
0 & 1&  i\gamma_1-2 & 1\\%
0 & 0 & 1 &  i\gamma_2-2  \end {array} \right). 
\end{equation}
The  $\PT$-symmetry ``phase diagram'' for matrix $H$ is shown in
Fig.~\ref{fig-PD}(c). It features three different
regions: unbroken $\PT$ symmetry where the spectrum is purely real
(black region), ``partially'' broken $\PT$ symmetry with two real and two complex eigenvalues (gray regions),
and ``fully broken'' $\PT$ symmetry where all eigenvalues are complex (white regions).
The boundaries separating different regions correspond to the exceptional points. 
The phase diagram also contains four \emph{triple points} where the
three regions meet. Another feature visible in the phase diagram is the   reentrant
$\PT$ symmetry (say after fixing $\gamma_1 =1.2$,  the $\PT$ symmetry
is broken at $\gamma_2=0$ and is restored when $\gamma_2$
increases).

In the case of arbitrary finite $N$,  characterization  of regions
of unbroken and broken $\PT$ symmetry is a non-trivial problem.
Detailed results on this front have been obtained for several particular cases. The
first case is a chain with \emph{alternating} gain and loss:
$\gamma_n = (-1)^{n+1}\gamma$, where $\gamma$ is a constant. Such a
chain can be considered as $N$ coupled identical dimers with
\emph{inter-dimer} coupling given by $\kappa$ and \emph{intra-dimer}
coupling given by  $\epsilon$.
The second case corresponds to a \emph{clustered} chain with two
segments, one consisting of lossy sites and the other consisting of
active sites, i.e., $\gamma_n = \textrm{sign}(n-1/2) \gamma$ [see
example in Fig.~\ref{fig-PD}(b)]. Systematic study of symmetry
breaking in these two chains with homogeneous linear coupling
($\kappa=\epsilon=1$) was performed by \textcite{KevrSIAM,BBA}. It was
found that $\PT$ symmetry is unbroken when the parameter $\gamma>0$
is below a threshold value $\gamma_{\PT}$. When
$\gamma>\gamma_{\PT}$ the $\PT$ symmetry is broken. For an open
alternating ($oa$) chain and alternating necklace ($an$) the
symmetry-breaking thresholds are
\begin{equation}
\label{eq:discr:oa}
\gamma^{(oa)}_\PT = \sin \frac{\pi}{2(2N+1)}, \quad
\gamma^{(an)}_\PT = \left \{
\begin{array}{ll}
0,&  N  \mbox{\ is even}\\
\sin\frac{\pi}{2N},&  N  \mbox{\ is odd}
\end{array}\right.,
\end{equation}
while for the open clustered ($oc$) chain  and  clustered necklace
($cn$) only the asymptotic results at $N\to\infty$ are available
\cite{BBA}:
\begin{eqnarray}
\label{eq:discr:oc}
\gamma^{(oc)}_{\PT} &=& 8.95(2N+1)^{-2}+O((2N+1)^{-3}), 
\\
\gamma^{(cn)}_\PT &=& 2.77 N^{-2}+O(N^{-3}). 
\end{eqnarray}
In all these cases, the region of unbroken $\PT$-symmetry shrinks to zero as
$N\to \infty$.

Another well-studied case is a chain with a $\PT$-symmetric
\emph{impurity}, i.e., having two ``defect'' sites
$\gamma_{d}=-\gamma_{1-d} = \gamma$ for some integer $d$ (with all
other sites having no gain or loss). For such a chain $\gamma_{\PT}
= 1$ if $d=1$ or $d=N$, and $\gamma_{\PT} \propto N^{-1}$ if $1<d<N$
\cite{JinSong09,Joglekar10,Scott11}.

Estimates for
$\PT$-symmetry breaking thresholds for chains with arbitrary configurations of sites with gain and losses  can be
obtained from the perturbation theory for $\PT$-symmetric operators
\cite{CaGraS}. For example, for an open chain with homogeneous
coupling, we separate matrix $H$ into real and imaginary parts,
i.e., $H=H_0+iG$ where $H_0$ is self-adjoint, and a diagonal matrix
$iG$ defines a non-self-adjoint perturbation. Then, by
Theorem~\ref{theor:GaGraS},
the spectrum of $H$ is real if
\begin{equation}
\label{eq:discr:unbroken}%
\max_{1\leq n\leq N}|\gamma_n|  \leq   4\sin^2\left( {\pi}{(4N+2)^{-1}}\right).
\end{equation}
At large $N$, this bound behaves as $\sim \pi^2(2N+1)^{-2}$, which is close to (\ref{eq:discr:oc}) obtained for the open clustered chain.

\textcite{PZK}   also found a sufficient condition for broken $\PT$ symmetry in a disordered open chain with homogeneous coupling: for $\epsilon=\kappa=1$, $\PT$ symmetry is broken if  $\sum_{n=1}^N \gamma_n^2 > 2N-1$.
This condition  is sharp for $N=1$.  For the quadrimer case ($N=2$),
this  condition ensures that $\PT$ symmetry is broken outside the
circle $\gamma_1^2+\gamma_2^2=3$. It is interesting to notice that
the triple points $T_j$ in the diagram of Fig.~\ref{fig-PD}(c) lie
exactly on this circle. Thus the condition is not sharp since the
regions with broken $\PT$ symmetry can also be found inside the
mentioned circle.

\subsubsection{Infinite lattices with unbroken $\PT$ symmetry}
\label{subsec:discr:infinite}

The results (\ref{eq:discr:oa})--(\ref{eq:discr:unbroken}) for the
finite chains indicate that $\PT$ symmetry tends to be more fragile
as the number of sites in the network increases, and the
$\PT$-symmetry breaking threshold can approach zero in the limit $N
\to \infty$. This indeed happens in  chains with $\PT$-symmetric disorder \cite{Bendix09}
and in an infinite chain with homogeneous coupling and alternating gain and loss \cite{Dmitriev}.
\textcite{Pelinovsky13} demonstrated that this situation is quite
general and can be encountered in different examples of infinite
$\PT$-symmetric networks with extended gain and loss. There are
however situations where an infinite linear
lattice has a nonzero $\PT$ symmetry breaking threshold. Some known examples are listed below

\paragraph{An open chain with alternating coupling and dissipation.} If
$\gamma_n = (-1)^n\gamma$ and alternating
coupling is characterized by the parameters $\kappa$ and $\epsilon$, as in Fig.~\ref{fig-PD}(a), the linear spectrum is real if $\gamma\leq \gamma_\PT =|\kappa-\epsilon|$ and is complex otherwise \cite{Dmitriev}.

\paragraph{An open chain with embedded defect.} Consider now an infinite   chain from Fig.~\ref{fig-PD}(a)  with homogeneous coupling $\kappa=\epsilon=1$  and an embedded finite $\PT$-symmetric \emph{defect} in a form of a dimer, i.e., $\gamma_0=-\gamma_1\ne 0$, and $\gamma_n=0$ if $n\not \in\{ 0, 1\}$. Then  $\PT$ symmetry is unbroken if
$|\gamma_{0,1}|\leq \gamma_\PT=\sqrt{2}$ \cite{SDSK12,SSDK12, KevrSIAM}. Notice that this $\PT$-symmetry breaking threshold
$\gamma_\PT=\sqrt{2}$ is larger than that for the isolated dimer
($\gamma_\PT=1$). It is also possible to consider a defect in a form of several adjacent dimers, but the domain of the unbroken $\PT$ symmetry  gradually shrinks   as the ``width'' of the defect grows.
%
%

A different case  when a pair of impurities $\pm \gamma$ are separated by one or
several conservative sites was addressed by \textcite{Bendix09}, who found that the $\PT$-symmetry breaking threshold becomes exponentially small as the distance $2d$ between the impurities tends to infinity:
$\gamma_{\PT} \approx e^{-d}$.

\begin{figure}
    \includegraphics[width=0.8\columnwidth]{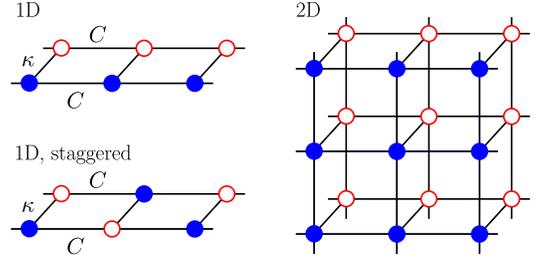}%
    \caption{(Color online) 1D   and 2D   arrays of identical $\PT$-symmetric dimers. Inter- and intra-dimer couplings are characterized by $\kappa$ and $C$. Open red and filled blue circles correspond to the sites with gain and losses, respectively.
        Based on ideas of \textcite{Bendix10} (1D),    \textcite{CLFLM14} (2D), and \textcite{DKM15} (1D, staggered). }
    \label{fig:discr:railway}
\end{figure}

\paragraph{Array of dimers.}
If gain and loss are alternating, $\gamma_n=(-1)^n\gamma$, then the
open chain in Fig.~\ref{fig-PD}(a) can be viewed as an array of
identical $\PT$-symmetric dimers, where the active site of each
dimer is coupled to the lossy site of the adjacent one.
\textcite{Bendix10} proposed another assembling of $\PT$-symmetric dimers in an array
which possess unbroken $\PT$ symmetry even if all couplings are equal. Such an array is illustrated  in Fig.~\ref{fig:discr:railway}(1D). This setup is described by the system
\begin{eqnarray}
\label{discr-1DdNLS}
\begin{array}{l}
i\dot{u}_n = i\gamma u_n + \kappa v_n + C(u_{n-1} + u_{n+1}) - \chi |u_n|^2u_n,
\\[1mm]
i\dot{v}_n = -i\gamma v_n + \kappa u_n + C(v_{n-1} + v_{n+1})- \chi |v_n|^2v_n,
\end{array}
\end{eqnarray}
where $\kappa>0$ and $C>0$ describe inter- and intra-dimer
couplings, and $\chi$ is the coefficient used to account the effect  of the Kerr nonlinearity. In the linear case ($\chi=0$), the condition of unbroken $\PT$ symmetry for (\ref{discr-1DdNLS})
 is  $\gamma\leq \gamma_{\PT}= \kappa$, i.e., $\PT$-symmetry breaking does not depend on $C$ \cite{Bendix10}. The nonlinear system with $\chi \ne 0$ supports propagation os solitons \cite{SMDK11}.
 Inverting gain and loss in a half of the chain produces a   model with the \emph{domain wall}. Scattering of linear waves in the latter system was studied by \textcite{SDMK12}.

\textcolor{black}{For a \textit{staggered} modification shown in Fig.~\ref{fig:discr:railway}(1D, staggered) with the orientations of the dimers alternating between the adjacent sites, $\PT$ symmetry is unbroken if   $\gamma\leq \gamma_{\PT}= \kappa-2C$, where $\kappa, C, \gamma>0$ \cite{DKM15}.}

\subsubsection{Array of metamaterial dimers}
\label{sec:1D_metamaterial}

Now we turn to a different type of an infinite $\PT$-symmetric chain
-- the model of 1D metamaterials which consists of
$\PT$-symmetric dimers of  SRRs discussed in Sec.~\ref{sec:metadimer} [see
Fig.~\ref{fig:metamat}(a,b)].  Assuming that $n$ enumerates the dimers $(q_{2n-1},q_{2n})$ in the
array, where odd and even SRRs have loss and gain, respectively
[Fig.~\ref{fig:metamat}(b)], and neglecting electric coupling, the
model describing the metamaterial can be written in the following
dimensionless form \cite{LazTsir13}
\begin{eqnarray}
\label{eq:1d_metamaterial}
\begin{array}{r}
\lambda^\prime\ddot{q}_{2n}+\ddot{q}_{2n+1}+\lambda \ddot{q}_{2n+2}+ q_{2n+1}+\gamma \dot{q}_{2n+1}
\\
+\alpha q_{2n+1}^2+\beta q_{2n+1}^3=\epsilon_0\sin(\Omega \tau),
\\
\lambda\ddot{q}_{2n-1}+\ddot{q}_{2n}+\lambda^\prime \ddot{q}_{2n+1}+ q_{2n}-\gamma \dot{q}_{2n}
\\
+\alpha q_{2n}^2+\beta q_{2n}^3=\epsilon_0\sin(\Omega \tau),
\end{array}
\end{eqnarray}
where $\lambda^\prime$ and $\lambda$ describe inter-  and
intra-dimer couplings. It is assumed the driving force is the same
for all SRRs and has frequency $\Omega$, and  the nonlinearity is
generalized to include quadratic terms.

The dispersion relation for the array (\ref{eq:1d_metamaterial}) in
the linear limit ($\alpha=\beta=0$) and in the absence of the
driving force ($\epsilon_0=0$) is readily found from the substitution
$(q_{2n-1},q_{2n})\sim (A,B)\exp[i(2nk-\Omega\tau)]$ with constants
$A$ and $B$~\cite{WA13}:
\begin{equation*}
\Omega_k^2=
\frac{2-\gamma^2\pm\sqrt{\gamma^4-4\gamma^2+4(\lambda-\lambda^\prime)^2+16\lambda\lambda^\prime\cos^2k}}{ 2[1-(\lambda-\lambda^\prime)^2-4\lambda\lambda^\prime\cos^2k]}.
\end{equation*}
At $\lambda^\prime=0$ (i.e., when all dimers are decoupled), this formula
recovers the eigenfrequencies of $\PT$-symmetric SRR
dimer~(\ref{eq:meta_eigen_dimer}). In the case of equal coupling,
$\lambda=\lambda^\prime$, $\PT$ symmetry of the array is broken. The
most unstable mode
is the one with $k=\pi/2$ corresponding to out-of-phase oscillating
dimers (i.e., with $\pi$-phase shift for the nearest SRRs). Thus
$\PT$-symmetry breaking  occurs at
$\gamma_{\PT}=\sqrt{2-2\sqrt{1-(\lambda-\lambda^\prime)^2}}$ and
$\gamma_{\PT}\sim |\lambda-\lambda^\prime|\to0$ as
$\lambda^\prime\to\lambda$. Domains of broken and unbroken $\PT$-symmetric phase are shown in Fig.~\ref{fig:metamat}(c).

\textcite{WA13}   considered the dynamics of metamaterial dimers
in the long-wavelength and small-amplitude limits and derived
coupled-mode equations supporting Bragg soliton solutions (see
Sec.~\ref{sec:bragg}).

\subsection{Nonlinear stationary  modes}
\label{subsec:nonlin_stat_mode}

An important class of solutions of Eq.~(\ref{eq:discr:dyn_main}) consists of nonlinear stationary modes. They have the form $\bq(z)=e^{-i E z} \bw$, where $E$ is a real parameter and
$\bw$ is a $z$-independent vector solving the algebraic system
\begin{equation}
\label{discr-stationary}
E \bw = H \bw + F(\bw) \bw.
\end{equation}
Notice that the equality $F(\bw) = F(\bq)$ follows from (\ref{eq:discr:F}).

\subsubsection{Exact solutions and the main features of nonlinear modes}
\label{subsec:discr:exact}

In some simple cases (like dimer or quadrimer models) solutions of
system (\ref{discr-stationary}) can be found explicitly. For the
$\PT$-symmetric dimer (\ref{opt:coupled_ode_nl}), this system is
\begin{eqnarray}
\label{eq:discr:dnls-dimer-stat}
\begin{array}{l}
E w_0 = \kappa w_1 - i \gamma  w_0   +   |w_{0}|^2 w_0,\\
E w_1 = \kappa w_0 + i \gamma  w_1   +   |w_{1}|^2 w_1.
\end{array}
\end{eqnarray}
Using polar representations $w_{0,1} = A_{0,1} e^{i\phi_{0,1}}$, $A_{0,1} \geq 0$, one can see that in addition to the trivial
solution $A_0=A_1=0$, other nonlinear modes are determined by the relations 
\begin{equation}
\label{eq:discr:dimer-sol}
A_0^2 = A_1^2 = E   \pm \sqrt{\kappa^2-\gamma^2}, \quad 
\sin(\phi_1 - \phi_0) = \gamma/\kappa.
\end{equation}
This result reveals several features. First, there exist two
branches of  nonlinear modes  (corresponding  to ``$+$'' and ``$-$''
signs in (\ref{eq:discr:dimer-sol})) when $ \gamma < \kappa$. At the
exceptional point   
$\kappa=\gamma$, these
two branches coalesce, and above the phase transition ($\gamma>\kappa$),
nonlinear modes do not exist. Below the phase transition, these modes
constitute two \emph{continuous families}: even if parameters of the
model $\gamma$ and $\kappa$ are fixed, one still can find a
continuous set of solutions by varying the free parameter $E$.
Amplitudes of the nonlinear modes tend to infinity as $E$ increases.
On the other hand, amplitudes of the modes vanish at $E=\pm
\sqrt{\kappa^2-\gamma^2}$, which corresponds to the eigenvalues of
the linear problem. Therefore both solution families bifurcate from
the linear limit. Any nonlinear mode
is $\PT$ invariant (up to a gauge transformation $w \to w^{i\theta}$), i.e., $\PT w = w$, which in the case at hand means $w_1 = w_0^*$.

\emph{Linear stability} of a nonlinear mode is examined by linearizing (\ref{eq:discr:dyn_main}) in the vicinity of the stationary solution $\bw$ and evaluating eigenvalues of the resulting linear problem. For the dimer model, the branch corresponding to the ``$-$'' sign is stable only if $E^2 \leq 4(\kappa^2-\gamma^2)$, while the ``$+$'' branch is always stable \cite{LiKev}.

The example of a dimer showcases several prototypical properties of
discrete nonlinear modes, but not all of them.
Indeed, for quadrimer models ($N=2$)
nonlinear modes (including stable ones) can exist even if $\PT$ symmetry of the underlying linear system is broken. Furthermore, there exist nonlinear modes that do not bifurcate from the linear limit and in the limit $E \to \infty$ some families of nonlinear modes can disappear \cite{ZK_12_PRL, LiKev, KevrSIAM,LKMG,LZKK}.

\subsubsection{Nonlinear modes as continuous families and isolated points}
\label{subsec:discr:nonlinPT}

As emphasized above, nonlinear modes of the $\PT$-symmetric
nonlinear dimer exist as continuous families. This property is
typical for conservative systems, but is quite unusual for systems
with gain and loss. On the other hand, $\PT$-symmetric arrays can
also feature properties of dissipative systems \cite{AA05}, i.e.,
they can admit another type of solutions which are \emph{isolated
    points}. For such solutions, $E$ is no longer a free parameter but
is determined by the balance between dissipation and gain, i.e., by
the system parameters. This feature can be illustrated in a
$\PT$-symmetric dimer with \emph{nonlinear} gain and loss
\cite{Miroshnichenko}
\begin{eqnarray}
\label{eq:discr:nonlinPT}
\begin{array}{l}
i\dot{q}_0 = \kappa q_1 - i \gamma  q_0  + i\Gamma |q_0|^2 q_0  +   |q_{0}|^2 q_0,
\\[1mm]
i\dot{q}_1 = \kappa q_0 + i \gamma  q_1  - i\Gamma |q_1|^2 q_1  +   |q_{1}|^2 q_1,
\end{array}
\end{eqnarray}
where in addition to linear gain and loss ($\gamma\geq 0$) one also
has $\PT$-symmetric nonlinear gain and loss characterized by
$\Gamma\geq0$. Looking for stationary nonlinear modes in the form of
$q_{0,1}=A_{0,1}e^{i\phi_{0,1}}e^{-iEz}$, one can classify possible
solutions into three groups \cite{Miroshnichenko, Duanmu13,CHCZLM15}:

\paragraph{Continuous families of $\PT$-symmetric solutions.} In this case  $E$ is a free parameter,
$A_0^2+A_1^2 \ne \gamma/ \Gamma$, $A_0=A_1$ are determined from the
equation
\begin{equation*}
(1+\Gamma^2)A_0^4-2(E+\gamma\Gamma)A_0^2+\gamma^2+E^2-\kappa^2=0,
\end{equation*}
and phases $\phi_{0,1}$ can be computed from $A_0$.

\paragraph{Isolated asymmetric solutions.} For solutions of this type, $E= \gamma/ \Gamma$,
$A_0^2\ne A_1^2$, and $A_0^2+A_1^2 = \gamma/\Gamma$. Notice that
these solutions have no counterparts  in the   dimer model without
nonlinear dissipation ($\Gamma=0$), because they require $\gamma/\Gamma >0$ and  exist only  due to the competition between
linear and nonlinear dissipation and gain.

\paragraph{Isolated $\PT$-symmetric solutions.} In this case, $E$ is determined from the equation
$( 2\Gamma E - \gamma)^2/(2\kappa \Gamma)^2 + \left(
{\gamma}/{2\kappa}\right)^2=1$, and $A_0^2 =  A_1^2$, $A_0^2+A_1^2 =
\gamma/\Gamma$. These modes exist only if both linear and nonlinear
gain and loss are present.

A similar classification of solutions can be elaborated for quadrimers
with nonlinear gain and loss~\cite{LKM14}.

\subsubsection{Continuous families of discrete solitons}

In a general case, families of nonlinear modes and their stability
analysis require numerical treatments. However, in some limiting
cases the problem can be investigated by means of an asymptotic
expansion or by the technique of analytical continuation.

\paragraph{Bifurcation from the linear limit.}
\label{sec:discr:fam:linear} If the underlying linear problem
possesses a real eigenvalue, one can search for a family of
nonlinear modes of (\ref{discr-stationary}) bifurcating from this
eigenvalue. Let us consider the case when $\tE$ is a {\em simple}
real eigenvalue of $H$ and $\tbw$ is the corresponding eigenvector,
i.e., $H\tbw=\tE\tbw$.  Up to a phase multiplier $e^{i\varphi}$ (see
discussion in Sec.~\ref{sec:1}), the eigenvector
can be chosen to be  {$\PT$ invariant}, i.e.,  $ \PT\tbw=\tbw$.
This property implies that the product $\langle \tbw^*,
\tbw\rangle$ is real.

In the vicinity of the \textit{linear limit} nonlinear modes
bifurcating from the eigenstate $\tbw$  can be searched using the
perturbation expansion~\cite{ZK_12_PRL}
\begin{equation}
\label{discr-expan1} \bw = \epsilon\tbw +\epsilon^3\bw^{(3)} +
o(\epsilon^3), \quad   E  = \tE +
\epsilon^2 E^{(2)} + o(\epsilon^2),
\end{equation}
where $\epsilon$ is a small real parameter, $\epsilon\ll 1$.
Coefficients $\bw^{(3)}$ and $E^{(2)}$ of
the expansions are to be determined.

Substituting expansions (\ref{discr-expan1}) into Eq.~(\ref{discr-stationary})
and noticing from Eq.~(\ref{eq:discr:F}) that $F(\bw) = \epsilon^2F(\tbw) +
O(\epsilon^3)$, one obtains the relation:
$E^{(2)}\tbw  = (H-\tE)\bw^{(3)} + F(\tbw)\tbw$. This equation
allows one to compute
$E^{(2)} =   {\langle\tbw^*, { F}(\tbw)\tbw
    \rangle }/{{\langle \tbw^*, \tbw\rangle}}$.~\cite{ZK_12_PRL,ZK_13_JPA}
Bifurcation of a family of nonlinear modes is possible only if $E^{(2)}$ is real. Hence, bearing in mind that $\langle \tbw^*,
\tbw\rangle$ is real, one obtains a necessary condition for a family of nonlinear modes to bifurcate from a simple eigenstate $\tE$:
$
\IM \langle \tbw^*, {F}(\tbw)\tbw \rangle = 0.
$
If the nonlinearity is $\PT$ symmetric, i.e., (\ref{eq:discr:PTreqnonlin}) holds,
then this condition is satisfied automatically. 

Validity of formal perturbation expansions (\ref{discr-expan1}) for
the particular case of the open chain with Kerr nonlinearity was
proven by \textcite{KevrSIAM} using the standard Lyapunov--Schmidt
method.  \textcite{DoSi15} developed a more general analysis and proved existence of nonlinear stationary modes bifurcating from  a simple eigenvalue in  systems with antilinear symmetry.
Expansions (\ref{discr-expan1}), however, must be modified if the
eigenvalue $\tilde{E}$ is not simple.   Nonlinear modes bifurcating
from {\em semi-simple} eigenvalues (i.e., eigenvalues with equal algebraic and
geometric multiplicities) were addressed by \textcite{ZK_13_JPA,
    LZKK}. Bifurcations from double and triple eigenvalues
(with algebraic multiplicity larger than the geometric one)
were studied by \textcite{ZK_13_JPA}.

\paragraph{Anticontinuum limit.}

Existence of continuous families can also be rigorously proven
in the so-called anti-continuum limit (ACL), developed in the
seminal work of \textcite{MA94} for a conservative dNLS equation. The method consists
in continuation from the limit of strong nonlinearity where the
linear coupling can be neglected and exact solutions of effectively
decoupled oscillators can be found. The solution family is
constructed using analytical continuation.

This idea can be extended to infinite $\PT$-symmetric arrays \cite{KPZ} as well as to finite $\PT$-symmetric chains and $\PT$-symmetric defects embedded in finite chains \cite{KevrSIAM,PZK}. As an
example, let us consider the system (\ref{eq:discr:dnls-general})
with all coupling coefficients $c_n$ scaled to unity and zero
boundary conditions $q_{-N}=q_{N+1}=0$. Substitution $q_n = w_n
e^{-i(E-2)z}$  yields  a system of  algebraic equations ($-N+1\leq n
\leq N)$:
\begin{eqnarray}
\label{eq:discr:stationary:general}
Ew_n = w_{n+1} + w_{n-1} + i \gamma_n w_n+
\nonumber \\
+[(1-\chi_n)|w_n|^2 + \chi_n |w_{1-n}|^2] w_n,
\end{eqnarray}
with  $w_{-N}=w_{N+1}=0$. Looking for $\PT$-invariant stationary
modes, $\PT w = w$, we obtain the relation $w_n=w^*_{1-n}$, which
reduces Eq.~(\ref{eq:discr:stationary:general}) to the system of $N$
equations  ($1 \leq n \leq N$)
\begin{equation}
\label{eq:discr:dnls-stat-reduct}
E w_n = w_{n+1} + w_{n-1} + i \gamma_n w_n + |w_{n}|^2w_n,
\end{equation}
with boundary conditions $w_0 = {w}_1^*$ and $w_{N+1} = 0$. Notice
that coefficients $\chi_n$ are not present in Eqs.~(\ref{eq:discr:dnls-stat-reduct}).

To enable the consideration of ACL,  we rescale variables as  $E =
1/\delta$ and ${ w_n} = {  W_n}/\delta^{1/2}$ with  $\delta\geq 0$,
and
rewrite (\ref{eq:discr:dnls-stat-reduct}) in the form 
\begin{equation}
\label{eq:discr:anti-continuum}
(1 - |W_n|^2) W_n = \delta( W_{n+1} + W_{n-1} + i \gamma_n W_n),
\end{equation}
where the boundary conditions now read $W_0 = {W}_1^*$ and $W_{N+1}
= 0$. In the limit   $\delta\to 0$ (i.e., $E\to\infty$),
Eqs.~(\ref{eq:discr:anti-continuum}) become decoupled and can be
solved analytically. The solutions obtained for $\delta=0$ can
then be analytically continued to the $\delta >0$ case by the
implicit function theorem.
Using this approach,
\textcite{PZK} proved that if the coefficients
$\gamma_1, \gamma_2, \ldots, \gamma_N$
satisfy  constraints
\begin{equation}
\label{eq:discr:condition} \left|\sum_{n=K}^N \gamma_n \right|<1
\quad \mbox{for all } K = 1,2,\ldots, N,
\end{equation}
then  in the limit $E\to \infty$ there exist $2^N$ unique  $\PT$-invariant
nonlinear modes  such that
\begin{equation}
\label{eq:discr:bound-nonlocal} \left| |w_n|^2 -  E \right| \leq C
\quad \mbox{for each } n = 1,2, \ldots, N,
\end{equation}
where $C$ is a positive $E$-independent constant.

The modes described by Eq.~(\ref{eq:discr:bound-nonlocal})
are characterized by unbounded amplitudes $|w_n|^2$ at all sites $n
= 1,2, \ldots, N$ when $E\to \infty$. They, however, do
not exhaust all possible nonlinear modes, and for a proper choice of
the coefficients $\gamma_1, \gamma_2, \ldots, \gamma_N$ one can also
construct  $2^M$ ($M=1,2, \ldots, N-1$) solutions whose amplitudes
in the limit of $E\to\infty$ grow unbounded only at $2M$ central
sites but vanish for other $2N-2M$ sites, i.e.,
\begin{equation*}
\begin{array}{ll}
\left| |w_n|^2 -  E \right| \leq  C  &\mbox{for all } n = 1,2, \ldots, M,\\[2mm]
|w_n|^2  \leq   C E^{-1} &\mbox{for all } n = M+1, M+2, \ldots, N,
\end{array}
\end{equation*}
where again  $C$ is a   positive constant which     does not depend on $E$.

\textcite{PZK} proved that under certain (not very
restrictive) conditions on coefficients $\gamma_1, \gamma_2,\ldots,
\gamma_N$, system (\ref{eq:discr:stationary:general}) admits altogether
$2^{N+1}-2$ $\PT$-invariant stationary solutions (unique up to a
gauge transformation) for all sufficiently large $E$.

ACL can also be used to classify linear stability of nonlinear modes
in the limit $E\to\infty$. Stability can be affected by the choice
of nonlinear coefficients $\chi_n$ [recall that $\chi_n$ do not
enter the stationary system (\ref{eq:discr:dnls-stat-reduct})].
\textcite{PZK} showed that $2^N$ nonlinear modes that exist in the
limit $E\to\infty$ under conditions (\ref{eq:discr:condition})
contain exactly one spectrally stable mode if $\chi_n<1/2$ for all
$n=1, 2, \ldots$ and exactly two spectrally stable modes if
$\chi_n=1/2$ for all $n$.

\begin{figure}
    \includegraphics[width=\columnwidth]{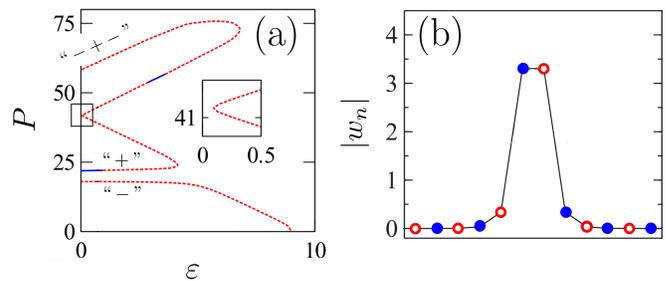}%
    \caption{(Color online)  (a) A diagram describing bifurcations of discrete $\PT$-symmetric solitons from the ACL which corresponds to $\epsilon=0$. Solid blue and dashed red fragments correspond to stable and unstable solitons, respectively.
        (b) Profile of an unstable soliton on the homogeneous lattice $\epsilon =\kappa$. Open (red) and filled (blue) circles correspond to the sites with gain and losses, respectively. For both panels,   $\kappa=1$,   $\gamma=0.1$ and $\mu=10$.  Adapted from \textcite{KPZ}.}
    \label{fig:discr:discr_sol}
\end{figure}

The ACL approach can also be used in infinite $\PT$-symmetric
lattices to prove existence of discrete solitons, i.e., nonlinear
modes satisfying boundary conditions $|w_n|\to 0$ as $|n| \to
\infty$ \cite{KPZ}. Consider an infinite network
(\ref{eq:discr:dnls-general}) with Kerr nonlinearity ($\chi_n=0$),
alternating gain and loss $\gamma_n = (-1)^{n}\gamma$, and
alternating coupling constants: $c_n=\kappa$ for even $n$ and
$c_n=\epsilon$ for odd $n$ ($n$ runs through all integers from
$-\infty$ to $\infty$). In this case, the ACL can be introduced as
the limit   $\epsilon\to 0$, when the infinite chain decouples into
a set of identical $\PT$-symmetric dimers.
Each dimer bears either the trivial
zero solution or one of the two non-zero solutions defined by
Eq.~(\ref{eq:discr:dimer-sol}). In the simplest situation, only one
dimer is exited (with a nonzero amplitude) in the ACL, while all
other dimers have zero amplitudes. Using the implicit function theorem, one can prove that
the obtained configuration can be continued analytically from the
limit $\epsilon=0$ to $\epsilon>0$. The obtained solution for $\epsilon>0$
represents a discrete $\PT$-symmetric soliton. Depending on the choice of sign
in  Eq.~(\ref{eq:discr:dimer-sol}), one can construct two types of
solutions, termed below as ``$+$'' solitons and ``$-$'' solitons.
These solitons can be continued to finite values of $\epsilon$
numerically, up to a certain threshold value of $\epsilon$ at which the
Jacobian matrix of the implicit function theorem becomes degenerate and further continuation is not
possible. An example of the bifurcation diagram is shown in
Fig.~\ref{fig:discr:discr_sol}(a), where the branch of ``$-$'' solitons
terminates at some critical value of $\epsilon$, and power $P=\sum_n
|w_n|^2$ of solitons vanishes. The branch of ``$+$'' solitons merges
with another branch of solitons designated as ``$-+-$''. In the ACL,
solitons of the ``$-+-$'' branch reduce to the configuration where
all decoupled dimers bear zero amplitude, except for three
consecutive dimers, the central one having amplitude
(\ref{eq:discr:dimer-sol}) with the ``$+$'' sign and the two others
having the ``$-$'' sign. Solutions from the ``$+$'' branch are
stable for sufficiently small $\epsilon$, but lose stability at
$\epsilon=\kappa-\gamma$, i.e., at the point of $\PT$ symmetry breaking.
However, the (unstable) discrete solitons can be continued to the
region of broken $\PT$ symmetry, and even up to the case where
$\kappa=\epsilon=1$ when the coupling becomes homogeneous
[Fig.~\ref{fig:discr:discr_sol}(b)].  {Stable} solitons in the infinite
chain with homogeneous coupling can be also found if the chain includes only a
finite number of sites with gain and loss (i.e., a $\PT$-symmetric
defect) \cite{KevrSIAM}. If the defect consists of only two sites,
localized modes can be obtained analytically
\cite{ZCHCLM,Dmitriev11}.

\subsubsection{Discrete compactons}
\label{subsec:compacton}

\begin{figure}
    \includegraphics[width=\columnwidth]{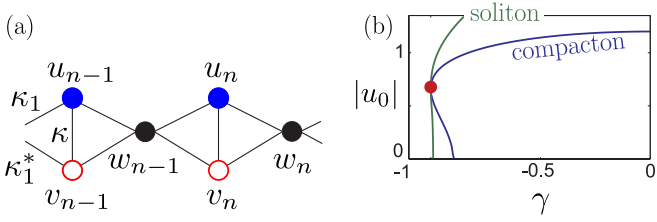}%
    \caption{(Color online) (a) Array of waveguides that supports discrete compactons.
        (b) Bifurcation diagram of branches of  compactons and solitons for $\kappa=1$,  $\tilde{\kappa}=0.25$, and  $\chi=1$.
        The point where the branches intersect is indicated by a dot.
        Adapted from \textcite{YuKo13}.}
    \label{fig:disc_compacton}
\end{figure}

{\em Compactons} were introduced by \textcite{RoHym93,Rosenau94} as excitations whose field is concentrated on
a finite support and is {\em exactly} zero outside this region.
Such objects cannot exist in systems with linear
dispersion. However linear dispersion can be completely suppressed
in specially designed $\PT$-symmetric arrays of waveguides
\cite{YuKo13}. Indeed let us consider an infinite network  shown in
Fig.~\ref{fig:disc_compacton}(a).  It consists of lossy
($u_n$), active ($v_n$), and conservative ($w_n$) waveguides. The
coupling coefficients are $\kappa$ (real and positive) and
$\kappa_1=\tilde{\kappa} e^{i\phi/2}$ with real $\tilde{\kappa}$ and
$\phi$. Gain and loss are described by the single parameter
$\gamma$. The array is governed by the dynamical system
\begin{eqnarray}
\label{main-conservative}
\begin{array}{l}
\label{eq_u}
i\dot{u}_n=\kappa v_n-i\gamma u_n+\kappa_1(w_{n-1}+w_{n})+ \chi|u_n|^2u_n,
\\[1mm]
\label{eq_v}
i\dot{v}_n=\kappa u_n+i\gamma v_n+\kappa_1^*(w_{n-1}+w_{n})+ \chi|v_n|^2v_n,
\\[1mm]
i\dot{w}_n=
\kappa_1 (u_n+u_{n+1})+\kappa_1^*(v_{n}+v_{n+1})+\chi |w_n|^2 w_n,
\end{array}
\end{eqnarray}
where $\chi$ is the coefficient of Kerr nonlinearity.

The linear dispersion relation is obtained by the ansatz $(u_n,v_n, w_n)=(u,v,w) e^{i(bz-kn)}$ with  $\chi=0$:
\begin{equation*}
b^3-(\kappa^2-\gamma^2)b
-8\tilde{\kappa}^2 [b\cos \phi+\gamma\sin\phi+\kappa] \cos^2(k/2)  =0.
\end{equation*}

In order to have a (linear) compacton, there must exist a dispersion
branch where the propagation constant $b$ is independent of $k$.
This is possible if $\gamma=-\kappa\sin\phi$. Then there is a $k$-independent
dispersion branch $b=-\kappa\cos\phi$ which describes the dipole
mode with $w_{dip}=0$ and
\begin{equation}
\label{discr:linear_compacton}
u_{dip}=\alpha v_{dip}, \,\, \alpha=\frac{\kappa^2-\tilde{\kappa}^2(1+i\cos k)}{\tilde{\kappa}^2(1+i\cos k)-\kappa^2\cos\phi}e^{i\phi}.
\end{equation}
This mode corresponds to the excitation of only one ``cell'' (say at
$n=0$), i.e., represents a linear compacton.

In the presence of nonlinearity, one can construct a continuous family of nonlinear solutions bifurcating from the   dipole mode (\ref{discr:linear_compacton}).  The nonlinear solutions persist as compactons if one follows along the line   $\gamma=-\kappa\sin\phi$. Bifurcation diagram for nonlinear $\PT$-symmetric  compactons   with $\chi=1$
is illustrated in Fig.~\ref{fig:disc_compacton}(b). Moving along
the bifurcation curve, one arrives at another bifurcation point
(indicated by a red dot) where the compacton branch intersects a
branch of conventional nonlinear $\PT$-symmetric modes. At this
intersection point the compacton and soliton coexist.

\subsubsection{Vortices in closed arrays}
\label{subsec:disc_vort}

\begin{figure}
    \includegraphics[width=\columnwidth]{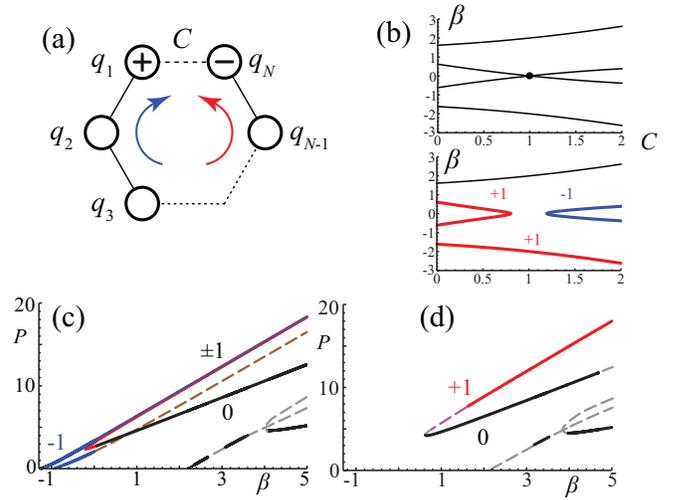}%
    \caption{(Color online) (a) Array of $N$ waveguides with one active ($n=1$) and one lossy ($n=N$) waveguide. Phase circulation direction is indicated by anti-clockwise (red, $m>0$) and clockwise (blue, $m<0$) arrows. (b) Linear propagation constants
        $\beta$ {\it vs.} $C$ for a conservative ($\gamma = 0$, upper panel) and $\PT$-symmetric ($\gamma = 0.2<\gamma_{\PT}$, lower panel) ring of $N=4$ waveguides. Degenerate vortex modes occur at the intersection marked by the black dot. Curves labeled  with ``$+1$'' and ``$-1$''  consist of the modes with the respective TC. Stable (unstable) nonlinear modes of $N=3$ ring are shown with solid (dashed) lines for (c) $C = 1.3, \gamma = 0.2<\gamma_{\PT}$ and (d) $C= 1.3, \gamma = 0.6>\gamma_{\PT}$.     TCs are indicated next to the curves, $m= +1$ in red (purple), $m=-1$ in blue (brown), and $m=0$ in black (grey).   Adapted from \textcite{Leykam13}.}
    \label{fig:disc_vortex}
\end{figure}

A discrete circular array of $N$ waveguides representing a system
with rotational symmetry supports vortex modes~\cite{DeDeFer11}.
These objects are characterized by the phase $\sim\exp(i2\pi m n/N)$
with $m=\frac{1}{2 \pi}\sum_{n=1}^N \text{Arg}( q_n^* q_{n+1})$
being the topological charge (TC)  and $n=1,2\dots N$ being the
waveguide number. The charge-flipping transformation
$m\leftrightarrow -m$ can be viewed as complex conjugation or time
reversion. Discrete vortices persist in arrays with embedded
$\PT$-symmetric defects similar to the one illustrated in
Fig.~\ref{fig:disc_vortex}(a) \cite{Leykam13}. Interplay between
nonlinearity and gain-loss breaks the $\PT$ symmetry and thus
degeneracy of the vortex modes.

Propagation of the monochromatic fields $q_n(z)$ ($n=1,...,N$) is   governed by the system ($\chi = \pm 1$)
\begin{eqnarray}
\label{main}
\begin{array}{l}
i \dot{q}_1 + C q_N + q_2 - i \gamma q_1 + \chi | q_1|^2 q_1 = 0,
\\
i \dot{q}_{n} + q_{n-1} + q_{n+1} + \chi |q_n|^2 q_n = 0,
\\
i \dot{q}_N + C q_1 + q_{N-1} + i \gamma q_N + \chi | q_N |^2 q_N = 0,
\end{array}
\end{eqnarray}
subject to the cyclic boundary conditions $q_{n+N} = q_n$.

The equations for linear stationary solutions  $q_n=w_ne^{i\beta z}$
can be cast in the general form $ \beta \bw  = H\bw$, where $H=H_0 +
i\gamma H_{1}$, $H_0$ is a matrix describing the array without
dissipation and loss, and $H_1$ has the only nonzero entries $(H_1)_{11}=(H_1)_{NN}=\gamma$. $H$ is $\PT$ symmetric, and the $\PT$-symmetry
breaking threshold is $\gamma_\PT=|C-1|$ \cite{Leykam13,SDSK12}.
Spectrum of the linear problem for a ring with  $N=4$ is illustrated in
Fig.~\ref{fig:disc_vortex}(b). At $\gamma=0$ vortex modes only exist
at $C = 1$, and they are degenerate (i.e.,  the modes with opposite TCs have the same energy). At $\gamma>0$ the degeneracy is broken
and vortex branches with $|m| = 1$ appear.

Typical families of nonlinear modes in an  array with  $N=3$ and $C>1$ are
shown in Fig.~\ref{fig:disc_vortex}(c) and (d) for unbroken and
broken $\PT$ symmetries, respectively. For $\gamma<\gamma_{\PT}$
nonlinear $m=-1$ modes bifurcate from the linear modes, and a pair
of $m=+1$ vortices is created from a saddle-node bifurcation [Fig.~\ref{fig:disc_vortex}(c)]. For $\gamma>\gamma_{\PT}$ the $m=-1$ modes are
destroyed, while the saddle-node bifurcation for $m=+1$ remains [Fig.~\ref{fig:disc_vortex}(d)].

Finally we notice that stable vortices as well as lifting of their degeneracy was also reported in continuous azimuthally modulated  $\PT$-symmetric rings~\cite{KarKonTor}.

\subsubsection{Solitons and vortices in coupled dNLS equations}

\begin{figure}
    \includegraphics[width=\columnwidth]{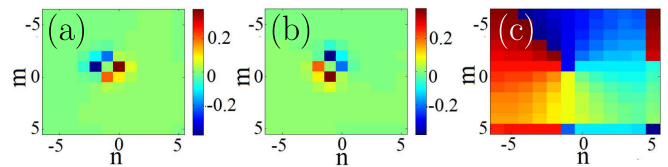}%
    \caption{(Color online) Real (a) and imaginary (b) parts, and the phase structure (c), of field $u_{m,n}$
        for a typical stable on-site $\PT$-symmetric vortex soliton with $(C, \kappa, \gamma)=(-0.03,-1, 0.4)$.  From \textcite{CLFLM14}.}
    \label{fig:disc_2Dvortex}
\end{figure}

$\PT$-symmetric dimers in an infinite array (\ref{discr-1DdNLS})
are described by two coupled 1D dNLS equations, whose soliton solutions
were reported by \textcite{SMDK11}.
A natural extension of that model is an infinite 2D array of coupled
dimers, i.e., in a plane as shown schematically in
Fig.~\ref{fig:discr:railway}(b). This leads to a model of coupled
2D dNLS equations \cite{CLFLM14}
\begin{equation*}
\begin{array}{l}
i\dot{u}_{n,m} =  \phantom{+} i\gamma u_{n,m} + \kappa v_{n,m} + C\Delta^{(2)} u_{n,m}
- |u_{n,m}|^2u_{n,m},\\
i\dot{v}_{n,m} = -i\gamma v_{n,m} + \kappa u_{n,m} + C\Delta^{(2)} v_{n,m}
-  |v_{n,m}|^2v_{n,m},
\end{array}
\end{equation*}
where $\Delta^{(2)}$ is a 2D second-order difference operator, i.e.,
$\Delta^{(2)}  u_{n,m} =  u_{n-1,m}+u_{n+1,m}+ u_{n,m-1} +
u_{n,m+1}-4 u_{n,m}$. If $\gamma < \kappa$ (i.e., $\PT$ symmetry is
unbroken), the system admits solutions with
$v_{n,m}=e^{i\delta}u_{n,m}$ where
\begin{equation}
\label{discr:delta}
\delta = -\arcsin(\gamma/\kappa)  \quad \text{or } \delta = \pi + \arcsin(\gamma/\kappa),
\end{equation}
and $u_{n,m}$ satisfies a single conservative 2D dNLS equation
\begin{equation*}
i\dot{u}_{n,m}  =     \kappa \cos(\delta) u_{n,m}   + C\Delta^{(2)} u_{n,m}-|u_{n,m}|^2u_{n,m}.
\end{equation*}
Each nonlinear mode of the latter equation yields two solutions of
the original system corresponding to two different $\delta$ in
(\ref{discr:delta}). A conservative 2D dNLS
equation admits on- and off-site vortex solitons characterized by
nontrivial phase circulations along a closed contour
\cite{Kevrekidis,MK01}. Their counterparts in the  2D
$\PT$-symmetric system were considered by \textcite{CLFLM14}, who
found that off-site vortices are unstable for almost any $C$, while
on-site vortices
can be stable in a wide range of parameters. An example of a stable
$\PT$-symmetric vortex is shown in Fig.~\ref{fig:disc_2Dvortex}.

\subsection{Nonlinear dynamics of $\PT$-symmetric arrays}
\label{sec:dim_nonlin_dyn}

\subsubsection{Conservative {\it vs.} dissipative  dynamics  }
\label{sec:cons_vs_dissip}

$\PT$-symmetric dynamical systems generally 
do not conserve energy, which  allows them  to possess unbounded solutions
(which are forbidden in conservative discrete lattices). The unbounded growth is a typical scenario of evolution of an unstable  $\PT$-symmetric stationary mode subjected to a small initial perturbation. Alternatively, unstable $\PT$-symmetric modes  break up
into long-lived transient 
structures, but typically do not evolve to an attractor (in this way $\PT$-symmetric dynamics has some features of conservative and Hamiltonian systems). If the initial conditions
correspond to a slightly perturbed stable nonlinear mode, then the
evolution also resembles that in a conservative system, i.e., the
amplitude of the perturbation remains nearly constant. Numerical
evidences of such behavior can be found in
\textcite{LiKev,PZK,LZKK,LKMG} for finite lattices, and in
\textcite{ZCHCLM, KevrSIAM} for infinite chains. Explanation of this
behavior might stem from the symplectic structure of the linear
operator that describes evolution of small perturbations of
stationary states~\cite{ABSK} (see Sec.~\ref{sec:coupler:basic}
below). \textcolor{black}{Despite absence of energy conservation, $\PT$-symmetric systems can conserve other quantities \cite{Ramezani} and admit a Hamiltonian representation. A variety of completely integrable Hamiltonian $\PT$-symmetric dimers were reported by \textcite{Barash_Hamilt,BPD15,BG14}}. In fact, some of such systems have been known much earlier outside the domain of the $\PT$ symmetry \cite{JCAH93,JC93}.

The best studied case corresponds to a finite
$\PT$-symmetric open chain with Kerr nonlinearity and alternating
gain and loss. \textcite{KevrJPA} proved that solutions of the
underlying configuration exist globally (i.e., do not blow up in
finite time) for any initial condition. At the same time, there
exist initial conditions that evolve to exponentially growing
solutions, even if $\PT$ symmetry of the underlying linear system is
unbroken [see (\ref{eq:discr:oa})].  More results on nonlinear
dynamics of $\PT$-symmetric oligomers (including categorization of
different dynamical scenarios) can be found in
\textcite{LiKev,LKMG,LZKK,LKFRK,Duanmu13,XuKevr14,DAmbroise12,Rodrigues13,Dmitriev11,DAmbroise14,ZCHCLM,SSDK12}.

Dynamics of solitons in infinite $\PT$-symmetric chains was
discussed by \textcite{SMDK11, Dmitriev}. Scattering on a
$\PT$-symmetric defect embedded in an infinite conservative chain
was studied in \textcite{DAmbroise12,
    Dmitriev11,DAmbr14ladder,ZCHCLM, SSDK12}.

We also mention a possibility for the existence of conserved
quantities in $\PT$-symmetric networks with an arbitrary number of
sites.  Recall that any linear $\PT$-symmetric (and hence
$\p$-pseudo-Hermitian) system admits an integral of motion
$Q=\langle \p q, q \rangle$ (see Sec.~\ref{subsec:pseudoHermit}).
This quantity   is not conserved in a nonlinear system
(\ref{eq:discr:dyn_main}) with generic nonlinearity $F(\bq)$.
However if the nonlinear operator $F(\bq)$ is
\emph{pseudo-Hermitian}, i.e.,
\begin{equation}
\label{eq:FpsH}
F^\dag(\bq) = \p F(\bq)\p \quad \mbox{ for any}\hspace{0.1cm} \bq,
\end{equation}
then the nonlinear system (\ref{eq:discr:dyn_main}) also   conserves the same quantity
$Q$~\cite{ZK_13_JPA}. This observation allows  to construct nonlinear
arrays with at least one integral of motion. Notice that if $\p  F^*(\bq)  = F(\bq) \p$
for all $\bq$ and $F(\bq)$ is a symmetric matrix,
i.e., $F^T(\bq) = F(\bq)$ for any $\bq$, then (\ref{eq:FpsH})
automatically holds.

\subsubsection{$\PT$-symmetric dimer}
\label{sec:PTdimer}

Dynamics of the nonlinear  dimer model (\ref{eq:discr:dimer-chi})
can be conveniently described using Stokes variables 
\begin{equation}
\label{eq:discr:spin}
\begin{array}{ll}
S_0=|q_0|^2+|q_1|^2,
& \quad
S_1=q_0 q_1^*+q_0^*q_1,
\\[1mm]
S_2=i(q_0 q_1^*- q_0^*q_1),
& \quad
S_3=|q_1|^2-|q_0|^2,
\end{array}
\end{equation}
which satisfy the identity $S_0^2=S_1^2+S_2^2+S_3^2$. From system
(\ref{eq:discr:dimer-chi}) one obtains
\begin{equation}
\label{eq:disr:Stokes:dyn}
\begin{array}{ll}
\dot{S}_0= 2\gamma  S_3,
& \quad
\dot{S}_1= (1-2\chi)  S_2 S_3,
\\ 
\dot{S}_2= 2  S_3 {-}(1-2\chi) S_1S_3,
& \quad
\dot{S}_3= 2\gamma S_0  {-} 2S_2.
\end{array}
\end{equation}

\paragraph*{Conserved quantities, integrability and Hamiltonian structure.}

\textcite{Ramezani} discovered that the $\PT$-symmetric dimer (\ref{eq:discr:dimer-chi}) with $\chi=0$
admits two integrals of motion. 
Indeed, using the new variable $r=\sqrt{S_1^2+S_2^2}/2$,
the first conserved quantity is found to be
$\rho = \sqrt{r^2-S_1 + 1}$.  The second constant of motion, $J$, is obtained from the relation
$2\rho\sin[(J-S_0)/(2\gamma)] = S_1-2$.
\textcite{PS13} used the integrals $\rho$ and $J$ to construct the phase portrait and to classify the behavior of all solutions of the system.
Further, \textcite{Barash_Hamilt} found that the dimer model admits a Hamiltonian representation, and  the Hamiltonian
\begin{eqnarray}
H=-2(\sqrt{\rho^2+1+2\rho\sin\theta}\cosh P_\theta + \gamma\theta)
\end{eqnarray}
is  expressed in terms of polar coordinates $\rho$ and $\theta$,
defined as $\rho\sin \theta = {S_1}/{2}-1$ and $\rho\cos
\theta=-{S_2}/{2}$, and momentum $P_\theta$, defined by the
relations $2r\sinh P_\theta = S_3$ and  $2r\cosh P_\theta = S_0$
[since $\rho$ is a conserved quantity, the conjugate momentum
$P_\rho$ does not enter the Hamiltonian]. The Hamiltonian equations
read
\begin{eqnarray*}
    \dot{\theta}=\frac{\partial H}{\partial P_\theta}=-S_3, \  \dot{P_\theta}=-\frac{\partial H}{\partial \theta}=2\left(\!\gamma+\frac{\rho\cos\theta}{r}\cosh P_\theta\!\right),
    \\\dot{\rho}=\frac{\partial H}{\partial P_\rho}=0,\quad
    \dot{P_\rho}=-\frac{\partial H}{\partial \rho}=2\,\frac{\rho+\sin\theta}{r}\cosh P_\theta.\hspace{1cm}
\end{eqnarray*}

The  $\PT$-symmetric dimer (\ref{eq:discr:dimer-chi}) with
$\chi=1/2$ is also integrable~\cite{PZK}. In this case substitution
$q_{0,1}=p_{0,1}\exp\left[\frac{1}{2i}\int (|p_0|^2 +
|p_1|^2)dz\right]$ transforms the model into a linear system:
$i\dot{p}_0 = -i\gamma p_0 + p_1$, 
$i\dot{p}_1 = i\gamma p_1 + p_0$,
meaning that all solutions are bounded for $\gamma<1$ and generically unbounded for $\gamma\geq 1$.

\paragraph*{Global existence, bounded and unbounded solutions.}

Turning back to the general model (\ref{eq:disr:Stokes:dyn}) with
arbitrary $\chi$, one concludes that solutions for any initial
condition exist globally as Gronwall's inequality implies that
$S_0(z) \leq S_0(0) e^{2 \gamma |z|}$ for all $z$.

For $\chi=0$ and $\gamma<1$,   there exist sufficiently small
initial conditions with globally bounded solutions \cite{KevrJPA}.
On the other hand, for $\chi \ne 1/2$ the system admits infinitely
growing solutions, even in the case of unbroken $\PT$ symmetry
\cite{KevrJPA, PZK}.  For $\chi=0$,  \textcite{BJF13} found an exact
unbounded solution:
$q_{0,1}=\exp\left\{  \mp \gamma(z-z_0) - i/\gamma\sinh[2\gamma(z-z_0)] \right\}$,
where $z_0$ is a free parameter.  For $\gamma>1$,  all trajectories
are generically unbounded, except for initial conditions that lie on
the stable manifold of the saddle point $q_0 = q_1 = 0$
\cite{BJF13}.

\paragraph*{Unidirectional propagation.}

A linear $\PT$-symmetric coupler displays non-reciprocal
behavior characterized by the field growth in the two arms (see
Sec.~\ref{subsec:optics}). The nonlinearity changes the situation leading to effectively
{\em unidirectional} light propagation \cite{SXK10, Ramezani}, i.e.,
to a light diode functionality. More specifically, if the
nonlinearity coefficient $\chi$ exceeds some threshold value
$\chi_{th}$, the output radiation almost entirely concentrates in
the arm with gain [the $q_1$ component in
(\ref{opt:coupled_ode_nl})] independent of which arm the input
radiation is applied; while for the nonlinearity below the critical
value, the output radiation is distributed between the two arms. The
critical nonlinearity, $\chi_{th}=4-2\pi\gamma$, was estimated from
the heuristic argument that for energy exchange to occur, there must
exist a maximum of $S_0$, i.e., a point $z$ where relations
$\dot{S}_0=0$ and $\ddot{S}_0<0$ hold simultaneously
\cite{Ramezani}. Description of possible evolution scenarios can be
found in~\textcite{SXK10}.

\subsubsection{Two coupled nonlinear oscillators}

Now we turn to nonlinear generalizations of the coupled oscillator
model (\ref{mech:two_oscil}). These studies were initiated
by~\textcite{CuKeSaK}, who considered periodic orbits of the model
\begin{equation*}
\ddot{x}+2\gamma\dot{x}+x+2\kappa y+x^3 =0,
\quad
\ddot{y}-2\gamma\dot{y}+y+2\kappa x+y^3 =0,
\end{equation*}
and showed that under the so-called rotating wave approximation,
this model is reduced to the nonlinear $\PT$-symmetric dimer
(\ref{opt:coupled_ode_nl}).

Another way to generalize the linear coupled oscillator model
(\ref{mech:two_oscil}) is based on a nonlinear extension of the
Hamiltonian (\ref{Hamilt_mecahnical}):
\begin{equation}
H=pq+\gamma(yq-xp)+(1-\gamma^2)xy
+\kappa (x^2+y^2)+\sum_{n,m} g_{nm} x^ny^m,
\end{equation}
which preserves the same Hamiltonian structure as described in
Sec.~\ref{subsec:mechan} for the linear case. By choosing
$g_{nm}=\delta_{n,1} \delta_{m,3}+\delta_{n,1} \delta_{m,3}$,
one obtains the relation between momenta and velocities given by
(\ref{velocity_momenta}), as well as the following dynamical
equations~\cite{BG14}:
\begin{eqnarray}
\label{eq:PT_oscil_complex}
\begin{array}{l}
\ddot{x}+2\gamma\dot{x}+x+2\kappa y+x^3+3xy^2 =0,
\\
\ddot{y}-2\gamma\dot{y}+y+2\kappa x+y^3+3yx^2 =0.
\end{array}
\end{eqnarray}
This system leads to a new integrable nonlinear $\PT$-symmetric
dimer. It was derived by~\textcite{BG14} using a multiple-scale
perturbation expansion under the scaling $2\kappa=3K\epsilon^2$ and
$2\gamma=\Gamma\epsilon^2$, where $\epsilon\ll 1$ is a small
parameter, and $K, \Gamma=O(1)$. Looking for a solution of
(\ref{eq:PT_oscil_complex}) in the form of perturbation expansions
$x=\epsilon x_1+\epsilon^3 x_3+\cdots$ and $y=\epsilon
y_1+\epsilon^3 y_3+\cdots$, and using the scaled time variables
$T_{2n}=\epsilon^{2n} t$ ($n=0,1,...$), one obtains the leading
order solution as
$x_1=\sqrt{ K}q_0e^{iT_0}+\textrm{c.c.},\quad y_1=\sqrt{K}q_1e^{iT_0}+\textrm{c.c.}$,
where $q_{0,1}$  solve the equations
\begin{equation}
\label{eq:discr:dimer_general}
\begin{array}{l}
{i\dot{q}_0+q_1+\left(|q_0|^2+2|q_1|^2\right)q_0+q_1^2q_0^*=-i \Gamma q_0},
\\[1mm]
{i\dot{q}_1+q_0+\left(|q_1|^2+2|q_0|^2\right)q_1+q_0^2q_1^*=\phantom{+}i\Gamma q_1}.
\end{array}
\end{equation}
Here the overdot denotes the derivative with respect to $\tau=3K
T_2/2$, and gain and loss are characterized by $\Gamma=\gamma/(3K)$.
System (\ref{eq:discr:dimer_general}) is Hamiltonian,
i.e., can be obtained from the equations of motion  $i\dot{q}_0 =-{\partial H}/{\partial q_1^*}$ and
$i\dot{q}_1=-{\partial H}/{\partial q_0^*}$, with the Hamiltonian
$H$ as
\begin{equation}
\begin{array}{c}
H=\left(|q_0|^2+|q_1|^2\right)\left(1+q_0^*q_1+q_0q_1^*\right)+i\Gamma\left(q_1^*q_0-q_0^*q_1\right).
\end{array}
\end{equation}
The system also conserves the quantity $S_1 = q_0q_1^*+q_0^*q_1$ and
is therefore integrable.

\textcolor{black}{Another interesting feature of system
    (\ref{eq:discr:dimer_general}) is the existence of stable nonlinear
    stationary modes for any value of $\Gamma$, even for $|\Gamma|>1$
    when   stable propagation of linear waves is not possible. Another
    modification of a $\PT$-symmetric dimer with   a similar property
    was reported by \textcite{CMKXS14}. Moreover, it is possible to find
    a family of dimers for which \textit{all} nonlinear trajectories
    remain bounded, irrespectively of the value of the gain-loss
    coefficient $\Gamma$ \cite{BPD15}. This phenomenon can be termed as
    nonlinearity-induced $\PT$-symmetry restoration.}

\subsubsection{Scattering on a $\PT$-symmetric defect}
An infinite conservative lattice with a $\PT$-symmetric defect
supports propagation of linear modes  which undergo scattering by the
defect~\cite{Dmitriev11, SSDK12}. The simplest case corresponds to a $\PT$-symmetric dimer
embedded in a conservative lattice. It can be represented
schematically by a chain in Fig.~\ref{fig-PD}(a), where all $\gamma_n=0$ except for
$\gamma_{1}=\gamma$, and all coupling constants are equal:
$\kappa=\epsilon=1$, except for  the one between $q_0$ and $q_1$ which is equal to some constant
$C$.
If all couplings are  equal, i.e., $C=1$, the transmission and reflection coefficients of a plane wave in the
linear limit $\chi=0$ incident from the left are
\begin{equation*}
T(k) = \frac{2i e^{-ik} \sin  k}{e^{-2ik}+\gamma^2-1}, \quad R(k) = \frac{-\gamma^2+2\gamma\sin k}{e^{-2ik}+\gamma^2-1},
\end{equation*}
where real $k$ is the Bloch wave number of the incident wave.

\begin{figure}
    \includegraphics[width=\columnwidth]{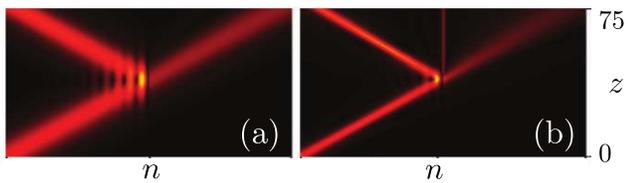}%
    \caption{(Color online) Two scenarios of discrete soliton scattering on a $\PT$-symmetric dimer defect. Panels (a) and (b) feature the same model parameters, but different amplitudes of the incident discrete soliton: $A=0.2$
        {\it vs.}  $A=0.5$. In both panels, $n$ runs from  $n=-75$ to $n=75$.
        Adapted from \textcite{SSDK12}.}
    \label{fig-discr-scatt}
\end{figure}

Since $|T(k)|^2$ and  $|R(k)|^2$ can be larger than unity, the
reflected and/or transmitted waves can be amplified after the
scattering. This property is verified in the nonlinear case as well.
Numerical study  shows that the
$\PT$-symmetric dimer defect can substantially amplify the incident
soliton \cite{SSDK12}. It was also found that soliton scattering can
occur without [Fig.~\ref{fig-discr-scatt}(a)]  or with
[Fig.~\ref{fig-discr-scatt}(b)] excitation of an internal
localized mode, depending on the amplitude of the incident soliton. Defect modes localized on a nonlinear $\PT$-symmetric
dimer were described in \textcite{ZCHCLM}.

Asymmetric scattering of left and right incident plane waves by
nonlinear  $\PT$-symmetric defects  embedded in a linear
conservative infinite chain was considered by
\textcite{DAmbroise12}. Scattering by a nonlinear defect embedded in
a linear ladder configuration, similar to that shown in
Fig.~\ref{fig:discr:railway}(a), was described
by~\textcite{DAmbr14ladder}.

\subsubsection{$\PT$-symmetric dimers with varying parameters}

A practically relevant issue is the management of nonlinear systems
by means of varying parameters. \textcite{DAmbroise14} investigated
numerically the effect of time-periodic gain on dynamics of the
nonlinear dimer (\ref{opt:coupled_ode_nl}) with periodic coupling
$\kappa=V_0 + V_1\cos(\omega z)$.
In the linear limit, such a system is characterized by the presence
of parametric resonance, and its long-term linear behavior is
determined  by the Floquet multipliers. On the plane of parameters
$(V_0,V_1)$  one can distinguish regions of stable and unstable
dynamics. The inclusion of nonlinearity    significantly affects the
dynamics, i.e., the same initial data can be bounded in the linear
case and unbounded in the nonlinear case and {\it vice versa}.
Effect of 
varying gain-loss profile was investigated by
\textcite{Horne2013, BDFP15}.

Another question of practical relevance is the effect of
\emph{random} modulations of system parameters which preserve the
$\PT$ symmetry only in average. Considering the $\PT$-symmetric
dimer (\ref{opt:coupled_ode_nl}) with coupling $\kappa+K(z)$ and
gain-loss coefficients $\gamma+\Gamma_{1,2}(z)$, where $K(z)$ and
$\Gamma_{1,2}(z)$ are delta-correlated white noises, \textcite{KZ14}
demonstrated that the statistically averaged intensity of the field
in the coupler grows exponentially. The growth occurs independently
of whether the average $\PT$ symmetry is broken or not, but the broken $\PT$ symmetry boosts the growth rate.

Stability regions for the model of coupled oscillators
(\ref{mech:two_oscil}) with periodically modulated gain-loss
coefficients were investigated by \textcite{PLT14}.

\subsection{\textcolor{black}{Observation of $\PT$-synthetic  solitons}}
\label{sec:synthetic_nonlin}
Now we turn to the experimental observation of   solitons in $\PT$-symmetric synthetic lattices reported by \textcite{Wimmer2015}. The experimental setting was briefly described in Sec.~\ref{subsec:optics} (see Fig.~\ref{fig:synthetic1}). It  is modeled by the nonlinear map (\ref{eq:synthetic1}), meaning that the temporal (i.e., evolution) coordinate ($m$ in this case) is also discrete.

\begin{figure}
    \includegraphics[width=\columnwidth]{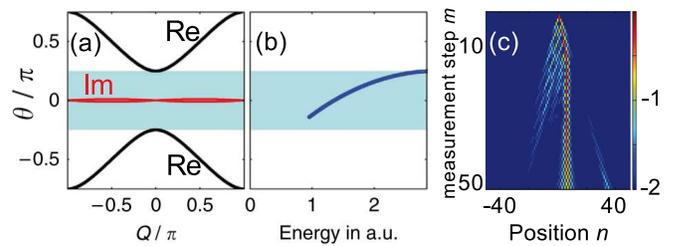}%
    \caption{(Color online)  Solitons in locally $\PT$-symmetric lattices.  (a) The linear dispersion relation (\ref{eq:synth_lin_loc}) with $G=1.1$. Black and red lines with labels ``Re'' and ``Im''  correspond to real and imaginary part of $\theta$, respectively. (b)  Family of solitons   in gap of the spectrum (shaded blue domain) on the diagram $\theta$ {\it vs} $E$. (c)   Formation and propagation  of a discrete soliton as the input power $P \approx 120$~mW. The colorbar shows $\log_{10}$(intensity).      Adapted from~\textcite{Wimmer2015}.  }
    \label{fig:synthetic2}
\end{figure}

\paragraph*{Discrete solitons in lattices with local symmetry} [see Fig.~\ref{fig:synthetic1}(a)] are described by the model (\ref{eq:synthetic1})  with $G_u=1/G_v=G$ and $\phi_n=0$. Neglecting the nonlinearity in the map (\ref{eq:synthetic1}) and using the ansatz $(u_n^m,v_n^m)\sim e^{i(Qn+\theta m)}$, one obtains the linear dispersion relation
\begin{eqnarray}
\label{eq:synth_lin_loc}
\theta=\pm \arccos\left[\cos\left(Q-({i}/{2})\ln G\right)/\sqrt{2}\right]
\end{eqnarray}
illustrated   in Fig.~\ref{fig:synthetic2} (a). Although the imaginary part of $\theta$ is not exactly zero in the infinite lattice, numerical results of \textcite{Wimmer2015}  show that in the finite lattice  the eigenvalues remain real as long as the gain does not exceed a certain critical value.

When nonlinearity is taken into account, increasing of the input power leads to a formation of a discrete soliton, as shown in   Fig.~\ref{fig:synthetic2}(c).
A stationary soliton is of the form $(u_n^m,v_n^m)=(U_n,V_n)e^{i\theta m}$ and can be characterized by the energy $E=\sum_n\left(|U_n|^2+|V_n|^2\right)$. One can identify a family of stationary solitons which can be visualized on  the plane $(\theta, E)$, see  Fig.~\ref{fig:synthetic2}(b).

\begin{figure}
    \includegraphics[width=\columnwidth]{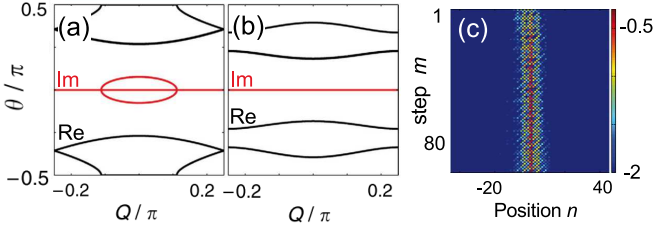}%
    \caption{(Color online)  Band-gap spectrum  (\ref{eq:synth_total_spectr})  for (a) the broken $\PT$ symmetry at $G=1.4$, $\phi_0=0$, and (b) unbroken  $\PT$ symmetry at $G=1.4$, $\phi_0=0.4\pi$. Black and red lines  with labels ``Re'' and ``Im''  show real and imaginary parts of $\theta$.  (c) Formation of a  broad single-hump soliton for $G=1.4$, and $\phi_0=-0.4\pi$.   The colorbar shows $\log_{10}$(intensity). 
        Adapted     from~\textcite{Wimmer2015}.  }
    \label{fig:synthetic3}
\end{figure}

\paragraph*{Discrete solitons in lattices with global $\PT$ symmetry} [see Fig.~\ref{fig:synthetic1}(b)] are described by the map (\ref{eq:synthetic1}) where the phase alternates as $n$ varies according to the following rule: $\phi_n=\phi_0$ for mod$(n+3,4)<2$, and $\phi_n=-\phi_0$ otherwise. The spectrum of the linear lattice in this case is determined by the equation
\begin{eqnarray}
\label{eq:synth_total_spectr}
\cos(4Q)=3\cos^2(2\theta)+8\cosh\left(\ln G\right)\cos(\phi_0)\cos(2\theta)
\nonumber \\
+ \cosh\left(2\ln G\right)-4\sin^2(\phi_0). \quad\quad
\end{eqnarray}
The examples of the numerical solution of the obtained equation are shown in Fig.~\ref{fig:synthetic3}(a,b). Formation and propagation of a discrete soliton  at large intensity of a broad Gaussian   pulse applied to the network input are shown in Fig.~\ref{fig:synthetic3}(c).

\section{$\PT$-symmetric coupled
    NLS equations}
\label{sec:4}

\subsection{The model and  its  basic  properties}
\label{sec:coupler:basic}

Generalization of discrete $\PT$-symmetric networks is given by {\em
    distributed couplers}, modeled  by two linearly coupled NLS
equations with gain and loss
\begin{equation}
\label{eq:NLS_gen}
\begin{array}{l}
i\psi_{1,z} =-\psi _{1,xx} - \kappa \psi_2+i\gamma \psi _{1}+\left( \chi|\psi _{1}|^{2}+\tchi |\psi _{2}|^{2}\right)
\psi _{1} ,
\\[1mm]
i\psi_{2,z} =-\psi_{2,xx}-\kappa \psi_1-i\gamma  \psi_{2}+\left( \tchi |\psi_{1}|^{2}+\chi|\psi_{2}|^{2}\right)
\psi _{2},
\end{array}
\end{equation}
where all coefficients are real and we assume $\kappa\geq 0$ and $\gamma \geq 0$.
%
Following optical terminology, the terms with $\chi$ and $\tchi$ are
referred to as self-phase modulation (SPM) and cross-phase
modulation (XPM), respectively.

The model (\ref{eq:NLS_gen}) is $\PT$-symmetric with the parity operator
$\p=\sigma_1$. This means that for a given solution
$\psi_{1,2}(x,z)$ of (\ref{eq:NLS_gen}) there exists a solution
$\psi_1^{(\PT)}(x,z) = \psi_2^*(x,-z)$ and $\psi_2^{(\PT)}(x,z) =
\psi_1^*(x,-z)$.

Using substitution $\psi_{1,2}\sim e^{ikx-ibz}$, we obtain the
dispersion relation of the underlying linear system ($\chi=\tchi=0$)
as
$b=k^2\pm\sqrt{\kappa^2-\gamma^2}$. 
Thus $\PT$ symmetry is unbroken for $\gamma \leq \kappa$, and
$\gamma=\gamma_\PT=\kappa$ is the exceptional point.

In the context of optical applications, model (\ref{eq:NLS_gen})
with $\chi=0$ was introduced  by~\textcite{Driben1} for constant
gain-and-loss coefficient $\gamma$, by~\textcite{AKOS} for a
$\PT$-symmetric defect with localized $\gamma=\gamma(z)$, and by
\textcite{Driben2} for periodic $\gamma(z)$ and $\kappa(z)$.
\textcite{PZK14}  proved that the Cauchy problem for
(\ref{eq:NLS_gen}) has a unique global solution in the energy space
$(\psi_1,\psi_2)\in H^1(\mathbb{R})\times H^1(\mathbb{R})$, with the
$H^1$ norm defined by $\|\psi\|_{H^1}^2 = \int_\mathbb{R}(|\psi|^2 +
|\psi_x|^2)dx$. This global existence
however does not rule out the possibility of indefinitely growing
total $H^1$ norm, $\|\psi_1\|_{H^1} + \|\psi_2\|_{H^1}$.
In the particular case of
$\chi=\tchi$, Eqs.  (\ref{eq:NLS_gen}) represent a $\PT$-symmetric
extension of the exactly integrable  model introduced by \textcite{Manakov}. In
this case, the system can be conveniently treated in terms of
integral Stokes variables [cf.~(\ref{eq:discr:spin})]
\begin{eqnarray*}
    \label{eq:integral_Stokes}
    S_0=\int_{\mathbb{R}}\!\left(|\psi_1|^2+ |\psi_2|^2\right)\! dx, \ S_1=\int_{\mathbb{R}}\!\left(\psi_1^*\psi_2+ \psi_2^*\psi_1\right)\! dx,
    \\
    S_2=i\int_{\mathbb{R}}\!\left(\psi_1^*\psi_2- \psi_2^*\psi_1\right)\! dx, \
    S_3=\int_{\mathbb{R}}\!\left(|\psi_1|^2- |\psi_2|^2\right)\!dx,
\end{eqnarray*}
which for $\chi=\tchi$ solve equations
\begin{equation*}
\label{eq:int_Stokes}
\dot{S}_0=2\gamma S_3,\quad\dot{S}_1=0,\,\, \dot{S}_2=  {-}2\kappa S_3,\,\, \dot{S}_3=2\gamma S_0+2\kappa S_2,
\end{equation*}
where the overdot stands for the derivative with respect to $z$.
Thus the model conserves two quantities: $S_1$ and $C=\kappa
S_0+\gamma S_2$, which allows one to obtain a general solution
\begin{eqnarray}
\label{eq:S0_dynam}
S_0=\kappa C/\omega^2+A_1\cos(2\omega z) +A_2\sin(2\omega z),
\end{eqnarray}
with $\omega=\sqrt{\kappa^2-\gamma^2}$ and constant $C$ and
$A_{1,2}$. Hence the total power $S_0(z)$ is bounded for $\gamma
<\kappa$ and generically unbounded otherwise.  On the basis of numerical
simulations, \textcite{PZK14} also  conjectured  that the $H^1$ norm
of all solutions in the  system with $\chi=\tchi$ is also bounded
for $\gamma < \kappa$.

Let us also point out that
if $\gamma \leq \kappa$,  then substitution 
\begin{equation}
\psi _{2} = e^{i\delta }\psi _{1}, \, \delta=\arcsin\gamma/\kappa \mbox{ or } \delta=\pi -\arcsin\gamma/\kappa
\label{seq:PT-NLS_subst}
\end{equation}%
reduces (\ref{eq:NLS_gen}) to a single conservative NLS equation for
function $\psi _{1}(x,z)$ \cite{Driben1,ABSK,BDKM}:
\begin{equation}
i \psi _{1,z}=-\psi _{1,xx} +\left( \chi+\tchi \right) |\psi _{1}|^{2}\psi _{1}-
\kappa \cos (\delta) \psi _{1}.
\label{eq:PT-NLS_reduced}
\end{equation}%


\subsection{Modulational instability}
\label{sec:bright_coupled_NLS}

System (\ref{eq:NLS_gen}) admits a solution in the form of a carrier
wave (CW) background~\cite{BKM}
\begin{equation}
\psi_{j}^{\mathrm{cw}}\!=\rho e^{ikx-ibz+i(-1)^{j}\delta /2},
~~b=k^{2}+\rho ^{2}(\chi
+\tchi )-\cos \delta ,
\label{eq:pl-wave-cur}
\end{equation}%
where 
$k$ and $\rho$ are constants. 
To study its linear stability, we use the standard substitution,
\[
\psi _{j}=\psi _{j}^{\mathrm{cw}}+\rho \left(\eta _{j}e^{-i(\beta z-\kappa
    x)}+\nu_{j}^*e^{i(\beta^*z-\kappa x)}\right) e^{ikx-ibz},
\]%
with  $|\eta _{j}|,\,|\nu _{j}|\ll  {1} $. The linearization gives 
two branches $\beta_{1,2}(\kappa)$  
of the stability eigenvalues:
\begin{equation}
\beta _{1}
= 2k\kappa \pm \kappa \sqrt{\kappa ^{2}+2\rho
    ^{2}(\chi+\tchi )},  \label{omega1}
\end{equation}
\begin{equation}
\beta _{2}
= 2k\kappa \pm \sqrt{\left( \kappa ^{2}+2\cos
    \delta \right) \left( \kappa ^{2}+2\cos \delta +2\rho ^{2}(\chi-\tchi )%
    \right) },
\label{omega2}
\end{equation}%
which feature several sources of modulational instability (MI). One
source of MI stems from Eq.~(\ref{omega1}) and corresponds
to
\begin{equation}
\chi +\tchi <0.
\label{cond1}
\end{equation}%
This is the MI  due to long-wavelength excitations;  it is not influenced by gain and loss and is present also
in a conservative system of NLS equations without linear coupling.
Another source of MI stems from Eq.~(\ref{omega2}):
\begin{equation}
\cos \delta <\max \{0,\rho ^{2}(\chi -\tchi)\},  \label{inst2}
\end{equation}%
and arises due to the linear coupling and is significantly affected by the gain and loss.

\begin{figure}
    \begin{center}
        \includegraphics[width=0.95\columnwidth]{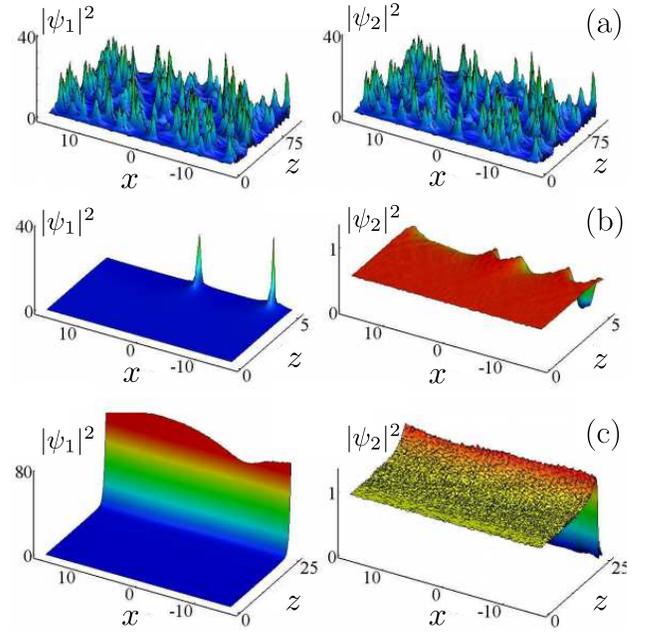}
    \end{center}
    \caption{(Color online)  Intensity evolution of the field components $|\psi_{1}|^{2}$ and
        $|\psi_{2}|^{2}$ (left and right columns) of the plane wave with
        $\rho =1.604$,  {$\chi=0.5$, $\tilde{\chi} =-1$} (a), $\rho =0.76$, $\chi=-1.5$, $\tchi =1$ (b), 
        and $\rho =0.98$, $\chi=0.25$, $\tchi =1$ (c). For all panels, $k=0$, $\delta =\pi /4$. Adapted from \textcite{BKM}.
    }
    \label{fig:mi}
\end{figure}

Different origins of the MI manifest themselves through different dynamical scenarios illustrated in Fig.~\ref{fig:mi}. 
In Fig.~\ref{fig:mi}(a), XPM nonlinearity is focusing, and we
observe a ``standard'' scenario of MI which is very similar to its
Hamiltonian counterpart where the power is distributed between the
two waveguides. If the XPM is defocusing, but condition (\ref{cond1})
is satisfied, one observes relatively fast power transfer from the
lossy waveguide to the active one, accompanied by fast growing peaks
[Fig.~\ref{fig:mi}(b)]. Such behavior is induced by the focusing
SPM, and therefore is not significantly changed even when one passes
from the domain of parameters (\ref{cond1}) [Fig.~\ref{fig:mi}(b)]
to the one defined by (\ref{inst2}) [not shown in
Fig.~\ref{fig:mi}]. The third distinctive scenario of the MI takes place when both XPM and SPM are defocusing [Fig.~\ref{fig:mi}(c)]. In this case  the MI
occurs only due to the imbalance between the gain and loss, induced by the nonlinearity and
resulting in nearly homogeneous growth (decay) of the field in the
waveguide with gain (loss) respectively.



\subsection{Bright solitons}
\label{sec:bright_coupled_NLS_eq}

When a CW background is unstable, a system can admit solitonic solutions.
Using substitution  (\ref{seq:PT-NLS_subst}), one can find a one-soliton exact solution  which for $\tchi=0$ and $\chi=-2$  reads  \cite{Driben1}
\begin{equation}
\label{eq:bright_sol_coupled_NLS}
\psi_1^{\mathrm{s}}\!=e^{i\delta-ibz }\frac{a}{\cosh(ax)}, ~~ \psi_2^{\mathrm{s}}\!=e^{-ibz}\frac{a}{\cosh(ax)},
\end{equation}
where the propagation constant $b=-a^2-\cos\delta$, amplitude
$a>0$, and we set $\kappa=1$.  Equation ~(\ref{eq:bright_sol_coupled_NLS}) actually describes two types of
solutions which can be termed {\em symmetric} ($0<\delta<\pi/2 $)
and {\em antisymmetric} ($\pi/2<\delta<\pi $) in accordance with the
respective conservative limits $\delta=0$ and
$\delta=\pi$~\cite{Stegeman}. Bright solitons in presence of SPM and XPM, i.e., for nonzero
$\chi$ and $\tchi$, were considered by \cite{BDKM}.

\textcite{Driben1} found that the symmetric soliton is stable for
\begin{eqnarray}
\label{eq:nls_lin_stab_bright}
a^2<a_{max}^2=2\sqrt{1-\gamma^2}/3,
\end{eqnarray}
which agrees with the known result for the conservative case
$\gamma=0$~\cite{Stegeman}. At $\gamma>0$, condition
(\ref{eq:nls_lin_stab_bright}) can be obtained from the linear
stability analysis \cite{ABSK} which starts with the substitution
\begin{equation*}
\psi_1=\psi_1^{\mathrm{s}} + e^{i\delta-ibz } (p+q)/{\sqrt{2}},
~~
\psi_2=\psi_2^{\mathrm{s}} + e^{-ibz} (p-q)/{\sqrt{2}},
\end{equation*}
where $p=\RE\left(p_1(x)e^{\mu z}\right)+i \RE\left(p_2(x)e^{\mu
    z}\right)$, $q=\RE\left(q_1(x)e^{\mu z}\right)+i
\RE\left(q_2(x)e^{\mu z}\right)$, $\mu=\nu-i\omega$, and  $p_{1,2}$,
$q_{1,2}$ are complex functions. The linearization with respect to
$p$ and $q$ gives the eigenvalue problem
\begin{equation}
\label{eq_bright_eigen_prob_a}
(L-\cos\delta)\bp+2\gamma  J \textbf{q}=\mu J\bp,
\quad
(L+\cos\delta) \textbf{q} = \mu J\textbf{q},
\end{equation}
where $\bp=(p_1,p_2)^T$, $\textbf{q}=(q_1,q_2)^T$, $J = -i\sigma_2$ [see (\ref{Pauli})],
\begin{equation*}
\label{eq:NLS_L}
L=\left(
\!\begin{array}{cc}
L_+ & 0 \\0 &L_-
\end{array}\!
\right),
~~
L_\pm=-\frac{d^2}{dx^2}-b-(4\pm 2)|\psi_2^{\mathrm{s}}|^2.
\end{equation*}
In fact, the stability analysis reduces to the second equation in  (\ref{eq_bright_eigen_prob_a}), since the first equation has a bounded solution $\bp$ for any bounded $\bq$ and $\mu \neq 0$.
The obtained symplectic eigenvalue problem pertains to Hamiltonian
evolution, and thus scenarios of evolution of instabilities are
expected to be characteristic of conservative systems, despite the
presence of gain and loss.

For the soliton (\ref{eq:bright_sol_coupled_NLS}), by introducing
$X=ax$, $\lambda=\mu/a^2$ and $\eta=\cos\delta/a^2$, the second
equation in (\ref{eq_bright_eigen_prob_a}) can be rewritten as
\begin{eqnarray}
\label{eq_eigen_nls_brigh_one_solit}
(L_-+\eta)(L_++\eta)q_1=-\lambda^2 q_1,
\end{eqnarray}
where
$
L_\pm=-d^2/dX^2+1-(4\pm 2)\sech^2X.
$
For the symmetric soliton (\ref{eq:bright_sol_coupled_NLS})  with
$\cos\delta>0$, the lowest eigenvalue of $L_-$ is zero, and  hence
$L_-+\eta$ is positive definite and the inverse $(L_-+\eta)^{-1}$
exists. The symmetry of the eigenvalue problem
(\ref{eq_eigen_nls_brigh_one_solit}) implies that if $\lambda$ is an
eigenvalue, so are $-\lambda$ and $\pm\lambda^*$. Hence stability of
the solitons requires that the minimal eigenvalue
expressed through the Rayleigh quotient
\begin{eqnarray*}
    -\lambda^2=\min\{{\langle q_1,(L_++\eta)q_1\rangle}/{\langle q_1,(L_-+\eta)^{-1}q_1\rangle}\}
\end{eqnarray*}
must be positive. This occurs if the lowest eigenvalue of the
operator $L_++\eta$, i.e., $\nu=-3+\eta$, is positive. Recalling the
definition of $\eta$ one recovers (\ref{eq:nls_lin_stab_bright}).

Numerical studies of anti-symmetric solitons show that all such
solutions are unstable \cite{Driben1,ABSK}. However lifetimes of the
solitons with small amplitudes are exponentially long so for some
purposes they can be regarded as stable.  Dynamics of unstable
solitons, either symmetric or anti-symmetric, is divided into two
asymptotic regimes: unbounded growth and formation of breathers.
Unbounded growth typically occurs when amplitudes of unstable
solitons are sufficiently large.

While bright solitons (\ref{eq:bright_sol_coupled_NLS}) correspond
to the most fundamental localized excitations in nonlinear
$\PT$-symmetric couplers, \textcite{LLM14} demonstrated that the
model also supports stable propagation of ``super-solitons''
\cite{NMMPG08}, i.e.,  localized excitations consisting of many
identical solitons which feature Newton-cradle-like dynamics. Dynamics and stability os 2-soliton solution, i.e. the input $\psi_1(z=0)=e^{i\delta} \psi_2(z=0)=\frac{2a}{\cosh(ax)}$, in a $\PT$-symmetric coupler, as well as switching of a 2-soliton initial pulse applied to only one of the arms were considered by~\cite{DM12}.


\subsection{Breathers}
\label{sec:breathers_coupled_NLS}

In numerical studies of soliton stabilization at the exceptional point 
(see~\ref{sec:solit_except}), \textcite{Driben2}  found breathers featuring  persistent irregular oscillations.
In further studies of bright solitons \textcite{ABSK}, 
found two main scenarios of development of instabilities: unbounded
growth of a soliton amplitude and emergence of periodic
breather-like excitations. This naturally poses a question on the
existence of breathers in $\PT$-symmetric coupled NLS equations
(\ref{eq:NLS_gen}). We address this question following
\textcite{BSSDK}, for the case of $\chi=-2$, $\tchi=0$, and $\kappa=1$, in
(\ref{eq:NLS_gen}). First, let us  observe that the global rotation
\begin{equation}
\label{glob_rot}
\left(\!\begin{array}{c}
q_1\\q_2
\end{array}\right)=U \left(\begin{array}{c}
\psi_1\\ \psi_2
\end{array}\!\right),~~ U=\frac{1}{2\cos\delta}\left(\!\begin{array}{cc}
e^{i\delta} &-1 \\ e^{-i\delta} &1
\end{array}\!\right),
\end{equation}
where $\delta$ is defined by  (\ref{seq:PT-NLS_subst}),
transforms (\ref{eq:NLS_gen}) into a new system of two  NLS-like
equations without any linear dissipation and coupling.
In
the limit of small amplitudes, $|q_{1,2}| \ll 1$, the new system
decouples into two linear equations: $iq_{j,z}=- q_{j,xx}\pm \cos (
\delta) q_{j}$, where $j=1,2$.
Thus at small amplitudes one can look for a solution of the
nonlinear problem in the form of a  multiple-scale expansion
\begin{eqnarray}
\label{multiscale}
q_j=\sqrt{\epsilon}e^{(-1)^ji\cos(\delta) z}\left(A_j+\epsilon A_j^{(1)}+\cdots \right),
\end{eqnarray}
where $A_j^{(n)}$ depend on $Z=\epsilon z$, $X=\sqrt{\epsilon}x$, as
well as the rest of scale variables $Z_n=\epsilon^{n+1} z$ and
$X_n=\epsilon^{n+1/2}x$ with $n=1,2,...$.
Substituting (\ref{multiscale}) into the nonlinear equations for
$q_{1,2}$, collecting all terms with the same power of $\epsilon$
and eliminating the secular terms, one obtains
\begin{eqnarray}
\label{syst_nls_rwa}
i A_{j,Z}+ A_{j,XX}+2\left(|A_j|^2+2|A_{3-j}|^2\right)A_j=0,
\end{eqnarray}
where $j=1,2$. This system has two obvious solutions: $(A_1,0)$ and $(0,A_2)$ which   in terms of  the original functions $\psi_{1,2}$ yield   antisymmetric and symmetric bright solitons.
System  (\ref{syst_nls_rwa}) also  admits   vector solitons with $A_1=A_2$.
Taking  the latter solution and inverting rotation  (\ref{glob_rot}),  one obtains a breather
\begin{eqnarray}
\label{eq:breathre}
\left(\!\begin{array}{c}
\psi_1\\ \psi_2
\end{array}\!\right)=\frac{2 a\exp(i a^2 z) }{\sqrt{3}\cosh(ax)}
\left(\!\begin{array}{c}
\cos\left[\cos(\delta) z\right]\\ i\sin\left[\delta +\cos(\delta) z\right]
\end{array}\!\right),
\end{eqnarray}
where the amplitude $a$ must be small enough ($a\sim  \epsilon^{1/2}
\ll 1$).  The frequency of the breather in the leading order is determined by
$\cos(\delta)=\sqrt{1-\gamma^2}$.

\begin{figure}
    \begin{center}
        \includegraphics[width=\columnwidth]{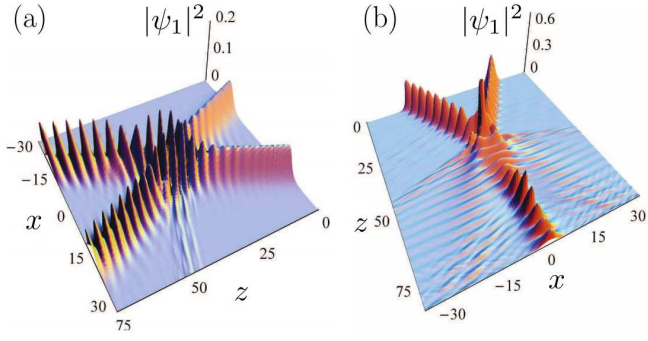}
    \end{center}
    \caption{(Color online)  (a) Collision of anti-symmetric (initially left) and symmetric (initially right) solitons for $\gamma=0.5$ and $a=0.3$.
        A pair of breathers emerge from the collision. (b) Inelastic collision of two breathers for $\gamma=0.3$ and $a=0.3$.
         Adapted from~\textcite{BSSDK}.}
    \label{fig:bretaher}
\end{figure}

Linear stability analysis and numerical simulations indicate that
breathers are stable.  Moreover, breathers appear to be rather
common objects which are excited when unstable solitons break up
\cite{ABSK}, after interaction of symmetric and antisymmetric
solitons [Fig.~\ref{fig:bretaher}(a)], and after interaction of a
soliton with a defect~\cite{BHHK}. Numerical studies of interaction
of breathers show appreciable inelastic effects as illustrated in
Fig.~\ref{fig:bretaher}(b)~\cite{BSSDK,RSD14}.

\subsection{Solitons in couplers with varying parameters}
\label{sec:varying_param}

\subsubsection{Stabilization of a soliton at an exceptional point}
\label{sec:solit_except}

The exceptional point of the linearized ($\chi=\tchi=0$) system
(\ref{eq:NLS_gen}) corresponds to $\gamma_\PT=\kappa$. It is evident
from the substitution (\ref{seq:PT-NLS_subst}) that at this point
symmetric and antisymmetric solitons merge into the same solution
with $\psi_2=i\psi_1$  which appears to be unstable. However, by
introducing simultaneous periodic modulations of the gain, loss and
coupling, i.e., by considering the model
\begin{eqnarray}
\label{eq:nls_driben}
\begin{array}{l}
i \psi_{1,z} =-\psi _{1,xx}+f(z)(\psi_2+i\psi _{1})+ \chi|\psi _{1}|^{2}    \psi _{1},
\\ 
i\psi_{2,z} =-\psi_{2,xx}+f(z)(\psi_1-i \psi_{2})+ \chi |\psi_{2}|^{2} \psi _{2} ,
\end{array}
\end{eqnarray}
where $f(z)$ describes the modulations, the soliton can be stabilized. 
This fact was established by \textcite{Driben2} who studied numerically
the case $f(z)=f_0\sin(2\pi z/L)$, where $f_0$ and $L$ are the
amplitude and period of the modulation. It was found that the
soliton can indeed be stabilized, and can even become an attractor
with a significantly broad basin.

Existence of stable solitons at the exceptional point was also
reported by \textcite{LiXie14} for the case where the coupling
constant is modulated periodically whereas the dissipation is
constant (or {\it vice versa}).

\subsubsection{Interaction of a soliton with an exceptional point}
\label{sec:solit_except_BHKK}

\begin{figure}
    \begin{center}
        \includegraphics[width=\columnwidth]{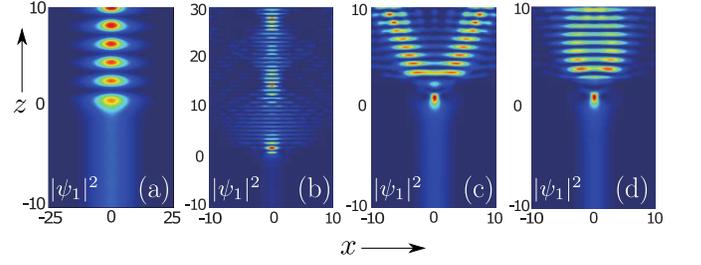}
    \end{center}
    \caption{(Color online) (a) Symmetric soliton passing the defect
        at $\kappa_0=2$, $\ell=1$ (in this case $\ell_{cr}\approx 7$).
        Interaction of anti-symmetric soliton with the defect at $\kappa_0=4$, $\ell=1.1$ (b) $\ell=2.2$ (c) and $\ell=2.7$
        (d) (in this case $\ell_{cr}\approx 3.4$). In (b) broadening is repeated with period $\approx 10$ while the breather period $\approx 0.8$.
        In all panels the  initial conditions are taken in the form (\ref{eq:bright_sol_coupled_NLS}) with $a=1/\sqrt{2}$, the defect is centered at $z=0$, and $\kappa_{min}=1$.
        Only the first component $|\psi_1|^2$ is shown; behavior of $|\psi_2|^2$ is similar. Adapted from~\textcite{BHHK}.}
    \label{fig:SolitExcept}
\end{figure}

By modulating the coupling constant
one can implement a situation where the coupled waveguides have
parameters corresponding to the exceptional point or to the broken
$\PT$ symmetry only at a single point or on a finite segment of the
propagation distance. Such a localized modulation of the coupling
can be referred to as {\em coupling defect}. Following
\textcite{BHHK}, now we consider the interaction of a bright vector
soliton with a coupling defect of the form
$\kappa(z)=\kappa_0-(\kappa_0-\kappa_{min})e^{-z^2/\ell^2}$, where
$\kappa_{min}$ and $\kappa_0$ are the minimum and maximum of the
coupling, and $\ell$ is a defect-length parameter. We consider system
(\ref{eq:NLS_gen}) in the absence of XPM ($\tilde{\chi}=0$), set $\chi=-1$, and take
$\gamma=1$ without loss of generality. The introduced
defect is centered at $z=0$, where the strength of coupling
$\kappa(0)=\kappa_{min}$ is the weakest. Thus when $\kappa_{min}=1$,
the exceptional point is achieved at $z=0$. Far from the defect the
NLS equations are homogeneous, and thus one can consider incidence of a
soliton initially given by (\ref{eq:bright_sol_coupled_NLS}) on the
defect. Numerical simulations revealed various scenarios visualized
in Fig.~\ref{fig:SolitExcept}.

If the defect length $\ell$ exceeds some critical value
$\ell_{cr}(\kappa_{min})$ (which depends on $\kappa_{min}$), then the
soliton energy grows without bound after interaction with the
defect, i.e., the soliton cannot ``overcome'' the defect. If $\ell$
is below $\ell_{cr}(\kappa_{min})$, then after passing the defect a
symmetric soliton is transformed into a breather propagating along
the homogeneous coupler [Fig.~\ref{fig:SolitExcept}(a)].
When an anti-symmetric soliton interacts with the coupling defect,
possible scenarios include  emergence of (quasi-)breathers with periodic broadening of the shape
[Fig.~\ref{fig:SolitExcept}(b)], splitting of a soliton into two
outgoing breathers [Fig.~\ref{fig:SolitExcept}(c)], as well as
splitting of the soliton into two breathers which after some
distance start moving towards each other [Fig.~\ref{fig:SolitExcept}(d)].

\subsubsection{Soliton switching by a $\PT$-symmetric defect}
\label{sec:solit_switch}

Now we turn to propagation of a soliton in a coupler having constant
coupling but localized gain-loss defects.  The problem is modeled by
\begin{eqnarray}
\label{eq:nls_management-except}
\begin{array}{l}
i \psi_{1,z}=-\psi_{1,xx}- \psi_2 - i \gamma_1(z)\psi _{1}- |\psi _{1}|^{2} \psi_{1}
\\
i \psi_{2,z} =-\psi_{2,xx} - \psi_1-i \gamma_2(z)\psi_{2} -  |\psi_{2}|^{2} \psi _{2},
\end{array}
\end{eqnarray}
where coefficients $\gamma_1(z)$ and $\gamma_2(z)$ are arbitrary so
far.

\begin{figure}[t]
    \begin{center}
        \includegraphics[width=0.48\columnwidth]{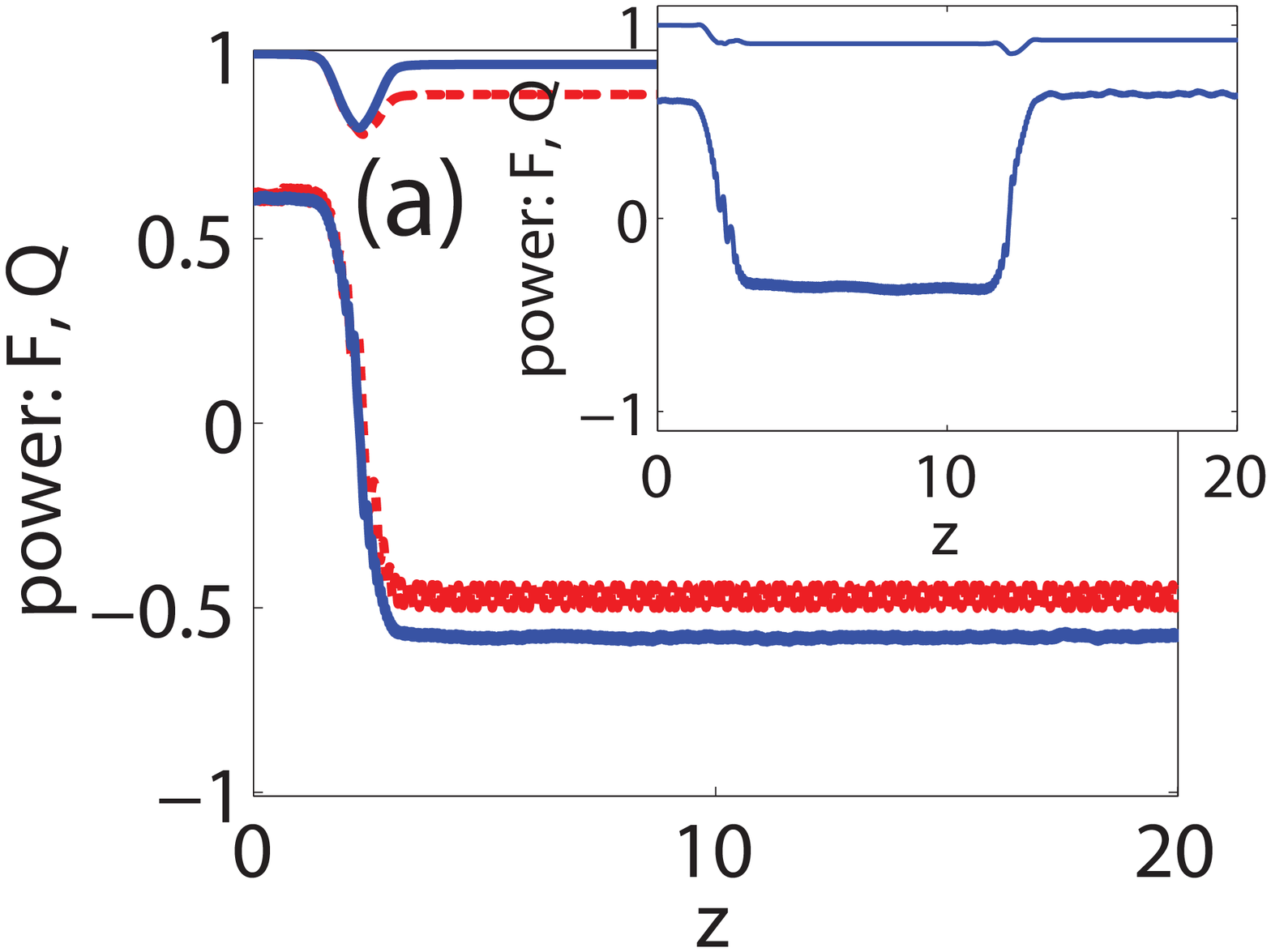}
        \includegraphics[width=0.48\columnwidth]{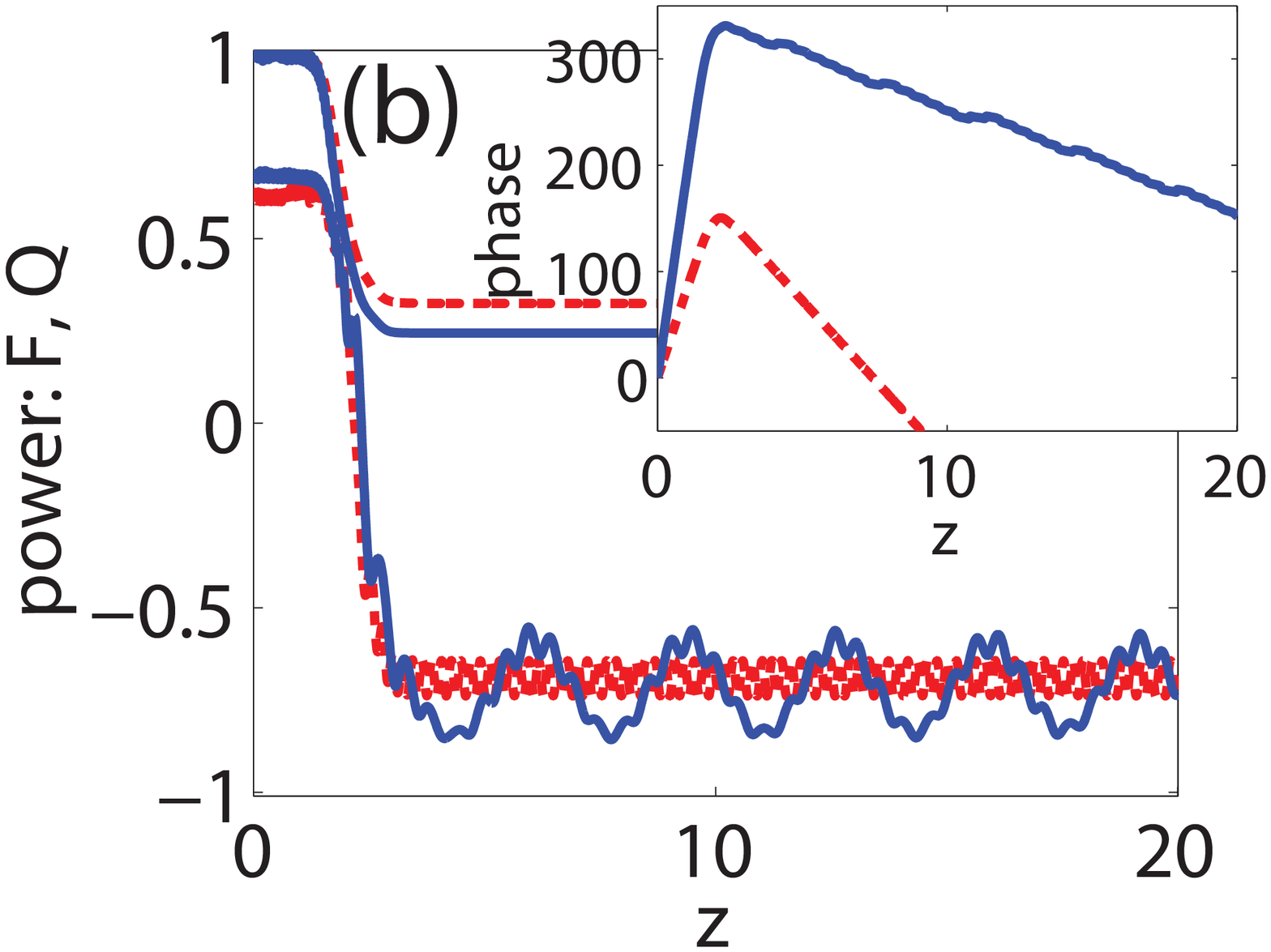}
    \end{center}
    \caption{(a) Soliton switching between arms by the ${\cal PT}$-symmetric segment  $\gamma_2(z)= - \gamma_1(z)=\Gamma \left(   \arctan [5(z-z_a)] -  \arctan [5(z-z_b)] \right)^2$ with $\Gamma= 0.065$, $z_a=1.5$, and $z_b=3.0$.
        In the inset of (a) a gain-loss segment is added at $z \sim 10$. The parameters  were engineered in order to keep $Q(z) \approx 1$. (b) The same as in (a) but with only dissipative segment included in the first arm ($\gamma_2(z)\equiv 0$) and with $\Gamma=0.135$. The inset shows the relative phase.
        In both panels $\psi_1(0)=20\, \textnormal{sech}(10\sqrt{2} x) $ and  $\psi_2(0)=5\, \textnormal{sech}(5  x /\sqrt{2} )$ [i.e. $F(0)=3/5]$. Adapted from~\textcite{AKOS}}
    \label{fig:SolitSwitch}
\end{figure}

Soliton switching can be described by the Lagrangian approach (see
\textcite{ParFlor} for the conservative coupler). This approach
relies on the ansatz $\psi_j=A_je^{\phi_j}/\cosh(ax)$ [cf.
(\ref{eq:bright_sol_coupled_NLS})] where amplitudes $A_{1,2}$ and
phases $\phi_{1,2}$ are considered as slow functions of the
propagation distance $z$. The
Lagrangian equations are~\cite{AKOS}
\begin{subequations}
    \label{mODE}
    \begin{eqnarray}
    F_{z} &=& -\gamma (1-F^2)+ 2   \sqrt{1-F^2}\sin(\phi)   , \label{mODE1}\\
    \phi_z &=& \delta F Q -2 F  \cos(\phi)/\sqrt{1-F^2} ,\label{mODE3} \\
    Q_{z} &=& - \gamma_1 Q(1+F)-\gamma_2 Q(1-F). \label{mODE2}
    \end{eqnarray}
\end{subequations}
Here $F=(P_1-P_2)/(P_1+P_2)$ is the relative distribution of the
power [$P_{1,2}(z)=\int_{-\infty}^\infty |\psi_{1,2}|^2dx$] between
the two arms of the coupler, $Q(z)=(P_1(z)+P_2(z))/(P_1(0)+P_2(0))$
is the total power normalized to the input one,
$\phi=\phi_1-\phi_2$ is the relative phase,
and $ \delta= aP_0/3$.
Equations (\ref{mODE}) are similar to those  describing a coupler
operating in the stationary regime \cite{AKS}. Thus  one can expect
various types of dissipative dynamics of a soliton, reproducing
light propagation in the $x$-independent dimer model. One of such effects, the switching of a
soliton by a dissipative defect localized on the distance interval
$[z_a,z_b]$, is illustrated in Fig.~\ref{fig:SolitSwitch}(a). The
input soliton is mainly concentrated in the dissipative arm ($P_1\gg
P_2$), but the output is concentrated in the active arm, resembling
unidirectional propagation described in Sec.~\ref{sec:PTdimer}.
By adding a new defect
with inverted gain and loss the switching can be repeated [see the inset of
Fig.~\ref{fig:SolitSwitch}(a)].

Notice that  $\PT$ symmetry is not necessary for switching since it
can be achieved even in a purely dissipative coupler [Fig.
\ref{fig:SolitSwitch}(b)].
Moreover, decay of the total power can be strongly suppressed by
increasing the strength of the dissipative defect. This phenomena
can be termed to as {\em macroscopic Zeno effect} \cite{SKon10} which is  a
macroscopic (meanfield) manifestation of the well-known quantum Zeno
effect \cite{FaPa08, Daley14}.  Macroscopic Zeno effect in a BEC
subject to an ionizing electronic beam was experimentally observed
by \textcite{BaLott}.

\subsection{Rogue waves}%
\label{sec:rogue_coupled_NLS}

In the regime of MI, system (\ref{eq:NLS_gen}) supports another type
of localized excitations known as rogue waves~\cite{KPS09}. Here we are
interested in deterministic rogue waves, which are nonlinear
excitations propagating on a nonzero background and localized in space and in time. The simplest rogue-wave solution is the
Peregrine soliton of the NLS equation~\cite{Peregrine}. In order to
construct a counterpart of the Peregrine soliton in the
$\PT$-symmetric coupled NLS equations (\ref{eq:NLS_gen}), one can
employ the reduction
(\ref{seq:PT-NLS_subst})--(\ref{eq:PT-NLS_reduced}). Then if the
condition of the MI (\ref{cond1}) is satisfied, the exact Peregrine
soliton of Eqs.~(\ref{eq:NLS_gen}) reads \cite{BDKM}
\begin{eqnarray}
\psi _{j}(x,z) =\rho e^{(-1)^{j}i\delta /2-ibz}\times  \hspace{3cm}\nonumber \\
\left[ 1-\frac{4\left(1-2i\left( \chi+\tchi \right) \rho ^{2}z\right)}{1-2\left(
    \chi +\tchi \right) \rho ^{2}x^{2}+4\left( \chi+\tchi \right)
    ^{2}\rho ^{4}z^{2}}\right].
\label{eq:peregrine-cur}
\end{eqnarray}%
When $|z|\rightarrow \infty $ or $|x|\rightarrow \infty $, this
solution approaches the constant background given by
Eq.~(\ref{eq:pl-wave-cur}) with $k=0$.

\begin{figure}
    \includegraphics[width=\columnwidth]{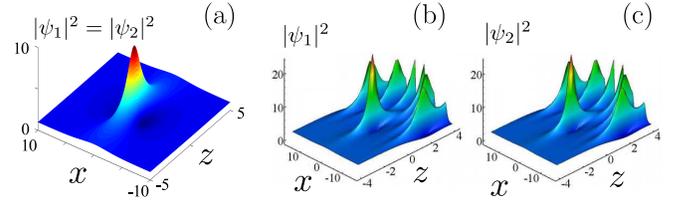}
    \caption{(a) Vector Peregrine soliton (\ref{eq:peregrine-cur})  with  $\rho=1$, $\chi=0.5$, $\tchi =-1$; (b) and (c) Intensities of the Peregrine solitons  for $\rho=1.604$, $\chi=0.5$, $\tchi =-1$, $\delta =\pi/4$, initiated with slightly perturbed initial conditions at $z=z_{ini}=-4$ The shown scenario of the evolution  corresponds to the scenario of MI shown in Fig.~\ref{fig:mi}(a). Adapted from \textcite{BDKM}.}
    \label{fig:Peregrine}
\end{figure}

The analytical solution (\ref{eq:peregrine-cur}) is  illustrated in Fig.~\ref{fig:Peregrine}.
Direct numerical simulations on the dynamics of Peregrine soltions subject to different initial conditions were performed by \textcite{BDKM}. In a generic situation due to the instability an emergence of a single peak follows by the development of the modulational instability, reflecting different scenarios corresponding to different relations among the nonlinear parameters described in Sec.~\ref{sec:bright_coupled_NLS}. An example of such evolution is shown in  Fig.~\ref{fig:Peregrine}(b,c).

System (\ref{eq:NLS_gen}) also admits more complex rogue-wave
solutions called higher-order rogue waves. They can be readily
obtained from higher-order rogue waves of the NLS equation through
the reduction (\ref{seq:PT-NLS_subst})--(\ref{eq:PT-NLS_reduced})
\cite{AAS2009,DGKM2010,GLL2012,OY2012,DH14}.

\subsection{Dark solitons} 
\label{sec:dark_coupled_NLS}

Dark solitons in coupled $\PT$-symmetric NLS equations can exist if
the CW background is stable \cite{BKM}. Under substitution
(\ref{seq:PT-NLS_subst}) we again obtain Eq.
(\ref{eq:PT-NLS_reduced}), but now  we assume $\chi+\tchi>0$.
Dark-soliton solutions of the NLS equation are well
known~\cite{Tsuzuki,FaTa87}. In the particular case of zero
velocity, the respective soliton  (also known as black soliton) reads $\psi^{\mathrm{ds}}(x,z)=u_{0}(x)e^{-ibz}$,
where
\begin{equation}
u_{0}(x)=\rho \tanh \left( \rho \sqrt{(\tchi+\chi)/2}\ x\right).
\label{eq:ds}
\end{equation}

As in the case of bright solitons (see
Sec.~\ref{sec:bright_coupled_NLS_eq}), the linear stability of a slightly perturbed soliton (with  perturbations $\propto e^{i\lambda z}$)   is reduced to two separate eigenvalue problems~\cite{BKM}:
$
L_{1,2}\psi =\Lambda_{1,2}\psi 
$,
where $\Lambda=\lambda^2$, and the   and the operators are defined by
\begin{eqnarray*}
    \label{L12}
    L_{1} = (L_{+}-L)(L_{-}+\cos \delta ),  \label{L2}
    \,\,
    L_{2} = (L_{-}-\cos \delta )(L_{+}+L),
\end{eqnarray*}%
with $L \equiv 2\chi u_{0}^{2}-\cos \delta $ and
\begin{eqnarray*}
    L_{\pm } = -\frac{\partial ^{2}}{\partial x^{2}}-b+[(2\pm 1)\chi
    _{1}+\chi ]u_{0}^{2}.
\end{eqnarray*}
The dark soliton is linearly stable if all eigenvalues
$\Lambda_{1,2}$ are real and positive.
The eigenvalue problem for $L_{2}$ is the  well-studied
stability problem for the black soliton in the conservative defocusing medium.
It is known that   $L_{+}+L$ is positive definite, and $L_{-}-\cos \delta $ has only one negative
eigenvalue and one zero eigenvalue~\cite{Barash_stabil}. It is
also known  that the minimal eigenvalue of $L_{2}$ is
positive \cite{Chen_stabili}. Thus,  $L_{2}$ does not give
instability, and the analysis is reduced to the study of operator $%
L_{1}$.


Linear stability of $\PT$-symmetric dark solitons was studied by \textcite{BKM}.
Stable dark solitons are robust and their collision is almost
elastic. If the system has a  weak imbalance between
gain $\gamma_1$ and loss $\gamma_2$ in the two waveguides, dark
solitons can still survive for a long time.


\subsection{Generalized $\PT$-symmetric coupled NLS equations}
\label{sec:generalized_couples}

The above studies do not exhaust rich dynamics governed by coupled $\PT$-symmetric NLS equations, including in particular resonant mode interactions in general and peculiar four wave mixing, in particular~\cite{WSKT}. Furthermore, generlizations of the model itself is possible, which is considered below in this section.

\paragraph{Circular arrays.}

As in Sec.~\ref{sec:3} where a dimer model was generalized to
$\PT$-symmetric oligomers, two coupled $\PT$-symmetric NLS equations
can be generalized to an array of $N$ waveguides. \textcite{BBA}
studied NLS equations
assembled in open and closed $\PT$-symmetric arrays with alternating 
and clustered 
gain-loss configurations (see Sec.~\ref{subsec:open_closed}). Here
we consider the alternating closed (necklace) configuration, modeled
by
\begin{equation}
i \psi_{n,z}+\psi_{n,xx}+2|\psi_n|^2\psi_n+\psi_{n-1}+\psi_{n+1}
=2i(-1)^n\gamma\psi_n,
\end{equation}
with $n=1,...,2N$, under boundary conditions $\psi_{2N+1}=\psi_1$
and $\psi_{2N}=\psi_0$. The $\PT$-symmetry breaking threshold of
this system is not affected by the dispersive terms $\psi_{n,xx}$.
Therefore the linear waves are stable if $\gamma <\gamma^{(an)}_\PT$ in (\ref{eq:discr:oa}). [Note that in a system
with alternating dispersion, i.e., with alternating signs in front
of the second derivative, $\PT$ symmetry is always broken
\cite{Gupta14}].

If $\PT$ symmetry is unbroken, a solitonic solution can be searched
in the form
$\psi_n=e^{i\phi_n+ia^2z}a\sech(ax)$,
where phases $\phi_n$ are determined from the relations
\begin{eqnarray*}
    \label{eq:necklace_phases}
    e^{-i\varphi_{n-1}}+ e^{i\varphi_{n}}=2 i(-1)^n\gamma,\quad \varphi_n=\phi_{n+1}-\phi_n.
\end{eqnarray*}
These equations yield 
$\phi_n=(-1)^n\arcsin\gamma +\pi n+\phi$,
where $\phi$  is a constant phase (appearing due to the phase invariance of the system).

\paragraph{Multidimensional NLS equations and wave collapse.}
Another extension of the $\PT$-symmetric
coupler model is coupled $\PT$-symmetric multidimensional NLS equations with more general nonlinearities, i.e., %
\begin{eqnarray}
\label{eq:nls_multidim}
\begin{array}{l}
i\psi_{1z} =-\nabla ^{2}\psi _{1}- \kappa \psi_2 - F_1\left(|\psi _{1}|,|\psi _{2}|\right)
\psi _{1}+i\gamma \psi _{1},
\\ [1mm]
i\psi_{2z} =-\nabla^{2}\psi_{2}-\kappa \psi_1 - F_2\left(|\psi_{1}|,|\psi_{2}|\right)
\psi _{2}-i\gamma  \psi_{2},
\end{array}
\end{eqnarray}
where $x\in\mathbb{R}^N$, $\nabla=\left(\partial/\partial x_1,...\partial/\partial x_N\right)$, and $F_j(\cdot,\cdot)$ describe the nonlinearities.

Starting with the case of cubic nonlinearity $F_1 = \chi |\psi_1|^2
+  \tchi |\psi_2|^2$ and $ F_2 = \tchi |\psi_2|^2 +  \chi
|\psi_1|^2$, we recall that for focusing SPM and XPM
($\chi,\tchi\geq 0$), solutions of a single NLS equation with $N\geq
2$  suffer finite-time blowup for a
wide range of initial conditions~\cite{Sulem}, even in the presence
of linear dissipation~\cite{Tsutsumi}. For the $\PT$-symmetric
system (\ref{eq:nls_multidim}) with critical dimensionality $N=2$,
no exact result on the global existence or blow-up of the solution
is available for general coefficients $\kappa$, $\gamma$, $\chi$ and
$\tchi$. However, a sufficient condition for global existence can be
formulated in the particular case of $\gamma<\kappa$ and
$\chi=\tchi$. In this   case, extending the arguments presented in
Sec.~\ref{sec:coupler:basic} on the 2D case, one can show that there
exists a {\it priori} upper bound $S_{max} =\sup_z S_0(z) < \infty$
for the Stokes component
$S_0(z)=\|\psi_1\|_{L^2(\mathbb{R}^2)}^2+\|\psi_2\|_{L^2(\mathbb{R}^2)}^2$.
The value of $S_{max}$  depends on the initial conditions. Using
this fact,   \textcite{PZK14} showed that if $\chi=\tchi=1$ and the
initial conditions satisfy the requirement $S_{max}<\frac 12
\|R\|_{L^2}^2$, then a global solution in $H^1(\mathbb{R}^2)\times
H^1(\mathbb{R}^2)$ does exist. Here,   $R=R(x)$  is the (unstable)
Townes soliton \cite{Townes}, i.e., the localized positive solution
of the 2D stationary problem $\nabla^2R-R+R^3=0$.

Study of 2D bright solitons withing the framework of (\ref{eq:nls_multidim})  with focusing cubic
nonlinearity and defocusing quintic nonlinearity, i.e.,
$F_j=|\psi_j|^2-|\psi_j|^4$ ($j=1,2$), was
reported  by  \textcite{BM13}.  Using the substitution
(\ref{seq:PT-NLS_subst}), the system is reduced to a single 2D NLS
equation whose radially symmetric solutions provide the shapes for
$\PT$-symmetric solitons. It was found that
the 2D solitons can be dynamically stable for $\gamma<\gamma_C$,
where $\gamma_C$ is some critical value which depends on the coupling
strength. Solitons in a closed array of three 2D waveguides with the cubic-quintic nonlinearity were studied  by \textcite{FZK15}.

In the supercritical case $N\geq 3$, the $\PT$-symmetric coupled NLS
equations with cubic nonlinearity may undergo finite-time blow-up,
whose sufficient conditions  were
established by \textcite{DFKZ14}. Numerical studies reveal that the
model features  different evolution scenarios, including decay of
the initial pulses, growth of the solution in the active or lossy
component, or both.

\subsection{Localized modes in $\CPT$-symmetric BECs}
\label{sec:CPT_BEC}

Following~\textcite{KaKoZe14} we now turn to the NLS equations with
gain and loss and linear SO-type coupling discussed in
Sec.~\ref{subsec:linear-CPT-BEC}. The two-body interactions can be
approximated by almost equal nonlinear coefficients [in an
experiment the difference was less than 1\%~\cite{LJ-GS11}]. This leads to the
coupled GPEs $i\bPsi_t = H\bPsi - \chi(\bPsi^\dag\bPsi)\bPsi$, where
the linear Hamiltonian $H$ is given by (\ref{eq:SO_BEC_Hamilt}).

\begin{figure}
    \includegraphics[width=\columnwidth]{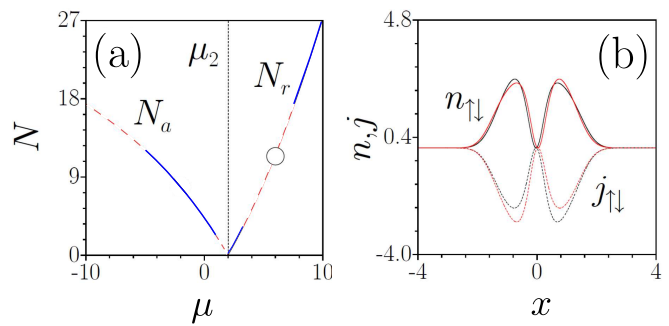}%
    \caption{ (Color online)
        Families of nonlinear modes of a SO-BEC in a parabolic trap $V(x)=\nu^2x^2/2$ bifurcating from $n=2$ (a) linear modes for $\kappa=1$, $\nu=2$, $\gamma=0.2$, and $\omega=0.5$.
        Solid blue (dashed red) curves correspond to stable (unstable) modes. Here $N=\int_{-\infty}^\infty \bPsi^\dag\bPsi dx$, and curves with labels $N_a$ and $N_r$ correspond to attractive and repulsive nonlinearities, respectively. Circle  in (a) correspond to a two-hump nonlinear mode  shown in (b).  Adapted from \textcite{KaKoZe14}.
    }
    \label{fig:SO_PT_nonlin}
\end{figure}

Stationary modes $\bPsi=e^{-i\mu t}\bpsi(x)$ can bifurcate from the
linear eigenstates.
From properties of the underlying linear system, one can identify
two fundamental (one-hump) nonlinear modes,  two two-hump modes,
etc. The SO-coupling induces nonzero currents $j_{\ua\da} =
\frac{1}{2i} \left(\Psi_{\ua\da}^{*}\frac{\partial
    \Psi_{\ua\da}}{\partial x}-\frac{\partial \Psi_{\ua\da}^{*}}{
    \partial x} \Psi_{\ua\da}\right)$, whose directions  in the
$|\ua\rangle$ and $|\da\rangle$ components coincide. At the same
time, directions of the currents in the two fundamental  modes are
opposite (the same being true for the two-hump modes, three-hump
modes, etc). Families of nonlinear modes [see
Fig.~\ref{fig:SO_PT_nonlin}] consist of alternating intervals of
stable and unstable segments, and stable nonlinear modes exist for
both attractive and repulsive nonlinearities.
Moreover, stable nonlinear modes exist even if the $\CPT$ symmetry
of the linear problem is broken.
$\CPT$ symmetry [specifically, the property
$\CPT\sigma_3=-\sigma_3\CPT$] implies that the nonlinear modes have
zero (pseudo-)magnetization: $M=\int_{-\infty}^\infty\bPsi^\dag\sigma_3\bPsi\, dx =0$.

\section{Nonlinear modes in complex  potentials}
\label{sec:5}

In this section,  we consider nonlinear modes supported by complex
potentials $U(x)$ in the NLS equation (\ref{opt:NLS}). More
specifically, we focus on potentials which are either localized,
$U(x)\to 0$, or unbounded, $U(x)\to \infty$, as $x\to\pm \infty$.
Stationary nonlinear modes in this equation are of the form $\Psi(x,
t)=\psi(x)e^{i\mu t}$, where $\mu$ is a real propagation parameter
and $\psi(x)$ solves the equation
\begin{equation}
\label{e:parabol:psi}
\psi_{xx} {-}U(x)\psi+g |\psi|^2\psi=\mu \psi,
\end{equation}
subject to the zero boundary conditions $\lim_{|x|\to \infty}\psi=0$.

If the underlying liner equation, i.e., Eq. (\ref{e:parabol:psi})
with $g=0$, admits a guided mode with a real propagation constant,
then a question of interest is the possibility for nonlinear modes
to bifurcate from that mode.

\subsection{Localized potentials}
\label{subsec:par:localized}

\paragraph*{Exact solutions.} Starting with localized potentials, we notice that linear spectra of some of them are available analytically \cite{Cooper95, Znojil2000}.
Moreover, many of such potentials admit exact expressions for
nonlinear modes.
The first known example corresponds to a $\PT$-symmetric Scarff~II
potential
\begin{equation}
\label{scarf}
U(x) =  {-}V_1 \sech^2x -  i V_2 \sech x \tanh x,
\end{equation}
with $V_1>0$ and $V_2\ne 0$, which is a complexification of the real
Scarff~II potential \cite{Cooper95}. Spectrum of (\ref{scarf}) was
found analytically by \textcite{Ahmed282,Ahmed287} through a
transformation of the corresponding Schr\"odinger equation to the
Gauss hypergeometric equation and by \textcite{BQ00,BQ02} using
complex Lie algebras. If $|V_2|<V_{cr}=V_1 + 1/4$, then the discrete
spectrum consists of a sequence of real eigenvalues.   At $|V_2|=V_{cr}$ the   real eigenvalues merge pairwise and split
into a complex-conjugate pairs as $|V_2|$ exceeds $V_{cr}$, i.e.,
$\PT$ symmetry becomes broken.

The nonlinear model
(\ref{e:parabol:psi})-(\ref{scarf}) admits an {\em exact} particular
solution for the focusing ($g>0$) and defocusing ($g<0$)
nonlinearities at $\mu=1$ \cite{Musslimani_1,SJZL11}:
\begin{equation}
\label{exact_scarf}
\psi =  \sqrt{\frac{V_1 - (V_2/3)^2 -2}{g}} \,\, \frac{\exp\left[i{(V_2/3)} \arctan(\sinh x)\right]}{ \cosh x}.
\end{equation}
In (\ref{exact_scarf}) it is assumed that parameters $V_{1,2}$ and
$g$ are chosen so that the expression under the radical is
positive.

The nonlinear mode (\ref{exact_scarf}) is $\PT$ symmetric, i.e.,
$\psi(x)=\psi^*(-x)$.
This mode belongs to a continuous family of localized
$\PT$-symmetric modes (fundamental solitons) which can be obtained
numerically by varying the propagation constant $\mu$ at fixed model
parameters $V_{1}$, $V_2$ and $g$. Numerical study of fundamental
and multipole solitons and their stability in the potential
(\ref{scarf}) was performed for focusing \cite{Musslimani_1} and
defocusing \cite{SJZL11,CHQ14} Kerr nonlinearities, as well as for
nonlocal nonlinearity \cite{SLZJ12}.

$\PT$-symmetric extension of the Rosen--Morse~II potential
$U(x) =  {-}V_1 \sech^2x + iV_2 \tanh x$ is another
potential which admits  explicit expressions for particular nonlinear
modes~\cite{Midya2013}. Notice that in this case only the real part
of the potential vanishes as $x\to \pm \infty$, while the imaginary
part approaches constant values.  The linear spectrum of this Rosen--Morse~II
potential was obtained by \textcite{Levai09}. Explicit expressions
for nonlinear modes in a more sophisticated potential  $U(x)=-V_1
\sech^2x +V_2^2 \sech^4x + 4iV_2 \sech^2 x \tanh x$ were reported by
\textcite{Musslimani_2} and generalized by \textcite{Midya2014,
    Khare12}. More generally, potentials allowing for exact solutions
can be constructed systematically using the ``inverse engineering''
approach~\cite{AKSY} which consists in assuming the given  field pattern and finding a potential shape sustaining such a pattern, or using the similarity
transformation~\cite{SerHas,PGTK} reducing a non-autonomous NLS
equation (with a potential $U(x,t)$ and time-depending coefficients)
to the autonomous NLS equation (\ref{opt:NLS}) \cite{DW14a,CDW14,
    DW14c,DWZ14}.

Exact particular solutions are also available in multi-dimensional
$\PT$-symmetric potentials [see also Sec.~\ref{sec:high_dim}],
including 2D and 3D versions of Scarff and Rosen--Morse
potentials~\cite{WDW14,DW14b,HC14,DW14c, DWZ14,WDW14_3D}.

One more example admitting explicit nonlinear solutions is
$\PT$-deformation of a parabolic potential \cite{MidyaGaussSoliton14} 
\begin{eqnarray}
\label{eq:parab_gauss}
U(x) = \Omega^2 x^2 - V_0e^{-2x^2} + i \gamma xe^{-x^2}.
\end{eqnarray}
Numerical study of linear spectrum for the   $\PT$-symmetric Gaussian potential [which corresponds to $\Omega=0$ in (\ref{eq:parab_gauss})]  was
performed by \textcite{Ahmed282}, and nonlinear modes were computed numerically
by \textcite{HMLYZH11,Jisha14}.

\textcite{KMT14} demonstrated that
stable solitons in the \emph{defocusing} nonlinearity can be found
in the absence of any real symmetric part, i.e., when
$U(x) = i \gamma xe^{-x^2}$, provided that the nonlinearity is spatially
modulated, and its profile grows rapidly enough as $x\to\pm \infty$,
e.g., $g=g(x) = -(g_1 + g_2 x^2) e^{x^2}$.  Existence of bright
solitons in such self-defocusing nonlinearity can be explained by
``nonlinearizability'' of the respective NLS equation at the soliton
tails.

\paragraph{Scattering on a $\PT$-symmetric defect.}

\begin{figure}
    \includegraphics[width=0.7\columnwidth]{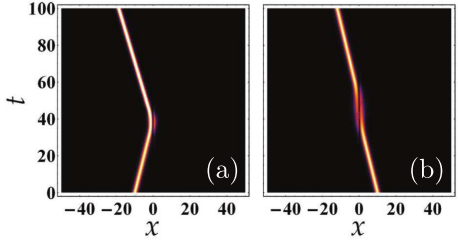}%
    \caption{(Color online) Scattering of a soliton by the potential $U(x)=-V_0^2\sech(V_0 x) + iW_0x\sech(V_0 x)$ with $V_0=W_0=-2$.
        (a) Almost perfect reflection of the soliton moving from the left. (b) Almost perfect transmission of the soliton moving from the right.
        In both cases the velocity of the incident soliton is $|v_0|=0.25$, and the initial position is at $|x_0|=10$. From  \textcite{AKAMBK13}.  }
    \label{fig-localized:unidir}
\end{figure}

A localized $\PT$-symmetric potential can be viewed as a defect
which scatters an incident wave. For certain cases, including the $\PT$-symmetric Scarf~II potential
(\ref{scarf}), the scattering data for the linear problem can be
found in an explicit form \cite{Cannata07,Levai01}.
In the nonlinear setting, one can
consider incidence of a soliton on a $\PT$-symmetric defect.
Numerical study of soliton scattering by (\ref{scarf}) reveals
several dynamical scenarios \cite{NNMF12,Karjanto15}. They include swinging and
self-trapping of the normally incident soliton, as well as
non-reciprocity of left- and right-incidence. Asymmetric evolution
of two simultaneously launched solitons on a 
$\PT$-symmetric defect was reported in \textcite{NNMFR13}. It was
also found by \textcite{AKAMBK13,AM14} that in a certain range of
the parameters of the potential and the incident soliton, one can
observe a unidirectional soliton flow, i.e., the soliton moving from
the left to right is almost perfectly reflected, while the soliton
moving in the opposite direction is almost perfectly transmitted
[see Fig.~\ref{fig-localized:unidir}]. Unidirectional transmission,
amplification and destruction of gap solitons (supported by a real
periodic potential) on a $\PT$-symmetric defect was also observed by
\textcite{ABS13}.

\subsection{Parabolic potential}
\label{subsec:parabol:parabol}

Now we turn to stationary nonlinear modes of
Eq.~(\ref{e:parabol:psi}) with a parabolic $\PT$-symmetric potential
(\ref{parabolic}).
Linear eigenmodes of this potential are given by ${\tmu}_n =
-(2n+1)$, $n=0,1, \ldots$, and the eigenfunctions $\tpsi_n(x)$ can
be expressed in terms of Hermite polynomials [see
Eqs.~(\ref{spect:parab})]. We first look for \emph{small-amplitude}
nonlinear modes bifurcating from the linear eigenstates
$\tpsi_n(x)$. In this case, the nonlinear modes can be constructed
by an asymptotic expansion
\begin{equation}
\label{eq:par:expans}
\begin{array}{l}
\psi_n(x)=\vep\tpsi_n + \vep^3 \psi_n^{(3)} + {o}(\vep^3), \\
\mu  =  {\tmu}_n +
g\vep^2\mu_n^{(2)} + o(\vep^2),
\end{array}
\end{equation}
where $\vep\ll 1$ is a small real parameter.
Substituting this expansion into Eq.~(\ref{e:parabol:psi}) and
collecting terms of order $\vep^3$ one obtains an equation for
$\psi_n^{(3)}$:
\begin{equation*}
(\psi_n^{(3)})_{xx}  -\tilde{\mu}_n \psi_n^{(3)}  - (x-i\alpha)^2\psi_n^{(3)} =  g\mu_n^{(2)}\tpsi_n- g  |\tpsi_n|^2 \tpsi_n.
\end{equation*}
The solvability condition (Fredholm alternative) for  this equation
requires its right-side term to be orthogonal to the kernel of the
adjoint operator in the left hand side, i.e., orthogonal to
$\tpsi_n^*$. This allows one to compute $\mu_n^{(2)}$ as
\cite{ZK_12_PRA,Yang14b}:
\begin{equation}
\label{eq:parabol:mu2}
\mu_n^{(2)} = \left. {\int_{-\infty}^\infty \tpsi_n^3(x)\tpsi_n^*(x)dx}\right/ {\int_{-\infty}^\infty \tpsi_n^2(x)dx}.
\end{equation}
For expansions (\ref{eq:par:expans}) to be meaningful, $\mu_n^{(2)}$
must be real [a similar constraint also arises for discrete systems
in Sec.~\ref{sec:discr:fam:linear}]. When $\alpha=0$, eigenfunctions
$\tpsi_n(x)$ are real-valued, hence coefficients $\mu_n^{(2)}$ are
positive for all $n$. When $\alpha\ne 0$, $\tpsi_n(x)$ are
complex-valued. However, parity of their real and imaginary parts
ensures that the coefficient $\mu_n^{(2)}$ is still real for any $n$
and $\alpha$. Thus for each $n$ one can identify a continuous family
of nonlinear modes $\psi_n(x)$ bifurcating from the $n$th linear
eigenstate $\tpsi_n(x)$.

\begin{figure}
    \includegraphics[scale=0.8]{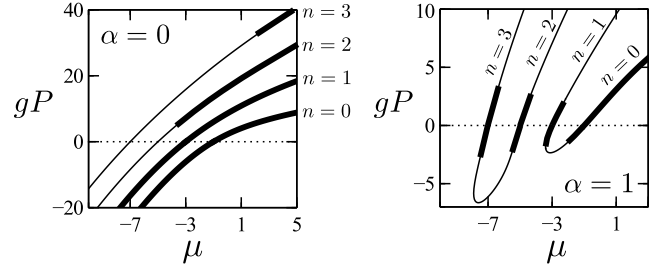}%
    \caption{Families of nonlinear modes in the conservative ($\alpha=0$) and  $\PT$-symmetric ($\alpha=1$) parabolic potential (\ref{parabolic}) and focusing ($g=1$) and defocusing ($g=-1$) nonlinearities. Bold segments on the power curves correspond to stable
        nonlinear modes. Adapted from \textcite{ZK_12_PRA}.
    } \label{fig-fams}
\end{figure}

\begin{figure}
    \includegraphics[scale=0.6]{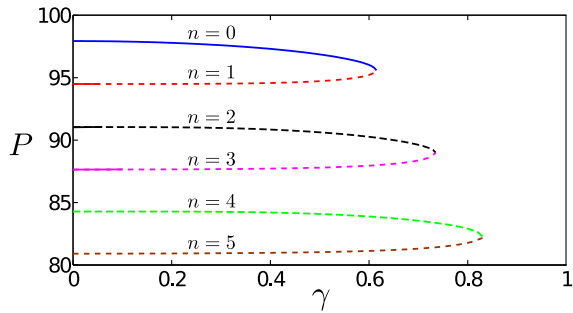}%
    \caption{(Color online) The power $P$ as a function of the strength $\gamma$
        of the imaginary potential in  (\ref{eq:parab_gauss}) with $V_0=0$.
        The bifurcation diagram shows the merging of different solution branches through saddle-node bifurcations,
        Solid (dashed) lines indicate dynamically stable (unstable) solutions.
        Here, $g=-1$, $\Omega\approx 0.07$ and  $\mu = - 3$.  Adapted from \textcite{Achil}.
    } \label{fig-achil}
\end{figure}

Further analysis of these nonlinear modes can be
performed numerically. The continuous families can be visualized as
curves $g P_n(\mu)$, where $P_n=\int_{-\infty}^\infty |\psi_n|^2dx$
is the power of the $n$th mode, see Fig.~\ref{fig-fams}. Each point
above (below) the axis $gP=0$ corresponds to a nonlinear mode under
focusing (defocusing) nonlinearity. A striking difference between
the two panels in Fig.~\ref{fig-fams} is \emph{coalescence} of
nonlinear modes bifurcating from different linear eigenstates, which
does not occur in the conservative parabolic potential ($\alpha=0$)
\cite{KAT01,KKRF05}, but becomes possible in its $\PT$-symmetric
counterpart ($\alpha\ne 0$) \cite{ZK_12_PRA}.  This coalescence can
be described in terms of a saddle-node bifurcation \cite{GP14}.
A similar scenario of collisions of nonlinear modes can be observed
if instead of the propagation constant $\mu$ one varies parameters of the
$\PT$-symmetric potential. Such collisions were observed in the
$\PT$-symmetric potential (\ref{eq:parab_gauss}) with $V_0=0$ and
varying values of $\gamma$ at a fixed propagation constant
\cite{Achil} [see Fig.~\ref{fig-achil}],
in a $\PT$-symmetric double-well potential \cite{Dast,Dast_b,CHDW12},
and in a slab waveguide with a piece-wise constant complex potential
\cite{TTA12}. In a way this behavior is reminiscent of linear $\PT$
phase transition: \textit{cf.} the $\PT$-symmetry breaking diagram
in Fig.~\ref{fig:spectrum_Bender}.

\subsection{Symmetry breaking of solitons}
\label{subsec:par:symm_break}

\begin{figure}
    \begin{center}
        \includegraphics[width=\columnwidth]{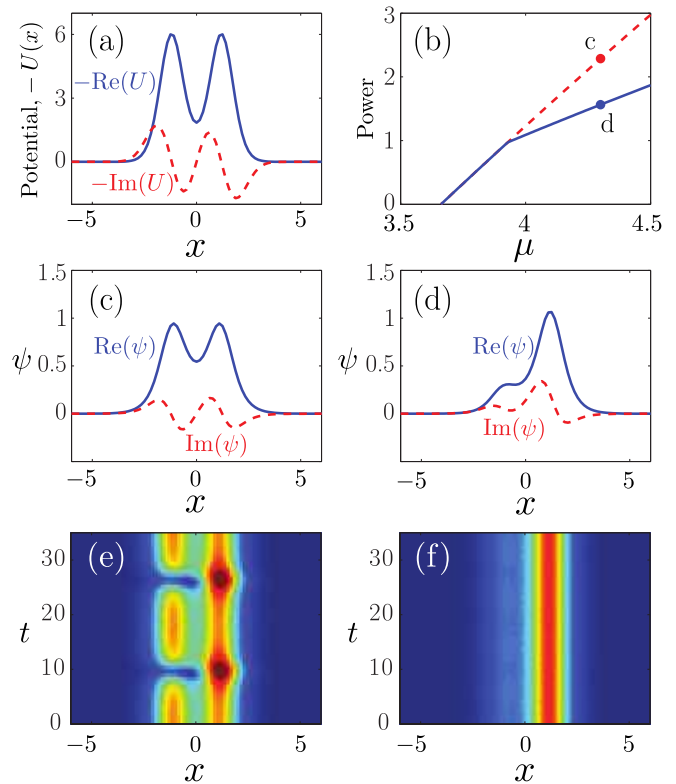}
        \caption{(Color online) (a) {Inverted plot} of the $\mathcal{PT}$-symmetric potential (\ref{e:VPT}) generated by (\ref{e:doublewell}) with $A_-=A_+=2$, $x_0=1.2$, $x_0=1.2$, and $\alpha=1$.
            (b) Power diagram for nonlinear modes with $g=1$ (solid blue: stable solitons;
            dashed red: unstable solitons). (c,d) $\mathcal{PT}$-symmetric and asymmetric solitons
            corresponding to the point `c'  and `d' on the power diagram
            with $\mu=4.3$; (e,f) Evolution of the solitons shown in (c,d) under $1\%$ random-noise
            perturbations. Adapted from \textcite{Yang14d}.
            \label{SB_fig1}}
    \end{center}
\end{figure}

The continuous families of nonlinear modes discussed above
are $\mathcal{PT}$-symmetric, i.e., satisfy $\PT \psi=\psi$ (up to a
phase shift $\psi \to \psi e^{i\theta}$, $\theta \in \mathbb{R}$).
This observation raises a question: can $\mathcal{PT}$-symmetric
systems admit continuous families of non-$\mathcal{PT}$-symmetric
solitons?
In conservative systems continuous families of asymmetric solutions
can exist due to symmetry-breaking bifurcations, where asymmetric
solitons bifurcate out from the base family of symmetric solitons as
the power ($L^2$-norm) of symmetric solitons exceeds a certain
threshold.
This symmetry-breaking usually occurs in a double- (or multi-) well
real potential or in a periodic potential
\cite{JW04,Kirr08,Saccetti09,AHY12,Yang2012, Malomed13}. However,
most of the studies of double-well and periodic $\PT$-symmetric
potentials \cite{Dast, Dast_b,Rodrigues13, cartar,
    Musslimani_1,NGY12,LHLD,MaMaRe,CHDW12} did not report non-$\PT$-symmetric
nonlinear modes with real propagation constants.
Indeed, continuous families of non-$\mathcal{PT}$-symmetric solitons
cannot be expected intuitively, since it is ``difficult'' for those
solitons to balance gain and loss. This intuition is supported by
mathematical analysis of \textcite{Yang14a}, who showed that for
non-$\mathcal{PT}$-symmetric soliton families to exist in a
$\mathcal{PT}$-symmetric potential, infinitely many nontrivial
conditions must be satisfied simultaneously, which is generically
impossible. However, in a generalized Wadati potential (\ref{VW})
\begin{equation}  \label{e:VPT}
U(x)= {-}[w^2(x)+\alpha w(x)+ i w'(x)],
\end{equation}
where $w(x)$ is a real and even function and $\alpha$ a real
constant, symmetry breaking of solitons can occur, and continuous
families of non-$\mathcal{PT}$-symmetric solitons
are possible
\cite{Yang14d}.   As an example, we consider
\begin{equation}
\label{e:doublewell} w(x)=A_-e^{-(x+x_0)^2}+A_+e^{-(x-x_0)^2},
\end{equation}
where $A_\pm$ and $x_0$ are constants.
When $A_-=A_+$, this function generates a $\mathcal{PT}$-symmetric
double-well potential $U(x)$
plotted in Fig.~\ref{SB_fig1}(a). The linear spectrum of this
potential is all-real (see Sec.~\ref{subsec:IST})  and contains
three positive isolated eigenvalues, the largest being $\approx
3.6614$. From this largest discrete eigenmode, a family of
$\mathcal{PT}$-symmetric solitons bifurcates out. Under focusing
nonlinearity ($g=1$), the power curve of this solution family is
shown in Fig.~\ref{SB_fig1}(b), and the soliton profile at the
marked point `c'
is displayed in
Fig.~\ref{SB_fig1}(c). At the
propagation constant $\mu_c\approx 3.9287$ of this base power
branch, a family of \emph{non-$\mathcal{PT}$-symmetric} solitons bifurcates
out. The power curve of this non-$\mathcal{PT}$-symmetric family is
also shown in Fig.~\ref{SB_fig1}(b). At the marked point `d' of the
bifurcated power branch, the non-$\mathcal{PT}$-symmetric solution
is displayed in Fig.~\ref{SB_fig1}(d). The most of the
energy in this soliton resides on the right side of the potential.

From Eq. (\ref{e:parabol:psi}) one can see that
if $\psi(x)$ is a solution, so is $\psi^*(-x)$. Thus for each of the
non-$\mathcal{PT}$-symmetric solitons $\psi(x)$ in
Fig.~\ref{SB_fig1}(b), there is a companion soliton $\psi^*(-x)$
whose energy resides primarily on the left side of the potential.
Thus this symmetry-breaking bifurcation is pitchfork-type.

Linear stability analysis shows that the base family of
$\mathcal{PT}$-symmetric solitons is stable before the bifurcation
point ($\mu<\mu_c$) and becomes unstable when $\mu>\mu_c$ due to the
presence of a real positive eigenvalue. However, the bifurcated
family of non-$\mathcal{PT}$-symmetric solitons is stable.
To corroborate these linear stability results, in
Fig.~\ref{SB_fig1}(e,f) we show direct simulations of soliton
evolutions under initial $1\%$ random-noise perturbations. It is
seen from Fig.~\ref{SB_fig1}(e) that the $\mathcal{PT}$-symmetric soliton in
Fig.~\ref{SB_fig1}(c) breaks up and becomes
non-$\mathcal{PT}$-symmetric. Upon further propagation, the solution
bounces back to almost $\mathcal{PT}$-symmetric again, followed by
another breakup. In contrast, Fig.~\ref{SB_fig1}(f) shows that the asymmetric
soliton in Fig.~\ref{SB_fig1}(d) is stable against perturbations.

It is noted that this symmetry-breaking bifurcation also occurs for
many other potentials of the form (\ref{e:VPT}), including periodic
potentials \cite{Yang14d}.

\subsection{Soliton families in asymmetric complex
    potentials}
\label{subsec:par:nonPT}

In a generic complex non-$\mathcal{PT}$-symmetric potential
[$U(x)\ne U^*(-x)$], continuous families of solitons are not
expected even if $U(x)$ has all-real linear spectra \cite{Yang14b}.
Indeed, let us
suppose that $U(x)$ is a complex
potential which admits a simple isolated real eigenvalue $\tilde{\mu}_n$, with the corresponding
localized eigenfunction $\tilde{\psi}_n(x)$.
If a soliton family bifurcates out from this linear eigenmode, then
in the small amplitude limit, one can expand these solitons into a perturbation series
analogous to (\ref{eq:par:expans}) and, following the analysis of Sec.~\ref{subsec:parabol:parabol}, we recover that the  bifurcation of the  continuous family from the real eigenvalue $\tilde{\mu}_n$ is possible only if the coefficient $\mu_n^{(2)}$ defined by Eq.~(\ref{eq:parabol:mu2}) is real.
As we saw in Sec.~\ref{subsec:parabol:parabol} on the example of a
parabolic potential, if $U(x)$ is $\PT$ symmetric, the reality of
$\mu_n^{(2)}$ is satisfied automatically. However, for a generic
complex potential $U(x)$ this condition is likely to fail.
Moreover, reality of $\mu_n^{(2)}$ is only the first condition for a
soliton family to bifurcate out from a linear mode. As we pursue
this perturbation calculation to higher orders, infinitely many more
nontrivial conditions would need to be met \cite{Yang14b}. This
shows that in a generic non-$\PT$-symmetric potential $U(x)$,
soliton families should not exist.

\begin{figure}
    \includegraphics[width=\columnwidth]{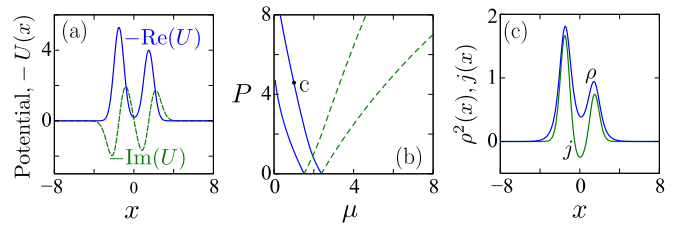}%
    \caption{(Color online) (a)  {Inverted} plot of the asymmetric
        potential $U(x)$ for $w(x)$ defined by (\ref{e:VPT})--(\ref{e:doublewell}) with $A_-=2.3$, $A_+=2$, $x_0=1.5$, and $\alpha=0$.
        (b) Power diagram for nonlinear modes.
        Solid blue (dashed green) lines correspond to defocusing (focusing) nonlinearity.
        (c) 
        Intensity $\rho^2$ and current $j=v\rho^2$ for the nonlinear mode  marked as ``c''  
        in panel (b). Adapted from \textcite{KZ14b}.}
    \label{fig-parabol:asymmetric}
\end{figure}

These generic arguments however do not apply if the system has
hidden symmetries. \textcite{FatkOL14} found numerically that
continuous families of nonlinear modes do exist in a one-hump
asymmetric Wadati potential (\ref{e:VPT}) generated by
$w(x)=\eta/\cosh[a(x) x]$, where $\eta$ is a real constant, and
$a(x)$ is a step function: $a(x)=a_-$ at $x<0$, and $a(x)=a_+$ at
$x>0$, with $a_{\pm}$ being real unequal constants. Explanation for
this observation given by~\textcite{KZ14b} consists in the existence
of an additional conservation law. Indeed, using the polar
representation for the stationary nonlinear mode $\psi(x)=\rho(x)
e^{i\int v(x)dx}$, we rewrite the stationary equation
(\ref{e:parabol:psi}) with potential (\ref{e:VPT}) in the
hydrodynamic form (without loss of generality, we assume $\alpha=0$
in this subsection)
\begin{equation}
\label{eq:nonPT:hydro}
\begin{array}{l}
\rho_{xx} - \mu \rho + w^2 \rho + g \rho^3 - v^2 \rho = 0,\\[1mm]
2\rho_x v + \rho v_x + w_x \rho = 0.
\end{array}
\end{equation}
It is straightforward to verify that these equations admit a
conserved quantity
\begin{equation}  \label{ICmotion}
I = \rho_x^2 + \rho^2(v+w)^2 - \mu \rho^2 +  {g}\rho^4 / 2,\quad  dI/dx\equiv 0.
\end{equation}
Due to existence of this integral of motion, for
each value of the propagation constant $\mu$, all localized nonlinear modes
can be identified through a solution of a system of {\em
    two} equations with {\em two} real unknowns (``shooting constants'') 
 which determine the asymptotic
behavior of $\psi(x)$ at $x\to\pm\infty$. This observation allows to
confirm the existence of continuous families of nonlinear modes
using a shooting-type argument: the number of constraints (matching
conditions at $x=0$) is equal to the number of available free
parameters. Using this approach, \textcite{KZ14b} found families of
nonlinear modes in an asymmetric complex double-hump potential
defined by (\ref{e:doublewell}) with $ A_-\ne A_+$, as illustrated
in Fig.~\ref{fig-parabol:asymmetric}.

{\color{black} Bifurcation of soliton families from linear modes in
asymmetric complex potentials was also studied analytically by
\textcite{NY16b}. Under a weak assumption, it was shown that the
stationary equation (\ref{e:parabol:psi}) admits a constant of
motion if and only if the complex potential $U(x)$ is of Wadati-type
(\ref{e:VPT}).
Using this constant of motion, the soliton equation
(\ref{eq:nonPT:hydro}) was reduced to a second-order equation for
the amplitude of the soliton. From this new soliton equation, it was
shown by perturbation methods that continuous families of solitons
bifurcate out from linear eigenmodes. It was also found that these
results hold not only for the cubic nonlinearity, but also for all
nonlinearities of the form $F(|\Psi|^2)\Psi$ in Eq. (\ref{opt:NLS}),
where $F(\cdot)$ is an arbitrary real-valued function.}

\section{Nonlinear waves in   periodic potentials}
\label{sec:periodic}
In this section, we review properties of solitons in 1D and 2D $\PT$-symmetric
periodic potentials. For the 1D case, we explore the NLS equation
(\ref{opt:NLS}) with a potential $U(x)\equiv U_{1D}(x)$, where
$U_{1D}(x)$ is a periodic and $\PT$-symmetric function, while in 2D
the model is
\begin{equation} \label{e:model}
i \Psi_t + \Psi_{xx} + \Psi_{yy}  {-}  {U}_{2D}(x,y)\Psi + g |\Psi|^2\Psi = 0,
\end{equation}
where ${U}_{2D}(x,y)$ is periodic in $x$ and $y$ and satisfies the
$\PT$ symmetry condition ${U}_{2D}^*(x,y) = {U}_{2D}(-x,-y)$.

Similar to their real-valued counterparts, $\PT$-symmetric periodic potentials feature band-gap spectra \cite{BDM99,Jones99}.
For many
familiar $\PT$-symmetric periodic potentials, it has been shown that
when the imaginary component of the potential is below a certain
threshold, then all the spectral bands lie on the real axis, and the spectrum of the potential is all-real. Above
this threshold, phase transition occurs, and complex eigenvalues
appear  \cite{Musslimani_1,Markis1,NGY12}. Sometimes, this threshold is
zero, meaning that complex eigenvalues exist for any imaginary
strength of the periodic potential \cite{Musslimani_2}.

In $\PT$-symmetric periodic potentials, special periodic solutions
can be found analytically~\cite{Musslimani_2,AKSY}. Continuous
families of bright solitons were found numerically by
\textcite{NGY12,ZL12,LHLD} in pure periodic potentials, and by
\textcite{WW11,LuZh} in a $\PT$-symmetric periodic potential with
local defects. Some of these soliton families bifurcate out from
edges of Bloch bands, while others do not. Above phase transition,
these solitons are all unstable; but below phase transition, they can
be stable in certain parameter regions. In addition to soliton
families, distinctive linear diffraction patterns were reported by
\textcite{Makris10, Regens}, and periodic bound states were reported
by \textcite{NZY12}. \textcolor{black}{Localization-delocalization transition of light propagating in quasi-periodic $\PT$-symmetric lattices was numerically obtained by \textcite{Hang2015}.}

In two dimensions, a distinctive pyramid
diffraction pattern was reported in both linear and nonlinear
regimes near phase transition \cite{NY13}. In addition to the above
results, other interesting phenomena, such as nonreciprocal Bloch
oscillations \cite{Longhi09b}, rectification and dynamical lozalization~\cite{KaVyKoTo}, and unidirectional propagation discussed in Secs.~\ref{subsec:optics} and \ref{sec:PTdimer},
have also been found in linear $\PT$-symmetric
periodic potentials.

\subsection{Linear spectrum of periodic potentials}

\paragraph*{1D lattice.}  Spectral properties  of  one-dimensional Schr\"odinger operators with complex periodic potentials have been intensively studied (for a recent review see \textcite{DM06,Markis}).  In particular,  \textcite{Gasymov1980} considered the Schr\"odinger operator  $H = -{d^2}/{dx^2} + U_{1D}(x)$, where
\begin{equation}
U_{1D}(x)= \sum_{n=1}^\infty u_n e^{inx}, \quad \sum_{n=1}^\infty |u_n| < \infty,
\end{equation}
is the $2\pi$-periodic potential  which becomes $\PT$ symmetric if all coefficients $u_n$ are real. It was proven that the   spectrum of $H$ is real and fills the  semi-axis $[0, \infty)$.  Eigenfunctions of $H$ constitute a complete basis in a properly defined  linear space. 

We illustrate  typical properties of linear periodic $\PT$-symmetric lattices using as an example the potential in the form
\begin{equation} \label{e:PTlattice}
U_{1D}(x) =  {-} V_0\left[ \cos^2(x) + i W_0 \sin(2x) \right].
\end{equation}
For this $\pi$-periodic lattice, $V_0$ characterizes the strength of
the real component of the potential, and $W_0\ne 0$ is the relative
magnitude of the imaginary component.

\begin{figure}
    \centering
    \includegraphics[width=0.45\textwidth]{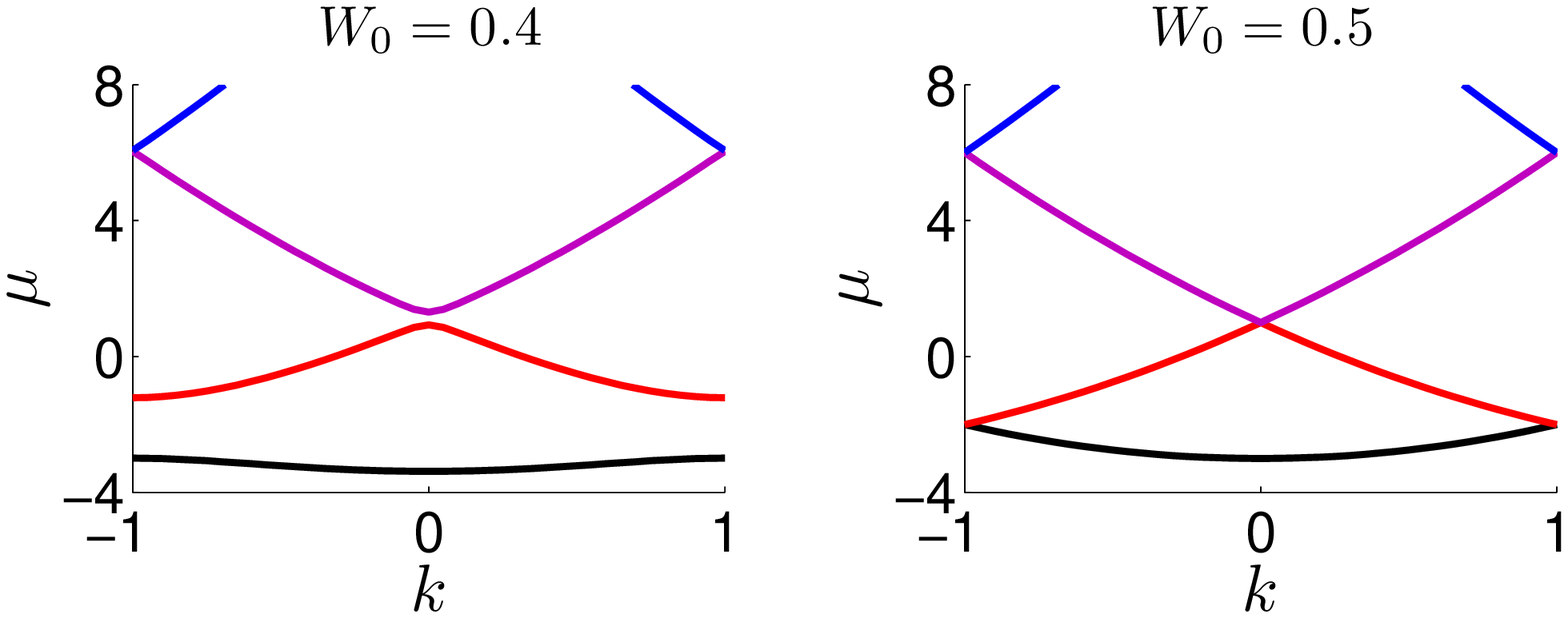}

    \vspace{0.5cm}
    \includegraphics[width=0.45\textwidth]{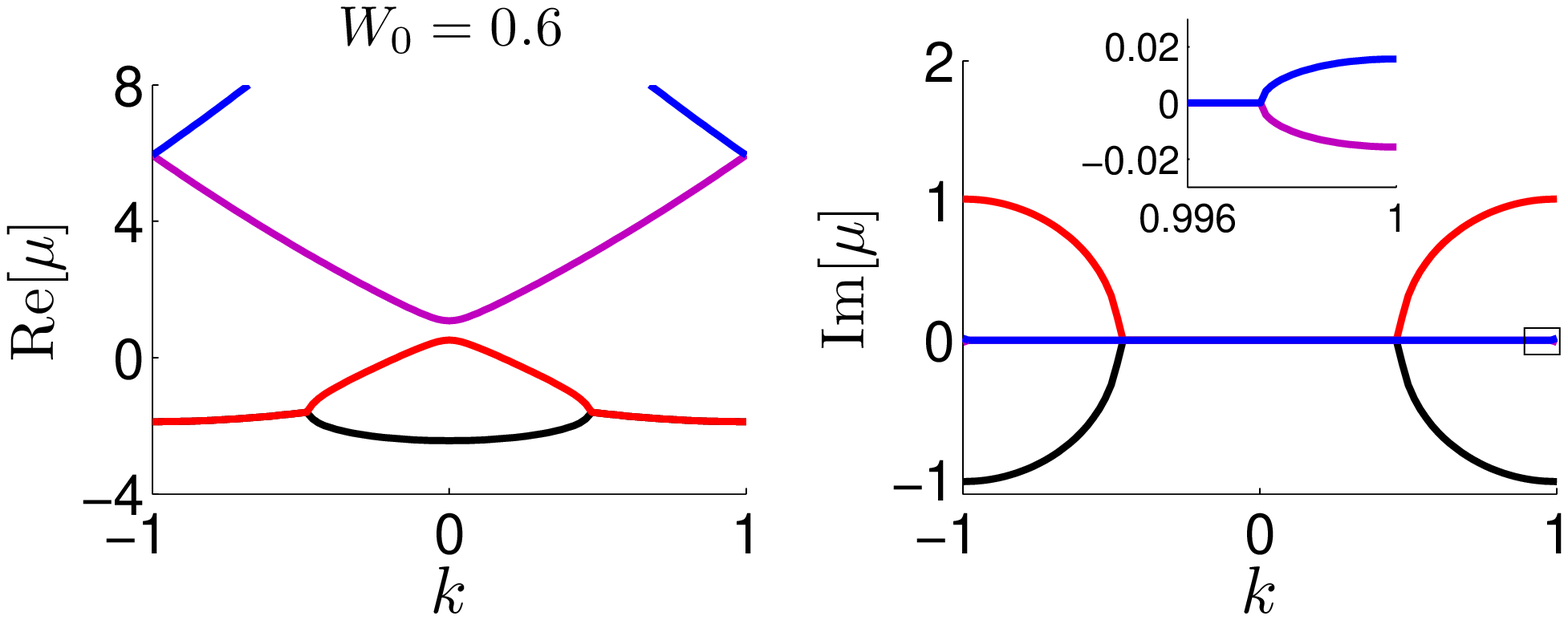}
    \caption{(Color online)  Diffraction relations of 1D $\PT$ lattice
        (\ref{e:PTlattice}) for three values of $W_0$ ($V_0=6$). The inset
        in the lower right panel is amplification of the small boxed region
        near $k=1$ and $\mbox{Im}[\mu]=0$ of the same panel. }
    \label{f:lattice_fig1}
\end{figure}

Linear modes of Eq. (\ref{opt:parabolic}) with potential
(\ref{e:PTlattice}) are
\begin{equation} \label{e:Bloch1D}
\Psi(x,t)=\psi(x)e^{-i\mu t}, \qquad  \psi(x)=p(x; k)e^{ikx},
\end{equation}
where $\psi(x)$ is a Bloch mode solving the eigenvalue problem
\begin{equation}
\mu \psi + \psi_{xx} + V_0 \left( \cos^2x + i W_0 \sin (2x) \right) \psi =
0,
\label{LinearMu}
\end{equation}
$p(x;k)$ is a $\pi$-periodic function in $x$, $k$ is the wavenumber in the irreducible Brillouin
zone (BZ) $-1\le k\le 1$, and $\mu$ is the propagation constant.
Function $\mu=\mu(k)$ is the  diffraction (or dispersive) relation,
and all admissible values of $\mu$ constitute the Bloch bands.
Dispersion  relations $\mu=\mu(k)$ for three values of $W_0$
are displayed in Fig.~\ref{f:lattice_fig1}. At $W_0=0.4$, $\PT$
symmetry is unbroken, and the Bloch bands are all-real and separated
by gaps. At $W_0=1/2$ (the exceptional point), all Bloch bands touch
each other and gaps disappear.
When $W_0>1/2$, $\PT$ symmetry is broken, and complex eigenvalues
appear in Bloch bands.

At $W_0 = 1/2$,
one can introduce the variable transformation $\xi= i \sqrt{V_0/2} \hspace{0.1cm} e^{i
    x}$ which reduces (\ref{LinearMu}) to the Bessel's equation,
\begin{equation}
\xi^2 \psi_{\xi\xi}+\xi \psi_\xi +\left(\xi^2-\mu-{V_0}/{2}\right)\psi=0,
\end{equation}
whose solution is $\psi(x) = J_{ k }( i \sqrt{{V_0}/{2}} \ e^{i
    x})$,
where $k = \pm\sqrt{\mu + {V_0}/{2}}$ \cite{BDM99,
    GraefeJones,NGY12,Berry98}. Using power-series expansion of the
Bessel function, the above solution can be rewritten as
$\psi(x)=e^{ikx}\tilde{p}(x;k)$, where $\tilde{p}(x;k)$ is a
$\pi$-periodic function of $x$. In order for this solution to be a
Bloch mode, $k$ should be real. Then if we also restrict $k$ to be
in the first BZ  $-1\le k\le 1$, then the exact diffraction relation
is $\mu = -{V_0}/{2} + (k+2m)^2$,
where $m=0, \pm 1, \pm 2 \dots$. This diffraction function matches
that shown in Fig. \ref{f:lattice_fig1} for $W_0=0.5$.  It shows
that all Bloch bands are real-valued. In addition, these Bloch bands
are connected either at the center ($k=0$) or edge ($k=\pm 1$) of
the BZ, where
$
\mu=-{V_0}/{2} + n^2,
$ and $n=0, 1, 2, \ldots.
$

We now consider the case where $W_0$ is near 1/2, i.e., $V_0(W_0 -
1/2) \equiv \epsilon \ll 1$.
In this case, Eq. \eqref{LinearMu} becomes
\begin{equation*} 
\left(\mu + {V_0}/{2}\right) \psi + \psi_{xx} + {V_0}/{2} \left( e^{2 i x} \right) \psi + \epsilon \, i \sin (2x)\psi =0.
\end{equation*}
Since complex eigenvalues
first appear near band
intersections
we only need to calculate the eigenvalue
at
$k=0$ and $\pm 1$, where the Bloch modes are $\pi$- or
$2\pi$-periodic. These solutions and the associated $\mu$ values can
be expanded as power series in $\epsilon^{1/2}$,
\begin{eqnarray}
\label{muexpand}
\begin{array}{l}
\mu = - {V_0}/{2}  +  n_0^2 + \epsilon^{1/2} n_1 + \epsilon
n_2 +
\ldots,
\\
\psi  = \psi_0 +  \epsilon^{1/2} \psi_1 + \epsilon \psi_2 +
\ldots,
\end{array}
\end{eqnarray}
where $n_0 = 0,1,2,\cdots$, and the coefficients
$n_1,n_2,n_3,\cdots$
are constants
shown in Table~\ref{tab:fonts}~\cite{NGY12}. When $n_0=1, 3$, the
coefficient $n_1$ or $n_3$ is imaginary, thus complex eigenvalues
bifurcate out simultaneously above the phase transition point
($\epsilon>0$). The imaginary part of these complex eigenvalues at
$n_0=3$ ($\sim\epsilon^{3/2}$) is much smaller than that at $n_0=1$
($\sim\epsilon^{1/2}$), and no complex eigenvalues bifurcate out
when $n_0=0, 2$. Continuing these calculations to higher $n_0$
values, one can find that the coefficient $n_{2m+1}$ is always
imaginary for $n_0=2m+1$, where $m=0, 1, 2, \cdots$. Thus complex
eigenvalues bifurcate out simultaneously from all odd values of
$n_0$ at the phase transition point $W_0=1/2$.

Table~\ref{tab:fonts} also shows that below the phase transition point
($W_0<1/2$, or $\epsilon<0$), the eigenvalue $\mu$
is real for all integers $n_0$.

\begin{table}
    \caption{\label{tab:fonts} Coefficients in the $\mu$ expansion
        (\ref{muexpand}).}
    \begin{center}
        \begin{tabular}{ | c | c | c|  c | }
            \hline
            $n_0$ & $n_1$ & $n_2$ & $n_3$ \\ \hline \hline
            0  &   0  & $ V_0/8 $& 0  \\ \hline
            1&   $ \pm i {V_0^{1/2}}/{2}$ & $ V_0/32$ &
            $\pm  i \hspace{0.04cm} \left( {V_0^{-1/2}}/{4} +{V_0^{3/2}}/{2^9} \right)$
            \\ \hline
            2 & 0 & $- {5V_0}/{48}$, ${V_0}/{48}$& 0 \\ \hline
            3 & 0 &  $-V_0/64$  & $\pm i { ~V_0^{3/2}}/{2^9}$ \\ \hline
            $N$ & 0 & $ - {V_0}/{8} (N^2 -1)^{-1}$  & 0 \\ \hline
        \end{tabular}
    \end{center}
\end{table}

\paragraph*{2D lattices.} In 2D, we focus on the separable potential
\begin{eqnarray}
\label{e:PTlattice2D}
{U}_{2D}(x,y) &=& U_{1D}(x) + U_{1D}(y),
\end{eqnarray}
whose linear spectrum can be obtained directly from the spectrum of
the 1D problem (\ref{e:PTlattice}). In this case,
$\mu={\mu}(k_1)+{\mu}(k_2)$ is the 2D diffraction relation, $k_1,
k_2$ are Bloch wavenumbers in the $x$ and $y$ directions which and are
located inside the first BZ, and
\begin{equation}
\Psi(x,y,t)=e^{ik_1x+ik_2y-i\mu t}p(x; k_1)p(y; k_2),
\end{equation}
where $p(x; k)$ is the 1D $\pi$-periodic function
as given in
(\ref{e:Bloch1D}).
This diffraction relation shows that complex eigenvalues appear in
2D $\PT$ lattice if and only if complex eigenvalues appear in the 1D
$\PT$ lattice (\ref{e:PTlattice}). Thus Bloch bands in the 2D
potential (\ref{e:PTlattice2D}) are all-real when $W_0\le 1/2$, and
a phase transition occurs at $W_0=1/2$ above which complex
eigenvalues arise.

\subsection{Solitons and their stability}

Solitons in $\PT$-symmetric periodic potentials exist as continuous
families \cite{Musslimani_1,NGY12,LHLD}. The simplest soliton
families are those that bifurcate out from edges of Bloch bands, and
they can be established analytically by exponential asymptotics
methods \cite{NY14}.
In addition to these simplest soliton families, an
infinite number of other soliton families were reported numerically  \cite{NGY12,LHLD}.

\paragraph*{Solitons in 1D latices.}

Let us consider
1D NLS equation~(\ref{opt:NLS}) with
$\PT$-symmetric periodic potential (\ref{e:PTlattice}).
Solitons are searched in the form $\Psi(x,t)=e^{-i\mu t}\psi(x)$,
where $\psi(x)$ is a stationary localized wavefunction solving
(\ref{e:parabol:psi}),
and $\mu$ is a real propagation constant. In full analogy with the
conservative case \cite{BraKon,Pelinovsky11,Yang2010}, exponentially
decaying soliton solutions (alias {\em gap solitons}) exist when
$\mu$ lies inside bandgaps of the underlying linear system. For
broken $\PT$ symmetry, all solitons are unstable since small tails
of solitons will be amplified. Thus below we only consider the
unbroken $\PT$ symmetry case where $W_0\le 1/2$.

\begin{figure}
    \centering
    \includegraphics[width=0.45\textwidth]{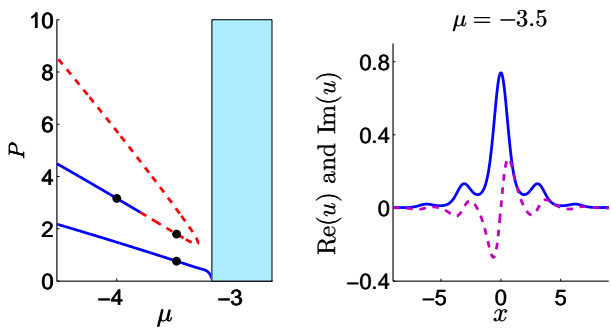}
    \caption{1D solitons in the semi-infinite gap of (\ref{e:PTlattice}) under focusing nonlinearity ($g=1$) for
        $V_0=6$ and $W_0=0.45$. (left) Power curves of these solitons;
        solid blue and dashed red lines represent stable
        and unstable solitons, respectively;
        the shaded region is the first Bloch band. (right)
        A fundamental soliton at $\mu = -3.5$ (marked by a dot on
        the lower curve of the left panel); the solid blue line is for the
        real part and dashed pink line for the imaginary part. }
    \label{PowerCurve1DFocusing}
\end{figure}

In Fig.~\ref{PowerCurve1DFocusing} (left) we illustrate two families of solitons
in the semi-infinite gap under focusing nonlinearity \cite{NGY12}. The lower power curve is for the fundamental solitons
which are $\PT$ symmetric,  i.e., $\psi^*(x)=\psi(-x)$,
and whose real parts possess a single dominant peak. The profile of such a soliton
at $\mu=-3.5$ is displayed in Fig. \ref{PowerCurve1DFocusing}
(right). This soliton family bifurcates out of the first Bloch band, and in the vicinity of the bifurcation the solitons can be described as     low-amplitude Bloch-wave
packets. The entire family of fundamental solitons   is linearly stable.

\begin{figure}
    \centering
    \includegraphics[width=0.5\textwidth]{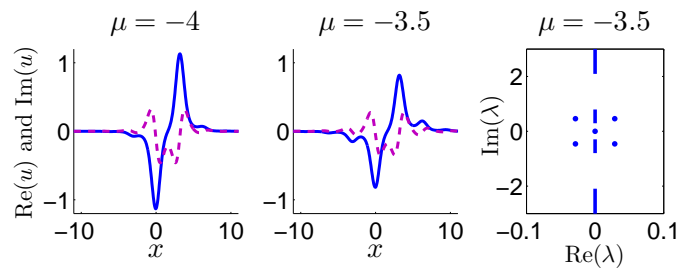}
    %
    \caption{(Color online) Left two panels: dipole solitons at the
        marked points of the lower power curve in
        Fig.~\ref{PowerCurve1DFocusing}. Right panel: linear-stability
        spectrum for the dipole soliton in the middle panel.}
    \label{Spectrum1DFocusing}
\end{figure}

The upper power curve in Fig. \ref{PowerCurve1DFocusing} consists of
dipole solitons. This power curve features double branches which
terminate through a saddle-node bifurcation before reaching the
first Bloch band. Profiles of two such solitons on the lower power
branch are displayed in Fig. \ref{Spectrum1DFocusing} (left two
panels). The real parts of these dipole solitons possess two
dominant peaks of opposite sign. This however does not violate $\PT$
symmetry, as due to phase invariance we have that
$\phi(x)=\psi(x)e^{i\pi/2}$ is a $\PT$-symmetric solution.

Dipole solitons are linearly stable only in a certain portion of
their existence region. Specifically, only dipole solitons on the
lower branch with $\mu \le \mu_a \approx -3.8$ are stable (see Fig.
\ref{PowerCurve1DFocusing} (left)). For dipole solitons in this
region, their spectra are entirely imaginary. At $\mu = \mu_a$,
stability switching occurs where a quadruple of complex eigenvalues
bifurcate off the edge of the continuous spectrum (see Fig.
\ref{Spectrum1DFocusing}, right panel). Within this unstable region,
there is a second eigenvalue bifurcation at $\mu\approx -3.4$ of the
lower branch (near and on the left side of the power minimum) where
a pair of real eigenvalues bifurcate out from zero.

Beside these soliton families, the model also admits other types of
solitons such as truncated-Bloch-mode solitons \cite{LHLD} which are
stable in certain parameter regimes. Stable dissipative solitons exist at the surface between homogeneous Kerr medium and a truncated lattice (\ref{e:PTlattice}) supported by the linear dissipation~\cite{HMZGK12}.

\begin{figure}
    \centering
    \includegraphics[width=0.24\textwidth]{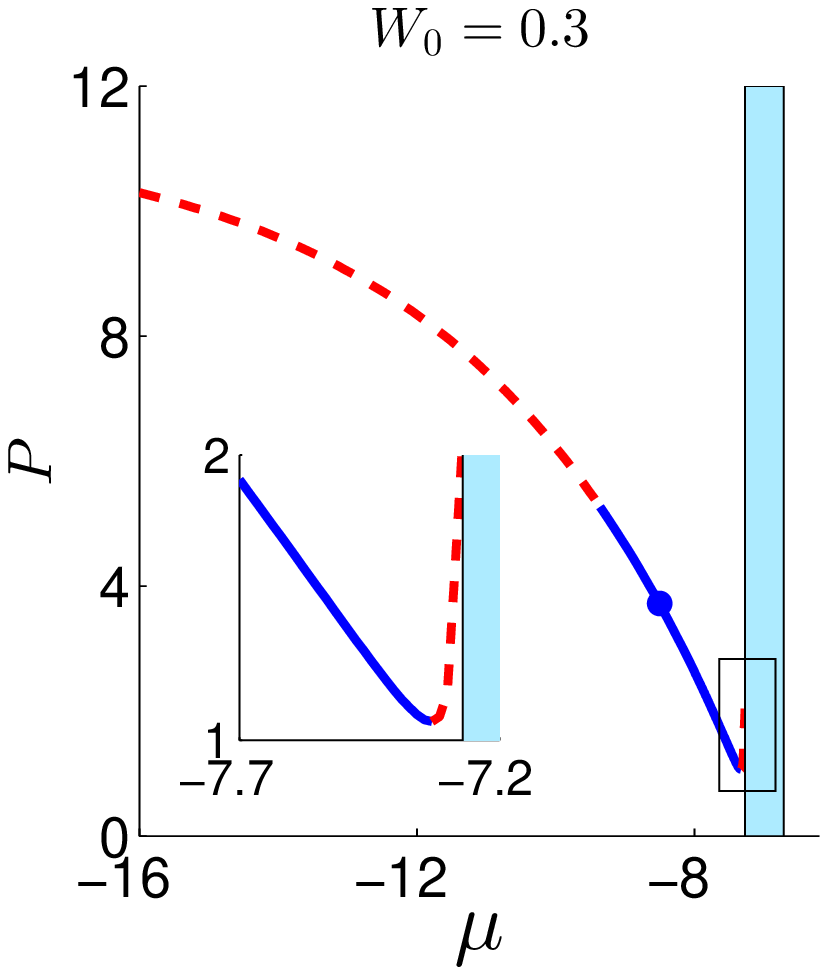}
    \includegraphics[width=0.22\textwidth]{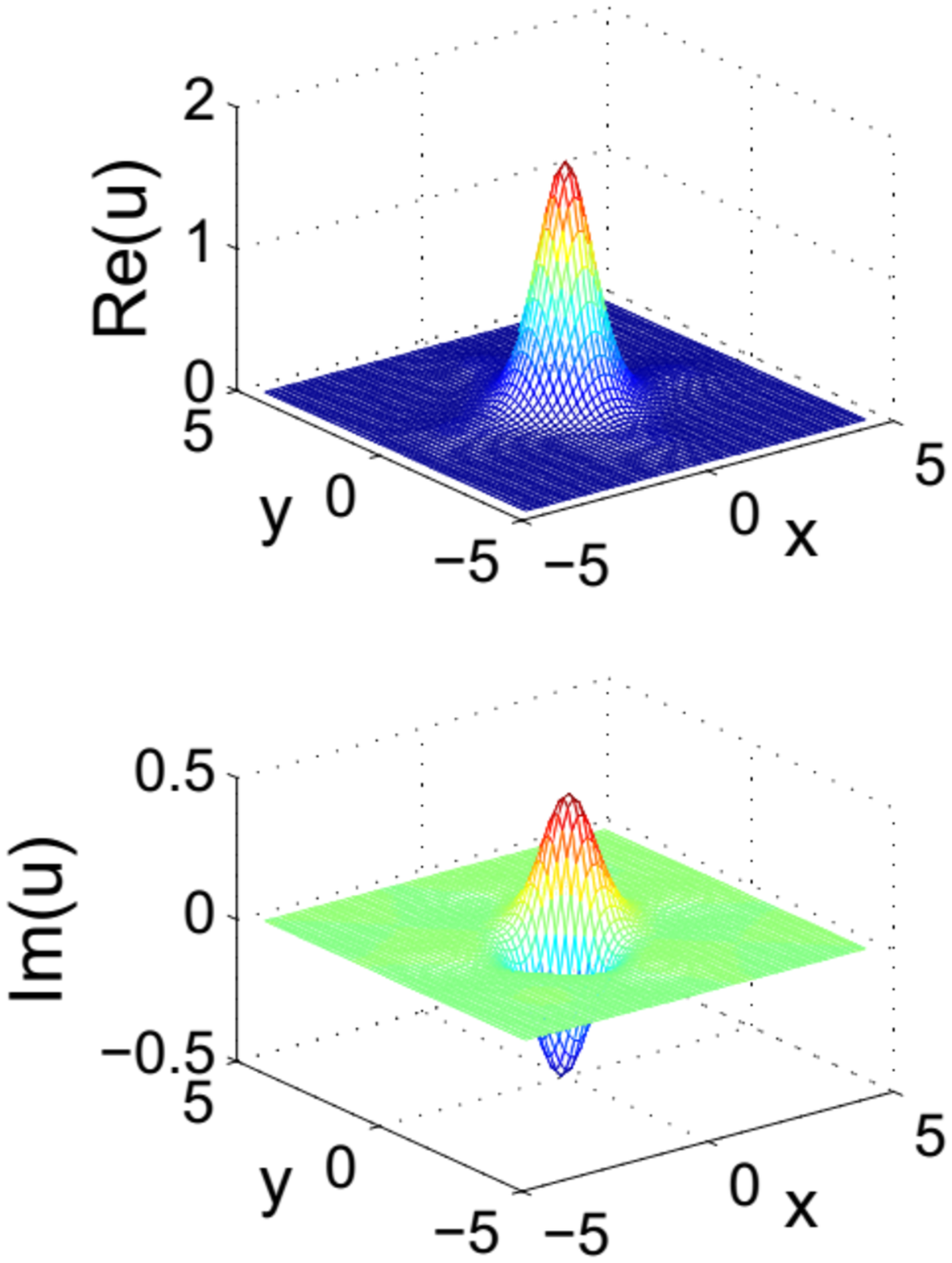}
    \caption{(Left) Power curve of fundamental 2D solitons in
        the semi-infinite gap under focusing nonlinearity ($g=1$) for
        $V_0=6$ and $W_0=0.3$. The inset is amplification of the power curve near the first Bloch band
        in the same panel. Solid blue lines indicate stable solitons, while dashed red lines indicate unstable solitons.
        (Right) Real and imaginary parts of the soliton $\psi(x,y)$ at $\mu=-8.5$ (marked by a dot on the power curve). }
    \label{PowerCurve2D}
\end{figure}

\paragraph*{2D solitons.}
Solitons and their stability in 2D  $\PT$-symmetric
periodic potentials (\ref{e:PTlattice2D}) have also been studied \cite{NGY12}. These solitons are of the form
$\Psi(x,y,t)=e^{-i\mu t}\psi(x,y)$.
Figure \ref{PowerCurve2D} (left panel) shows  fundamental 2D solitons in the semi-infinite
gap under focusing nonlinearity ($g=1$).
Similar to the conservative case,
there exists a threshold power ($L^2$ norm) necessary for the
existence of such solitons. The profiles of the solitons possess
$\PT$ symmetry, $\psi^*(x,y)=\psi(-x,-y)$, and their real parts have
a single dominant peak [Fig. \ref{PowerCurve2D}(right)]. These
fundamental solitons are stable only in a finite $\mu$-interval,
even though their existence region is infinite.
For large negative values of $\mu$, the instability is due to a
quadruple of complex eigenvalues, whereas for $\mu$ values near the
first band, the instability is due to a pair of real eigenvalues.

Beside the fundamental solitons in Fig. \ref{PowerCurve2D}, other
types of solitons such as vortex solitons and multipole solitons
have also been reported in 2D $\PT$-symmetric lattices \cite{LZSMLL14,RWWH14,WSRZMMH15,ZWLHH13}.

\paragraph*{Nonlinear periodic solutions \textcolor{black}{and constant-intensity waves}.}

As we have mentioned earlier, a $\PT$-symmetry threshold can be
zero. This fact, however, does not prohibit existence of nonlinear
periodic solutions, which do not require the propagation constant to
belong to a bandgap of the linear spectrum. For instance, phase
transition for the $\PT$-symmetric periodic potential
$U_{1D}(x)=-[V_0\sin^2x+3iW_0\sin x]$,
where $V_0$ and $W_0$ are real constants, has zero threshold
(complex eigenvalues appear in Bloch bands for any nonzero  $W_0$).
However, in the presence of focusing nonlinearity, it admits a
stationary $x$-periodic solution found in exact form by
\textcite{Musslimani_2},
\begin{equation*}
\Psi(x,t)=\sqrt{V_0+W_0^2}\cos x \hspace{0.02cm} e^{iW_0\sin x -i\mu t}, \quad \mu=1-V_0,
\end{equation*}
provided that $V_0>-W_0^2$. These periodic solutions may   be
stable, even though the periodic potential is above phase transition
\cite{LPRS13}. Examples of stable periodic solutions in defocusing
medium can also be found \cite{AKSY}.

\textcolor{black}{ The complex Wadati potentials
    $U_{1D}(x)=-[w^2(x)+iw_x(x)]$ (see also Sec.~\ref{subsec:par:symm_break} and \ref{subsec:par:nonPT})
    support exact constant-intensity solutions
    $   \Psi= A \exp[-i\int w(x)dx +igA^2 t]$,
    where $A$ is a real amplitude \cite{MMCR15}.    These   solutions can be  generalized  to the 2D case, where the $\PT$-symmetric potential reads $U_{2D}(x,y)=-|{\bf W}|^2-i\nabla\cdot {\bf W}$, where ${\bf W}=(W_1,W_2)$ is a vector potential constrained by the condition $(W_1)_y=(W_2)_x$. Then  the 2D exact  solutions have the form
    $\Psi(x,y,t)= A \exp[-i\int_C {\bf W}\cdot d{\bf r} +igA^2 t]$,
    where $C$ is a smooth open curve between any two points in the $(x,y)$-plane. The constant-amplitude solutions   were explored by  \textcite{MMCR15} in the context of modulational instability.}

\subsection{Nonlinear dynamics near phase transition point}
\label{sec:nonlin_dyn}

Beyond the question of stationary modes, a more general question is
how an initial wave evolves in a $\PT$-symmetric periodic potential.
Here we review linear and nonlinear dynamics of wavepackets near the phase
transition point, investigated analytically by
\textcite{NZY12,NY13}, and describe new phenomena such as wave blowup,
periodic bound states and linear or nonlinear pyramid diffraction
patterns. 

\paragraph*{1D dynamics.}
We first consider the model (\ref{opt:NLS}) with potential
(\ref{e:PTlattice}), which in this subsection is rewritten as
$U(x) = U_{1D}(x) = \textcolor{black}{-}V_0^2 \left[ \cos(2x) + i
W_0 \sin(2x) \right]$.
For this form of the potential, phase transition occurs at $W_0=1$.
At this phase transition point, the diffraction relation is $\mu =
(k + 2m)^2$, where $k$ is in the first BZ $k\in [-1,1]$, and $m$ is
any nonnegative integer (see Fig.~\ref{f:lattice_fig1} for
$W_0=1/2$).

At $k = 0$ and $\pm 1$, adjacent Bloch bands intersect each other.
At these intersection points, Bloch solutions are degenerate and
$\pi$- or $2\pi$-periodic in $x$. Posed as an eigenvalue problem for
$\Psi = \phi(x) e^{-i \mu t}$ in the linear
Eq.~(\ref{opt:parabolic}), we get $L \phi  = -\mu \phi$, where $L \equiv
\partial_x^2  {-} U(x)|_{W_0=1}$. Then at these
band-intersection points, the eigenvalues are $\mu = n^2$, where $n$
is any positive integer. These eigenvalues all have geometric
multiplicity 1 and algebraic multiplicity 2, thus there exists a
generalized eigenfunction $\phi^\textrm{g}$ satisfying $\left( L+\mu \right)
\phi^\textrm{g} = \phi$ [for construction of the complete basis at an exceptional point see \cite{GraefeJones,Gasymov1980}].

To study nonlinear dynamics of wave packets near
the phase-transition point (i.e., $W_0\sim 1$), one can use the
asymptotic expansion $\Psi =  e^{-i n^2 t} [\epsilon A(X,T) \phi(x)
+ \epsilon^2 \psi_1 + \ldots]$, where
$A(X,T)$ is an envelope of the degenerate Bloch mode $\phi(x)$,
$X=\epsilon x$, $T=\epsilon t$ are slow variables, and
$0<\epsilon\ll 1$. From the multiple-scale perturbation analysis one
finds that near $n=1$
and $W_0=1-c\epsilon^2$ where $c$ is a constant, the envelope is governed by a nonlinear
Klein-Gordon (KG) equation,
\begin{equation}
A_{TT} - 4A_{XX}  +\alpha A+ \gamma  |A|^2A = 0.
\label{Eq:NLWn}
\end{equation}
Here $\alpha=c V_0^4/2$ and $\gamma =\frac{2 g}{\pi}
\int_{-\pi}^\pi |\phi|^2 \phi^{2} dx$ are real constants. At higher
$n$ values, similar envelope equations can be derived under
appropriate scalings of $W_0(\epsilon)$.

Envelope dynamics in the nonlinear KG equation (\ref{Eq:NLWn})
proves to closely mimic the corresponding wavepacket dynamics in the
original $\mathcal{PT}$ model (\ref{opt:NLS}).

\begin{figure}
    \vspace{0.3cm}
    \includegraphics[width=0.9\columnwidth]{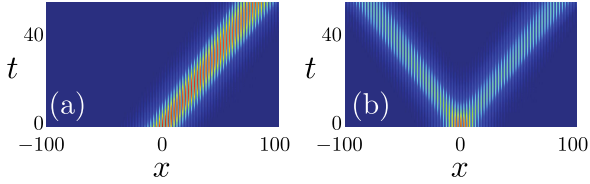}
    \caption{
        (a) Linear unidirectional wavepacket and (b) linear wavepacket
        splitting at the phase-transition point
        $W_0=1$. Other parameters are $\epsilon=0.1$, $\mu=1$ and $V_0=\sqrt{6}$.
    } \label{Fig:Diffraction}
\end{figure}

First, at the phase-transition point ($c=0$), solutions for the left
and right propagating waves (resembling solutions of the wave
equation $A_{TT} - 4A_{XX}=0$) can be found in the original linear
$\mathcal{PT}$ model (\ref{opt:NLS}) with $g=0$. Two examples are
shown in Fig.~\ref{Fig:Diffraction}. The spreading-shelf solutions
in this wave equation were reported experimentally by
\textcite{Regens}.

\begin{figure}
    \includegraphics[width=8.1cm]{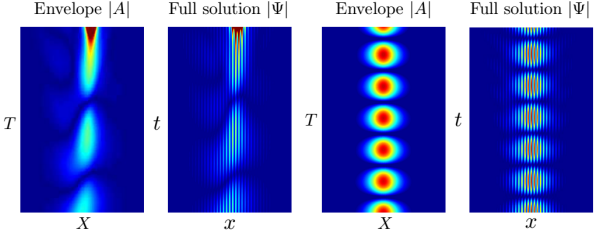}
    \vspace{-0.2cm} \caption{(Color online) Nonlinear wavepacket solutions below the
        phase-transition point ($c=1$) under self-defocusing nonlinearity.
        From left to right: a blowup solution in the envelope equation and
        the full equation; a periodic bound state in the envelope equation
        and the full equation. } \label{Fig:NLExample}
\end{figure}

Next we consider envelope solutions in the KG equation
(\ref{Eq:NLWn}) with self-defocusing nonlinearity ($\gamma <0$) near
the lowest band-intersection point ($n=1$). \textcite{NZY12}
numerically found that below the phase-transition point ($c>0$)
envelope solutions blow up to infinity
[first panel in Fig.~\ref{Fig:NLExample}].
In the full model (\ref{opt:NLS}), similar growing solutions were
found and displayed in Fig. \ref{Fig:NLExample} (second panel). In
the full model, this blowup may eventually be suppressed, but that
is already beyond the asymptotic regime of the KG model
(\ref{Eq:NLWn}).
Under self-defocusing nonlinearity Eq.~(\ref{Eq:NLWn}) also admits
breather-like solutions shown in the third panel of
Fig.~\ref{Fig:NLExample}, as well as
stationary solitons $A(X,T) =F(X) e^{i \omega T}$ with $\omega\in
[-\sqrt{\alpha},\sqrt{\alpha}]$. The corresponding breather solution
in the full model is shown in the fourth panel of
Fig.~\ref{Fig:NLExample}.

\small
\normalsize

If the nonlinearity is self-focusing ($g=1$), envelope solutions do
not blow up, periodic bound states cannot be found, and stationary
solitary waves do not exist in the envelope equation. In this case,
breathers as well as nonlinear diffracting solutions similar to the
linear diffracting pattern reported in \textcite{Makris10} can be
found. Dynamics near the breaking point appears to be rich even in the linear limit, allowing in particular for the resonant mode conversion~\cite{Vysloukh}.

\paragraph*{2D dynamics.}

\begin{figure}
    \begin{center}
        \includegraphics[height = 1.6in]{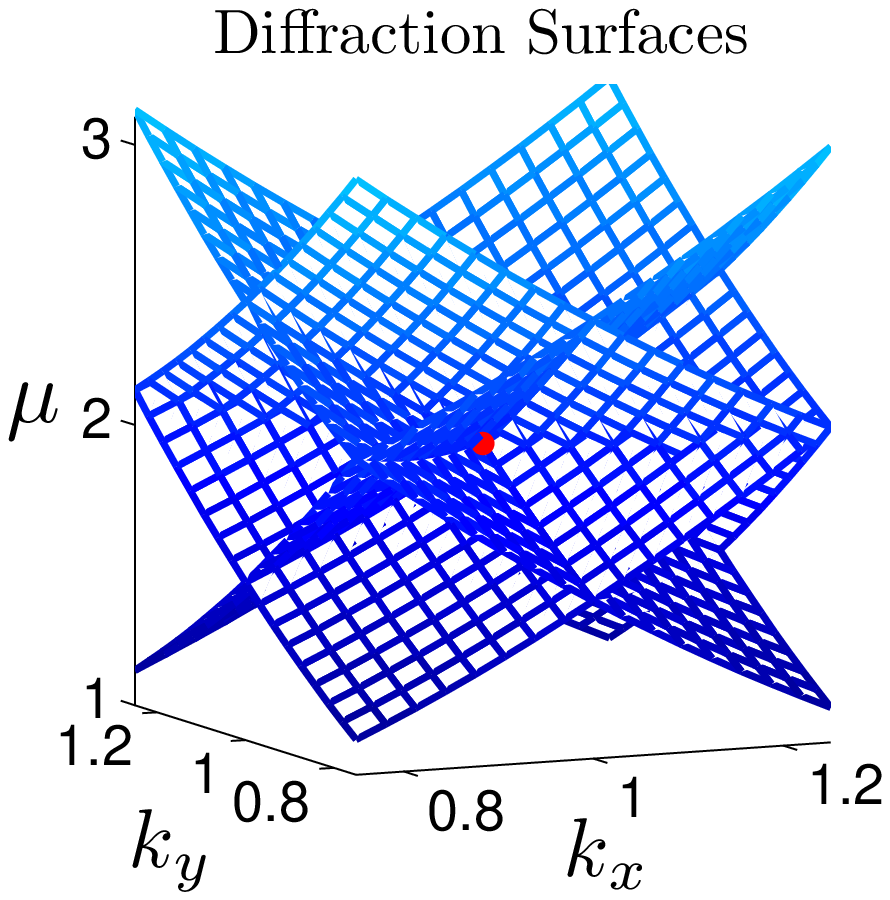} \hspace{0.2cm}
        \includegraphics[height = 1.6in]{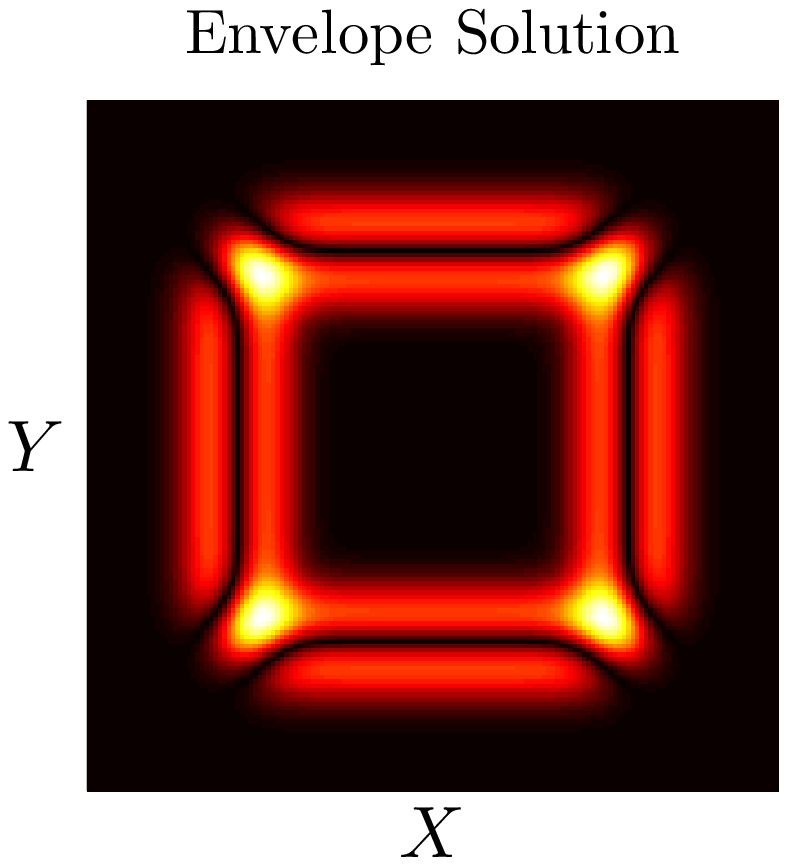}

        \caption{(Color online)  (Left) Diffraction relation near intersection point $(k_x,
            k_y, \mu) = (1,1,2)$ (marked by red dot). (Right) Linear pyramid
            diffraction of initial Gaussian envelope at phase transition point
            in the envelope equation (\ref{Eq:EnvelopeA}).  }
        \label{Fig:Dispersion}
    \end{center}
\end{figure}

Next we consider dynamics of wave packets in a 2D $\PT$-symmetric
periodic potential near the phase transition point \cite{NY13}. The
mathematical model is taken as (\ref{e:model}) with $U_{2D}(x,y)=
U_{1D}(x) + {U}_{1D}(y)$, where $U_{1D}(x)$ is the same periodic
potential used in the 1D dynamics above. At the phase transition
point $W_0=1$, the linear diffraction relation reads $\mu = (k_x +
2m_1)^2 + (k_y + 2m_2)^2$, where $(k_x, k_y)$ are Bloch wavenumbers
in the first BZ $-1 \le k_x, k_y \le 1$, and $m_{1,2}$ are
nonnegative integers. The most complex degeneracies occur at points
$k_x = 0, \pm 1$ and $k_y = 0, \pm 1$, where the diffraction surface
intersects itself four-fold as illustrated in Fig.
\ref{Fig:Dispersion}. When a linear Bloch wave $\Psi = \phi(x,y)
e^{-i \mu t}$ is chosen at one of these degeneracies, $\phi(x,y)$
satisfies an eigenvalue equation $L \phi = -\mu \phi$, where
$L=\nabla^2  {-} U(x,y)|_{W_0=1}$, $\mu = n_1^2+n_2^2$, and
$(n_1,~n_2)$ are any pair of positive integers.

We will conduct the analysis at the
lowest intersection point, $\mu =2$. The perturbation expansion for
the wave packet near this point is $\Psi = \epsilon^{\frac{3}{2}} e^{-i \mu t}\left(A(X,Y,T) \phi^{01}(x,y)+ \epsilon \psi_1  + \ldots\right)
$,
where $\phi^{01}$ is the Bloch mode at the point $(k_x, k_y,
\mu)=(1,1, 2)$, $(X,Y,T) =(\epsilon x, \epsilon y, \epsilon t)$ are
slow variables, and $0< \epsilon \ll 1$. Near the phase-transition
point we express $W_0 = 1 - \eta \epsilon^2/V_0^2$, where $\eta$
measures the deviation from phase-transition. Through a perturbation
calculation we obtain \cite{NY13}
\begin{align} \label{Eq:EnvelopeA}
\partial_T^4 A &- 8 (\partial_X^2 + \partial_Y^2) \partial_T^2  A + 16 (\partial_X^2-\partial_Y^2)^2A  \notag \\
& + \alpha \partial_T^2A+ i\tilde{g}  \partial_T \left(|A|^2A
\right) =0,
\end{align}
where $\alpha=2V_0^2\eta$, and $\tilde{g}$ is a real constant. Equation
\eqref{Eq:EnvelopeA} reveals important physical features, which are
demonstrated below using the initial conditions
\begin{equation}
\label{Eq:ic}
A= A_0 e^{-(X^2+Y^2)}, \, A_T = A_{TT} = 0, \, \partial_T^3 A = -
\ri \tilde{g} |A|^2A
\end{equation}
in the envelope equation (and corresponding initial conditions in
the full Eq.~\eqref{e:model}). Further, we take $V_0^2 = 6$,
$\epsilon = 0.1$, and $\eta=0$ or $1$ (at or below phase transition,
respectively), which yields $\alpha = 12\eta$ and $\tilde{g}\approx
7.3g$.

In the linear limit $g=0$ and at the phase transition point $\alpha
= 0$, Eq.~\eqref{Eq:EnvelopeA} becomes linear and is readily solved.
Its general solution corresponds to an expanding square wave front
propagating with speeds $\pm 2$ in both $X$ and $Y$ directions,
which is termed {\em pyramid diffraction}. For the initial
conditions (\ref{Eq:ic}), this pattern is illustrated in Fig.
\ref{Fig:Dispersion}.

\begin{figure}
    \begin{center}
        \vspace{0.2cm}
        \includegraphics[height = 1in]{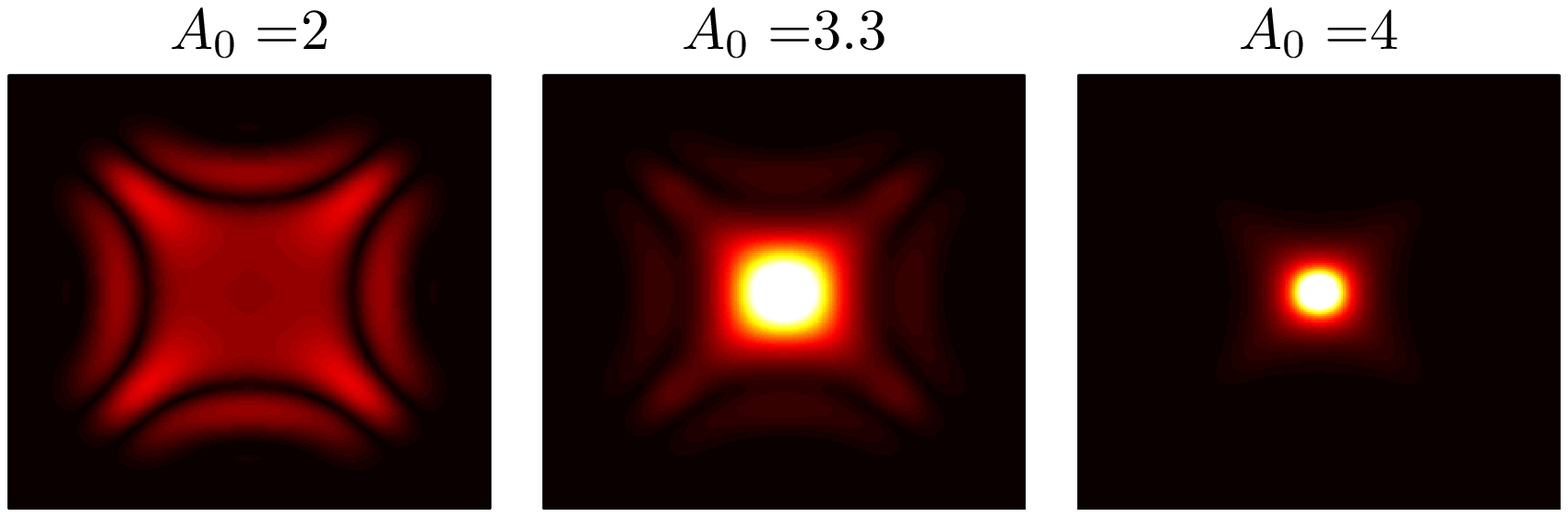}

        \vspace{0.15cm}
        \includegraphics[height = 1in]{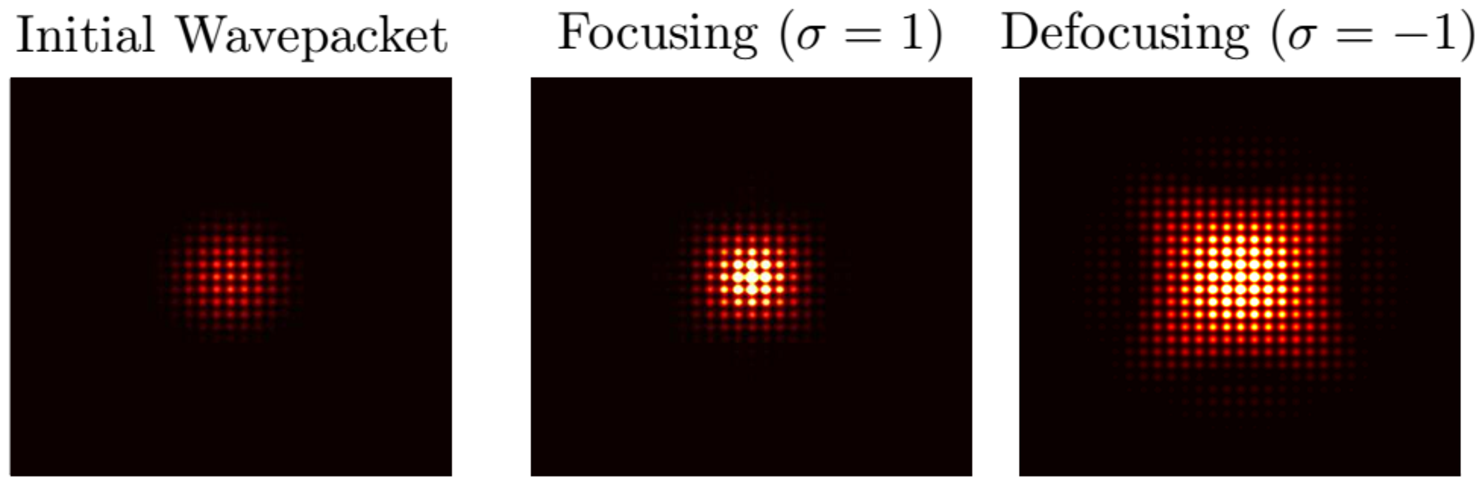}
        \caption{(Color online) Nonlinear dynamics of wave packets below
            phase transition. Upper row: envelope solutions in
            (\ref{Eq:EnvelopeA}) at $T\approx 2$ for three values of $A_0$ in
            (\ref{Eq:ic}). Lower row: solutions of the full equation
            (\ref{e:model}) for the initial wavepacket with $A_0=6$ (left) at
            later times under focusing (middle) and defocusing (right)
            nonlinearities. } \label{Fig:NonlinearEvolve}
    \end{center}
\end{figure}

In the presence of nonlinearity ($\tilde{g}\approx 7.3g$) and below
phase transition, the wave packet diffracts away if its initial
amplitude is below a certain threshold value, as displayed in the
upper left panel of Fig. \ref{Fig:NonlinearEvolve}. If the initial
amplitude is above this threshold, the envelope solution blows up to
infinity in finite time. For example, with the initial condition
(\ref{Eq:ic}), the envelope solution in (\ref{Eq:EnvelopeA}) blows
up when $A_0>3.2$ [Fig. \ref{Fig:NonlinearEvolve} (upper middle and
right panels)]. Remarkably, this blowup is independent of the sign
of nonlinearity, a fact which is clear from the envelope equation
(\ref{Eq:EnvelopeA}), since a sign change in $\tilde{g}$ can be
accounted for by taking the complex conjugate of this equation. In
the full equation (\ref{e:model}), it was  confirmed that similar
growth occurs for both signs of the nonlinearity as well (see Fig.
\ref{Fig:NonlinearEvolve}, lower row) at initial stages of
evolution, although the finite-time blowup is ruled out in the
defocussing medium at longer times.

\subsection{Nonlinear $\PT$-symmetric lattices}
\label{subsec:nonlin_lat}

So far, we have considered nonlinear models whose linear parts obey
$\PT$ symmetry. Now we turn to wave propagation in a
\emph{nonlinear} $\PT$-symmetric lattice governed by the equation
\begin{equation}
\label{eq:periodic:purelynonlin}
i \Psi_t + \Psi_{xx}  + g[1 +  U_{NL}(x)]|\Psi|^2\Psi = 0,
\end{equation}
where $U_{NL}(x)=U_{NL}^*(-x)$ is a $\PT$-symmetric nonlinear
potential. Equation (\ref{eq:periodic:purelynonlin}) can be
considered  as a $\PT$-symmetric deformation of conservative
nonlinear lattice models studied intensively during the last decade
[see \textcite{KMT11} for a review]. This model with a periodic
$\PT$-symmetric potential $U_{NL}(x) = V_0\cos( x) + i W_0\sin( x)$
was introduced by \textcite{AKKZ}. It   supports continuous
families of stable solitons (detailed linear stability analysis was
performed by \textcite{ZKK}). Interestingly, stable solitons can
be found even if periodic modulation of the real part of the
potential is absent, i.e., $V_0=0$.
Stable solitons also exist in non-periodic nonlinear landscapes, such as $U_{NL}(x) =  iW_0\tanh x$ or  $U_{NL}(x) = i W_0 x$.

Further natural generalization is  a model of combined linear and nonlinear $\PT$-symmetric lattices \cite{Hemixed}:
\begin{equation}
i \Psi_t + \Psi_{xx}   {-}  U_L(x)\Psi + g[1 +  U_{NL}(x)]|\Psi|^2\Psi = 0,
\end{equation}
where $U_L(x)=U_L^*(-x)$ and $U_{NL}(x)=U_{NL}^*(-x)$. The presence
of both linear and nonlinear modulations enriches the problem and
makes it possible to consider a general case where linear and
nonlinear modulations are different from each other and the special
case where the two lattices are identical \cite{Hemixed}. Stable gap
solitons can be found in both cases, as well as in the case of
real-valued functions $U_{NL}(x)$
\cite{HJMLC12,HM12,ML13}. It is of interest to consider in-phase modulations,
i.e., $U_L(x) = U_{NL}(x)$ \cite{Hemixed} and out-of-phase modulations, i.e., $U_L(x) = -U_{NL}(x)$.
The latter type of modulation can support stable fundamental and multi-pole solitons whose counterparts in in-phase lattices are unstable \cite{HLD13}.

\subsection{Solitons in generalized lattice models}

The basic $\PT$-symmetric nonlinear models described above allow for
numerous generalizations accounting for more complex forms of
nonlinearities and periodic lattices, as well as for multi-component
situations. Such generalized models also support a variety of
solitons whose properties were studied systematically. Here we
provide a succinct review of the available results [some
generalizations were also summarized by \textcite{HeMal13}].

\paragraph*{Vector mixed-gap solitons.} A model describing two incoherently coupled fields ($\Psi_{1,2}$)
in a $\PT$-symmetric lattice $U(x)$
\begin{equation*}
i\frac{\partial \Psi_{1,2}}{\partial t}+\frac{\partial^2 \Psi_{1,2}}{\partial x^2} {-} U(x)\Psi_{1,2} + (|\Psi_1|^2 + |\Psi_2|^2)\Psi_{1,2}=0,
\end{equation*}
was studied by \textcite{Kartashov}. Since the model does not
include linear coupling between the two fields, it supports  the
so-called mixed-gap solitons characterized by different propagation
constants for the two field components: $\Psi_{j}(x, t)=e^{-i\mu_{j}
    t}\psi_{j}(x)$, where $\mu_1 \ne \mu_2$ lie in different gaps of the
$\PT$-symmetric potential.
A further generalization of this model accounting for both linear
and nonlinear periodic lattices was considered by
\textcite{ZCSHL14}.

\paragraph*{Generalized nonlinearities.} A $\PT$-symmetric lattice model with a more general form of nonlinearity is
\begin{equation}
\label{eq:periodic:nonlocal}
i\Psi_t+\Psi_{xx}- U(x)\Psi  +  F(x, |\Psi|^2) \Psi=0.
\end{equation}
The case of nonlocal nonlinearity corresponds to $F(x, |\Psi|^2)
=g(x) \int_{-\infty}^\infty K(x-y)|\Psi(x)|^2dy$, where $K(x)\geq 0$
is a kernel function describing nonlocal properties of the medium,
and $g(x)$ is the nonlinear coefficient. Stable nonlocal gap
solitons were found for both self-focusing [$g(x) \equiv 1$]
\cite{LJZS12} and self-defocusing [$g(x) \equiv -1$]
\cite{ZLWH13,JABA14} nonlocal nonlinearities. Spatially modulated
nonlocal nonlinearity with a periodic function $g(x)$ was also
considered \cite{YHLX12}. ``Accessible solitons'' \cite{SM97} in a
strongly nonlocal 2D $\PT$-symmetric medium were reported by
\textcite{ZBH12}.

Saturating nonlinearity $F(x, |\Psi|^2) = g |\Psi|^2 / (1+|\Psi|^2)$
also supports gap solitons \cite{CZHL14,HJ15}.
Multistable solitons in $\PT$-symmetric lattices in the presence of
cubic-quintic nonlinearity $F(x, |\Psi|^2)=g_1|\Psi|^2 + g_2
|\Psi|^4$ were reported by \textcite{LLD12}. \textcite{HM13} studied
soliton propagation in the cubic-quintic Ginzburg-Landau model with
a $\PT$-symmetric lattice.

\paragraph*{$\PT$-symmetric superlattices.}
Superlattices are combinations of several periodic potentials with
different periods. \textcite{ZWZLH11} considered a $\PT$-symmetric
superlattice with the potential $U(x)$ having real and imaginary
parts: $V(x)=\vep \sin^2(x+\pi/2) + (1-\vep)\sin^2(2(x+\pi/2))$ and
$W(x) = W_0\sin(2x)$, and found that such a superlattice supports
stable gap solitons. The extension of this study to the case where
both real and imaginary parts of the complex potential are
dual-periodic superlattices was addressed by \textcite{WLZZH14}.

\paragraph*{Defect solitons.}
Periodic lattices or superlattices locally perturbed by a defect can
support defect solitons. As an example, one can consider a periodic
lattice whose real part is given as $V(x) = \cos^2(x)[1+\vep
f_D(x)]$, where $f_D(x)$ is a localized function and the coefficient
$\vep$ controls the defect strength \cite{WW11}. If the imaginary
part is an unperturbed periodic function, say, $W(x)=W_0\sin(2x)$,
then the resulting $\PT$-symmetric defect lattice is known to
support stable defect solitons under self-focusing \cite{WW11,HH12}
and self-defocusing \cite{HH13a} nonlinearities. Further
developments in this directions include defect solitons in
superlattices \cite{LuZh,HLMGH12, WHSZH14,HH13c,FGSZL14},  as well
as in nonlocal \cite{HLMGH12,HMLZH12,FGSZL14},  saturable
\cite{HH13b}, and $\PT$-symmetric \cite{WHSZH14,WHZZLH12}
nonlinearities. Defect solitons in 2D $\PT$-symmetric lattices were
reported by \textcite{XSCCLMH14}.

Finishing this review of generalized $\PT$-symmetric lattice models,
we also mention other possibilities such as chirped (quasi-periodic)
$\PT$ lattices \cite{CYCMLW13} and effects of higher-order
diffraction on $\PT$ models \cite{GSMZD14}.

\subsection{Bragg solitons}
\label{sec:bragg}

The NLS-type models considered above do not account for waves
reflected from the periodic structure, i.e., they are only valid
away from the Bragg resonance. If a stop gap is relatively narrow and
the carrier wave frequency falls inside that gap, the interference
of forward and backward propagating waves
\begin{equation}
\label{eq:bragg1}
E=E_f(z,t)e^{i(\beta_0z-\omega_0 t)}+E_b(z,t)e^{-i(\beta_0z+\omega_0 t)}
\end{equation}
must be considered.   Here $\omega_0$ is the carrier-wave frequency, and
$\beta_0=n_0\omega_0/c$ is the unperturbed propagation constant.
Then under Kerr nonlinearity the medium supports the so-called {\em
    Bragg solitons}~\cite{ChJo89,AcWab89}. Bragg solitons persist also
in the case of a periodic $\PT$-symmetric grating and are described
by the coupled equations \cite{Miri2012a}:
\begin{equation}
\label{Bragg_syst}
\begin{array}{c}
\displaystyle
\frac{i}{v}\frac{\partial E_{f}}{\partial t}+ i \frac{\partial E_{f}}{\partial z} +(\kappa +g)
E_b 
+ \gamma \left(|E_f|^2+2|E_b|^2\right)\! E_f=0,
\\[2mm]
\displaystyle
\frac{i}{v}\frac{\partial E_{b}}{\partial t} -i\frac{\partial E_{b}}{\partial z} +(\kappa -g)
E_f 
+ \gamma \left(2|E_f|^2+|E_b|^2\right)\! E_b=0,
\end{array}
\end{equation}
where $v=c/n_0$, $\kappa$ is the coupling arising from the real Bragg grating
itself, $g$ is the antisymmetric coupling arising from the complex
$\PT$-symmetric potential, and $\gamma$ is a nonlinear coefficient.

In the linear limit ($\gamma=0$), the substitution $E_{f,b} =
\Psi_{f,b} e^{i(Kz-i\Omega t)}$ yields the dispersion relation
$\Omega^2 = v^2 (K^2+\kappa^2-g^2)$. Thus $\PT$ symmetry is unbroken
($\Omega$ is real for any $K$) if $g\leq \kappa$; the case
$g=\kappa$ corresponds to the exceptional point; and $\PT$ symmetry
is broken if $g>\kappa$. When $\PT$ symmetry is unbroken, the
nonlinear model (\ref{Bragg_syst}) supports traveling soliton
solutions
\begin{equation}
E_{f,b} = \pm \alpha \sqrt{\frac{\kappa_\rho}{2\gamma}}\Delta^{\mp 1}\sin(\sigma)\sech\left(\theta \mp \frac{\sigma}{2}\right)  e^{i\eta(z,t)},
\end{equation}
where
$\kappa_\rho = \sqrt{\kappa^2-g^2}$, $\theta = \kappa_\rho \sin(\sigma) (z-mvt)/\sqrt{1-m^2}$,
$m=(1-\Delta^4)/(1+\Delta^4)$, and $\Delta$, $\sigma\in (0,\pi)$ are
free parameters. The amplitude $\alpha$ and phase $\eta(z,t)$ can
also be found in analytical form \cite{Miri2012a}.

The nonlinear model (\ref{Bragg_syst}) also supports plane wave
solutions whose modulational instability was classified by
\textcite{Sarma14} in different parameter regimes. Families of more
general traveling wave solutions (including bright solitons in
forward waves and dark solitons in backward waves) were
reported by \textcite{Gupta14a}.

\section{$\PT$-symmetric $\chi^{(2)}$ media}
\label{sec:quadratic}

In this section, we consider $\PT$-symmetric optical media with
quadratic [i.e., $\chi^{(2)}$] nonlinearity. Our main concern is the
existence and stability of nonlinear modes. We have seen before that
the existence of continuous solution families requires not only
balance between linear gain and loss, but also specific form of
nonlinearity. This balance can be achieved either in spatially
extended systems, or in linearly coupled multicomponent systems.
$\chi^{(2)}$ media appear as a special case since they arise due to
frequency conversion and intrinsically have two components having
``different nonlinearities''.
In addition, they do not allow linear coupling between the two
components due to different frequencies.
Therefore $\PT$-symmetric optics of quadratic media can be developed
either on the basis of coupled extended systems where one (or each)
component is subject to gain and loss, or as a combination of (at
least two) $\PT$-symmetric models where the first two components are
linearly coupled with the second two components. Below we consider
the respective examples.

\subsection{Quadratic media with $\PT$-symmetric potentials}
\label{subsec:chi2_loc_mod}

The dimensionless mathematical model for quadratic media with
$\PT$-symmetric potentials reads:
\begin{subequations}
    \label{eq:ch2_cont}
    \begin{eqnarray}
    \label{eq:ch2_cont1}
    i{ q_{1,z}}&=&-{ q_{1,xx}}+ V(x)
    q_{1}+2q_{1}^{\ast }q_{2},
    \\[1mm]
    \label{eq:ch2_cont2}
    i{ q_{2,z}} &=&-(1/2){q_{2,xx}}+
    2[\tilde{V}(x)+\beta]
    q_{2}+q_{1}^{2},
    \end{eqnarray}%
\end{subequations}
where $q_1$ and $q_2$ are the fundamental-frequency (FF) and
second-harmonic (SH) fields, $\beta$ is the mismatch parameter, and
$V(x)$, $\tilde{V}(x)$ are $\PT$-symmetric potentials.
Stationary localized solutions of (\ref{eq:ch2_cont}) are searched
in the form $q_1=w_1(x)e^{ibz}$ and $q_2=w_2(x)e^{2ibz}$, where $w_1, w_2$ can be
required to obey the symmetries $\{w_{1}(x),w_{2}(x)\}=\{w_{1}^{\ast
}(-x ),w_{2}^{\ast }(-x )\}$ or $\{w_{1}(x ),w_{2}(x
)\}=\{-w_{1}^{\ast }(-x ),w_{2}^{\ast}(x)\}$. In the so-called
cascade limit corresponding to large $\beta$ for which the
approximate solution $w_2\approx - w_1^2/2\beta$ of
(\ref{eq:ch2_cont2}) is valid~\cite{SteHaTo96}, Eq.
~(\ref{eq:ch2_cont1}) for the FF is reduced to the stationary NLS
equation (\ref{e:parabol:psi}) with a $\PT$-symmetric potential.
This suggests the existence of localized modes in the model
(\ref{eq:ch2_cont}) with $\PT$-symmetric potentials. Such modes
indeed were found \cite{Moreira1} in the case of the Scarf II potential (\ref{scarf})
for the FF and $\tilde{V}= 2V_1/\cosh^2x$ for the SH.
On the other hand, exact $\sech$-shaped or cnoidal-shaped solutions
can be found by ``inverse engineering'' \cite{AU14,TVAAK15}.

Another case where stationary modes in $\PT$-symmetric quadratic
media were found~\cite{Moreira2} corresponds to the periodic
$\PT$-symmetric potential $V(x)$ given by (\ref{e:PTlattice}) in the
FF and a real periodic potential
$\tilde{V}(x)=\tilde{V}_1\cos^2(2x)$ in the SH. The resulting
periodic model supports families of gap solitons which
bifurcate from band edges of the underlying linear model. Such
bifurcations are always characterized by vanishing fields in the FF
($w_{1}\rightarrow 0$), while the asymptotics of SH may be
different.
More precisely, there exist three possibilities for bifurcation of
gap solitons from band edges.

\begin{figure}
    \includegraphics[width=\columnwidth]{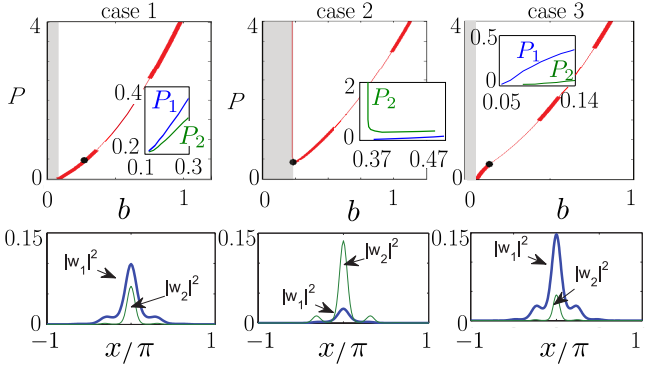}%
    \caption{(Color online) Upper panels: families of fundamental solitons in the semi-infinite gap.
        Left, center and right panels correspond to cases 1, 2, and 3, with  $\beta=-0.316$, $\beta=0$ and $\beta=-0.5134$ respectively.
        Thick and thin lines correspond to stable and unstable solitons. Shaded regions denote bands of
        the FF and/or SH. Insets show powers of the FF ($P_{1}$) and SH ($P_{2}$) fields close to the band-edge. Lower panels: intensities of stable fundamental solitons  indicated by black circles in the respective upper panels:  $b=1.25$ (case 1),  $b=1.43$ (case 2), and $b=1.21$ (case 3).
        Other parameters are $V_0=\tilde{V}_0=2$ and $W_0 =0.35$.
        Adapted from \textcite{Moreira2}.}
    \label{fig:chi_2_pt_pot}
\end{figure}

\emph{Case 1}: Both components are of the same order, i.e.,
$w_{2}\sim w_{1}$ and $w_{1}\rightarrow 0$.
In this case (see left column of Fig.~\ref{fig:chi_2_pt_pot}), in the vicinity of the bifurcation
the nonlinearity is negligible, and both components
are governed by linear equations: the FF
is described by a solution of the stationary
equation~(\ref{LinearMu}) with the pair ($\mu$, $\psi$) replaced by ($b$, $w_1$), i.e., $w_1\sim \psi$,  while  $w_2\sim \tilde{\psi}$, where
$\tilde{\psi}$ solves the linear Mathieu  equation with the potential
$\tilde{V}(x)$. For the existence of a gap soliton, the propagation constant
$b$ should belong to the gaps of both FF and SH, which requires a
non-empty overlap of the stop gaps of both components; such an overlap
can be termed as a \textit{total gap}.  Moreover, the adopted
scaling implies that band edges of FF and SH should
coincide exactly. This constraint makes this case uncommon, although in
practice it can always be achieved by adjusting the mismatch
parameter $\beta$.

\emph{Case 2}: The SH field remains finite: $w_{2}=\mathcal{O}(1)$,
and $w_{1}\rightarrow 0$. In this case (see middle column of
Fig.~\ref{fig:chi_2_pt_pot}),
the vicinity of the total gap must coincide with the respective SH
gap edge. Then, while the FF component is vanishing
($w_{1}\rightarrow 0$) as $b$ approaches the total gap edge, the
amplitude of the SH $w_{2}$ remains finite and its width increases
(i.e., the SH in this limit becomes delocalized). This explains
divergent powers of the SH $P_2$ (shown in the inset in the middle panel of Fig.~\ref{fig:chi_2_pt_pot}) and, respectively, the divergence  of the total
power.

\emph{Case 3}: The SH amplitude scales as the square of the FF
amplitude: $w_{2}=O(w_{1}^{2})$, and $w_{1}\rightarrow 0$. This case
(see right column of Fig.~\ref{fig:chi_2_pt_pot}) takes place when
an edge of the total gap, from which gap solitons bifurcate, is determined by
one of the respective edges of the gap of the FF. Then in the
small-amplitude limit the SH is determined by the field distribution
in the FF and the total power vanishes at the bifurcation point.


\subsection{$\PT$-symmetric coupler with quadratic nonlinearity}
\label{subsec:chi2_coupler}

A model of a $\PT$-symmetric coupler with $\chi^{(2)}$ nonlinearity, introduced by \textcite{LZKKA}, reads:
\begin{equation}
\label{eq:ch2_disc}
\begin{array}{ccl}
i\dot{u}_{1} &=& k_1 u_2-2u_1^* v_1 + i\gamma_1 u_1,\\
i\dot{v}_{1} &=& k_2 v_2-u_1^2-\beta v_1 + i\gamma_2 v_1,\\
i\dot{u}_{2} &=& k_1 u_1-2u_2^* v_2 - i\gamma_1 u_2,\\
i\dot{v}_{2} &=& k_2 v_1-u_2^2-\beta v_2 - i\gamma_2 v_2.
\end{array}
\end{equation}
Here two modes
propagate in each waveguide: the FF $u_j$ and the SH $v_j$, and $j=1,2$ enumerates the coupler arm. Linear
coupling between two FFs (two SHs) is described by $k_1$ ($k_2$),
the gain (loss) strength in the arms is given by $\gamma_{j}$, and
$\beta$ is the mismatch parameter.

Stationary modes of (\ref{eq:ch2_disc}) are searched in the form
$(u_1,v_1, u_2, v_2)^T= e^{-i\Lambda bz} \bw$, where $\Lambda=\diag
(1,2,1,2)$, $b$ is the propagation constant, and
$\bw=(w^{(1)},w^{(2)},w^{(3)},w^{(4)})^T$ solves the stationary
nonlinear problem $E\Lambda\bw = H\bw - F(\bw) \bw$, where
matrix $H$ describes the linear part of system (\ref{eq:ch2_disc})
and matrix function $F(\bw)$ has $F_{1,2}=
2[w^{(1)}]^*$, $F_{2,1}= w^{(1)}$,  $F_{3,4}= 2[w^{(3)}]^*$, and
$F_{4,3}= w^{(3)}$ with other entries being zero. $H$ is $\PT$ symmetric with
$\p=\sigma_1\otimes\sigma_0$.

From a physical point of view  (\ref{eq:ch2_disc}) represents a
coupler with each of the arms guiding two modes; from a mathematical
point of view, it is a quadrimer similar to one considered in
Sec.~\ref{sec:discrete}. Eigenvalues of $H$ are given by
\begin{equation*}
\tilde{b}_{1,2}=\pm\sqrt{k_1^2-\gamma_1^2}, \quad \tilde{b}_{3, 4}
=\frac12\left(-\beta\pm\sqrt{k_2^2-\gamma_2^2}\right),
\end{equation*}
and its   eigenvectors  are:
\begin{equation*}
\begin{array}{ll}
\tbw_1=(e^{i\theta_1},0,e^{-i\theta_1},0)^T, & \tbw_2=i(e^{-i\theta_1},0,-e^{i\theta_1},0)^T,
\\
\tbw_3=(0,e^{i\theta_2},0,e^{-i\theta_2})^T, & \tbw_4=i(0,e^{-i\theta_2},0,-e^{-i\theta_2})^T,
\end{array}
\end{equation*}
where $\theta_{1,2} = \frac{1}{2}\arctan\left({\gamma_{1,2}}/\sqrt{k_{1,2}^2 - \gamma_{1,2}^2}\right)$.
If $\PT$ symmetry is unbroken, i.e.,  $|\gamma_{1,2}| < k_{1,2}$,
then    eigenvectors $\tbw_j$ are  $\PT$ { invariant}, i.e., $\PT
\tbw = \tbw$.

In the nonlinear problem, one looks for nonlinear modes obtained by
continuation from the linear eigenvectors [see
Sec.~\ref{sec:discr:fam:linear}]. Notice that $F(\tbw_{3,4})=0$,
which means that eigenvectors $\tbw_{3,4}$ also solve the nonlinear
problem (\ref{eq:ch2_disc}). As a result, one can construct
nonlinear modes of {\em two} different types. For the first type, one
looks for nonlinear continuation of $\tbw_{1,2}$ in the form of
expansions $\bw_j = \vep \tbw_j + \vep^2 \textbf{W}_j+\ldots$ and
$b_j=\tb_j+\vep b_j^{(2)}+\ldots$, where coefficients $\textbf{W}_j$
and $b_j^{(2)}$ ($j=1,2$) are computed from the solvability
conditions at higher orders. This is the ``standard'' expansion: in
the linear limit ($\vep=0$), the power of nonlinear modes
$P=|w^{(1)}|^2+|w^{(3)}|^2+2(|w^{(1)}|^2+|w^{(3)}|^2)$ (which
corresponds to the Manley-Rowe invariant of the conservative coupler
with $\gamma_{1,2}=0$) vanishes. For modes of the second type, which
bifurcate from $\tbw_{3,4}$, the expansions are of the form $\bw_j =
\alpha_j\tbw_j+\vep \textbf{W}_j+\ldots$,
$b_j=\tb_j+\vep^2b_j^{(2)}+\ldots$, where $\alpha_j$ ($j=3,4$) are
(generically nonzero) coefficients. In this case, at $\vep=0$ the
mode amplitudes do not vanish. Admissible values of the coefficients
$\alpha_j$ are found from compatibility conditions that arise from
the underlying expansions. Generally speaking there exist two
admissible values of $\alpha_j$, thus $\tbw_3$ and $\tbw_4$ admit
{\em two} continuous families of nonlinear modes \cite{LZKKA}.

\begin{figure}
    \includegraphics[width=\columnwidth]{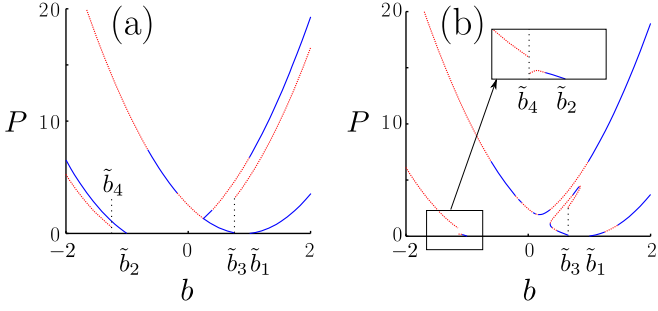}%
    \caption{(Color online)  Families of
        nonlinear modes in $\chi^{(2)}$ coupler (\ref{eq:ch2_disc}) for   $k_1=1$, $k_2 =2$, $\beta =
        0.5$, and  $\gamma_{1,2}=0$ [panel (a)];    $\gamma_1=0.1$ and $\gamma_2=0.9$  [panel (b)].
        Solid blue and dashed red segments correspond to stable and unstable modes respectively.
        Adapted from \textcite{LZKKA}.}    \label{fig:chi2_discrete}
\end{figure}

Numerical results are illustrated in
Fig.~\ref{fig:chi2_discrete}. In both  conservative and
$\PT$-symmetric cases there are two solution families
bifurcating from the eigenstates $\tbw_{1}$ and $\tbw_{2}$. These
bifurcations take place at the limit $P=0$. Bifurcations
of nonlinear modes from $\tbw_{3,4}$ occur at finite $P$ (in some
cases these values of $P$ are
quite small and hardly  distinguishable from $P=0$ on the scale of
the figure). In the conservative case, either $\tbw_3$ or $\tbw_4$
gives birth to {\em one} physically distinct family. In the $\PT$-symmetric case
{\em two} distinct solution families originate from either of
$\tbw_3$ and $\tbw_4$.

\section{Partial $\PT$ symmetry}
\label{sec:high_dim}

Multi-dimensional complex potentials considered so far obey the $\PT$ symmetry with the canonical $\p$ operator (\ref{p_reflect}) resulting in inversion of all spatial variables. Now we discuss situations where the complex potential is
not $\PT$ symmetric in this sense but is \emph{partially $\PT$-symmetric} meaning that the
potential is invariant under complex conjugation and reflection in a
\emph{single} spatial direction. Such
potentials can still admit all-real spectra and support continuous families
of solitons~\cite{Yang14c,KarKonTor}.

The mathematical model we use is a 2D NLS equation (\ref{e:model}) with a
potential $U_{2D}(x, y)$ which is partially $\PT$ symmetric with respect to $x$:
\begin{equation} \label{e:PPTcondition}
U_{2D}^*(x,y)=U_{2D}(-x,y).
\end{equation}
No symmetry is assumed in the $y$ direction.

First, we argue that the linear spectrum of a partially
$\PT$-symmetric potential can be all-real.
If the potential is separable, i.e., $U_{2D}=U_{2D}^0 = V_1(x)+V_2(y)$,
then the partial $\PT$ symmetry condition (\ref{e:PPTcondition})
implies that $V_1^*(x)=V_1(-x)$, $V_2^*(y)=V_2(y)$,
and its eigenvalues are $\lambda=\Lambda_1+\Lambda_2$,
where $\Lambda_j$ are eigenvalues of 1D potentials
$V_j(x)$. Since $V_1(x)$ is
$\PT$ symmetric, its eigenvalues $\Lambda_1$ can be all-real. Since
$V_2(y)$ is strictly real, i.e. the respective Hamiltonian is Hermitian, its eigenvalues $\Lambda_2$ are all-real
as well. Thus eigenvalues $\lambda$ of the separable potential
$U_{2D}(x,y)$ can be all-real.

Next, we consider a separable potential with all-real spectra perturbed
by a localized potential $ U_{2D}^p(x,y)$: $U_{2D}=U_{2D}^0+\epsilon U_{2D}^p$,
where $\epsilon$ a small real parameter, and both $U_{2D}^0$ and
$U_{2D}^p$ satisfy the partial $\PT$ symmetry
(\ref{e:PPTcondition}). Since $U_{2D}^p$ is localized, continuous
spectrum  of the perturbed potential $U_{2D}$ coincides with that of
$U_{2D}^0$ and is thus all-real. Regarding isolated eigenvalues of
$U_{2D}$, they can be shown to be real as well by a perturbation
calculation \cite{BenderJones08,Yang14c}.

Partially $\PT$-symmetric potentials possess some typical feature of
standard $\PT$-symmetric potentials, such as phase transition in the
linear case and existence of continuous families of stationary solitons in the nonlinear case. 
One can show that, from each simple real discrete
eigenvalue $\tmu_n$ of a partially $\PT$-symmetric potential, a
continuous family of solitons bifurcates out under both focusing and
defocusing nonlinearities. Indeed, introducing small-amplitude
expansions for nonlinear modes bifurcating from a linear eigenvalue
$\tmu_n$ with eigenfunction $\tpsi_n(x,y)$, similar to those in
Eqs.~(\ref{eq:par:expans}), one can show that bifurcations are
governed by the coefficient $\mu_n^{(2)}$ defined by
Eq.~(\ref{eq:parabol:mu2}), where the integrals should be taken over
$dxdy$.
For real eigenvalue $\tmu_n$, its eigenfunction $\tpsi_n$ inherits
the partial $\PT$ symmetry of the potential. Thus both integrals in
the expression for $\mu_n^{(2)}$ are real, so $\mu_n^{(2)}$ is also
real.
Pursuing the  perturbation calculation to higher orders, one can
construct a perturbation solution to all powers of $\epsilon$, and
thus a continuous family of solitons, parameterized by $\mu$,
bifurcates out from the linear eigenmode $(\tmu_n, \tpsi_n)$.

\begin{figure}
    \begin{center}
        \includegraphics[width=\columnwidth]{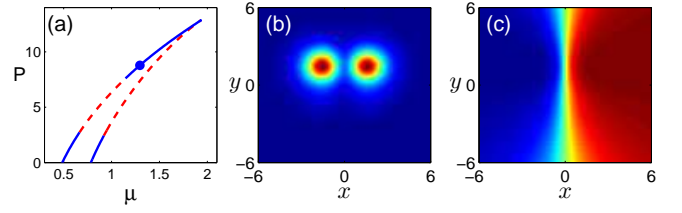}
        \caption{(Color online) (a) power diagram of soliton families in potential (\ref{e:Vexample}) under focusing nonlinearity
            (solid blue segments are stable and dashed red unstable); (b,c) amplitude and phase fields of the soliton at the marked point of the power curve.
            Adapted from \cite{Yang14c}. \label{PPT_fig2} }
    \end{center}
\end{figure}

The existence of soliton families can be verified numerically. For
this purpose, we take the partially $\PT$-symmetric potential%
\begin{eqnarray}
\label{e:Vexample}
U_{2D}=
-3\left(\!e^{-|\br-\br_+|^2}+e^{-|\br-\br_-|^2}\!\right) \quad\quad\quad\quad\quad\quad\quad\quad\quad\quad
\nonumber \\
-2\left(\!e^{-|\br+\br_+|^2}+e^{-|\br+\br_-|^2}\!\right)
+i\gamma\left(-\! e^{-|\br+\br_-|^2}+e^{-|\br+\br_+|^2}
\right. \nonumber \quad\\ \left.
-2e^{-|\br-\br_+|^2}+2e^{-|\br-\br_-|^2}
\!\right), \  \mbox{where} \  \br_\pm=(\pm x_0, y_0).
\qquad\quad
\end{eqnarray}
For  $\gamma=0.1$ and $x_0=y_0=1.5$, this potential has  three discrete eigenmodes,  from each of which a soliton family bifurcates out. Soliton families bifurcated from the
first and second linear eigenmodes of the potential under focusing
nonlinearity ($g=1$) are displayed in Fig.~\ref{PPT_fig2}(a).
Interestingly, these two power curves are connected through a fold
bifurcation and have an upper bound. The profile of a stable soliton
on the power curve is displayed in Fig.~\ref{PPT_fig2}(b,c).

Results of numerical linear stability analysis show that
most solitons of the upper power branch are stable.  This is
surprising, since in conservative potentials solitons on the upper
power branch are generally less stable.
The increased stability of the upper power branch
here is due to the complex partially $\PT$-symmetric potential
(\ref{e:Vexample}), which stabilizes solitons at higher powers.

Solitons in Fig. \ref{PPT_fig2} are partially $\PT$-symmetric as the
underlying potential (\ref{e:Vexample}) itself.
In special classes of partially $\PT$-symmetric potentials, symmetry
breaking of solitons can occur, where families of
non-partially-$\PT$-symmetric solitons can bifurcate out from the
base branch of partially-$\PT$-symmetric solitons \cite{Yang15}.
This situation is analogous to 1D $\PT$-symmetric potentials (see Sec.~\ref{subsec:par:symm_break} and
\textcite{Yang14d}).

\section{Spectral singularities}
\label{sec:lasing_absorption}

\subsection{Spectral singularities in the linear theory}
\label{se:lasing}

A non-Hermitian Hamiltonian at an exceptional point does not admit a complete bi-orthonormal basis and  cannot be diagonalized. Alternatively, completeness of the basis is lost if the continuum spectrum features a {\em  spectral singularity}~\cite{Naimark60},  which may  be relevant for
physical applications~\cite{Most2009b,MosMeDe2009,Most2011a,Most2011b,Longhi09a,Longhi09b,Longhi10} (see also~\cite{Most2015} for a recent review).

Consider an eigenvalue problem $H\tpsi_k=k^2\tpsi_k$
for the Schr\"odinger operator (\ref{Schrod}) with a localized
complex potential: $\lim_{x\to\pm\infty}|U(x)|=0$ (the tilde stands
for eigenstates of the linear problem). The associated Jost
solutions $\tpsi_{k\pm}(x)$ are defined by their asymptotics:
$\tpsi_{k\pm}(x)\sim e^{\pm ikx}$  {for $x\to\pm\infty$},
while an arbitrary eigenfunction of the continuous spectrum has  asymptotics
$
\tpsi_k(x)\to A_\pm e^{ikx}+B_\pm e^{-ikx}
$
at $x\to\pm\infty$.
The constants, $A_\pm$ and $B_\pm$ are not independent: the link among them defines the transfer matrix $M(k)$
\begin{eqnarray}
\label{eq_transfer}
\left(A_+, B_+ \right)^T=M(k) \left(
A_-,  B_- \right)^T.
\end{eqnarray}
The Wronskian of the Jost solutions
$
W[\tpsi_{k-},\tpsi_{k+}]=2ikM_{22}(k)
$
is $x$-independent  and $\det M(k)=1$. If $M_{22}(k_*)=0$ for some
real $k_*\ne 0$, then the number $k_*^2>0$ is said to be \emph{a
    spectral singularity} \cite{MosMeDe2009}.
The explicit form of the Wronskian implies that linear dependence of
the Jost solutions is necessary and sufficient for the spectral
singularity~\cite{Most2013a}. Notice that bound states of $H$, if
any, are also defined by the roots of $M_{22}(k)$ which, however,
are located in the upper half-plane of the complex $k$: Im$(k)>0$.

Spectral singularities are known for a number of particular cases,
including the step-like potential
(\ref{pot:step_like})~\cite{Most2009b,Most2011a,Most2014},
$\delta$-function potentials~\cite{MosMeDe2009}, $\PT$-symmetric
Scarff II potentials~\cite{Ahmed09}, and special types of periodic potentials~\cite{Gasymov1980,Longhi10}.

Consider now  a monochromatic wave incident on the potential $U(x)$
either from the left or from the right and suppose that there exists a real ${k}_*$ solving  $M_{22}({k}_*)=0$, i.e., there exists a spectral singularity. Relations (\ref{opt:tr})
among the scattering characteristics still hold. Thus at the  spectral
singularity the reflection and transmission coefficients diverge,
behaving like zero-width resonances~\cite{Most2009b}.
It follows from (\ref{eq_transfer}) that there exists a solution
with $A_-({k}_*)=B_+({k}_*)=0$. This is a solution where only radiation propagating away
from the potential exists, i.e., it means that the potential operates
as a laser~\cite{Longhi10,Most2011a}. 

On the other hand, if a medium
allows for a zero  of $M_{11}(\tilde{k}_*)=0$ with real
$\tilde{k}_*$,  then there exists a solution with
$A_+(\tilde{k}_*)=B_-(\tilde{k}_*)=0$ meaning absence of radiation
propagating away from the potential. Such a medium operates as a
coherent perfect absorber (CPA), which was predicted
by~\textcite{ChoGeCaSto10} and observed experimentally
by~\textcite{WaCho11}. $(\tilde{k}_*)^2$ is referred to as time-reversed spectral singularity~\cite{Longhi11,Most2015}. Realization of a CPA using a pair of passive resonators was reported by \textcite{Sun20014}. The authors observed complete absorption of light when the system was in (passive) $\PT$-symmetric phase; however the complete absorption was not observed when the $\PT$-phase was broken.

In a generic case, there may exist either spectral singularities or time-reversed spectral singularities, or both with ${k}_*\neq \tilde{k}_*$. However, if  a complex potential is $\PT$ symmetric, then  for each $k^*$ there exist $\tilde{k}_*={k}_*$ giving origin to so-called self-dual spectral singularity. Such a potential  can operate either as a laser or as an absorber, i.e., as a CPA-laser~\cite{Longhi10,ChoGeSto11}. \textcolor{black}{Furthermore,  the interplay between $\PT$ symmetry and Fano resonances can result in  singularities emerging from the coincidence of two independent singularities and having highly directional responses~\cite{Ramezani2014}.}


\subsection{Spectral singularities of a nonlinear layer}

Generalization of spectral singularities to nonlinear media is
justified by at least two reasons. First, in a realistic system
infinite transmission or reflection coefficients are the idealization,
and regularizing mechanisms must be involved, nonlinearity being one
of them. Second, the nonlinearity is expected to become a dominating mechanism at large field amplitudes.

Effects of nonlinearity on spectral singularities were considered
by~\textcite{Most2013a,Most2013b,Most2014} for the nonlinear
eigenvalue problem
\begin{equation}
\label{eq:nonlin_eigen}
H^{nl}\psi_k=k^2\psi_k,
\,\,\, H^{nl}\psi\equiv - \psi^{\prime\prime} + U(x)\psi+F(\psi^{\prime}, \psi,x)\psi,
\end{equation}
where $\psi(x)$ is a complex-valued function, $U(x)$ is a rapidly decaying complex potential, and $F(\psi^{\prime}, \psi,x)$ describes the nonlinearity, which is  confined to the interval $[0,1]$:  $F(\psi^{\prime}, \psi,x)\equiv 0$ for $x<0$ and $x>1$. 
In this case one can still exploit Jost solutions  $\tpsi_{k\pm}$  
of the underlying linear problem 
and define a nonlinear spectral singularity  as
follows~\cite{Most2013a}: {\em a positive real number $k^2$ is a
    spectral singularity of $H^{nl}$ if there exists a solution $\psi_k$
    of the nonlinear problem (\ref{eq:nonlin_eigen})
    such that $\lim_{x\to\pm\infty}\psi_k(x)=C_{\pm}\tpsi_{k\pm}(x)$, where $C_{\pm}$ are complex numbers}. %

Continuity of the field and its derivative at the boundaries of the
nonlinear medium requires
\begin{subequations}
    \label{eq:continuity}
    \begin{eqnarray}
    \label{eq:continuity_0}
    \psi_k(0)=\psi_{k-}(0),\quad  \psi_k^\prime(0)=\psi_{k-}^\prime(0),
    \\
    \label{eq:continuity_1}
    \psi_k(1)=\psi_{k+}(1), \quad \psi_k^\prime\!(1)=\psi_{k+}^\prime(1).
    \end{eqnarray}
\end{subequations}
This suggests  an algorithm for obtaining spectral
singularities~\cite{Most2013a,Most2013b}. Consider a solution
$\psi_{k-}(x)$ [or $\psi_{k+}(x)$] of (\ref{eq:nonlin_eigen}) on the
semi-axis $x\in[0,\infty)$ [or $x\in(-\infty,1]$] with the
``initial'' conditions (\ref{eq:continuity_0}) [or
(\ref{eq:continuity_1})]. Compute $k$ ensuring the asymptotics
$\psi_{k}(x)\sim C_+e^{ikx}$ as $x\to\infty$ [or $\psi_{k}(x)\sim
C_-e^{-ikx}$ as $x\to-\infty$]. If the problem has a real solution for
either of $\psi_{k}(x)$, the respective $k$ then yields a spectral
singularity $k^2$.

Let us consider an example of a non-$\PT$-symmetric potential
$U(x)=\zeta\delta(x-a)$, where $\zeta$ is a complex number, $a\in(0,1)$ is the position of the
defect inside the nonlinear layer, and Kerr nonlinearity $F \equiv
\chi|\psi|^2$ with real $\chi$ for $x\in(0,1)$ and $F \equiv 0$
otherwise \cite{Most2013a}. Assuming  $\chi|\psi|^2$ to be small,
one finds that a nonlinear spectral singularity occurs at $k$ satisfying
$
\zeta\approx 2ik+ {i\chi|A_-|^2} (e^{2ik(1-a)}+e^{2ika}-2)/(2k)
$
[at $\chi=0$ one recovers the linear spectral singularity~\cite{MosMeDe2009}].
The field intensity can be found from this expression:
\begin{eqnarray}
\label{eq:filed_int_singular}
\chi|A_-|^2\approx - {k\mbox{Re}(\zeta)}\,/\, ({\cos[k(1-2a)]\sin k}),
\end{eqnarray}
which is valid for $\mbox{Re}(\zeta)\ll 1$ and $k\neq\pi m,\, \pi(m+1/2)/(1-2a)$.
The obtained nonlinear spectral singularities obey parity symmetry, i.e., they are invariant under the transformation $a\to 1-a$; they are amplitude-dependent and sensitive to location of the defect $a$. One also observes that the right hand side of (\ref{eq:filed_int_singular}) has a minimal value, which means that no singularities arise if the amplitude is below a certain threshold.

A nonlinear $\PT$-symmetric bi-layered structure with
potential (\ref{pot:step_like}) was considered by
\textcite{Most2014}. It was found that  for $V_0=1$  the lossy layer
results in a decrease of the lasing threshold of the gain.  On the
other hand, when $V_0-1\gg \gamma$ the threshold value of the
gain-loss coefficient, considered as a function of the real part of
the refractive index, has a minimum  (for a homogeneous active layer
the lasing threshold decreases with $V_0$).

From these examples, one can conclude that while the cubic
nonlinearity makes the spectral singularity amplitude-dependent, it
does not regularize the scattering characteristics. Similar
conclusions can be reached from the study of step-like potentials
(\ref{pot:step_like})~\cite{Most2013b,Most2014}. However,
regularization of the spectral singularity is indeed possible. It
was obtained by \textcite{LiGuAg14}  for a $\PT$-symmetric
bi-layered structure with saturable nonlinearity modeled by
\begin{eqnarray}
\label{eq:spectr_sing_saturable}
\frac{d^2\psi}{dx^2}+k^2\xi(x)\frac{(\delta +i)\psi}{1+\delta^2+\alpha|\psi|^2}=0,
\end{eqnarray}
where $\xi(x)=-1$ for $x\in(-L,0)$ and $\xi(x)=1$ for $x\in(0,L)$,
and $\delta$, $\alpha$ are real constants.

The described algorithm of obtaining nonlinear spectral
singularities relies on the solution of the scattering problem where
the output radiation is fixed, rather than on the solution of a
problem with the fixed amplitude of the incident wave. These are two
different statements of the nonlinear scattering problem, referred
to as the {\em fixed-output} and {\em fixed-input} problems [see
e.g.~\textcite{KonVaz} and references therein]. The fixed-input
problem manifests multistability phenomenon, while the fixed-output
problem does not. In the numerical study of transmission-coefficient
dependence on the input intensity, bi-stability was reported
by~\textcite{LiGuAg14}.

\section{$\PT$ symmetry in Klein-Gordon models}
\label{sec:KG}
$\PT$ symmetry can be introduced in Klein-Gordon (KG) models which
in the conservative case read
\begin{eqnarray}
\label{KG}
u_{tt}-u_{xx}+f(u)=0,
\end{eqnarray}
where $f(u)$
is a nonlinear function of the field $u(x,t)$. The two celebrated
examples are the sine-Gordon (SG) equation given by $f(u)= \sin(u)$
and the $\phi^4$ model given by $f(u)=2(u^3-u)$.
Now the $\PT$ symmetry is defined by the
transformation $(x,t)\to (-x,-t)$. To preserve this symmetry, gain
and loss can be introduced by adding a spatially inhomogeneous
dissipative term $\gamma(x)u_t$ with $\gamma(x)$ being an odd
function: $\gamma(x)=-\gamma(-x)$~\cite{DFKSS14}. This yields the
model
\begin{eqnarray}
\label{PT-KG}
u_{tt}-u_{xx}+\gamma(x)u_t+f(u)=0,
\end{eqnarray}
where lossy and gain domains correspond to regions with
$\gamma(x)>0$ and $\gamma(x)<0$, respectively.

For   $\PT$-symmetric KG model (\ref{PT-KG}), the Galilean
invariance is broken. Therefore, unlike its conservative
counterpart (\ref{KG}), the model does not admit traveling-wave solutions.
Nevertheless, stationary solutions are not affected by the dissipation and gain and are given by $u=\phi(x)$, where $\phi(x)$ solves the equation $\phi_{xx}=f(\phi)$. The most interesting solution is the
kink, which is a topological object given by $\phi(x)=4\arctan
(e^{x-x_0})$ in the SG model, and $\phi(x)=\tanh (x-x_0)$ in the $\phi^4$
model ($x_0$ stands for the position of the kink center).

To study linear stability of these kinks, one substitutes
$u(t,x)=\phi(x)+v(x)e^{\lambda t}$,   with $v(x) \ll
1$, into Eq.~(\ref{PT-KG}) and obtains a linear eigenvalue problem
\begin{eqnarray}
\label{eq:spectra_KG}
\lambda^2 v+\lambda\gamma(x)v-v_{xx}+f^\prime(\phi)v=0.
\end{eqnarray}
\textcite{DFKSS14} performed general analysis of
(\ref{eq:spectra_KG}) as well as a numerical study of stability for
$\gamma(x)=\epsilon xe^{-x^2/2}$, where the constant $\epsilon$
characterizes the strength of gain and loss. The main findings for
SG and $\phi^4$ kinks can be summarized as follows. The linear
stability spectrum is unaffected by $\gamma(x)$ (except for a possible shift
of the discrete spectrum along the imaginary axis) if the kink
center $x_0$ coincides with the boundary between the domains with gain and loss,
i.e., with $x=0$. If however the kink center is shifted to the lossy or to the gain
gain region, then the kink becomes spectrally stable and unstable,
respectively. \textcolor{black}{Behavior of kinks in $\PT$-symmetric SG and $\phi^4$ models can be also described by means of a generalized collective coordinate method which was developed by \textcite{Kevrekidis14} on the basis of a proposition of \textcite{Galley13} who suggested  an approach to  formulation of the Lagrangian and Hamiltonian dynamics of generic non-conservative systems.}

The conservative SG equation is also known to admit a breather solution $\displaystyle{\phi=4\arctan\frac{\sigma \cos[a(t-t_0)]}{a\cosh[\sigma (x-x_0)]}}$,
where $\sigma=\sqrt{a}$, $0<a<1$, $x_0$ is the center  of the breather, and $t_0$ is a constant.
\textcite{LuKeCu14}
addressed the existence and stability of breathers in the $\PT$-symmetric SG model.
Unlike kinks, breathers are always affected by the gain and loss
because of their time-periodic nature. This in particular makes
their persistence in the $\PT$-symmetric model possible only if they
are centered at the boundary between the gain and loss. Numerical
analysis
shows that even for a small amplitude of the gain and loss $|\gamma(x)|$
the breather becomes unstable through a Hopf bifurcation.
It was also found that if a breather is initially centered at the
lossy side, then it will decay away. If, however, the breather is
initially shifted toward the gain region, then its energy will grow
until a pair of a kink and an anti-kink is nucleated.

\section{$\PT$-deformations of nonlinear equations}
\label{sec:integrable}
The nonlinear models we considered so far were constructed by adding
nonlinear terms (physically, by accounting for nonlinear
interactions) to linear models with complex potentials, dissipation,
or gain. In this section, we address another possibility, where the
so-called {\em $\PT$-deformation} (alias $\PT$-extension) is
performed by extending purely real coefficients of  a nonlinear equation  to the complex plane.

\subsection{Deformed KdV equation}

The idea was formulated by~\textcite{BBCF} and can be described as
follows. Consider wave dynamics governed by the KdV equation
\begin{eqnarray}
\label{eq:KdV}
u_t+uu_x+u_{xxx}=0.
\end{eqnarray}
Now under parity transformation ($x\to - x$), $u$ also has to
change its sign: $u\to-u$. Since we are dealing with a classical
Hamiltonian system, the time reversal operator $\T$ results in the
change $t\to-t$ with simultaneous change $u\mapsto-u$. Then in order to
continue $u$ from the real axis to the complex plane, one can ``borrow''
from the quantum mechanics the rule of changing $i\to -i$ when applying
time reversal. This suggests to introduce a $\PT$-symmetric
extension of the KdV equation as~\cite{BBCF}
\begin{eqnarray}
\label{eq:KdV-PT}
u_t-iu(iu_x)^\varepsilon+u_{xxx}=0, \quad \varepsilon\in\mathbb{R}.
\end{eqnarray}
Obviously, both equations (\ref{eq:KdV}) and (\ref{eq:KdV-PT}) are
invariant under the $\PT$ transformation. At $\epsilon=1$,
Eq.~(\ref{eq:KdV-PT}) reduces to (\ref{eq:KdV}). Other physically
relevant cases include $\varepsilon=0$, which leads to the
dispersive equation $v_t+v_{xxx}=0$ with $v=e^{-it}u$,
and $\varepsilon=3$, which leads to the nonlinear equation $u_t-u(u_x)^3+u_{xxx}=0$ \cite{Fushchych91,BBCF}.

Another $\PT$-deformation of the KdV equation can be obtained using the
Hamiltonian formulation. The original KdV equation (\ref{eq:KdV})
can be written in a Hamiltonian form as 
\begin{eqnarray}
\label{eq:KdV_Hamilt_eq}
u_t=\frac{\partial }{\partial x}\frac{\delta H}{\delta u(x)}=\{u,H\},
\end{eqnarray}
where
\begin{eqnarray}
\label{eq:KdV_Hamilt}
H(t)=\int_{-\infty}^{\infty} \cH(x,t) dx, \quad \cH(x,t)=\frac 12 u_x^2+u^3.
\end{eqnarray}
\textcite{Fring07} considered a $\PT$-symmetric generalization of the Hamiltonian density $\cH(x,t)$ as
\begin{eqnarray}
\label{eq:KdV_Halilt_gen}
\cH(x,t)=-{(1+\varepsilon)^{-1}} (iu_x)^{\varepsilon+1} +u^3,
\end{eqnarray}
which satisfies the relation $\cH(u(x))=\cH^*(u(-x))$. The latter
property ensures that the energy on each symmetric interval $[-a,a]$
is real:
\begin{eqnarray}
\label{eq:class_spectrum}
E=\int_{-a}^a\cH(u(x))dx=
\int_{-a}^a\cH^*(u(x))dx=E^*.
\end{eqnarray}
Equation   (\ref{eq:KdV_Hamilt_eq}) with the Hamiltonian
(\ref{eq:KdV_Halilt_gen}) yields another $\PT$ deformed KdV
equation~\cite{Fring07}
\begin{equation}
\label{eq:PT_deformed_gen}
u_t-6uu_x+i\varepsilon(\varepsilon-1) (iu_x)^{\varepsilon-2}u_{xx}^2+\varepsilon (iu_x)^{\varepsilon-1}u_{xxx}=0.
\end{equation}

While the physical relevance of models like~(\ref{eq:KdV-PT}) or
(\ref{eq:PT_deformed_gen}) for arbitrary values of the deformable
parameter $\varepsilon$ remains an open question, the systems themselves possess
some interesting properties which justify the attracted interest.
One of them is the existence of integrals of motion, which is a
nontrivial issue for nonconservative systems. In particular, model
(\ref{eq:KdV-PT}) with  $\varepsilon=3$ admits two integrals of
motion \cite{BBCF}
\begin{eqnarray}
\label{eq:integrals_pt-kdv}
I_{\pm}=\int dx\!\!\int\limits_{0}^{2^{1/3}u(x,t)}ds\left[\Bi(s)\pm \sqrt{3}\Ai(s)\right],
\end{eqnarray}
where $\Ai(\cdot)$ and $\Bi(\cdot)$ are the Airy functions. Quantity  $I_-$ is strictly positive (when $u(x,t)$ is not identically zero) and therefore can be   interpreted as the energy.

For other $\PT$-deformations of the KdV equation, as well as for
examples of their solutions, see~\cite{CaFrBa11,BaFr08}.

\subsection{Deformed Burgers equation}

\textcite{CaFr12} studied a $\PT$-deformed
Hopf equation
\begin{eqnarray}
\label{PT-Burgers}
u_t-if(u)(iu_x)^\varepsilon=0,
\end{eqnarray}
where $f(u)$ is a well behaved function, and  $\varepsilon$ is a real
rational number. For $\varepsilon=1$, Eq.~(\ref{PT-Burgers}) reduces
to the real-valued Hopf equation
\begin{eqnarray}
\label{eq:Hopf}
w_t+f(w)w_x=0.
\end{eqnarray}
This deformation extends the  earlier results of~\textcite{BeFein2008}
on  the $\PT$-deformation $v_t-iv(iv_x)^\varepsilon=0$
of the inviscid Burgers' equation $w_t+w w_x=0$.

The
$\PT$-deformed Hopf equation (\ref{PT-Burgers}) can be obtained from
its original version (\ref{eq:Hopf}) for arbitrary rational
$\varepsilon$ through an explicit map, i.e., change of variables
\cite{CaFr12,CF08}. For the particular case of $f(w)=w^n$, the map
reads $w \mapsto \sqrt[n]{\varepsilon u(iu_x)^n}$.
This direct mapping allows for the straightforward analysis of wave dynamics in the deformed model on the basis  of the knowledge about  the ``seed'' real nonlinear equation. 
We illustrate this on the example of shock formation
\cite{BeFein2008,CaFr12}. Suppose the initial condition to the Hopf
equation (\ref{eq:Hopf}) is $w(x_0,0)=w_0(x_0)$, and consider a
characteristic, i.e., a curve in the plane $(x,t)$ for which
$w(x,t)= w_0(x_0)$. The characteristic has the form $x=f(w_0)t+x_0$.
At the point of gradient catastrophe two characteristics cross and
$w_x$ tends to infinity. Then by computing
\begin{eqnarray}
w_x=w_0^\prime(x_0)\frac{dx_0}{dx}=
\frac{w_0^\prime(x_0)}{1+t\left(df(w_0)/dx_0\right)}
\end{eqnarray}
and utilizing the above map, one finds that for $f(w)=w^n$ the
earliest time for shock formation in the $\PT$-deformed Hopf
equation (\ref{PT-Burgers}) is
\begin{eqnarray}
\label{eq:shock_time}
t_s^w= -\varepsilon^{-{1}/{n}}\left[\frac{d}{dx_0}\left(u_0^{1/n}\left(iu_{x_0}\right)^{(\varepsilon-1)/n} \right)\right]^{-1}.
\end{eqnarray}
Requiring the time $t_s^w$ to be real, one finds that this condition is
satisfied under the replacement $u_0\to i^\alpha\tilde{u}_0$, where
$\tilde{u}_0\in\mathbb{R}$, $\alpha=(4m\pm 1)n/\varepsilon$, and
$m\in\mathbb{Z}$. Thus for certain combinations of $\varepsilon$ and
$n$ one can not observe shock wave formation for real solutions of
the deformed equation (\ref{PT-Burgers}).
On the other hand, the deformed model offers other possibilities for singularities of the solutions to occur \cite{CaFr12}.
Such possibilities  correspond to a curvature catastrophe. For the inviscid Burgers' equation ($f(w)=w$)
this phenomenon stems from the  relation
$w_x=i\varepsilon(iu_x)^{\varepsilon-2}\left[u_x^2+(\varepsilon
-1)uu_{xx}\right]$, meaning that the shock of the $u$-field
($u_x\to\infty$) always corresponds to the shock of $w$. In the
meantime, the converse is not necessarily true because
$w_x\to\infty$ can also occur when $u_{xx}\to\infty$.

For other $\PT$-symmetric deformations of the Hopf equation see
also~\textcite{Zhenya2008}.

\subsection{Deformed short pulse equation}

$\PT$-deformation of yet another model, the short pulse equation
$u_{xt}=u+\frac{1}{2}(u^2u_x)_x$~\cite{ScWa2004}, was constructed
by~\textcite{Zhenya2012}:
\begin{eqnarray}
\label{SPE_PT}
i[(iu_x)^\sigma]_t=u+bu^m+ic[u^n(iu_x)^\varepsilon]_x.
\end{eqnarray}
Here parameters $b$, $c$, $\sigma$, $n$, $m$, and $\varepsilon$ are
all real. For their specific choices  Eq.~(\ref{SPE_PT}) allows for
soliton, kink, or compacton solutons.

\subsection{Nonlocal NLS equation}
\label{subsec:nonloc_NLS}

The NLS equation (\ref{eq:NLS}) belongs to the so-called Abblowitz-Kaup-Newell-Segur (AKNS) scheme~\cite{AKNS}, allowing for obtaining a wide class of equations integrable by the inverse scattering transform. More specifically, NLS equation (\ref{eq:NLS}) with $g=2$ is a particular case of the more general integrable system  \cite{AS81,NMPZ} 
\begin{equation}
\label{eq:AKNS}
i\psi_t+\psi_{xx}+2\phi\psi^2=0, \quad
i\phi_t-\phi_{xx}-2\psi\phi^2=0,
\end{equation}
subject to the reduction $\phi(x,t)=\psi^*(x,t)$.

\textcite{AM13} considered yet another reduction, $\phi(x,t)=\pm \psi^*(-x,t)$ leading to the equation with nonlocal nonlinearity
($\sigma=\pm 1$)
\begin{equation}
\label{sec:PT-NLS}
i\psi_t(x,t)=\psi_{xx}(x,t)+ 2\sigma \psi(x,t)\psi^*(-x,t)\psi(x,t).
\end{equation}
Notice that in this equation, the nonlinear term can be represented as $F(\psi)\psi$, where  $F(\phi) = 2\sigma \phi(x,t)\phi^*(-x,t)$.  For any $\psi$, the nonlinearity satisfies the identity $(F(\psi))^* = \p F(\psi)\p$, and thus commutes with $\PT$ and can be termed as $\PT$ symmetric or $\p$-pseudo-Hermitian in the sense of the definition (\ref{eq:FpsH}). In the discrete case this property guarantees the existence of at least one integral of motion. For the continuous model  (\ref{sec:PT-NLS}), one can find an infinite number of conserved quantities. The first one is given by (\ref{eq:int_Q}), and the second and
third ones read
\begin{equation}
\label{sec:nls_int_2}
Q_2=\int_{-\infty}^{\infty}\!\left[\psi_x(x,t)\psi^*(-x,t)+\psi(x,t)\psi_x^*(-x,t)\right]dx,
\end{equation}
\begin{equation}
\label{sec:nls_int_3}
Q_3=\int_{-\infty}^{\infty}\!\left[\psi_x(x,t)\psi_x^*(-x,t)-\sigma\psi^2(x,t)\psi^{*2}(-x,t)\right]dx.
\end{equation}
The one-soliton solution for  (\ref{sec:PT-NLS}) reads
\begin{eqnarray}
\label{eq:PT_NLS_solit}
\psi(x,t)=-\frac{2(\eta+\bar{\eta})e^{i\bar{\theta}}e^{-4i\bar{\eta}^2t}e^{-2\bar{\eta}x}}{1+e^{i(\bar{\theta}+\theta)}e^{4i(\eta^2-\bar{\eta}^2)t}e^{-2(\bar{\eta}+\eta)x}},
\end{eqnarray}
where $\eta,\bar{\eta}\, (>0)$, $\theta$ and $\bar{\theta}$ are
constants.

Being a particular case of the model (\ref{eq:AKNS}),  Eq. (\ref{sec:PT-NLS}) is only the first equation in a  hierarchy of integrable nonlocal models. Furthermore, it
can be generalized to the vectorial case and to include a wider class
of symmetries through the reductions of the type  $\phi(x,t)=\psi^*(\epsilon_1x,\epsilon_2 t)$, where $\epsilon_1$ and $\epsilon_2$ take values $\pm1$~\cite{Zhenya2015}.

Furthermore, one can construct a discrete analog of (\ref{sec:PT-NLS}), which was reported by \textcite{AbMu14}.

\section{Conclusions and perspectives}
\label{sec:conclusion}
In this article, we have reviewed recent progress on nonlinear wave
dynamics in $\PT$-symmetric systems. We have shown that the
interplay between nonlinearity and $\PT$ symmetry creates a host of
new phenomena which sets nonlinear $\PT$-symmetric systems apart
from traditional conservative or dissipative systems. For instance,
even though $\PT$ systems contain gain and loss and are dissipative
in nature, they admit continuous families of nonlinear modes and integrals of motion --
properties which are common in conservative systems but rare in
dissipative systems. $\PT$-symmetry breaking of nonlinear modes in
certain types of $\PT$ systems is another surprising property which
is highly non-intuitive. Stabilization of nonlinear modes in
$\PT$-symmetric systems above phase transition is a fascinating
property as well.

Most of the materials reviewed in this article are on theoretical
aspects of $\PT$-symmetric systems. \textcolor{black}{But a number of
    experimental validations of the main concepts as well as several
    practical applications were also described. Since $\PT$ symmetry is
    prevalent in a wide range of physical systems, further experimental
    studies are expected to continue. A reason $\PT$ symmetry can be
    physically useful is that it allows for overcoming losses while
    still preserving guidance properties of the system. In addition,
    under $\PT$ symmetry, the gain and loss can be varied, thus paving
    the way for optimal and flexible control of wave-guiding systems.
    Exciting applications of $\PT$ symmetry have already appeared.
{\color{black} They include optical switches, unidirectional
    reflectionless
    $\PT$-symmetric metamaterials at optical frequencies,
    single-mode $\PT$-symmetric micro-ring lasers,
    CPA-lasers and phonon lasers.}
    In those applications, the effects of
    nonlinearity can be an important issue. For instance, it is well
    known that lasing is an intrinsic nonlinear process. Thus studies of
    nonlinear effects in those emerging applications are important open
    questions. }

We anticipate that nonlinear $\PT$ systems may find further
applications in the near future, especially because the paradigm is
relevant practically to all branches of contemporary physics. We
also expect growing interest in this subject from the mathematical
community, which is justified by the novelty and beauty of
properties of $\PT$-symmetric models and, more
    generally, of non-Hermitian systems.

\begin{acknowledgements}

We are indebted to our colleagues F. Kh. Abdullaev, Y. V. Bludov, C. Hang, G. Huang, Y. V. Kartashov, P. G. Kevrekidis, B. A. Malomed, D. E. Pelinovsky, and Z. Yan for fruitful collaboration and for enlightening discussions on several topics of the present review.
The work of VVK and DAZ was supported by the FCT (Portugal) through
the grant PTDC/FIS-OPT/1918/2012, and
the work of JY was supported in part by AFOSR (Grant USAF
9550-12-1-0244) and NSF (Grant DMS-1311730).
\end{acknowledgements}

%

\end{document}